\documentclass{article}
\pdfoutput=1

\usepackage{varioref}
\usepackage{amsmath,amsfonts,amsthm,%
  mathrsfs}
\usepackage[normalem]{ulem}
\usepackage{multirow,color,graphics}
\usepackage{jheppub}
\usepackage{enumerate}
\usepackage{fancyvrb}
\usepackage{verbatim}
\usepackage{wrapfig,bbm}

\newtheoremstyle{statement}
  {\topsep} %
  {\topsep} %
  {\it} %
  {} %
  {} %
  {:} %
  {.5em} %
  {} %
\theoremstyle{statement}
\newtheorem{statement}{{\bf Statement}}

\newtheorem{property}{{\bf Property}}

\newcommand{\hstretch}{\rule{0pt}{0pt}\hspace{\stretch 1}\rule{0pt}{0pt}}

\preprint{
  \begin{minipage}{0.2\linewidth}
    NORDITA-2014-60 \\
    TCDMATH 14--05
  \end{minipage}
}

\newcommand{\ifpicture}[1]{#1}

\numberwithin{equation}{section}

\newcommand{\fprefix}[1]{#1-}

\newcommand{\fct}[1]{{\fprefix{#1}}function}
\newcommand{\fcts}[1]{{\fct{#1}s}}
\newcommand{\Yfcts}{\fcts{Y}}
\newcommand{\Yfct}{\fct{Y}}
\newcommand{\Qfcts}{\fcts{Q}}

\newcommand{\bQfcts}{\fcts{\(\bQ\)}}
\newcommand{\mQfcts}{\fcts{\(\mQ\)}}

\newcommand{\Tfcts}{\fcts{T}}
\newcommand{\bTfcts}{\fcts{\(\bT\)}}
\newcommand{\sys}[1]{{\fprefix{#1}}system}
\newcommand{\Ysys}{\sys{Y}}
\newcommand{\Pfcts}{\fcts{\(\bP\)}}
\newcommand{\Tsys}{\sys{T}}
\newcommand{\Tfct}{\fct{T}}
\newcommand{\Pmsys}{\sys{$\bP\mu$}}
\newcommand{\Qosys}{\sys{$\bQ\omega$}}
\newcommand{\Qsys}{\sys{Q}}

\newcommand{\ie}{i.e.}

\newcommand\secref[1]{section~\ref{#1}}
\newcommand\Secref[1]{Section~\ref{#1}}
\newcommand\appref[1]{appendix~\ref{#1}}

\newcommand\figref[1]{figure~\ref{#1}}

\newcommand\equref[1]{equation~\eqref{#1}}
\newcommand\Equref[1]{Equation~\eqref{#1}}
\newcommand\IM{{\rm Im}\,}

\newcommand\es{\emptyset}

\newcommand{\bal}{\begin{align}}
\newcommand{\eal}{\end{align}}
\newcommand{\beq}{\begin{equation}}
\newcommand{\eeq}{\end{equation}}
\newcommand\beqa{\begin{eqnarray}}
\newcommand\eeqa{\end{eqnarray}}
\newcommand\bea{\begin{array}}
\newcommand\eea{\end{array}}

\newcommand{\eq}[1]{(\ref{#1})}

\newcommand{\bB}{{\bf B}}

\newcommand{\su}{\mathfrak{su}}
\renewcommand{\sl}{\mathfrak{sl}}
\newcommand{\psu}{\mathfrak{psu}}
\newcommand{\gl}{\mathfrak{gl}}
\newcommand{\uuu}{\mathfrak{u}}

\newcommand{\Gl}{\mathsf{GL}}

    \newcommand{\nn}{\nonumber}

    \newcommand{\neqa}{\nonumber\end{eqnarray}}
    \newcommand{\la}[1]{\label{#1}}

\def\o{{\omega}}
\def\O{{\Omega}}
\def\a{{\alpha}}

\def\[{\left[}
\def\]{\right]}
\def\l{\lambda}
\def\e{\epsilon}

\def\a{\alpha}
\def\b{\beta}

\def\<{\langle}
\def\>{\rangle}

\def\i2{\frac{i}{2}}

\def\bQ{{\bar Q}}
\def\cQ{{\cal Q}}

\def\<{\langle}
\def\>{\rangle}
\def\cF{{\cal F}}

\def\cT{{\cal T}}

\def\i2{\frac{i}{2}}

\def\1h{\hat 1}
\def\2h{{\hat 2}}
\def\3h{{\hat 3}}
\def\4h{{\hat 4}}

\def\be{\begin{eqnarray}}
\def\ee{\end{eqnarray}}
\def\no{\nonumber}

    \def\CA{{\cal A}}

    \def\CF{{\cal F}}

    \def\<{\left\langle\,}
    \def\>{\, \right\rangle}
    \def\[{\left[}
    \def\]{\right]}

    \def\D{{\rm D}}
    \def\bT{{\bf T}}
    \def\disc{\,{\rm disc}\,}
    \def\sT{{\mathscr T}}
    \def\wT{{{\mathbb{T}}}}

\def\clap#1{\hbox to 0pt{\hss#1\hss}}

\def\wTl{\underleftarrow{\mathbb{T}}}
\def\wTr{\underrightarrow{\mathbb{T}}}

\newcommand{\ps}{{\bf p}}
\newcommand{\qs}{{\bf q}}
   \def\gl{{\mathfrak{gl}}}
   \def\su{{\mathfrak{su}}}
   \def\sl{{\mathfrak{sl}}}

\def\hbZ{{\widehat{ Z}}}

\renewcommand{\Re}{{\rm Re}\,}
\renewcommand{\Im}{{\rm Im}\,}

\usepackage{amssymb}
\usepackage{subfigure}

\usepackage{tikz}

\def\e{\epsilon}
\def\abadm{{ABA-diagram}}

    \def\bT{{\bf T}}
    \def\bQ{{\bf Q}}
    \def\bP{{\bf P}}
    \def\bPt{\tilde{\bf P}}
    \def\bQt{\tilde{\bf Q}}
    \def\wT{{\mathbb{T}}}
    \def\wQ{{\mathbb{Q}}}

\def\mQ{{\bQ}}
\newcommand{\bosQ}[1]{{\bQ_{\check #1}}}
\newcommand{\bosQup}[1]{{\bQ^{\check #1}}}
\newcommand{\hbosQ}[1]{{\hat\bQ_{\check #1}}}
\newcommand{\hbosQup}[1]{{\hat\bQ^{\check #1}}}
\newcommand{\tbosQ}[1]{{\tilde\bQ_{\check #1}}}
\newcommand{\tbosQup}[1]{{\tilde\bQ^{\check #1}}}
\newcommand{\bosQs}[2]{{\bQ_{\check #1}^{#2}}}
\newcommand{\hbosQs}[2]{{\hat\bQ_{\check #1}^{#2}}}
\newcommand \bos \check

    \def\CF{{\cal F}}
\def\hbZ{{\widehat{ Z}}}

\DeclareMathOperator{\Pf}{Pf}

\DeclareMathOperator{\Tr}{Tr}
\DeclareMathOperator{\tr}{tr}

\newcommand{\sm}{{-}}

 \newcommand{\gQ}{Q}

\makeatletter
\def\maketag@@@#1{\@Domain\hbox{\m@th\normalfont#1}}
\def\Domain#1{\def\@Domain{\llap{$#1$\quad}}}
\Domain{}
\makeatother

\definecolor{mblue}{rgb}{0,0,0}
\definecolor{mred}{rgb}{0,0,0}

\newcommand{\fQ}{{\mathcal Q}}

\newcommand{\neqfour}{{\ensuremath{\mathcal{N}=4}}}

  \title{Quantum spectral curve\\ for  arbitrary state/operator in AdS$_5$/CFT$_4$
}
\author[a,b]{Nikolay Gromov}
\author[c,d,e]{~~~Vladimir Kazakov}
\author[f]{~~~S\'ebastien Leurent}
\author[g,h]{~~~Dmytro Volin}

\affiliation[a]{Mathematics Department, King's College London,
The Strand, London WC2R 2LS, UK.}
\affiliation[b]{St.Petersburg INP, Gatchina, 188 300, St.Petersburg,
  Russia.}
\affiliation[c]{LPT, \'Ecole Normale Superieure, 24, rue Lhomond 75005
  Paris, France.}
\affiliation[d]{Universit\'e Paris-VI, Place Jussieu, 75005 Paris,
  France.}
\affiliation[e]{School of Natural Sciences, Institute for Advanced Study, Princeton, NJ08540, USA
}
\affiliation[f]{Institut de Math\'ematiques de Bourgogne, UMR 5584 du CNRS, Universit\'e de
  Bourgogne, 9 avenue Alain Savary, 21000 DIJON, France.}
\affiliation[g]{Nordita
KTH Royal Institute of Technology and Stockholm University,\\
Roslagstullsbacken 23, SE-106 91 Stockholm, Sweden
}
\affiliation[h]{School of Mathematics, Trinity College Dublin, College
  Green, Dublin 2, Ireland.}

\abstract{
We give a derivation of quantum spectral curve (QSC) - a finite set of Riemann-Hilbert equations for exact spectrum of planar \neqfour{} SYM theory proposed  in our recent paper\\ {\it Phys.Rev.Lett.}112~(2014). We also generalize this construction  to {\it all}  local single trace operators of the theory, in contrast to the TBA-like approaches worked out only for a limited class of states. 
We reveal a rich  algebraic and analytic structure of the QSC in terms of a so called Q-system -- a finite set of Baxter-like Q-functions. This new point of view on the finite size spectral problem is shown to be completely compatible, though in a far from trivial way,  with  already known exact equations (analytic Y-system/TBA, or FiNLIE).
We use the knowledge of this underlying Q-system  to demonstrate how the
classical finite gap solutions and the asymptotic Bethe ansatz emerge from our formalism in appropriate limits. }
\usepackage{graphicx}

\makeatletter
  \def\tagform@#1{\tagstuff \maketag@@@{(#1)\@@italiccorr}}
\makeatother

\newcommand{\tagstuff}{}

\definecolor{drkgreen}{RGB}{0,150,0}

\definecolor{drkred}{RGB}{150,0,0}

\def\fip{\cite{Gromov:2011cx}}

\begin{document}

 \maketitle

\newpage

\section{Introduction}
\label{sec:intro}

The discovery and exploration of integrability in the planar AdS/CFT correspondence has  a long and  largely
successful history \cite{Beisert:2010jr}. In the two most advanced examples of 4-dimensional
\neqfour{} SYM and 3-dimensional ABJM model it became possible \cite{Gromov:2009tv} to study the planar spectrum
of anomalous dimensions of some simple but non-protected single trace operators at any 't Hooft coupling \(\lambda\). The computations were efficiently done in various limits and also numerically, with
sufficiently high precision  \cite{Gromov:2009zb,Frolov:2010wt,LevkovichMaslyuk:2011ty},
by means of an explicit but immensely complicated Thermodynamic Bethe Ansatz (TBA) formalism  \cite{Gromov:2009bc,Bombardelli:2009ns,Arutyunov:2009ur}.

One should admit that the complexity of the TBA-like equations appeared to be in a  stark contradiction with the elegant integrability concept for the spectrum of these beautiful maximally super-symmetric gauge theories.
Fortunately, the situation was not hopeless as some signs of hidden simplicity started to emerge here and there.
The system of integral nonlinear TBA equations was known to have a reformulation in terms of a simple and universal
 Y-system \cite{Gromov:2009tv,Gromov:2009bc} supplied  by
 a relatively simple analytic data  \cite{Cavaglia:2010nm}.
 Furthermore, the Y-system, an infinite system of nonlinear functional equations is
 equivalent to the integrable Hirota bilinear equation (T-system) \cite{Gromov:2009tv} which by itself was known to be integrable.
 The integrability of the latter would imply that it can be rewritten in terms of a finite number of Q-functions of the spectral parameter --
 analogues of the Baxter polynomials in the studies of integrable spin chains \cite{Gromov:2010vb,Gromov:2010km}.
This venue
was explored in our paper \cite{Gromov:2011cx} where, using this discrete classical  integrability and an important analyticity input, a {\it finite} system of non-linear integral equations (FiNLIE), somewhat reminding in spirit the Destri-De Vega equations, was formulated and successfully tested numerically.

But even this finite FiNLIE system, which allowed for some numerical tests and even for the calculation of  8 loop Konishi anomalous dimension \cite{PhysRevLett.109.241601,Leurent:2013mr}, 
was still quite complicated and tricky for the practical use,
 though already conceptually simpler then the infinite system of TBA equations.
It was clear that behind all these quite mysterious analytic structures a much simpler truth  should be hidden.
In our opinion, to a great extent this truth was unveiled in \cite{Gromov:2013pga} where we proposed a simple finite set of non-linear Riemann-Hilbert equations which we called the quantum spectral curve (QSC) of the AdS/CFT correspondence.
Due to their simplicity, the equations have found numerous applications in the practical calculations:
They were successfully applied to the analysis of weak coupling
expansion in the $\sl(2)$ sector (Konishi up to 9 loops!) \cite{Dima-talk,Dima-toappear}
as well as at strong coupling \cite{Gromov:2013pga}, for the slope and curvature functions for twist-2 operators at any coupling and pomeron intercept at strong coupling
\cite{Gromov:2014bva}. Recently, the QSC was also found for the ABJM model in \cite{Cavaglia:2014exa}, which was used \cite{Gromov:2014eha} to make a well-grounded conjecture for
the interpolation function \(h(\lambda)\) entering  numerous physically relevant quantities such as cusp anomalous dimension and magnon dispersion relation
in this theory.

The name QSC is justified by the fact that this system of equations reveals a natural generalization of the classical spectral curve of superstring sigma model on AdS\(_5\times\)S\(^5\) background \cite{Beisert:2005bm}  -- the AdS counterpart of the \neqfour{} SYM. Namely, the 8 basic
 Q-functions \(Q_j(u)\) entering the QSC equations should be closely related to the exact wave function of the theory (in  separated variables). in particular
 in the quasi-classical regime they take a familiar  quantum mechanical form \(Q_j(u)\simeq e^{ i\int^up_j(u)du}\)
 where the roles of the momentum and  of the coordinate are played  by the quasi-momenta \(p_j(u)\) and the spectral parameter \(u\), respectively.
 Notably, the exact quantum \(Q_j\)'s have an analytic structure which,  on their defining Riemann sheet, is very similar to their classical counterparts: there are only two branch points  at \(u= \pm \frac{\sqrt{\l}}{2\pi}\,\,\)
 forming a cut (that we will call Zhukovsky cut) which can be uniformized by
  Zhukovsky map \(u=\frac{\sqrt{\l}}{4\pi}(x+\frac{1}{x})\) -- an important element of the whole construction. One of our main findings is that their analytic continuation to another
 sheet is governed by the mondromy matrices \(\mu\) or \(\omega\) which are entangled into a closed system with \(Q_j\) themselves.
 This system takes a form of very concise and elegant Riemann-Hilbert type equations containing all the necessary dynamical information  for the spectral problem.
 Depending on the choice of basic Q-functions within the construction we call such closed system of equations as \(\bP\mu\)- or \(\bQ\omega\)-system.

To uncover the complete algebraic and analytic structure of the AdS/CFT spectral problem  we discuss the so called Q-system -- the full set of 256 Q-functions of the problem being Pl\"ucker coordinates for a set of Grassmanians in \(\mathbb{C}^8\) and thus related to each other by the Pl\"ucker bilinear identities (which are often called QQ-relations). Pl\"ucker identities allow one to express all the Q-functions \(Q_{I|J}\), where \(I,J\subset{\{1,2,3,4\}}\) are two ordered subsets of indices,  through the basis of 8 one-index \(Q_j\)-functions. In that sense, the  analytic structure of all Q-functions is completely fixed by the basic ones, having only a single Zhukovsky cut on a certain, defining sheet. The multi-index Q-functions already have infinite sequences of Zhukovsky cuts, spaced by \(i\), on all sheets, as a result of solution of these Pl\"ucker relations.
In this language, the \(\bP\mu\)-system (or \(\bQ\omega\)-system) is reformulated as a certain
symmetry, or rather a morphism mapping the Q-functions analytic in the upper half-plane (i.e. having there no Zhukovsky cuts) to the other ones analytic in the lower half-plane.   One can even invert the logic and (almost) completely fix the Riemann-Hilbert equations defining QSC by the requirement of existence of such symmetry. Such a Q-system supplied by the analyticity structure will be called the {\it analytic} Q-{\it system}, and it should be considered as a rightful successor of the analytic Y-system.
In other words, our present paper in addition to a firm closed \(\bP\mu\)-system (or \(\bQ\omega\)-system)
presents a new point of view on the AdS/CFT spectral problem: the quantum integrability amounts to reducing the whole problem to the construction of a  Q-system with certain rather remarkable analytic properties.

     The claims of this paper are based of course on  solid derivations from  TBA and analytic Y-system.
 For that purpose, we derived {\bf in \secref{sec:QSCY}} from the analytic Y-system the existence of   8 one-index Q-functions having only one Zhukovsky cut on the defining sheet.  The monodromy equations allowing the analytic continuation of various Q-functions through  Zhukovsky cuts are also induced from the analytic Y-system. The emergence of the Q-system with the announced analyticity properties is thus explicitly demonstrated. This section already contains the full set of QSC equations.

{\bf In  \secref{sec:YTQ}} we take an opposite point of view: we first formulate the \((4|4) \) graded Q-system and identify there  the relevant Q-functions  with the quantities  obtained from the analytic Y-system in \secref{sec:QSCY}.
The correspondence appears to be perfect.
At the end of \secref{sec:YTQ}, the exact finite size Bethe ansatz equations for  zeros of the Q-functions are derived.

The Bethe equations simplify in the large volume limit to the well-known Asymptotic Bethe Ansatz (ABA) equations of \cite{Beisert:2006ez,Beisert:2005fw}, as we demonstrate {\bf in \secref{sec:ABA}}.

{\bf In \secref{sec:QCA}} we derive the classical  limit of the quantum spectral curve and even, partially, the quasiclassical corrections, in full agreement with the known results for the  classical algebraic curve.

The {\bf \secref{sec:conclusions}} is devoted to the conclusions and prospects.

Some technical details are discussed in appendices.

\section{Notations and conventions}\la{secnot}
\label{sec:notations}

\subsection{Spectral parameter and Riemann sheets}
\label{sec:spectr-param-riem}

\begin{figure}
{\hstretch}\ifpicture{
 \subfigure[Function $f$ in the physical kinematics]
{
\includegraphics{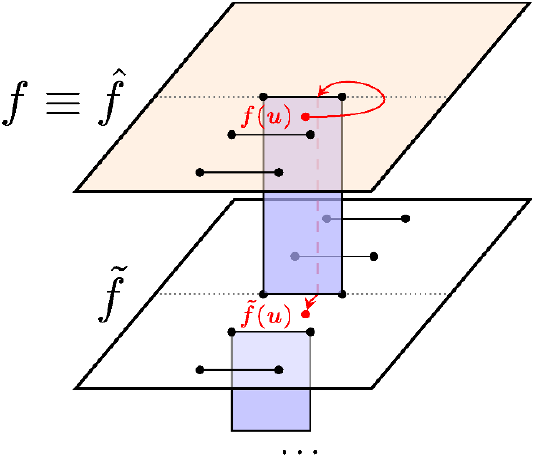}
\label{fig:HatTilde}}
{\hstretch}{\hstretch}   \subfigure[Function $\check f$ in the mirror kinematics]
{
\includegraphics{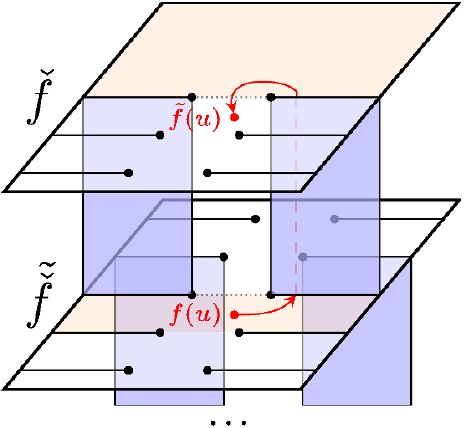}
\label{fig:TildeCheck}}}{\hstretch}
  \caption{Riemann sheets of the function \(f\): the notation
    \(f\) denotes the function $\hat f$ (left) which
    has infinite ladders of cuts except on the upper half plane of its
    main Riemann sheet. It coincides with $\check f$ (right) on
    this upper half plane. The red arrow 
    indicates a path to define the tilde transformation.
  }
\label{fig:HatTildeCheck}\end{figure}

In this article, we will use functions of the spectral
parameter \(u\), and denote
\begin{align}
  \label{eq:156}
  F^{[\pm n]}\equiv&F(u\pm\tfrac i 2 n)\,,&F^{\pm}\equiv F^{[\pm1]}.
\end{align}

Many such functions will actually be multi-valued analytic functions, and we will refer to
the two specific arrangements of branch cuts as to the ``mirror'' and the ``physical''
kinematics. In general, functions will have branch points at positions
\(\pm 2g+i\frac {n}2,\quad n\in \mathbb{Z}\), where \(g=\tfrac {\sqrt \lambda}{4\pi}\). Each function can
have many branch points\footnote{There will be a general constraint on
the values of \(n\) giving the position of the branch points of a
function: for each function, all the \(n\)'s have the same parity.} corresponding to different values of \(n\).
The branch points will be connected by cuts parallel to the real axis,
which will either be ``short'', {\ie} of the form \([-2g+i\frac
{n}2,2g+i\frac {n}2]\), or ``long'', {\ie} of the form \(]-\infty+i\frac
{n}2,-2g+i\frac
{n}2]\cup[2g+i\frac {n}2,\infty+i\frac {n}2[\). The mirror sheet of a
function denoted as \(\check F\) is a sheet where all cuts will be
long,
whereas the physical sheet denoted as \(\hat F\) is a sheet where all cuts will be
short.\footnote{When the
  sheet is clear from the context, we will simply write \(F\) instead of
  \(\hat{F}\) or \(\check{F}\).}

We will sometimes somewhat  abusively write equalities
between functions which do not have the same cut structure. In that
case, we mean that equalities hold slightly above the real axis, {\ie}
when \(0<\mathrm{Im}(u)<1/2\). For instance we can define the functions \(\check x(u)=\tfrac u{2g}+i
\sqrt{1-\tfrac{u^2}{4g^2}}\) and \(\hat x(u)=\tfrac u{2g}+
\sqrt{\tfrac u{2g} -1}\sqrt{\tfrac u{2g} -1}\) which have a cut on the
real axis. By our convention, when \(\mathrm{Im}(u)>0\) one has \(\hat
x(u)=\check x(u)\), whereas when \(\mathrm{Im}(u)<0\) one has \(\hat
x(u)=1/\check x(u)\). We will hence write
\begin{equation}
  \label{eq:19}
  \check x=\hat x\,,\ \ \ \ \check x^-=1/\hat x^-\,.
\end{equation}

The analytic continuation\footnote{This analytic continuation is
  performed along a path which encloses \(\pm 2g\) but no other branch
  point. As usual in the analysis of AdS/CFT, we assume that the cuts
  are of square-root type, {\ie} that  this
  analytic continuation gives the same result for clockwise or
  anticlockwise continuation, and that enclosing \(2g\) or \(-2g\) gives
  the same outcome.} of a function \(F\) around the branch point at
position \(\pm 2 g\) is denoted as \(\tilde F\). For instance, \(\widetilde{\check x}(u)=\tfrac u{2g}-i
\sqrt{1-\tfrac{u^2}{4g^2}}=1/\check x(u)\). Note that for example
\(\widetilde{\hat x^{[+2]}}=\hat x^{[+2]}\) because \(x^{[+2]}\) has no
branch points at position \(\pm 2 g\) (it only has branch points at
position \(\pm 2g-i\)).

The discontinuity of a function on its cut on the real axis will be
denoted as \(\disc(F)\equiv F-\tilde F\).

As an illustration of the above notations, one can consider a
function  \(f\) which will appear in
\appref{sec:find-texorpdfstr-tex}.  It is one of the functions having
the most complicated analyticity 
properties: it is analytic in the upper half-plane, and has infinitely
many Zhukovsky cuts in the lower half-plane. We will
conventionally denote this function as \(f \equiv \hat f\), which means
that we use the physical kinematics if no check symbol is explicitly
written. Some Riemann sheets of this function are illustrated in the physical kinematics (see
\figref{fig:HatTilde}) and in the mirror kinematics (see
\figref{fig:TildeCheck}). Notice that $\check f$ coincides
with $f$ if $\mathrm{Im}(u)>0$, and that $\tilde f$
coincides with $\check f$ if $-1<\mathrm{Im}(u)<0$. In this
figure, the ``tilde''
transformation, {\ie} the analytic continuation around the branch
point, is illustrated by the path along the red line.

We will often use the abbreviations for some functions and even for Q-systems of functions: UHPA and LHPA, which means upper half-plane analytic and lower half-plane analytic, respectively.

\subsection{Multi-indices and sum conventions}
\label{sec:multi-indices-sum}

In this article, we will use tensor objects with upper and lower
indices taking values in \(\{1,2,3,4\}\). We use Einstein's sum
convention, {\ie} we sum over repeated
indices. For instance we have \(\mu_{a,b}\bP^b\equiv \sum_{b=1}^4
\mu_{a,b}\bP^b\).

We will use the Levi-Civita tensors, {\ie} the completely antisymmetric
tensors \(\e^{abcd}\) and \(\e^{ab}\) such that \(\e^{1234}=1\) and \(\e^{12}=1\).

We call ``multi-index'' an ordered set of indices for various Q-functions. It will be
denoted by a capital letter. We
will use the same sum convention for multi-indices. For
instance\footnote{In the example of the expression \(\e^{A,B}\mu_B\),
  one has to know from the context that \(B\) has two indices, {\ie} that
  \(\mu\) is a matrix. Then one can deduce that \(\e^{A,B}\mu_B\equiv \sum_{1\le c,d\le 4}
\epsilon^{Acd}\mu_{cd}\).} we
have \(\e^{AB}\mu_B\equiv \sum_{1\le c,d\le 4}
\epsilon^{Acd}\mu_{cd}\). In this expression, if
\(A=(a,b)\), then \(\e^{Acd}\) simply denotes \(\e^{abcd}\).

If  the indices are not summed over, they will be denoted as \(a_0\), \(b_0\),
etc, as in \equref{AABB0}. Similarly, if a multi-indexed is
not summed over, it will be denoted as \(A'\), \(B'\), etc, as in
\equref{Hodgedef}.

In the rest of this article, we will sometimes use the following matrices:
\begin{equation}
\eta_{ij}=
\left(
\begin{array}{cccc}
0 & 1 & 0 & 0 \\
-1 & 0 & 0 & 0 \\
0 & 0 & 0 &+1 \\
0 & 0 & -1 & 0 \\
\end{array}
\right)\,,\ \ \
\eta^{ij}\equiv (\eta^{-1})_{ij}=
\left(
\begin{array}{cccc}
0 & -1 & 0 & 0 \\
1 & 0 & 0 & 0 \\
0 & 0 & 0 &-1 \\
0 & 0 & +1 & 0 \\
\end{array}
\right)\,,\\
\end{equation}
\begin{equation}\label{etachi}
\chi_{ab}=-\chi^{ab}=\begin{pmatrix}0 & 0 & 0 & 1 \\
0 & 0 & -1 & 0 \\
0 & 1 & 0 & 0 \\
-1 & 0 & 0 & 0 \\
\end{pmatrix}\end{equation}

\section{Quantum Spectral Curve from Analytic Y-system}
\label{sec:QSCY}

\subsection{Inspiration from TBA}
\label{sec:QSCTBA}

{\Ysys} and TBA equations have played an important role for the comprehensive resolution of the spectral problem of the planar AdS\(_5\)/CFT\(_4\) correspondence.  We will relate this old formulation of the spectral problem to
our language of Quantum Spectral Curve (QSC).
In this section we summarize the main steps leading to the {\Pmsys}, which is one of the ways to define the QSC, leaving the details for the \appref{sec:quant-sprect-curve}. Then we depart from the {\Pmsys} and derive an alternative description -- {\Qosys}.
The role of this section is to demonstrate, in somewhat schematic but hopefully inspiring way, the origins of our QSC approach. In the next section, we will reveal a more general underlying algebraic and analytic structure of QSC in terms of the analytic Q-system.

\subsubsection{TBA equations as a set of functional equations}
\begin{figure}
 \centering
\ifpicture{
\includegraphics{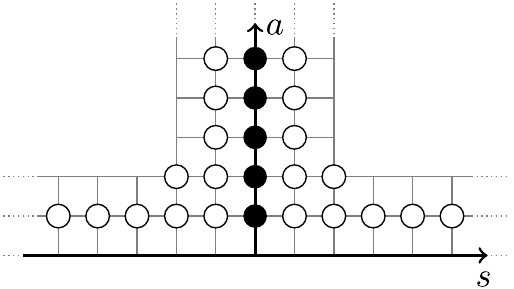}}
 \caption{\textbf{The AdS\(_5\)/CFT\(_4\) T-hook}: the domain for the variables \(a\) and \(s\) in the {\Ysys} (circles) and the Hirota equation (all nodes of the underlying grid).}
 \label{fig:Thook}
\end{figure}

TBA equations were written as an infinite set of nonlinear integral equations on {\Yfcts} with rather complicated kernels.
Despite its complexity they were suitable for the first numerical analysis of the spectrum for short operators
in the work \cite{Gromov:2009zb}. However, these equations look extremely complicated for analytic study.
The {\Ysys}, conjectured even before TBA \cite{Gromov:2009tv} as the solution of spectral AdS/CFT problem, already points out the possibility of considerable simplifications of TBA. It states that the {\Yfcts}, the same as those entering the TBA,
satisfy  the  specific functional equations  in a special, T-hook
domain, shown in \figref{fig:Thook}. This Y-system reads as follows
\begin{equation}\label{Ysystem}
Y_{a,s}(u+i/2)Y_{a,s}(u-i/2)=\frac{(1+Y_{a,s+1}(u))(1+Y_{a,s-1}(u))}{(1+1/Y_{a+1,s}(u))(1+1/Y_{a-1,s}(u))}\,.
\end{equation}
Although equations (\ref{Ysystem}) already contain a big part of the information about the spectrum, one should supplement them with an extra input to establish the full equivalence with TBA. The {\Yfcts} have infinitely many cuts \((-\infty,-2g+i n/2]\cup [2g+i n/2,+\infty)\) for some integers \(n\). And this missing part of the information concerns the behaviour of the {\Yfcts} w.r.t. the analytic continuation under these cuts.

The analytic continuation properties can be summarized in three discontinuity relations \cite{Cavaglia:2010nm,Balog:2011nm}.
As an example, one  of these relations simply states that
\begin{equation}
 \disc\log\left[Y_{1,1}^{[2n]}Y_{2,2}^{[2n]}\prod_{a=1}^n\left(1+Y_{a,0}^{[2n-a]}\right)\right]=0,\qquad n\ge 1\,,
\label{Italian-1}
\end{equation}
where \(\disc\) denotes a discontinuity on such cut {\ie} \(\disc f\equiv f(u+i0)-f(u-i0)\). The Y-system together with these analyticity relations can be called {\it analytic Y-system}.

It was noticed in \cite{Gromov:2011cx}  that these relations can be further simplified by introducing {\Tfcts}
related to {\Yfcts} as follows
\begin{equation}\label{eq:14}
Y_{a,s}=\frac{T_{a,s+1}T_{a,s-1}}{T_{a+1,s}T_{a-1,s}}\;.
\end{equation}
This relation does not define {\Tfcts} uniquely leaving a certain
gauge freedom: different sets of {\Tfcts} could produce the same
Y's. 
In the language of {\Tfcts} \eq{Ysystem} becomes
\begin{equation}\label{Hir}
T_{a,s}^+T_{a,s}^-=T_{a+1,s}T_{a-1,s}+T_{a,s+1}T_{a,s-1}\;.
\end{equation}
which is the so-called discrete Hirota equation. But most importantly, the discontinuity relations of \cite{Cavaglia:2010nm} turn out to be very simple statements about the analyticity of {\Tfcts}.
In particular, as was shown in \cite{Gromov:2011cx}, \eq{Italian-1} is equivalent to a statement of existence of a special gauge choice
 for {\Tfcts}, denoted as \(\bT\) in \cite{Gromov:2011cx}, such that
\begin{eqnarray}\label{bTprop}
\bT_{a-1,0}\;,\;\bT_{a,\pm 1}\;,\;\bT_{a+1,\pm 2}\;\;\text{have no cuts inside the strip}\;\;\;\;|{\rm Im}\;u|<\frac{a}{2}
\end{eqnarray}
which additionally obey the ``group theoretical'' constraints\footnote{
In the classical limit, the T-functions reduce to characters of the
\(\psu(2,2|4)\) symmetry algebra \cite{Gromov:2010vb,Gromov:2010km}, and the integers \(a\) and \(s\) label
certain representations of \(\psu(2,2|4)\). In this limit, the
conditions \eqref{bT-gauge} reduce to group-theoretical statements
about these representations \cite{Gromov:2011cx}.

In the general case, although the interpretation in terms of
representations is much less clear than in the classical limit,
\eqref{bT-gauge} provides the generalization, in terms of T-functions,
of the properties of \({\rm {\rm PSU}}(2,2|2)\) characters.}
\begin{equation}\label{bT-gauge}
\bT_{2,\pm 3}=\bT_{3,\pm 2}\;\;,\;\;\bT_{0,0}^+=\bT_{0,0}^-\;\;,\;\;\bT_{0,s}=\bT_{0,0}^{[+s]}\;.
\end{equation}
Similarly, the meaning of the two remaining  discontinuity relations of
\cite{Cavaglia:2010nm} becomes clear in another  choice of the gauge of {\Tfcts} which we denote
 \(\wT\), defined through the previous one as
\begin{equation}\label{eq:95}
  \wT_{a,s}\equiv (- 1)^{a\, s }\bT_{a,s}\left(\bT_{0,0}^{[+a+s]}\right)^{\frac{a-2}{2}}.\end{equation}
Firstly, it is obvious from the definition that \(\wT_{0,s}=1\). Secondly, the discontinuity relations strongly suggest  to choose a Riemann sheet with short branch cuts \([-2g+i n/2,2g+in/2]\)
in \fcts{\(\wT\)} instead of the initial long cuts. With this choice
\(\hat\wT_{1,s}\) have only two short cuts for \(|s|>0\)! To distinguish between different choices of the cuts we will put the hat over the functions
which we choose to have short cuts. Thirdly, in this gauge
 \(\hat\wT_{2,\pm s}=\hat \wT^{[+s]}_{1,\pm 1}\hat \wT^{[-s]}_{1,\pm 1}\) for \(|s|\ge2\).%

In the next subsection we will see that these rather simple properties of {\Tfcts} can be taken into account all together by a particular parameterization of {\Tfcts} in terms of a few Q-functions entering the \(\bP\mu\)-system.

\subsubsection{Emergence of  \texorpdfstring{{\Pmsys}}{P-mu system}}\label{sec:emergence}
We show in this subsection that it is possible to parameterize the {\Tfcts} in a particularly nice way so that all analyticity properties
 deduced from TBA and
described in the previous section are easily satisfied.
We start from the right band of the \(\wT\)-hook {\ie} \(s\geq a\). We know that \(\wT_{1,s}\) should have only two short cuts.
This fact can be reflected by the following parameterization
\begin{eqnarray}\label{Tp1p2}
{\mathbb T}_{1,s}&=&
\bP_1^{[+s]}
\bP_2^{[-s]}
-
\bP_2^{[+s]}
\bP_1^{[-s]}\;\;,\;\;s\geq 1
\end{eqnarray}
where each \(\bP_a\) has only one short cut on the real axis.
It is easy to convince oneself that together with
\begin{equation}\label{eq:12}
\wT_{0,s}=1\;\;{\rm and}\;\;\hat\wT_{2,\pm s}=\hat\wT_{1,\pm 1}^{[+s]}\hat\wT_{1,\pm 1}^{[-s]}\;\;,\;\;s\geq 2
\end{equation}
the ansatz \eq{Tp1p2} solves indeed  \eq{Hir} for \(s>a\) and has all required analytic properties.
Similarly, for negative \(s\) one can parameterize
\footnote{Note that we introduce here the ``contravariant'' upper
  indices \(3,4\) in contrast to the ``covariant'' indices \(1,2\) of
  the first two {\Pfcts}. The meaning of it, and its relation to the
  whole system of {\Qfcts} formally solving the {\Tsys} \eq{Hir} in
  the \(\wT\)-hook \cite{Gromov:2010vb}, will be clear later.}
 \begin{eqnarray} \label{eq:15}
{\mathbb T}_{1,s}&=&{\bP^{4[+s]}}
{\bP^{3[-s]}}
-{\bP^{3[+s]}}
{\bP^{4[-s]}}
\;\;,\;\;s<0\;.
\end{eqnarray}
where we introduced a pair \(\bP^3\), \(\bP^4\) with upper indices,
also having a single short Zhukovsky cut on the real axis. 

In what follows we will construct other {\Pfcts} $\bP^1$, $\bP^2$,
$\bP_3$ and $\bP_4$, and show that the analyticity of the {\Tfcts} is
the statement that these new {\Pfcts} also have a single Zhukovsky cut
on the real xis.

  The {\Tfcts} \eqref{Tp1p2},~\eqref{eq:15} alone do not allow to reconstruct all {\Tfcts}.
In fact we will need   only one additional quantity to build a complete parameterization of the T-hook.
Let us introduce a notation
\begin{equation}
\mu_{12}\equiv \left(\bT_{0,1}\right)^{\frac 1 2}\,,
\end{equation}
which is \(i\)-periodic due to \eqref{bT-gauge}   on the sheet with long cuts:
\begin{equation}\mu_{12}^{++}=\mu_{12}\;.\end{equation}

The claim is that any {\Tfct} (and hence any {\Yfct})
can be written in terms of \(5\) functions \(\bP_1\), \(\bP_2\), \(\bP^3\), \(\bP^4\) and \(\mu_{12}\).
Indeed, one simply has to use Hirota equation to find, one-by-one, all {\Tfcts}.
To exemplify the procedure let us find \(\bT_{2,1}\). For that we write the Hirota equation \eq{Hir} at \((a,s)=(2,2)\):%
\begin{equation}\label{hir22}
\wT_{2,1}\wT_{2,3}=\wT_{2,2}^+\wT_{2,2}^--\wT_{1,2}\wT_{3,2}\,.
\end{equation}
which holds in mirror kinematics (i.e. for long cuts).
Next, we can replace in the vicinity of the real axis \(\wT_{2,3}\) by \(\hat \wT^{[+3]}_{1,1}\hat \wT^{[-3]}_{1,1}\), \(\wT_{2,1}\)
is the same as \(\bT_{2,1}\) and due to the condition \(\bT_{2,3}=\bT_{3,2}\) we have
\(\wT_{3,2}=\wT_{2,3}\mu_{12}\). The first term in the r.h.s. should be treated with some care:
we should use~(\ref{eq:12}) \(\hat\wT_{2,2}=\hat \wT_{1,1}^{[+2]}\hat \wT_{1,1}^{[-2]}\),
but we should remember that \(\hat\wT_{2,2}\) and \(\wT_{2,2}\) coincide only inside the
analyticity strip \(-1/2<{\rm Im}\ u<1/2\). To avoid ambiguities we always assume that the
argument has a small positive imaginary part which means that \(\wT_{2,2}^+\) is outside the
analyticity strip and is not equal to \(\hat\wT_{2,2}^+\) but to its analytic continuation
\(\wT_{2,2}^+=\widetilde{\hat\wT_{2,2}^+}=\hat \wT_{1,1}^{[+3]}\widetilde{\hat \wT_{1,1}^{-}}\), which results in:
\begin{equation}\label{2444}
\wT_{2,2}^+\wT_{2,2}^-=\hat \wT_{1,1}^{[+3]}\widetilde{\hat \wT_{1,1}^{-}}\hat\wT_{1,1}^+\hat\wT_{1,1}^{[-3]}\,,
\end{equation}
where we will always denote by tilde the analytic continuation of a
function around the branch point \(u=2g\).
Thus \eq{hir22} gives \(\bT_{2,1}=\widetilde{\hat \wT_{1,1}^{-}}\hat\wT_{1,1}^+\sm\wT_{1,2}\mu_{12}\) or
\begin{equation}\label{bt21}
\bT_{2,1}=\left(\tilde\bP_1\bP_2^{[-2]}-\tilde\bP_2\bP_1^{[-2]}\right)
\left(\bP_2\bP_1^{[+2]}-\bP_1\bP_2^{[+2]}\right)
+\mu_{12}
\left(\bP_1^{[-2]}\bP_2^{[+2]}-\bP_1^{[+2]}\bP_2^{[-2]}\right)
\;.
\end{equation}
In the same way we find $\bT_{2,-1}$ in terms of $\bP^3,\bP^4$ and $\mu_{12}$.

Next, using Hirota at \((a,s)=(1,1)\) one finds \(\bT_{1,0}\), after that \(\bT_{2,0}\), etc.
The complexity of  expressions grows fast, but it is important to stress that in principle
one can explicitly recover all T- and {\Yfcts} in terms of \(\bP_1,\bP_2,\bP^3,\bP^4,\mu_{12}\) and their analytic continuations through the  cuts.
In the \appref{apA} we wrote a simple {\it Mathematica} code which allows to automatize this procedure.

So far our parameterization ensures all the nice properties of the right and of the left bands of \(\wT\)-hook, but  can we guarantee  the correct analytic properties
of the upper band? Namely,  we have to ensure that the analyticity strips of \(\bT\) are such as dictated by \eq{bTprop}. In particular, it must be that \(\bT_{2,1}\)
has no cut on the real axis or, in other words, that the analytic continuation around the point \(u=2g\) is trivial \(\bT_{2,1}-\tilde\bT_{2,1}=0\).
This can be written from \eq{bt21} as
\begin{equation}\label{formath1}
\left(\bP_1^{[-2]}\bP_2^{[+2]}-\bP_1^{[+2]}\bP_2^{[-2]}\right)\left(\mu_{12}-\tilde\mu_{12}-\tilde \bP_1\bP_2+\tilde \bP_2 \bP_1\right)=0\;.
\end{equation}
The first multiplier is \(\wT_{1,2}\) which cannot be zero, so we must require that
\begin{equation}\label{tmu12}
\tilde\mu_{12}=\mu_{12}+\bP_1\tilde \bP_2-\bP_2\tilde \bP_1\;.
\end{equation}
Assuming \eq{tmu12}, it is easy to see that \(\bT_{1,0}\) is
analytic on the real axis as it is given by
\beq\la{T10}
\bT_{10}=\mu_{12}(\mu_{12}+\bP_1\tilde \bP_2-\bP_2\tilde \bP_1)=\mu_{12}\tilde\mu_{12}\;.
\eeq
As we shall  soon see, another  restriction on $\bP$'s and $\mu_{12}$ comes from  equations on \(\bT_{2,0}\) and \(\bT_{3,1}\). At this point, for a better transparency we will first consider an important particular case of the left-right (LR) symmetric states \(Y_{a,s}=Y_{a,-s}\) and later generalize it to all states. Since $\wT_{1,s}=\wT_{1,-s}$ in this case, we can make the identification \(\bP^4=\bP_1\) and \(\bP^3=-\bP_2\) which is used in the next subsection.

\subsubsection{\texorpdfstring{{\Pmsys}}{P-mu system} in Left-Right--symmetric case}
To illustrate what could come out from this LR-symmetry condition let us write  down \(\bT_{20}\)  explicitly
\begin{align}
\label{eq:13}\bT^+_{20}=&\mu_{12}^2+\mu_{12}\left(-\bPt_1\bP_2+\bPt_2\bP_1-\bPt_1^{[2]}\bP_2^{[2]}+\bPt_2^{[2]}\bP_1^{[2]}\right)\\
\nn&+\left(\bPt_1\bP_2-\bPt_2\bP_1\right)\left(\bPt^{[2]}_1\bP^{[2]}_2-\bPt^{[2]}_2\bP^{[2]}_1\right)-\left(\bPt_1\bP^{[2]}_2-\bPt_2\bP^{[2]}_1\right)^2\;.
\end{align}
We observe the appearance of a new type of objects \(\bPt_a^{[2]}\)
which means that we first go under the cut\footnote{By contrast \(\widetilde{\bP_a^{[2]}}\) is
simply equal to \(\bP_a^{[2]}\) because \(\bP_a^{[2]}\) has no cut on the
real axis.} to get \(\bPt_a\) and then
analytically continue further shifting the argument by \(+i\).
Even though on its main defining sheet \(\bP_a\) had only one single
short cut, on the next sheet there will be a ``ladder'' of infinitely many cuts with branch points
at any \(\pm 2g+ i n,\;n\in{\mathbb Z}\). But, nevertheless, the analyticity properties \eqref{bTprop}  require that \(\bT^+_{20}\) does not have a cut on the real axis, so that
\beq\la{neweq}
\bT^+_{20}=\widetilde{\bT^+_{20}}\;.
\eeq
The r.h.s. the above equation will contain \(\widetilde{\bPt_1^{[2]}}\) and  \(\widetilde{\bPt_2^{[2]}}\) in addition to the usual terms. Schematically we can write \eq{neweq} equation as
\(
A_1 \widetilde{\bPt_1^{[2]}}+A_2\widetilde{\bPt_2^{[2]}}=A_0
\)
where the \(A_j\)'s contain only \(\bP_a,\bPt_a,\bP_a^{[2]},\bPt_a^{[2]}\) and \(\mu_{12}\). Analyticity condition on \(\bT_{31}^+\) also has a very similar structure, which schematically can be written as
\(
B_1 \widetilde{\bPt_1^{[2]}}+B_2\widetilde{\bPt_2^{[2]}}=B_0
\).
Thus the analyticity conditions for \(\bT_{20}^+\) and \(\bT_{31}^+\) can be considered as a system of two linear equations on two unknowns \(\widetilde{\bPt_1^{[2]}}\) and  \(\widetilde{\bPt_2^{[2]}}\).
The solution of this system is simpler than one would expect:
\begin{eqnarray}\label{generaln}
\widetilde{\bPt_1^{[2n]}}&=&\bPt_1^{[2n]}\frac{\tilde\mu_{12}}{\mu_{12}}+\bP_1^{[2n]}\frac{\bP_1\bP_2-\bPt_1\bPt_2}{\mu_{12}}
-\bP_2^{[2n]}\frac{\bP_1^2-\bPt_1^2}{\mu_{12}}\,,
\end{eqnarray}
where \(n=1\),
and a similar expression for \(\widetilde{\bPt_2^{[2n]}}\)  obtained by interchanging the indices \(1\leftrightarrow 2\) and changing the sign \(\mu_{12}\to -\mu_{12}\).
Amazingly, repeating this procedure for
\(\bT_{30}^{[2]}\) and \(\bT_{41}^{[2]}\) and requiring their analyticity at the real axis we find again \eq{generaln}, with \(n=2\) this time.
It is of course very appealing to assume
 that this condition must hold for any integer \(n\geq 1\) to ensure the required analyticity of \(\bT_{a,s}\) in the upper band (see \appref{sec:quant-sprect-curve} for a formal proof of this fact).
Thus the \equref{generaln} is all we need to close our system of spectral equations!

 In the rest of this section we will try to find an  aesthetically  more attractive form of the \equref{generaln}.
 For that we will need to introduce a few new objects. First, we
 introduce two auxiliary functions \(\bP_3\) and \(\bP_4\)\footnote{One
   should be careful not to confuse the new functions \(\bP_3\) and \(\bP_4\)
with the  functions \(\bP^{3},\bP^{4}\) of the left band which carry the upper indices.}
\begin{equation}\label{P3P4}
\bP_3\equiv\frac{1}{\mu_{12}}\bPt_1-\frac{\mu_{14}}{\mu_{12}}\bP_1
+\frac{\mu_{13}}{\mu_{12}}\bP_2\;\;,\qquad
\bP_4\equiv\frac{1}{\mu_{12}}\bPt_2-\frac{\mu_{24}}{\mu_{12}}\bP_1
+\frac{\mu_{23}}{\mu_{12}}\bP_2\,,
\end{equation}
where \(\mu_{14},\;\mu_{13},\;\mu_{24},\;\mu_{23}\) are some new \(i\)-periodic functions with long cuts.
Let us show that, with a convenient choice of functions \(\mu_{13}\),
\(\mu_{14}\), \(\mu_{23}\) and \(\mu_{24}\), the \equref{generaln} is nothing but the condition of analyticity of \(\bP_3\) in the upper half-plane.
Indeed, the analyticity condition for \(\bP_3\) reads
\begin{equation}\label{P3nod}
0=\bP^{[2n]}_3-\widetilde{\bP^{[2n]}_3}=
\left(
\frac{1}{\mu_{12}}\bPt_1^{[2n]}-\frac{\mu_{14}}{\mu_{12}}\bP_1^{[2n]}
+\frac{\mu_{13}}{\mu_{12}}\bP_2^{[2n]}
\right)
-
\left(
\frac{1}{\tilde\mu_{12}}\widetilde{\bPt_1^{[2n]}}-\frac{\tilde\mu_{14}}{\tilde\mu_{12}}\bP_1^{[2n]}
+\frac{\tilde\mu_{13}}{\tilde\mu_{12}}\bP_2^{[2n]}
\right).
\end{equation}
Note that in the r.h.s. there is no tilde over \(\bP_1^{[2n]}\) and  \(\bP_2^{[2n]}\)
as they have no cuts except for the real axis. We see that if we
choose \(\mu_{14}\) and \(\mu_{13}\) so that
\begin{subequations}
\begin{align}\label{m14}
\frac{\mu_{14}}{\mu_{12}}-\frac{\tilde\mu_{14}}{\tilde\mu_{12}}=&\frac{\tilde\bP_1\tilde\bP_2-\bP_1\bP_2}{\mu_{12}\tilde\mu_{12}}\,&
\frac{\mu_{13}}{\mu_{12}}-\frac{\tilde\mu_{13}}{\tilde\mu_{12}}=&\frac{\tilde\bP_1^2-\bP_1^2}{\mu_{12}\tilde\mu_{12}}\,,
\end{align}
which is of course always possible as so far the only condition on \(\mu_{14}\) and \(\mu_{13}\) was their periodicity\footnote{The equation \(\tilde f-f=g\) for a periodic function \(f\) and arbitrary distribution
\(g\) can be always solved, modulo an arbitrary regular periodic function, by \(f(u)=\frac{1}{2i}\int \coth(\pi (u-v))g(v)dv\).}, we recognize in
\eq{P3nod} the main result of this section \eq{generaln}!
 Similarly,
we obtain $\bP^{[2n]}_4-\widetilde{\bP^{[2n]}_4}=0$ by setting
\begin{align}
\frac{\mu_{24}}{\mu_{12}}-\frac{\tilde\mu_{24}}{\tilde\mu_{12}}=&\frac{\tilde\bP_2^2-\bP_2^2}{\mu_{12}\tilde\mu_{12}}\,,
&
\frac{\mu_{23}}{\mu_{12}}-\frac{\tilde\mu_{23}}{\tilde\mu_{12}}=&\frac{\tilde\bP_1\tilde\bP_2-\bP_1\bP_2}{\mu_{12}\tilde\mu_{12}}\,
\end{align}
\end{subequations}
and then we see that we can set $\mu_{23}=\mu_{14}$.

We have to justify the strange way the periodic coefficients \(\mu_{ab}\)
where introduced in \eq{P3P4}. First, we can use \eq{m14} and \eq{tmu12}
to get
\begin{eqnarray}\label{m1314}
\tilde\mu_{13}-\mu_{13}&=&\frac{\bP_1^2-\bPt_1^2}{\mu_{12}}
+\mu_{13}\frac{\bP_1\bPt_2-\bP_2\bPt_1}{\mu_{12}}\;\;,\notag\\
\tilde\mu_{14}-\mu_{14}&=&\frac{\bP_1\bP_2-\bPt_1\bPt_2}{\mu_{12}}
+\mu_{14}\frac{\bP_1\bPt_2-\bP_2\bPt_1}{\mu_{12}}\,\,.
\end{eqnarray}
Next, from \eq{P3P4} we can express \(\bPt_3\) and \(\bPt_4\), then by excluding from them \(\tilde\mu_{ab}\) with the use of \eq{m1314} we find
\begin{equation}\label{p3t}
\bPt_3=\frac{1}{\mu_{12}}\left(\bP_1+\bPt_2\,\mu_{13}-\tilde\bP_1\,\mu_{14}\right)
\,,\ \ \ \
\bPt_4=\frac{1}{\mu_{12}}\left(\bP_2+\bPt_2\,\mu_{23}-\bP_1\,\mu_{24}\right)
\end{equation}
from where, together with \eqref{P3P4},  it is easy to see that
\eqref{m1314} reduces to
\begin{equation}
\tilde\mu_{13}=\mu_{13}+\bP_1\tilde \bP_3-\bP_3\tilde \bP_1\;,\ \ \ \tilde\mu_{14}=\mu_{14}+\bP_1\tilde \bP_4-\bP_4\tilde \bP_1\;.
\end{equation}
Acting similarly for the other \(\mu_{ab}\) we find a universal relation
\begin{equation}\label{muab}
\tilde\mu_{ab}=\mu_{ab}+\bP_a\tilde \bP_b-\bP_b\tilde \bP_a\;
\end{equation}
which generalizes our previous relation for \(\mu_{12}\) \eq{tmu12}!
So far we had only \(\mu_{ab}\) with \(a<b\), we can define \(\mu_{ba}=-\mu_{ab}\).
The only missing \(\mu_{34}\) is then defined by
\begin{equation}\label{defmu34}
\mu_{34}\equiv\frac{1+\mu_{13}\mu_{24}-\mu_{14}\mu_{23}}{\mu_{12}}\,,
\end{equation}
so that it also satisfies \eq{muab}, as one can check by using \eqref{P3P4} and \eqref{p3t}. Finally, we can exclude \(\bPt_{1}\) and \(\bPt_{2}\)
from \eq{p3t} which then becomes:
\begin{equation}
\bPt_3=\mu_{34}\bP_1-\mu_{14}\bP_3+\mu_{13}\bP_4
\,,\ \ \
\bPt_4=\mu_{34}\bP_2-\mu_{24}\bP_3+\mu_{14}\bP_4\,,
\end{equation}
which also appears to be on equal footing with a similar equation
for \(\bP_1\) and \(\bP_2\) \eq{P3P4}.

To write these identities compactly in the matrix form we introduce a  matrix \begin{equation}\chi_{ab}=-\chi^{ab}=\begin{pmatrix}0 & 0 & 0 & 1 \\
0 & 0 & -1 & 0 \\
0 & 1 & 0 & 0 \\
-1 & 0 & 0 & 0 \\
\end{pmatrix}\end{equation} so that
\begin{equation}\label{bPta}
\bPt_a=\mu_{ab}\chi^{bc}\bP_c\;.
\end{equation}
Notice that by contracting the last equation with $\chi^{ad}\bP_d$, we set the r.h.s. to zero because \(\mu\) is anti-symmetric and obtain \(\bPt_a\chi^{ad}\bP_d=-\bPt_1\bP_4+\bPt_2\bP_3-\bPt_3\bP_2+\bPt_4\bP_1=0\),
as expected from \eq{muab} and from the equality
\(\mu_{23}=\mu_{14}\).

With the help of $\chi$, one can also write \eq{defmu34} in a covariant way as
\begin{equation}
\mu\chi\mu\chi=1\;.
\end{equation}

\subsubsection{\texorpdfstring{{\Pmsys:}}{P-mu system} General case}
For the general non-left--right symmetric case we proceed in exactly the same way.
This time \(\bT_{2,-1}\neq \bT_{2,1}\) and thus in principle we can get some new condition from its analyticity
at the real axis. In addition to \eq{tmu12}  coming from analyticity of \(\bT_{2,1}\) we find
\begin{equation}\label{tmu34}
\tilde\mu_{12}=\mu_{12}+\bP^3\tilde \bP^4-\bP^4\tilde \bP^3\;
\end{equation}
which implies that \(\tilde \bP\)'s are not all linearly independent
\begin{equation}\label{constr}
\bP^3\tilde \bP^4-\bP^4\tilde \bP^3=\bP_1\tilde \bP_2-\bP_2\tilde \bP_1\;.
\end{equation}
At the next step we require analyticity of \(\bT_{3,1}^+\), \(\bT_{2,0}^+\) and \(\bT_{3,-1}^+\).
This gives us \(3\) linear equations on \(\widetilde{\bPt_1^{\,\,[2]}},\widetilde{\bPt_2^{\,\,[2]}},\widetilde{\bPt^{3[2]}},\widetilde{\bPt^{4[2]}}\)
and the \(4^{\rm th}\) identity follows from \eq{constr}:
\(
\bP^{3[2]}\widetilde{\tilde \bP^{4[2]}}-
\bP^{4[2]}\widetilde{\tilde \bP^{3[2]}}
=
\bP_1^{\,\,[2]}\widetilde{\tilde \bP_2^{\,\,[2]}}-
\bP_2^{\,\,[2]}\widetilde{\tilde \bP_1^{\,\,[2]}}\;.
\)
Solving this linear system we get  a generalization of \eq{generaln}
\begin{eqnarray}\label{mastereq}
\frac{\widetilde{{\bPt_\alpha}^{\,\,[2n]}}}{\tilde\mu_{12}}&=&
\frac{{{\bPt_\alpha}}^{\,\,[2n]}}{\mu_{12}}-\bP^{4[2n]}\frac{\bP_\alpha\bP^3-\bPt_\alpha\bPt^3}{\mu_{12}\tilde\mu_{12}}
+\bP^{3[2n]}\frac{\bP_\alpha\bP^4-\bPt_\alpha\bPt^4}{\mu_{12}\tilde\mu_{12}}\;\;,\;\;\alpha=1,2\notag\\
\frac{\widetilde{\bPt^{{\dot\alpha}[2n]}}}{\tilde\mu_{12}}&=&
\frac{\bPt^{{\dot\alpha}[2n]}}{\mu_{12}}+{{\bP_1}}^{[2n]}\frac{\bP^{\dot\alpha}\bP_2-\bPt^{\dot\alpha}\bPt_2}{\mu_{12}\tilde\mu_{12}}
-{\bP_2}^{[2n]}\frac{\bP^{\dot\alpha}\bP_1-\bPt^{\dot\alpha}\bPt_1}{\mu_{12}\tilde\mu_{12}}\;\;,\;\;{\dot\alpha}=3,4\notag\\
\end{eqnarray}
Again, to solve the  \equref{mastereq} at once for all \(n>0\) we introduce
new auxiliary functions \(\bP^1,\;\bP^2\) and \(\bP_3,\;\bP_4\)
\begin{eqnarray}\label{P1234}
&&\bP^2\equiv+\frac{1}{\mu_{12}}\bPt_1-\frac{\mu_{14}}{\mu_{12}}\bP^4
-\frac{\mu_{13}}{\mu_{12}}\bP^3\;\;,\qquad
\bP^1\equiv-\frac{1}{\mu_{12}}\bPt_2+\frac{\mu_{24}}{\mu_{12}}\bP^4
+\frac{\mu_{23}}{\mu_{12}}\bP^3\,,\notag\\
&&\bP_3\equiv+\frac{1}{\mu_{12}}\bPt^4-\frac{\mu_{23}}{\mu_{12}}\bP_1
+\frac{\mu_{13}}{\mu_{12}}\bP_2\;\;,\qquad
\bP_4\equiv-\frac{1}{\mu_{12}}\bPt^3-\frac{\mu_{24}}{\mu_{12}}\bP_1
+\frac{\mu_{14}}{\mu_{12}}\bP_2\,.\notag\\
\end{eqnarray}
Provided that the \(i\)-periodic functions \(\mu_{ab}\) obey the relations
\begin{eqnarray}\label{tmu}
\frac{\tilde\mu_{\alpha4}}{\tilde\mu_{12}}-\frac{\mu_{\alpha4}}{\mu_{12}}=\frac{\bPt_\alpha\bPt^3-\bP_\alpha\bP^3}{\mu_{12}\tilde\mu_{12}}\;\;,\;\;
\frac{\tilde\mu_{\alpha3}}{\tilde\mu_{12}}-\frac{\mu_{\alpha3}}{\mu_{12}}=
\frac{\bP_\alpha\bP^4-\bPt_\alpha\bPt^4}{\mu_{12}\tilde\mu_{12}}\;\;,\;\;\alpha=1,2
\end{eqnarray}
the equations \eq{mastereq} are equivalent to the analyticity of the
newly defined \(\bP^{1},\;\bP^2,\;\bP_3\;,\bP_4\) on the upper
half-plane.
We also define \(\mu_{34}\) in the same way as it was defined before
in \eq{defmu34} for the symmetric case.

One should note that if we had started from requiring the analyticity of
\(\bT_{3,1}^-\), \(\bT_{2,0}^-\) and \(\bT_{3,-1}^-\), the same steps
would have shown that the {\Pfcts} defined by \eqref{P1234} are also
analytic on the lower half-plane. Additionally, one should show that  the equations \eqref{P1234} above real axis and below real axis are related by analytic continuation that avoids the short cut \([-2g,2g]\). This is not immediately obvious since the {\Pfcts} are defined as functions with short cuts and \(\mu\)-functions have long cuts, but the computation is quite straightforward\footnote{
    For instance, one gets for $u\in]-\infty,-2g[\cup]2g,+\infty[$ the jumps $P(u+i0)-P(u-i0)$:: ${\bP^2}^{[+0]}-{\bP^2}^{[-0]}= \tfrac 1 {\mu_{12}\tilde\mu_{12}}\left((\bP^3\bPt^4-\bP^4\bPt^3)\bPt_1+
(\bPt_1\bPt^3-\bP_1\bP^3)\bP^4+
(\bP_1\bP^4-\bPt_1\bPt^4)\bP^3\right)=0$.
}. Hence, all the {\Pfcts} have a single, short Zhukovsky
cut on the real axis.

At this moment, we can already  state that we found a full system of equations solving the spectral problem.
There is however a further simplification expected, like in the symmetric case.
We notice that, using \eq{P1234},\eq{tmu34},\eq{defmu34} and \eq{constr},  the \equref{tmu}  can be equivalently written in a more covariant form, similar to \eqref{muab} and \eqref{bPta} of the left-right symmetric case:
\begin{eqnarray}\label{Pmufinal}
\boxed{\tilde\mu_{ab}-\mu_{ab}=\bP_a\bPt_b-\bP_b\bPt_a
\;\;,\;\;
\tilde\bP_a=\mu_{ab}\bP^b\;\;,\;\;\bP_a\bP^a=0\;\;,\;\;{\Pf(\mu)=1\,,}
}
\end{eqnarray}
where \(\Pf(\mu)\equiv
\mu_{12}\mu_{34}+\mu_{23}\mu_{14}-\mu_{13}\mu_{24}\) is the Pfaffian of
the matrix \(\mu_{ab}\).

The equations \eqref{Pmufinal} represent a complete set of spectral equations
of planar \neqfour{} SYM for any local single-trace operator. In the next section we will describe in all details
how the global symmetry charges enter into the asymptotics of these functions, to completely fix the rules of the game.

A few comments are in order. Firstly, it is obvious, by the left-right symmetry of the {\Tsys} equations (not solutions!) that there should exist a supplementary system of equations, by the exchange of the lower and upper indices.
Such system can be deduced immediately from \eq{Pmufinal} by introducing
\(\mu_{ab}\mu^{bc}=\delta_a^c\) and it has the form
\begin{eqnarray}\label{PmufinalLR}
\tilde\mu^{ab}-\mu^{ab}=-\bP^a\bPt^b+\bP^b\bPt^a
\;\;,\;\;
\tilde\bP^a=\mu^{ab}\bP_b\;\;,\;\;\bP_a\bP^a=0\;\;.
\end{eqnarray}
Secondly,  this formulation of the spectral problem,  is rather appropriate for the  description of the degrees of freedom in the left and right bands of the \(\wT\)-hook (which can be related to the  \(S^5\) degrees of freedom of the string).
In this description, the formalism becomes very complicated (although valid of course!) once one tries to go to the upper band ({\ie} \(AdS^5\) part). Thus one may expect that there is an alternative description
which treats \(AdS_5\) in a way similar to \(S^5\) giving the analogue of the {\Pmsys} for the \(AdS_5\) degrees of freedom.  In the next sections we will
formulate such a system of equations called the {\Qosys}.

\subsection{Generalization and extension}
\label{sec:gener-extens}
\subsubsection{\texorpdfstring{{\Qosys}}{Q-omega system}}\label{sec:Qomega}
In this section we build an alternative set of spectral equations
which we call {\Qosys}. As we will discuss in \secref{sec:QCA},
{\Pfcts} are quantum analogs of the quasi-momenta in S\(^5\), whereas the
new {\bQfcts} correspond to the quasi-momenta in AdS\(_5\).

We shall construct {\bQfcts} from {\Pfcts} in the
following way: first we have to find a solution of the finite
difference equation for a 4-vector \(X_a\)
\begin{equation}\label{eqX}
X^{-}_{a}={U_{a}}^b X^+_b\;,\quad{\rm where}\;\;{U_a}^b\equiv \delta_a^b+\bP_a\bP^b \;.
\end{equation}
We can construct \(4\) solutions to this equation
in the form of a formal infinite product\footnote{Strictly speaking this product is divergent and an appropriate regularization is needed.
We use this formal solution just to illustrate that there are indeed
\(4\) independent analytic in the upper half-plane solutions.}\\
\(X_a=[{U^{[+1]}U^{[+3]}\dots]_{a}}^bX_b^\infty\)
which solves the \equref{eqX} for any constant (or periodic) vector \(X_b^\infty\). This shows that it is always
possible to construct \(4\) linearly independent solutions
which are analytic in the upper half-plane. We define these linearly independent analytic solutions as \(X_a={\cal Q}_{a|i}\)
where \(i=1,2,3,4\) labels these \(4\) solutions\footnote{The utility of such notation will become clear in the next section where these quantities will be interpreted as certain Q-functions.}, so that we formally have
\({\cal Q}_{a|i}=[{U^{[+1]}U^{[+3]}\dots]_{a}}^jM_{ji}\,\) for some constant matrix $M_{ji}$.

Let us define the 4 {\bQfcts} as follows
\begin{equation}\label{QP}
\bQ_i\equiv-\bP^a {\cal Q}^+_{a|i}\;\;\text{for}\;\;{\Im u}>0\;.
\end{equation}
To define $\bQ$'s for \({\Im u}<0\) one has to do an analytic continuation.
This definition has  very intriguing feature
which one can explore with our knowledge of the properties of \(\bP\)'s \eq{Pmufinal}: if we consider $\bQ$'s as the functions with long cuts, then these functions have only one cut on their first sheet, which nicely complements the property of $\bP$'s who are considered as functions with short cuts and have only one cut on their first sheet.  We will now derive this feature.

An immediate consequence of the definitions \eqref{eqX} and \eqref{QP}  is an important formula
\begin{equation}\label{QPQ}
{\cal Q}_{a|i}^+ - {\cal Q}_{a|i}^-=\bP_a\bQ_i\,.
\end{equation}
We can also introduce the matrix \(V=U^{-1}\), and note that due to
\(\bP^a\bP_a=0\) this inverse matrix is simply \({V_a}^b=
\delta_a^b-\bP_a\bP^b\).
From the equations \eqref{Pmufinal} or \eqref{PmufinalLR}, it is then
easy to derive some useful identities
for the \(U\)-matrix \begin{eqnarray}
\label{id1}
\tilde U_a^{\,\,b}=\delta_a^b+\bPt_a\bPt^b=\mu_{ac}{V_d}^c\mu^{db}\;\;&\Leftrightarrow&\;\;\boxed{\tilde U=\mu^{-1}U^{-t}\mu}\\
\label{id2}\tilde\mu^{ab}{V_b}^c=\tilde \mu^{ac}-\bPt^a\bP^c=\mu^{ac}-\bP^a\bPt^c={U_d}^a\mu^{dc}\;\;&\Leftrightarrow&\;\;
\boxed{\tilde \mu=
U^{t}\mu U}\\
\label{id3}\tilde\mu^{ab}{\tilde U_{b}}^{\,\,c}=\tilde\mu^{ac}+\bP^a\bPt^c
=\mu^{ac}+\bPt^a\bP^c=\mu^{ab}{U_b}^c
\;\;&\Leftrightarrow&\;\;
\boxed{\tilde \mu \tilde U=\mu U}
\end{eqnarray}
Where the superscript \(t\) denotes the transposed matrix and \(-t\) transposed and inverted.
We show now that \(\bQ_i\) defined in this way should have only one long cut on the real axis \((-\infty,-2g]\cup[+2g,+\infty)\).
First, we see that \(\bQ_i\) is analytic by construction in the upper half-plane.
We define \(\bQ_i\) on the sheet with long cuts. To analytically continue under the real axis in
the definition \eq{QP} we have to go under the cut of \(\bP^a\):
\begin{equation}\label{QPt}
\bQ_i=-\tilde\bP^a {\cal Q}^+_{a|i}=\bP_a\mu^{ab} {\cal Q}^+_{b|i}=\bP_a\mu^{ab}({U_b}^{c})^{[+2]} {\cal Q}^{[+3]}_{c|i}
\;\;\text{for}\;\;0>{\Im u}>-1\;.
\end{equation}
 In this form, it is not hard to see that the monodromy around the branchpoint \({-i\pm2g}\) is trivial. Indeed, we have
\begin{equation}
\widetilde{\bQ_i^{[-2]}}=
\bP^{[-2]}_a\tilde\mu^{ab}{\tilde U_b}^{\,\,c} {\cal Q}^{[+1]}_{c|i}
\end{equation}
and using \eq{id3} and the fact that \(\bP^{[-2]}_a\) and \({\cal Q}^{[+1]}_{d|i}\)are regular on the real axis we  see that
\(\widetilde{\bQ_i^{[-2]}}={\bQ_i^{[-2]}}\).
This procedure can be easily continued further and it is not hard to prove by induction
that there are no cuts in the lower half-plane.
For that let us show (by induction) that the general expression for the analytic
continuation of \(\bQ\) is given by:
\begin{subequations}
\label{eq:20}

  \begin{align}\label{QPn0}
    \bQ_i=&\bP_a [V^{t[2]}V^{t[4]}\dots V^{t[2n-2]}\mu]^{ab} {\cal
      Q}^{[2n-1]}_{b|i}\,,&-n+1>&{\Im u}>-n\;,\\ \intertext{and then rewrite it equivalently using the equation for \(\cal Q\)
\eq{eqX}:}
\label{QPn} \bQ_i=&\bP_a
    [V^{t[2]}V^{t[4]}\dots V^{t[2n-2]}\mu U^{[2n]}]^{ab} {\cal
      Q}^{[2n+1]}_{b|i}\,,&-n+1>&{\Im u}>-n\;.
  \end{align}
\end{subequations}
At \(n=1\) the equations (\ref{eq:20}) are nothing but the
relation (\ref{QPt}), and to go to the next strip \(-n>{\Im u}>-n-1\) we
have to go under the cut of \(U^{[+2n]}\) in (\ref{QPn}), replacing it
by \(\tilde U^{[+2n]}\). Using \eq{id1} we convert it into \(\mu
V^{t}\mu\) and get (\ref{QPn0}) in the next strip ({\ie} at level \(n+1\)), which
also gives (\ref{QPn}) in the next strip.
 This proves the general formula (\ref{QPn},\ref{QPn0}).
Next, we have to show that there is no branch cut at \(\Im u=-n,\,\, n>0\). This is again obvious due to \eq{id3} which
tells that the combination \(\mu U\) has a trivial monodromy and all the other factors are explicitly regular.

\paragraph{Defining \(\omega\)}
We define \(\omega\) as a counterpart of \(\mu\) for \(\bQ\), {\ie} it should appear in the relation for \(\tilde \bQ\):
\begin{equation}\label{PPQQ}
\tilde\bQ_i=-\tilde\bP^a {\cal Q}^\pm_{a|i}=\bP_b\mu^{ba} {\cal Q}^\pm_{a|i}\,,
\end{equation}
where one can note that the sign of the shift in \({\cal Q}^\pm_{a|i}\)
is irrelevant due to \eqref{QPQ}\footnote{ Indeed, we have \(\tilde\bP^a ({\cal Q}^+_{a|i}-{\cal
  Q}^-_{a|i})=\tilde\bP^a \bP_a \bQ_i=(\bP_a\mu^{ab}\bP_b)\bQ_i\) where
\(\bP_a\mu^{ab}\bP_b\) vanishes due to the antisymmetry of \(\mu\).
}.
To close the {\Qosys} we have to define \(\bQ^i\) with upper indices in
the same way\footnote{It may seem unnatural to choose the sign in
  (\ref{QP2}) opposite to the sign in (\ref{QP}). We actually see in
  \secref{sec:analyticQ} that these relations can be interpreted
  in the setup of the Q-system where such a choice of signs looks natural.}
as \(\bQ_i\) {\ie}
\begin{equation}\label{QP2}
\bQ^i\equiv \bP_a({\cal Q}^{{a|i}})^+\;\;,\;\;
{\cal Q}^{a|i}\equiv-\left({\cal Q}_{a|i}\right)^{-t}\,.
\end{equation}
This new \({\cal Q}^{a|i}\) satisfies a similar equation \({\cal Q}^{a|i-}={V_{b}}^a ({\cal Q}^{b|i})^+\) and, as a consequence, \(\bQ^i\) has the same analyticity properties
as \(\bQ_i\). One consequence of \eq{QP2} is
 \(\bP_a=-\bQ^i{\cal Q}^\pm_{a|i}\).
Using these identities we can get rid of \(\bP\) in \eq{PPQQ} and get
\begin{align}
\tilde\bQ_i=&-\bQ^j{\cal Q}^\pm_{b|j}\mu^{ba} {\cal Q}^\pm_{a|i}=-\bQ^j\omega_{ji}\;\;,\\
\label{mudef}\text{where}\quad\omega_{ji}\equiv&{\cal Q}^-_{b|j}\mu^{ba}{\cal Q}^-_{a|i}\,.
\end{align}
Solving the last relation for \(\mu^{ab}\) we obtain:
\begin{align}
\label{muomega} \mu^{ab}= ({\cal Q}^{a|i})^-({\cal Q}^{b|j})^-\,\omega_{ij}\,.
\end{align}
We can now show that \(\omega\) is $i$-periodic with short cuts:
 \begin{equation}
\hat\omega^{[+2]}-
\omega
={\cal Q}^{t+}\tilde\mu {\cal Q}^{+}-{\cal Q}^{t-}\mu {\cal Q}^{-}
={\cal Q}^{t+}\tilde\mu{\cal Q}^{+}- {\cal Q}^{t+}U^{t}\mu U {\cal Q}^{+}
=0\;,
\end{equation}
and also it has a very similar discontinuity relation
  \begin{equation}
    \label{eq:22}
    \begin{aligned}
      \tilde\omega_{ij}-\omega_{ij}=&({\cal
        Q}^+_{a|i}-\tilde\bP_a\tilde\bQ_i)\tilde\mu^{ab} ({\cal
        Q}^+_{b|j}-\tilde\bP_b\tilde\bQ_j) - ({\cal
        Q}^+_{a|i}-\bP_a\bQ_i)\mu^{ab} ({\cal
        Q}^+_{b|j}-\bP_b\bQ_j)\\=&{\cal Q}^+_{a|i}(\tilde
      \mu^{ab}-\mu^{ab}) {\cal Q}^+_{b|j}-{\cal Q}^+_{a|i}(\bP^a\tilde
      \bQ_j-\tilde\bP^a \bQ_j) +(\bP^b\tilde\bQ_i-\tilde\bP^b\bQ_i)
      {\cal Q}^+_{b|j}\\=&{\cal Q}^+_{a|i}(\tilde\bP^a \bP^b-\bP^a
      \tilde\bP^b) {\cal
        Q}^+_{b|j}+\bQ_i\tilde\bQ_j-\tilde\bQ_i\bQ_j-\tilde\bQ_i\bQ_j
      +\bQ_i\tilde\bQ_j\\=&\bQ_i\tilde\bQ_j-\tilde\bQ_i\bQ_j\;.
    \end{aligned}
  \end{equation}
Finally we have to show that the Pfaffian of \(\omega\) can be set to \(1\).
For that we notice that \({\rm det}U=1-\bP_a\bP^a=1\), which implies that \({\rm det}{\cal Q}^+={\rm det}{\cal Q}^-\)
{\ie} \({\rm det}{\cal Q}\) is a periodic function. We also know that \(\cal Q\) is analytic in the upper half-plane
which implies that due to the periodicity \({\rm det}{\cal Q}\) could not have cuts. Thus we can always normalize it
to be \(1\) (by rescaling \({\cal Q}_{a|i}\) and hence \(\bQ_i\)) without any effect for our construction.
In this normalization we thus must have \(\det\omega=\det\mu=1\) which also ensures, due to the manifest anti-symmetry of \(\omega\), that up to an irrelevant sign \(\Pf(\omega)=1\).
We finally can summarize the complete set of \(\bQ\omega\)-equations
\begin{eqnarray}\label{Qomegafinal}
\boxed{\tilde\omega_{ij}-\omega_{ij}=\bQ_i\bQt_j-\bQ_j\bQt_i
\;\;,\;\;
\tilde\bQ_i=\omega_{ij}\bQ^j\;\;,\;\;\bQ_i\bQ^i=0\;\;,\;\;\Pf(\omega)=1\;.}
\end{eqnarray}
and a similar system obtained from here by exchange of lower and upper indices:
\begin{eqnarray}\label{QomegafinalLR}
{\tilde\o^{ij}-\o^{ij}=-\bQ^i\bQt^j+\bQ^j\bQt^i
\;\;,\;\;
\tilde\bQ^i=\o^{ij}\bQ_j\;\;,\;\;\bQ_i\bQ^i=0\;\;.
}
\end{eqnarray}
and   \begin{equation}\label{mutmu}
\o_{ij}\o^{jk}=\delta_i^k\end{equation} following from the these two sets of equations.

We also note that, by an obvious help of the above equations, one has the following orthogonality properties
\be\label{ortoPP}
\bP_i\bP^i=0\,,\quad \bP_i\tilde\bP^i=0 \ \ \text{ and }\ \ \ \bQ_{i}\bQ^{i}=0\,,\quad \bQ_{i}\tilde \bQ^{i}=0\;.
\ee

In the next subsection we will show how the global charges of the theory enter into the construction.

 \begin{figure}
\ifpicture{
 \centering
\subfigure[{\Pmsys}]
{\includegraphics{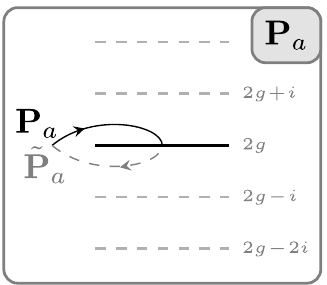}~\includegraphics{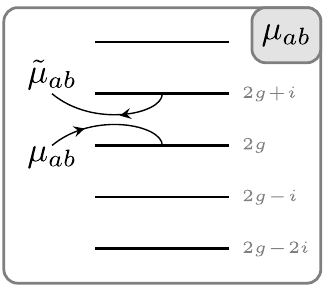}}\(\qquad\)
\subfigure[{\Qosys}]
{\includegraphics{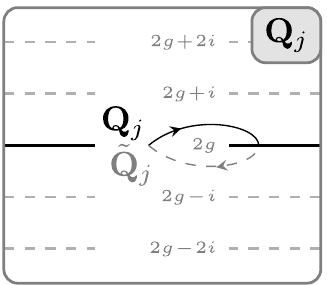}~\includegraphics{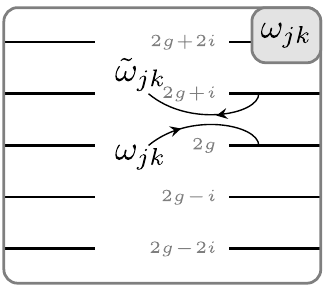}}
 \caption{Cuts structure of the \(\bP\mu\) and the {\Qosys}: the
   functions $\bP_a$ (resp $\bQ_j$) are analytic except on a short
   (resp long) Zhukovsky cut on the real axis. By contrast, 
$\mu_{ab}$ has infinite ladder of cuts and is $i$-periodic in the
mirror kinematics (hence it obeys $\tilde \mu_{ab}=\mu_{a,b}^{[+2]}$
in the physical kinematics). Similarly, $\omega_{jk}$ is periodic in
the physical kinematics.}
} \label{fig:CutsPmuQomega}
\end{figure}

\subsubsection{Regularity}
So far we only discussed the  structure of Zhukovsky cuts of the \(\bP\mu\)- and \(\bQ\o\)-functions, and we
did not  yet discuss a possible presence of other singularities, such as poles\footnote{An extra branch cut coming from infinity could arise in \(\bP\). This is however an artefact of the gauge normalization, and it can be always removed by what we call the $x$-rescaling \eqref{oneparameterxrescaling}. We do not discuss this cut in the following.}.
What we observed from the TBA equations on particular examples is that it is possible to construct
these functions so that they have no poles anywhere on their Riemann surfaces.
Of course, we hope that our construction has a broader applicability
domain than the TBA equations known only for a very limited set of states.
Thus we conjecture that in general \(\bP\mu\)-functions  have no poles for any  state/operator
, this will be referred to as the regularity requirement. The only additional singularity could be at infinity, and this we consider in detail in the next sections.
We found that the behavior at infinity is governed by the global symmetry changes of the state.

 Note that it is enough to ensure regularity of \(\bP\)'s and \(\mu\)'s
on their main sheet, because analytic continuation to the next sheets can be re-expressed as  algebraic
non-singular combinations of these functions on their main sheets via \eq{Pmufinal}. Also, inverse of $\mu$ encountered in \eqref{PmufinalLR} does not lead to possible singularities because the $\Pf(\mu)=1$ condition implies $\mu^{ab}=-\frac 12\e^{abcd}\mu_{cd}$.

The regularity requirement in a sense is very similar to those of Q-functions
of the spin chains. For the \(\su(n)\) Heisenberg spin chain
the spectral problem can be reformulated as the polynomiality,  in particular absence of poles, requirement of \(n\) Q-functions
(\(n\) independent solutions of the Baxter equations).

 The regularity of \(\bP\)'s and \(\mu\)'s also ensures the absence of poles in \(\bQ\)'s and \(\omega\)'s. Then it follows that our requirement will be also true in a broader context of all Q-functions in the Q-system defined in the next section.

\subsubsection{Asymptotics at large \texorpdfstring{$u$}{u}}
\label{sec:asymptotics-at-large}
Recall that the TBA approach, in its current state, is mostly restricted to the states of the rank-1
sector of the theory. The generalization to other states is a hard and generally unsolved problem. Our current approach gives a natural and concise   generalization of spectral equations   to all local operators of the theory. In our formalism such a generalization is very simple --
we only have to generalize a possible large \(u\) behavior of \(\bP\mu\) and \(\bQ\omega\)-functions!
At the same time in the TBA formulation consideration of general states require a complicated modification by
the extra ``driving'' terms, whose structure  is not known in general even for \(\sl(2)\) states. On the contrary, in our formalism this generalization is effortless.\footnote{We also expect the same system of equations to be applicable
for \(\beta\)-deformed case and for the integrable observables with boundary (like cusp anomalous dimension \cite{Correa:2012hh,Drukker:2012de} or DD-brane system \cite{Bajnok:2013wsa}).
Generalization to these cases can be done through modification of the asymptotic by relaxing power-like
behavior at infinity. These cases should be rather straightforward to understand.
}

In this section we conjecture that the  behavior at \(u\to\infty\) of all considered \(\bP\)- and {\bQfcts} is  governed by the global symmetry charges.
The situation is very similar to the classical spectral curve, formed by a set of
\(4+4\) functions, called quasi-momenta, \({(\tilde p_{ 1}(u)},\linebreak[1]\tilde p_{ 2}(u),\linebreak[1]\tilde p_{ 3}(u),\linebreak[1]\tilde p_{ 4}(u)|\hat p_{ 1}(u),\linebreak[1]\hat p_{ 2}(u),\linebreak[1]\hat p_{ 3}(u),\linebreak[1]{\hat p_{4}(u))}\).
At large \(u\) the quasi-momenta contain the information about  global charges of the classical solution so that\footnote{
These charges are related to the Dynkin labels
\([r_1,r_2,r_3]\) and  \([q_1,q_2,q_3]\)
of representations of \(\su(4)\) and \(\su(2,2)\) subalgebras by
 \(r_1=J_2-J_3\), \(r_2=J_1-J_2\), \(r_3=J_2+J_3\) , \(q_{1}=S_1+S_2\),  \(q_2=-\Delta-S_1\), \(q_3=S_1-S_2\); {\it cf.} \appref{sec:kac-dynkin-vogan}.}
\begin{eqnarray}
\left(\bea{c}
\tilde p_1\\
\tilde p_2\\
\tilde p_3\\
\tilde p_4
\eea
\right)
\simeq
\frac{1}{2u}
\left(\bea{c}
+J_1+J_2-J_3\\
+J_1-J_2+J_3\\
-J_1+J_2+J_3\\
-J_1-J_2-J_3
\eea
\right)
\;\;,\;\;
\left(\bea{c}
\hat p_1\\
\hat p_2\\
\hat p_3\\
\hat p_4
\eea
\right)
\simeq
\frac{1}{2u}
\left(\bea{c}
+\Delta-S_1+S_2\\
+\Delta+S_1-S_2\\
-\Delta-S_1-S_2\\
-\Delta+S_1+S_2
\eea
\right).
\end{eqnarray}

In the \secref{sec:QCA} we discuss in detail the classical limit of our system, but the output is very simple
\begin{eqnarray}
&&{\bf P}_a\sim \exp\left(-\int^u \tilde p_{a}(v)dv\right)\;\;,\;\;
{\bf P}^a\sim \exp\left(+\int^u \tilde p_{a}(v)dv\right)\,,\\
&&{\bf Q}^i\sim \exp\left(-\int^u \hat p_{i}(v)dv\right)\;\;,\;\;
{\bf Q}_i\sim \exp\left(+\int^u \hat p_{i}(v)dv\right)\,.
\end{eqnarray}
This simple insight tells us that the large \(u\) asymptotic of \(\bP\) and {\bQfcts} should contain the global charges in a very specific way\footnote{Compared to \cite{Gromov:2013pga}, we exchanged the notation $P_1\leftrightarrow P_2$ and $P_3\leftrightarrow P_4$ so as to have a natural relation to the Dynkin labels.}
\begin{eqnarray}\label{largeu2}
\bP_a\simeq A_a\, u^{-\tilde M_a}\,,\ \ \bQ_{i}\simeq B_i\,u^{\hat M_i-1}\,,\qquad   \bP^a\simeq A^a\, u^{\tilde M_a-1}\,,\ \ \bQ^{i}\simeq B^i\,u^{-\hat M_i}\,,
\end{eqnarray}
where
\begin{align}
\label{relMta}
\tilde M_a=&\left\{\frac{1}{2}
   (J_1+J_2-J_3+2),\frac{1}{2}
   (J_1-J_2+J_3),\frac{1}{2}
   (-J_1+J_2+J_3+2),\frac{1}{2}
   (-J_1-J_2-J_3)\right\}\\
\hat M_i=&\left\{\frac{1}{2} (\Delta -S_1-S_2+2),\frac{1}{2}
   (\Delta +S_1+S_2),\frac{1}{2} (-\Delta
   -S_1+S_2+2),\frac{1}{2} (-\Delta
   +S_1-S_2)\right\}
\label{M-ass}\end{align}
The shifts of powers by \(-1\) are of course not detectable in the classical limit, but they can be seen at weak coupling when the Asymptotic Bethe Ansatz (ABA) is applicable. The origin of shifts is similar to the known phenomena of length changing under duality transformations in ABA. We explain how to fully derive \eqref{largeu2} first at weak and then at finite coupling in \secref{sec:NRCGC}.

From the   \equref{QPQ}, replacing the \(\pm \frac{i}{2}\) shifts by \(1\pm \frac{i}{2}\partial_u\) at large \(u\),
we obtain \begin{eqnarray}\label{Qkjlargeu}
\fQ_{a|j}\simeq -i\, A_a\,B_j\,\frac{u^{-\tilde M_a+\hat M_j}}{-\tilde M_a+\hat M_j}\,.
\end{eqnarray}
The constants \(A^a\) and \(B^i\) are explicitly expressed through the \(\tilde M_a\) and \(\hat M_j\). The fastest way to find them is to plug the corresponding asymptotic expressions into
\eqref{QP} and get in this way a set of algebraic equations \(-1= i\,\sum_{a=1}^4\frac{A^aA_a}{\tilde M_a-\hat M_j}\)
which defines \(A^{a_0}A_{a_0}\) for each \(a_0\in\{1,2,3,4\}\). In the same way we compute \(B^{j_0}B_{j_0}\).
 The
final result is
\begin{eqnarray}\label{AABB0}
A^{a_0}A_{a_0}=i\frac{\prod\limits_{j}(\tilde M_{a_0}-\hat M_j)}{\prod\limits_{b\neq a_0}(\tilde M_{a_0}-\tilde M_b)}\,,\qquad
                B^{j_0}B_{j_0}=i\frac{\prod\limits_{a}(\hat M_{j_0}-\tilde M_a)}{\prod\limits_{k\neq
{j_0}}(\hat M_{j_0}-\hat M_k)}\;,\qquad a_0,j_0=1,2,3,4
\end{eqnarray}
(with no summation over \(a_0\) or \(j_0\) in l.h.s.!).

For completeness, let us also discuss the asymptotics of \(\mu\) and \(\omega\).
Our main assumption is that they have power-like asymptotic at infinity.\footnote{This
restriction becomes too strong in some cases. For example for analytic continuation in the Lorentz spin \(S\)
for twist two operators one could also have exponential factors. A similar situation is known to arise  for the twisted case and
for the boundary TBA case. But in all cases where we have  considered local  physical  operators in \neqfour{} SYM, this rule holds
and is compatible with TBA.}
First, \(\omega\) on the sheet with short cuts must be also periodic at large \(u\). The only power-like
periodic function is a constant. Thus \(\omega_{ij}\) at infinity becomes an antisymmetric matrix with the unit Pfaffian. By making the appropriate choice of the \(\bQ_i\) basis or, in other words, by the choice of \({\cal Q}_{a|i}\)
we can always choose it to be at \(u\to +\infty\) of the form
\begin{eqnarray}
\label{largeuomega}
&&\omega_{ij}\simeq\eta_{ij} \,,\quad \text{where }\,\,\eta_{ij}=
\begin{pmatrix}
0 & 1 & 0 & 0 \\
-1 & 0 & 0 & 0 \\
0 & 0 & 0 & 1 \\
0 & 0 & -1 & 0 \\
\end{pmatrix}\;,
\end{eqnarray}

Curiously at \(u\to-\infty\) one can find
\begin{eqnarray}
\label{largeuomegab}
&&\omega_{ij}\simeq \pm
\begin{pmatrix}
0 & e^{+i\pi\gamma} & 0 & 0 \\
-e^{+i\pi\gamma} & 0 & 0 & 0 \\
0 & 0 & 0 & e^{-i\pi\gamma}\\
0 & 0 & -e^{-i\pi\gamma} & 0 \\
\end{pmatrix}\;.
\end{eqnarray}
where \(\gamma=\Delta-\Delta_0\) is the anomalous dimension. It will be shown in  \secref{omega-ass}.

Finally, knowing the asymptotics of \({\cal Q}_{a|i}\)
we then define the asymptotics of \(\mu_{ab}\) through the identities \eqref{mudef} and \eqref{mutmu}      \begin{align}\hspace{6em}\label{muass}
&\mu_{12}=-\mu^{34}\simeq u^{\Delta-J_1}\,,&&\mu^{12}=-\mu_{34}\simeq u^{\Delta+J_1}\;,\no\\
&\mu_{13}=+\mu^{24}\simeq u^{\Delta-J_2-1}\,,&&\mu^{13}=+\mu_{24}\simeq
u^{\Delta+J_2+1}\,,\ \ \no\\
&\mu_{14}=-\mu^{23}\simeq u^{\Delta+J_3}\,,&&\mu^{14}=-\mu_{23}\simeq
u^{\Delta-J_3}\,.
\hspace{4em}
\end{align}

\section{Quantum Spectral Curve as an Analytic
  \texorpdfstring{{\Qsys}}{Q-omega system}}
\label{sec:YTQ}
In the previous section, we sketched out the derivation of the \(\bP\mu\)  and {\Qosys}s. Either of these systems defines the quantum spectral curve of AdS\(_5\times\)S\(^5\) duality  generalising the classical algebraic curve of the Metsayev-Tseytlin string sigma model \cite{Beisert:2005bm}. We departed from a well established and tested formalism based on the AdS/CFT {\Ysys} supplied with the analyticity constraints following from the TBA  and reduced it to  a significantly simpler set of Riemann-Hilbert equations on a finite number  of  functions of spectral parameter, \(\bP_a\),  and \(\mu_{ab}\)   (and a similar set  on  \(\bQ_i\) and \(\omega_{ij}\)) which turn out to have very transparent analytic properties.
Some of these functions, like \(\bP_1,\bP_2\) and \(\o_{ab}\),  were already familiar from our previous, FiNLIE construction of \cite{Gromov:2011cx}, whereas the others, like \(\bP_3,\bP_4,\bQ_i\) and \(\mu_{ab}\), seem to be new and somewhat mysterious.
 In this section, we will reveal  a nice mathematical structure emerging behind the Riemann-Hilbert equations allowing to interpret the \(\bP\mu\)  and {\Qosys}s as parts of a broader, Grassmannian object which we call the  {\Qsys}\footnote{In certain more mathematically-oriented literature, the name {\Qsys} is used for a different object: the  {\Tsys} in the character limit.
 We rather mean by that the system formed by Baxter-type Q-functions, which justifies the name.}.

This {\Qsys} obeys some generic, model independent algebraic properties  which are well known  from the analytic Bethe ansatz for the Heisenberg \(\gl( n|m)\)  spin chains. In the context of our current spectral problem, it is applicable in a domain of the
complex plane which is free from branch points, which is true in our construction for a sufficiently large (but finite) imaginary part of the
spectral parameter.
 We describe the setup for this {\Qsys} for the case of interest, the \(\gl(4|4)\) algebra, in sub\secref{sec:algebraic Q-system} of this section. In sub\secref{sec:analyticQ},  we  complement this algebraic construction by analytic properties of the underlying {\Qfcts} which are specific to the AdS/CFT integrability.  They will allow us to embed the  \(\bP\mu\)  and {\Qosys}s into the  full {\Qsys}.

\subsection{{\Qsys} -- general algebraic description}\label{sec:algebraic Q-system}

In the Heisenberg spin chains  the {\Qsys} appears as a set of Q-operators, or their eigenvalues -- Q-functions,   satisfying the Baxter-type functional equations appearing on various stages of  B\"acklund reduction (or equivalently, the nesting procedure \cite{KRNB,Kigm,PhysRevLett.74.816}) of the corresponding {\Tsys} \cite{Krichever:1996qd,Kazakov:2007fy,Tsuboi:1997iq}. It was pointed out in \cite{Krichever:1996qd} that, for a system with \(\su(n)\) symmetry, the set of all \(2^{n}\) {\Qfcts} can be identified with Pl\"ucker coordinates on finite-dimensional Grassmanians \(G^k_n\), $k=0,1,\ldots,n$, where each \(G^k_n\) is a collection of all  $k$-dimensional linear subspaces  of the vector space \(\mathbb{C}^n\). This Grassmanian construction can be also adopted to the case of superalgebras.  We exploit this fact
to build a comprehensive algebraic description of the {\Qsys}. We will focus on the case of  \(\gl( 4|4)\) relevant for the AdS$_5$/CFT$_4$ integrability\footnote{
The \(\gl( n|m)\) case of the generic Q-system will be considered in \cite{KLV-Qsystem} (see also the papers \cite{Krichever:1996qd,Tsuboi:2011iz}).}.
\subsubsection{Definition of {\Qsys}  and QQ-relations}
\label{sec:qq-relations}
The \(( 4|4)\) {\Qsys} is a set of \(2^{8}\) {\Qfcts} of the spectral
parameter \(u\) denoted \(Q_{A|I}\equiv
Q_{a_1a_2\dots|i_1i_2\dots}\), where each label \(A\) and \(I\) is a
multi-index from the set \(\{1,2,3,4\}\). The multi-index \(A\) will
be called bosonic and \(I\) -  fermionic. The {\Qfcts} are
antisymmetric with respect to permutation of bosonic or fermionic
indices: \(Q_{\ldots ab\ldots|\ldots ij\ldots}=-Q_{\ldots ba\ldots|\ldots ij\ldots}=Q_{\ldots ba\ldots|\ldots ji\ldots}\), so  we are dealing here with the anti-symmetric tensors, the elements of the linear space \(\Lambda(\mathbb C^4)\otimes\Lambda(\mathbb C^4)\). Note that this space does not have any anti-commuting variables,  ``bosons'' and ``fermions'' is just a terminological convention here.

As everywhere in this article, we use the standard notations for the shifts in the imaginary direction: \(Q^{[n]}\equiv Q(u+\frac{i\,n}{2})\), \(Q^{\pm}\equiv Q^{[\pm 1]}\).

\paragraph{Defining Pl\"ucker's QQ-relations.} The {\Qsys} can be defined by a set of
so-called Pl\"ucker's  QQ-relations \cite{Kazakov:2007fy,Gromov:2010km} (see also \cite{Tsuboi:1998ne,Gromov:2007ky}), which is a set of bilinear  constraints on
various  Q-s:
 \begin{subequations}\label{definingQQ}
    \begin{align}
       \label{QQbb}
       \gQ_{A|I}\gQ_{A ab|I} &=\gQ_{A a|I}^{+} \gQ_{A b|I}^{-}-
       \gQ_{A a|I}^{-}
       \gQ_{A b|I}^{+}\,,\\
       \label{QQff}
       \gQ_{A|I}\gQ_{A|I ij} &=\gQ_{A|I i}^{+} \gQ_{A|I j}^{-}-
       \gQ_{A|I i}^{-} \gQ_{A|I j}^{+}\,,\\
       \label{QQbf}
       \gQ_{A a|I}\gQ_{A|I i} &= \gQ_{A a|I i}^{+}\gQ_{A|I}^{-}-
       \gQ_{A|I}^{+} \gQ_{A a|I i}^{-} \,.
     \end{align}
\end{subequations}

The first two exchange two indices of the same type (of grading) and they are usually called bsonic QQ-relations \cite{Pronko:1998xa,Gromov:2007ky}. The last one exchanges two indices of different types and usually is called fermionic QQ-relation \cite{Kazakov:2007fy,Tsuboi:1998ne}. They naturally appeared in the chain of
B\"acklund transformations for the integrable Heisenberg \(\gl( n|m)\)    spin chains \cite{Kazakov:2007fy}\footnote{They arise very naturally in the context of interpretation of {\Qfcts} as components of exterior forms. This point of view, together with the proofs of various relations given in this section, will be developed in the forthcoming publication of 3 of the current authors \cite{KLV-Qsystem}.}.
All  other  relations among {\Qfcts} follow from these QQ-relations. In this sense we will call them the defining relations.

The defining QQ-relations enjoy the gauge symmetry
\footnote{It induces a particular case of the gauge transformations \eqref{gaugeapp} of the related {\Tsys} which will be defined in \appref{sec:solution-t-system}.}
\begin{eqnarray}\label{rescalingsym}
Q_{A|I}\to \frac{g^{[+(|A|-|I|+1)]}}{g^{[-(|A|-|I|+1)]}}\,Q_{A|I}\,,
\end{eqnarray}
which we  fix in this article by imposing  the overall normalization of the {\Qsys} as follows
 \begin{equation}
\label{Qes}
\gQ_{\emptyset}\equiv\gQ_{\emptyset|\emptyset}=1\,.
\end{equation}
Two explicit examples of relations \eqref{definingQQ} are
\begin{subequations}
\begin{eqnarray}
  \label{premainQQbf}      Q_{a|\es}Q_{\es|i}&=&Q_{a|i}^+-Q_{a|i}^-\,,\\
\label{qqexam}  \gQ_{\emptyset|i}\gQ_{ab|i}
&=&\gQ_{a|i}^{+}
\gQ_{b|i}^{-}-
\gQ_{a|i}^{-}
\gQ_{b|i}^{+}\,.
\end{eqnarray}
\end{subequations}
If we replace \(i\to \es\) in  the r.h.s. of the second one the reader may recognize in it the Wronskian of the Baxter equation for homogeneous \(\su(2)\)
spin chain where \(Q_{12|\es}=u^L\).
The Bethe equations follow from it simply by imposing the polynomiality of all Q-functions entering there.

\subsubsection{A complete basis for parameterization of all {\Qfcts\ }}\label{subsec:Qbasis}

Due to the QQ-relations \eqref{QQbb}-\eqref{QQbf}, all multi-index {\Qfcts} can be expressed through a basis of 8 (rank of the superalgebra) {\Qfcts}.
A natural basis is given by the
single-indexed {\Qfcts}
\(\gQ_{a|\emptyset}\,,\ \  \gQ_{\emptyset|i}\).
\paragraph{Determinant relations in components}

With the normalization (\ref{Qes}), all the Q-functions with only one type of indices
can be reduced in a simple way to the Q-functions with one index using defining QQ-relations.
These identities have the following determinant form:%
\begin{subequations}
\label{determinantQQ}
\begin{gather}
\label{eq:3}\gQ_{a_1\ldots a_k|\es}=k!~\gQ_{[a_1|\es}^{[k-1]}\gQ_{
  \vphantom{[}a_2|\es}^{[k-3]}\ldots \gQ_{
  a_k|\es]}^{[1-k]}=\det\left(\gQ_{a_m|\es}^{[k+1-2n]}\right)_{1\le m,n\le k}\,,\\
 \gQ_{\es | i_1\ldots  i_k}=k!~\gQ_{
   [\es | i_1}^{[k-1]} \gQ_{
   \vphantom{[}\es | i_2}^{[k-3]}\ldots \gQ_{
   \es | i_k]}
^{[1-k]}
=\det\left(\gQ_{\es | i_m}^{[k+1-2n]}\right)_{1\le m,n\le k}
\,,
\label{eq:5}
\end{gather}
\end{subequations}
where  \([\dots]\) stands for the standard anti-symmetrization of the indices.
Furthermore when the number of fermionic and bosonic indices is the same we can again write
the corresponding Q-function as a determinant of \(Q_{i|j}\) which is a simple consequence of the QQ-relations \eq{definingQQ} \cite{Tsuboi:2011iz,KLV-Qsystem}:
\begin{equation}\label{symdet0}
 Q_{a_1\ldots a_k|i_1\ldots i_k}=\det_{1\leq m,n\leq k} Q_{a_m|i_n}\,.
\end{equation}
This suggests that any Q-function can be expressed explicitly ({\ie} without infinite sums) in terms of \(Q_{a|\es}\), \(Q_{\es | i}\), and \(Q_{a|i}\).
Indeed such relations are known\footnote{Again, this is very easy to check starting from \eq{definingQQ}.  For general $\mathfrak{gl}(n|m)$ case, the formal proof will be given in \cite{KLV-Qsystem} (these relations can be recognised also inside sparse determinants used in \cite{Tsuboi:2011iz}).}:
\begin{subequations}\label{symdet}
\begin{gather}\label{eq:125}
 Q_{a_1\ldots a_{k+n}|i_1\ldots i_k}=
\tfrac{(n+k)!}{n!\,k!} 
 Q_{{
     [a_1\ldots a_n}|\emptyset}^{\vphantom{[]}}Q_{{
     a_{n+1}\ldots a_{n+k}]}|i_1\ldots i_k}^{[\pm n]}\,,\\\label{eq:126}
Q_{a_1\ldots a_{k}|i_1\ldots i_{k+n}}= (-1)^{n\,k} \tfrac{(n+k)!}{n!\,k!}
Q_{a_{1}\ldots
a_{k}|{
  [i_{1}\ldots i_{k}}}^{[\pm n]} Q_{{\emptyset|
i_{k+1}\ldots i_{k+n}]}}^{\vphantom{[]}}\,.
\end{gather}
\end{subequations}
For example,  of a particular importance for us  are the relations
\begin{eqnarray}\label{explicit1}
Q_{ab|\es }=\begin{vmatrix}Q_{a|\es}^+ & Q_{b|\es}^+ \\
Q_{a|\es}^- & Q_{b|\es}^- \\
\end{vmatrix},\ \
Q_{ab|ij}=\begin{vmatrix}Q_{a|i} & Q_{a|j} \\
Q_{b|i} & Q_{b|j} \\
\end{vmatrix}\,.
\end{eqnarray}
We will see that the functions  \(Q_{a|\es}\), \(Q_{\es | i}\),  and \(Q_{i|j}\) play an important role and they  will be extensively used in the following. They will be identified with the Q-functions with similar notations from the previous \secref{sec:QSCY}.
Note, however, that    \(Q_{a|i}\) are not independent and satisfy the QQ-relation (\ref{premainQQbf}):
 \begin{equation}
  \label{sport}
   \gQ_{a|i\vphantom{a|i}}^+-\gQ_{a|i\vphantom{a|i}}^-=\gQ_{a|\es}\gQ_{\es| i\vphantom{a|i\hat
i}}\,,
\end{equation}
which can be formally solved by
\begin{equation}\label{sportive}  \gQ_{a|i}=-\sum\limits_{n=1}^{\infty}\gQ_{a|\es\vphantom{a|i\hat
i}}^{[2n-1]}\gQ_{\es| i\vphantom{a|i\hat
i}}^{[2n-1]}+\cal{P}\;,
\end{equation}
where \(\cal{P}\) is an \(i\)-periodic function\footnote{Alternatively, the relation
   (\ref{sportive}) can be  solved as \begin{equation} \gQ_{a|i}=\sum\limits_{n=1}^{\infty}\gQ_{a|\es \vphantom{a|i\hat
i}}^{[1-2n]}Q_{\es|i}^{[1-2n]}+\cal{P'}\,,\end{equation}  which will be important for our discussion of the lower-half-plane versus  upper-half-plane analyticity in sub\secref{subsec:LHS/UHS}.}.
One should also properly regularise the above sum which is  in general divergent. For instance, one can differentiate it a sufficient number of times to make it convergent, and then integrate the result of summation back. Since we have to satisfy \eqref{sport} at the end, the only arising ambiguity in this procedure is fully accounted by \({\cal P}\) which is so far arbitrary.
One can restrict \(\cal{P}\) by requiring a certain asymptotics at infinity. In particular, the polynomial behaviour restricts  \({\cal P}\) to be a constant, and if $\fQ_{a|i}$ should decrease, one can put  \({\cal P}=0\) in the appropriate regularisation.  Later we will encounter other self-consistency requirements which will allow one to fix \(\cal{P}\) for all $\fQ_{a|i}$.

We see that all the {\Qfcts} can be found if \(8\) one-indexed {\Qfcts} are known.\footnote{It is known \cite{Gromov:2010km} that there are other choices of  the basis of \(8\) {\Qfcts} from which all other {\Qfcts} can be restored in terms of determinants, with no need to solve any difference equation like \eqref{sport}.}

\paragraph{Hasse diagram.}
\begin{figure}
\centering
\ifpicture{
\includegraphics{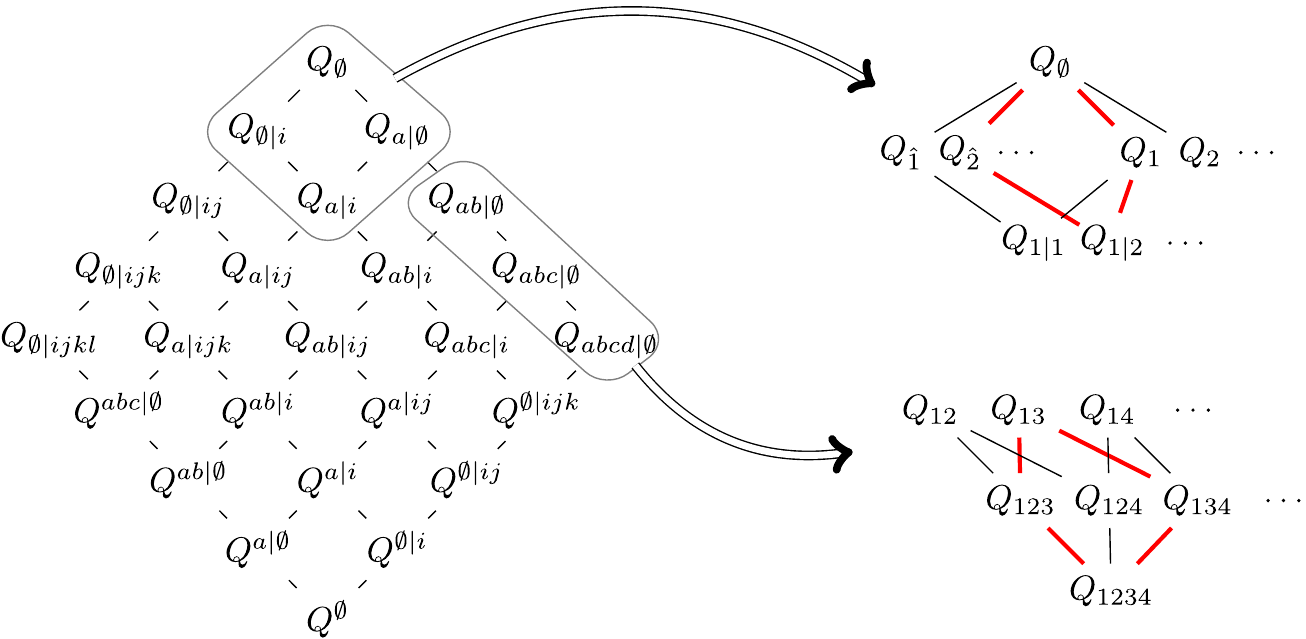}
}\caption{\label{fig:HasseDiagram}A projection of the Hasse diagram
  (left), where all Q-functions having the same grading (number of bosonic and
  fermionic indices) are identified.
A more precise picture (right) of some small portions of this diagram
illustrates the ``facets'' (red) corresponding to the QQ-relations
 \(Q_{13|\es}Q_{1234|\es}=Q_{123|\es}^+Q_{134|\es}^--Q_{123|\es}^-Q_{134|\es}^+
 \) and \(Q_{1|\es}Q_{\es|2}=Q_{1|2}^+Q_{\emptyset}^--Q_{1|2}^-Q_{\emptyset}^+\).}
\end{figure}

It is instructive to demonstrate the relations between {\Qfcts} in a graphical way, using the Hasse diagram of the \figref{fig:HasseDiagram},  originally proposed in \cite{Tsuboi:2009ud}
for \(2^{m+n}\) Q-functions of \(\gl(m|n)\) spin chains.

\subsubsection{Symmetries}\label{subsec:symmetry}

\paragraph{a) Gauge symmetry}
In the previous section, a gauge symmetry \eqref{rescalingsym} was
used to enforce the condition \(\gQ_{\emptyset}=1\). Actually, there
remains  only one more independent gauge  symmetry\footnote{Whereas Hirota equation has 4 gauge symmetries, only T-functions in certain gauges can be expressed in terms of Q-functions, see \appref{sec:wronparam}. On the level of Q-functions, only two gauge symmetries are possible.}:
\begin{align}
  \label{rescalingsym2}
  Q_{A|I}\to \frac{g^{[+|A|-|I|]}}{g^{[-|A|+|I|]}}\,Q_{A|I}\,.
\end{align}
We will see that this freedom is important in ensuring analytic
properties in the construction of T- and {\Qfcts}.

\paragraph{b) Hodge transformation.}
In the case of AdS/CFT {\Ysys} the left and right wings of the T-hook look symmetric.
In the language of the {\Qsys} the interchange of the wings is a nontrivial symmetry transformation which, in particular,  replaces Q functions by their Hodge-dual Q-functions. We define the Hodge-dual of the {\Qfcts} as\footnote{One should note that, as stated at the end of \secref{sec:multi-indices-sum}, there is no summation over the multi-indices $A'$ and $I'$.}
\begin{equation}\label{Hodgedef}
  \begin{aligned}
    Q^{A|I}\equiv&
    {(-1)^{|A'||I|}\e^{A'A}\e^{I'I}Q_{A'|I'}}{}\,,&\textrm{where
    }\{A'\}=&\{1,2,3,4\}\setminus \{A\},\\&& \{I'\}=&\{1,2,3,4\}\setminus
    \{I\},
  \end{aligned}
\end{equation}
where \(\e\) is a 4D Levi-Civita symbol. %
 In particular, the Hodge dual of \(Q_{\emptyset|\es}\)
is \(Q^{\emptyset|\emptyset}=Q_{\bar\emptyset|\bar\emptyset}\equiv Q_{1234|1234}\). In the case of  ${\rm AdS}_5/{\rm CFT}_4$ spectral problem,  one can impose
\(Q_{1234|1234}=1\), and we believe that this possibility reflects the unimodularity of the symmetry algebra $\psu(2,2|4)$%
\footnote{
This interpretation follows closely the argument of \cite{Gromov:2011cx}: the combination \({Q_{\bar\es|\bar\es}^-}/{Q_{\bar\es|\bar\es}^+}\)
is a ``quantum determinant'' of the monodromy matrix. Classically it is equal to \(1\) due to the unimodularity. At the quantum level it can be also shown to be \(1\) from TBA, it is essentially the consequence of \eqref{per1}.
This implies that \(Q_{\bar\es|\bar\es}\) is a periodic function. Since we are building the {\Qsys} so that Q-functions are analytic sufficiently high above the real axis and have power-like asymptotics,  we  conclude that \(Q_{\bar\es|\bar\es}\) is a constant which can be always scaled to \(1\).
}.
 From now on, we assume this restriction, which simplifies many relations, and otherwise continue our general consideration of the algebraic properties of {\Qsys}.

With this definition the Hodge-dual Q-functions satisfy exactly the same QQ-relations \eqref{definingQQ} and \eqref{determinantQQ}
with all indices raised.
Moreover,  the following formulae hold:
\begin{subequations}
\label{QQorto1}
\begin{align}\label{QQrel1}
Q_{\emptyset|i}=&
\, -Q^{a|\emptyset}Q_{a|i}^\pm\,,& Q^{\emptyset|i}=&
Q_{a|\emptyset}(Q^{a|i})^\pm\,,\\
\label{QQrel2}
Q_{a|\emptyset}=&-
Q^{\emptyset|i}Q_{a|i}^\pm\,,&  Q^{a|\emptyset}=&
Q_{\emptyset|i}(Q^{a|i})^\pm\,.
\end{align}
\end{subequations}
Furthermore, as a consequence  of QQ-relations and of the condition \(Q_{1234|1234}=1\), one can demonstrate the following orthogonality relations
\begin{align}\label{QQorto2}
Q^{a|i}Q_{a|j}=&-\delta^{i}{}_j&Q^{a|i}\,Q_{b|i}=&-\delta^a{}_b\,.
\end{align}
as well as
\begin{align}
\label{orto1}
Q^{a|\es}Q_{a|\es}=&0&Q^{\es|i}Q_{\es|i}=&0\,.
\end{align}

In conclusion, we see that the Hodge dual Q-functions satisfy the same set of QQ-relations as the original Q-functions.
Thus we can think of the Hodge transformation as of a symmetry of our system. In fact it has a clear physical meaning
in the \neqfour{} SYM -- %
it interchanges Left and Right wings of the {\Ysys} (up to a relabeling of the indices with a constant matrix \(\chi\)), see \secref{sec:particular-case-left}. That why we  call it the LR-symmetry.

\paragraph{c) H-symmetry.}
\label{sec:h-symmetry-gauge}
Let us finally discuss  some useful residual symmetries of the
 algebraic relations defined above. These relations
are invariant
under
a \(\mathsf{Gl}(4)\times\mathsf{Gl}(4)\) symmetry
of the {\Qsys},
\begin{eqnarray}
\label{Htransform}
 Q_{A|I}\to \sum_{\substack{|B|=|A|\\|J|=|I|}}(H_b^{[\, |A|-|I|\, ]})_A{}^{B}(H_f^{[\, |A|-|I|\, ]})_I{}^{J}Q_{B|J}\,,
\end{eqnarray}
 where \(H\) is an arbitrary \(i\)-periodic matrix. The notation \(H_{I}{}^{J}\) with multi-indices \(I,J\) means the tensor product of \(4\times 4\) matrices \(H\), explicitly: \(H_{I}{}^{J}\equiv H_{i_1}{}^{j_1}H_{i_2}{}^{j_2}\ldots H_{i_{|I|}}{}^{j_{|I|}}\). Some explicit examples are
\begin{eqnarray}
Q_{a|\es}\to (H_b^{\pm})_{a}{}^{c}\,Q_{c|\es}\,,\ \ Q_{a|i}\to (H_b)_{a}{}^{c}(H_f)_{i}{}^{j}\,Q_{c|j}\,,\ \ Q_{\bar\es|\bar\es}\to\det{H_b}\ \det{H_f}\,Q_{\bar\es|\bar\es}\,.
\label{explex}
\end{eqnarray}
\paragraph{Rescaling}
As a simple example of H-transformation consider
 diagonal $H$-matrices which simply generate the following rescaling
\be\label{cresc}
&&{  Q}_{a|\emptyset}\to\a_a\,{Q}_{a|\emptyset}\,\ \ \ {  Q}_{\emptyset|i}\to\beta_i\,{  Q}_{\emptyset|i}\,,
\no\\
&& {  Q}^{a|\emptyset}\to\frac 1{\a_a}\,{  Q}^{a|\emptyset}\,,\ \ {  Q}^{\emptyset|i}\to\frac{1}{\beta_i}{  Q}^{\emptyset|i}\,.
\ee
This $8$-parametric freedom is constrained by $Q_{\bar\es|\bar\es}=1$ which implies $\a_1\a_2\a_3\a_4\b_1\b_2\b_3\b_4=1$.
\subsection{{\Qsys} and AdS/CFT spectral problem}
\label{sec:analyticQ}

Above we discussed the general properties of {\Qsys}s. In this section we will demonstrate how
the construction of \secref{sec:QSCY} finds its natural description in terms of the {\Qsys}.
We will first identify various objects in the {\Qsys} language and then discuss their properties from this
new point of view.

\subsubsection{\texorpdfstring{$\bP$ and $\bQ$}{P and Q} as {\Qfcts}}
Despite their simplicity, \(\bP\mu\)- and {\Qosys}s may look rather mysterious. In particular, their derivation in \secref{sec:QSCTBA} contained a number of surprises on the way and it looked
to some extend like a magic trick. The {\Qsys} described in the previous subsection allows to unveil this mystery and inscribe the \(\bP\mu\)- and {\Qosys}s into a  mathematical framework related to the classical integrability.

Our claim is that \(\bP\)'s and \(\bQ\)'s naturally can be embedded into a {\Qsys} described above in the following way
\begin{eqnarray}
\label{eq:8}
\fQ_{a|\emptyset}\equiv\bP_a\,,\ \ \fQ_{\emptyset|i}\equiv \bQ_{i}\,;
\end{eqnarray}
furthermore we set
\begin{equation}
{\cal Q}_{\emptyset|\emptyset}={\cal Q}_{\bar \emptyset|\bar \emptyset}=1\,.
\end{equation}
The second of the above constraints, as we already mentioned, has in fact  a natural interpretation as a quantum unimodularity condition.

 This identification provides us with a set of \(2^8\) Q-functions, with very particular analyticity properties,  constitute what we call the {\it fundamental} {${\rm AdS}_5/{\rm CFT}_4$ \Qsys}\footnote{As concerns the main text, we alternatively use the names "fundamental \Qsys" and "analytic \Qsys". However, in \appref{app:relation-to-TBA} we introduce the "mirror \Qsys" which is not equivalent to the fundamental one but nevertheless  it has nice analytic properties and hence can be also called analytic.}.
 To distinguish it from the generic Q-functions we denote the fundamental {\Qsys} by the calligraphic font $\fQ$.

In this identification one should remember about one important subtlety. The QQ-relations are the finite difference relations
and thus they may have a tricky meaning when the functions are not single valued. To avoid this complication we
first note that our \(\bP\)'s and \(\bQ\)'s are free from branch points above the real axis.
We use this property to {\it define} all Q-functions in terms of \(\bP_a,\bQ_a\)  in the upper half of the complex plane.
More precisely, we first find \({\cal Q}_{a|b}\) recursively from \eq{sport}, which becomes
\begin{equation}
  \label{sport2}
   {\cal Q}_{a|i}^+-   {\cal Q}_{a|i}^-=\bP_{a}\bQ_{i}\,,
\end{equation}
so that it is analytic above the real axis. This can be formally done by taking \eq{sportive} with \({\cal P}\) being  some (generically infinite) constant%
. Next, we define all Q-functions explicitly in terms of
\(\bQ\)'s, \(\bP\)'s
 and \({\cal Q}_{a|b}\) via  identities
\eq{determinantQQ}. This ensures that all Q-functions are analytic and that the QQ-relations are satisfied sufficiently far above the
real axis.

Giving this construction we can start recognizing various object defined in \secref{sec:QSCY}.
We can already see that \eq{sport2} is exactly the \equref{QPQ} from \secref{sec:Qomega}. Furthermore, the general  identities
\eq{QQrel1} and \eq{QQorto2} related to Hodge duality are the same as \eq{QP} and \eq{QP2}!

What remains perhaps unclear  is the role of \(\mu\) and \(\omega\) in the {\Qsys} picture.
At the same time it looks unnatural that  in the above construction we gave a preference to the analyticity in the upper half of the complex plane w.r.t. the lower half.
As we shell see later these two problems are tightly related.

\subsubsection{\texorpdfstring{$\mu$ and $\omega$}{mu and omega} as linear combinations of {\Qfcts}}
The identification \eqref{eq:8} also allows to give an interesting interpretation to \(\mu\) and \(\omega\) as a certain combination
of Q-functions. To see this we start from \eq{muomega} and rewrite it, using \eq{explicit1} and the antisymmetry of \(\omega^{ij}\), as
\begin{align}
\label{muomega2} \mu_{ab}=
\frac{1}{2}{\cal Q}^-_{ab|ij}\;\omega^{ij}\;.
\end{align}
 \({\cal Q}_{ab|ij}\)  satisfies a curious relation  which is a consequence of QQ-relations:
\begin{eqnarray}\label{eqQijkl}
\fQ_{ab|ij}^+-\fQ_{ab|ij}^-=-(\delta_a^c\bP_b\bP^d-\delta_b^c\bP_a\bP^d)\fQ_{cd|ij}^\pm\,.
\end{eqnarray}
It follows from a similar relation for \({\cal Q}_{a|i}\) (following from \eq{QP} and \eq{QPQ})
\begin{equation}\label{QabPP}
{\cal Q}_{a|i}^+ - {\cal Q}_{a|i}^-=-\bP_a\bP^c {\cal Q}^+_{c|i}
\end{equation}
and from the fact
that \({\cal Q}_{ab|ij}\) can be built out of these two-index Q-functions via the determinant \eq{explicit1}.

Due to the $i$-periodicity of \(\hat\omega\) this identity allows to
translate \eq{muomega2} into a similar relation for \(\mu_{ab}\):
\begin{equation}\label{mubaxter}
\hat\mu_{ab}^{[+2]}-\hat\mu_{ab}=-(\delta_a^c\bP_b\bP^d-\delta_b^c\bP_a\bP^d)\hat\mu_{cd}\,.
\end{equation}
The last relation can be also understood solely within the {\Pmsys}
as a combination of mirror $i$-periodicity (with long cuts) and the discontinuity property of \(\mu_{ab}\)  \cite{Gromov:2013pga}:
\begin{equation}
\hat\mu_{ab}(u+i)=\tilde\mu_{ab}(u)=
\hat\mu_{ab}-(\delta_a^c\bP_b\bP^d-\delta_b^c\bP_a\bP^d)\,\hat\mu_{cd}=\hat\mu_{ab}(u)+\bP_a\tilde\bP_b-\bP_b\tilde\bP_a
\,.
\end{equation}
A possible interpretation of these identities is the following:
\(\mu_{ab}\) solves the finite difference matrix equation of the first order on 6 functions \eq{mubaxter}.
The finite difference \equref{mubaxter} in general has \(6\) linear independent solutions.
\eq{eqQijkl} tells us that these \(6\) independent solutions could be packed into an antisymmetric tensor which is nothing but
\({\cal Q}_{ab|ij}\) (where the blind indices \(i,j\) simply label 6 different solutions)! Thus \(\mu_{ab}\), as a particular solution, must be a linear combination of these \(6\) solutions with
some $i$-periodic coefficients. These coefficients are precisely the matrix elements of \(\omega^{ij}\).

Let us also note that \eqref{muomega2} can be inverted using \eqref{ortoPP}, and we can write
\be\label{omegamu2}
\omega_{ij}=\frac{1}{2}{\cal Q}^-_{ab|ij}\,\mu^{ab}\,,
\ee
which shows that $\omega_{ij}$ is also a linear combination of Q-functions with periodic coefficients $\mu^{ab}$. But now these coefficients are periodic on the Riemann sheet with long cuts.

\subsubsection{\texorpdfstring{$\mu$ and $\omega$}{mu and omega} as symmetry generators of {\Qsys}}\label{subsec:LHS/UHS}
\label{sec:discussionaxioms}

Apart of being, in a sense, Q-functions, as suggested by \eqref{muomega2} and \eqref{omegamu2}, $\mu$ and $\omega$ have another interesting role in the analytic structure of the fundamental {\Qsys}: they appear to be certain symmetries, or rather morphisms. Moreover, the very existence of these symmetries almost completely determines the QSC itself! We derive this point of view in this subsection, and elaborate on its interpretation in \secref{sec:difpoint}.

Let us summarize the logic we followed starting from \secref{sec:gener-extens}. One takes \(\bP_a,\bP^a\) of the \(\bP\mu\)-system and then defines \(\fQ_{a|i}\) as four independent solutions of \eqref{QabPP}. If we identify \(\fQ_{a|\es}=\bP_a\) and \(\fQ_{a|i}\)   with the appropriate Q-functions of a \(\gl(4|4)\) Q-system with \(Q_{\es|\es}=1\) then we can fully reconstruct all the other Q-functions and  prove that the obtained Q-system has  nice analytic properties. In particular,    \(\bQ_{i}\equiv\fQ_{\es|i}=-\bP^ a\fQ_{a|i}^+\) has a  single long cut,  etc. This Q-system can be also generated from \(\bP_a\) and \(\bQ_i\) alone, using the formulae of sub\secref{subsec:Qbasis}.

One of the advantages of this  construction was the fact that all Q's  of this Q-system were analytic   in the upper half-plane (we will abbreviate it as UHPA Q-system, contrary to the lower half-plane analyticity LHPA). We strove for this analyticity to avoid the problem of discussing various branches of the Riemann surface when solving QQ-relations. The UHPA was achieved by taking solutions \(\fQ_{a|i}\) of \eqref{QabPP} that are analytic in the upper half-plane including the real axis.

But obviously, one could instead generate a  solution of \eqref{QabPP} so as to make it LHPA.  Such a possibility can be easily seen by a slight modification of arguments in \secref{sec:gener-extens}. In $\fQ_{a|i}$, the index \(i\)  labels different solutions of  \eqref{QabPP} which are analytic in the upper half-plane. Let us label the new-type solutions, which are analytic in the lower half-plane, as \(\fQ_{a}|^{i}\):
\be\label{QabPP2}
(\fQ_{a}|^{i})^+-(\fQ_{a}|^{i})^-=-\bP_a\,\bP^b\,(\fQ_{b}|^{i})^+\,,\ \ \ \ \ \Im(u)<0\,.
\ee
By repeating the same steps, one  generates from here a new, LHPA Q-system, denoted in what follows by objects \(\fQ_{A}|^{I}\).~\footnote{It is our convention that the
  QQ-relations are not sensible to the position (upper/lower) of
  indices, hence they allow one to generate \(\fQ_{A}|^{I}\) from
  \(\fQ_{a}|^{i}\) defined above, from \(\fQ_{a}|^{\emptyset}\)
which coincides with \(\bP_a\) on the lower half-plane, and from
\(\fQ_\es|^i\equiv-\bP^a {\cal Q}^+_a|^i\). In
  the setup of this LHPA Q-system, the fact that fermionic ({\ie} $i,j,k,\dots$) indices are
  upper indices should not be confused with a Hodge transformation.} All other analytic  properties for this new Q-system are as nice as for the original, UHPA Q-system.

If the Q-system plays a fundamental role in our formalism, the two Q-systems should be related by some symmetry.
As one can see from \eqref{QabPP} and \eqref{QabPP2} \(\hat\fQ_{a|i}\) and \(\hat\fQ_{a}|^{i}\) are two complete sets of solutions of the same functional equation
if considered in the physical kinematics, with short cuts,  and thus should be linear combinations of each other with \(i\)-periodic coefficients.
We actually have already one \(i\)-periodic \(4\times 4\) matrix in
the physical kinematics -- \(\omega\). And, remarkably enough we can
always choose %
$\hat\fQ_{a}|^{i}$
such that
\begin{equation}\label{defahi}
(\hat\fQ_{a}|^{i})^-=\omega^{ij}\hat\fQ_{a|j}^-\,.
\end{equation}
More precisely, let us show that \((\hat\fQ_{a}|^{i})\) defined by \eqref{defahi}   is analytic for \(\Im(u)<1/2\).  First, using \eq{QabPP} we rewrite \eqref{defahi} slightly above the real axis as
\begin{equation}
(\hat\fQ_{a}|^{i})^-=\omega^{ij}\fQ_{a|j}^+-\omega^{ij}\bP_a\bQ_j=\omega^{ij}\fQ_{a|j}^+-\bP_a\tilde\bQ^i\,.
  \label{sport4}
\end{equation}
The discontinuity of the r.h.s. vanishes
\begin{equation}
\label{discomegadef}
(\omega^{ij}-\tilde\omega^{ij})\fQ_{a|j}^+-\bP_a\tilde\bQ^j+\tilde\bP_a\bQ^j=\left(\bQ^i\tilde\bQ^{j}-\bQ^{j}\tilde\bQ^i\right)\fQ_{a|j}^+-\bP_a\tilde\bQ^i+\tilde\bP_a\bQ^i=0\,,
\end{equation}
where on the last step we use \eq{QQrel2} which gives \(\bP_a=-\bQ^j{\cal Q}_{a|j}^+\) and \(\tilde\bP_a=-\tilde\bQ^j{\cal Q}_{a|j}^+\) (the second function is analytic on \(\mathbb{R}\)). Hence \(\disc(\fQ_{a}|^i)^-=0\).

Further on, using the periodicity of \(\omega\), one has (slightly above real axis)
\be
(\hat\fQ_{a}|^{i})^+-(\hat\fQ_{a}|^{i})^-=\omega^{ij}\left(\fQ_{a|j}^+-\fQ_{a|j}^-\right)=\omega^{ij}\bP_a\,\bQ_j=\bP_a\,\tilde\bQ^i\,.
\ee
One can analytically continue this equation to the lower half-plane by avoiding short cuts, to get
\be
(\hat\fQ_{a}|^{i})^+-(\hat\fQ_{a}|^{i})^-=\bP_a\,\bQ^i\,,\ \ \  \Im(u)<0\,.
\ee
The last equality, given that \(\bP_a\) and \(\bQ^i\) are analytic in the
lower half-plane, proves by recursion the desired analyticity for
\(\fQ_{a}|^{i}\). But more than that, it is an analog of
\eqref{premainQQbf}. Hence it tells us that we should identify
\(\fQ_{\es}|^{i}=\bQ^i\) in the lower half-plane. Therefore, the LHPA Q-system  can be thought
of as generated by a pair of LHPA single-indexed Q-functions \(\bP_a\),
\(\bQ^i\).

At this stage, the reason for notation $\fQ_{A}|^{I}$ becomes clear. Consider the H-rotation of fermionic indices by a constant H-matrix with $\det H=1$. While $\bQ_i$ transforms as $\bQ_i\to H_{i}^{j}\,\bQ_j$, the $\bQ^i$ should transform contra-variantly: $\bQ^i\to \bQ^j\,(H^{-1})^i{}_j$. Since we want to preserve the relation \eqref{defahi}, the H-rotation  can be applied  to both UHPA and LHPA Q-systems to transform one into another: we use the position of the indices to keep track of the appropriate covariance.

Generically, \(\omega\) defines an H-rotation in the physical kinematics (short cuts) that relates two systems
\begin{equation}\label{omULA}
\hat\fQ_{A}|^{I}=\left(\omega^{IJ}\right)^{[\,|A|+|I|-1\,]}\hat\fQ_{A|J}\,,
\end{equation}
where \(\omega^{IJ}=\omega^{i_1j_1}\,\omega^{i_2j_2}\,\ldots\,\omega^{i_kj_{k}}\), \(k=|I|=|J|\).

To complete the description of the symmetry exchanging UHPA and LHPA
Q-systems  we define the Hodge-dual
  \begin{equation}
    \begin{aligned}
      \label{HodgedefLHP}
      \fQ^{A}|_{I}\equiv&(-1)^{|A'|\,|I'|} \epsilon^{A'A}\epsilon_{II'}\,\
      \fQ_{A'}|^{I'}\,,&\textrm{where
    }\{A'\}=&\{1,2,3,4\}\setminus \{A\},\\&& \{I'\}=&\{1,2,3,4\}\setminus
    \{I\}.
    \end{aligned}
\end{equation}
Note that the ordering $II'$ in \(\e_{II'}\) is opposite to \eqref{Hodgedef} which reflects the change of the covariance of the object.

It is immediate to see that in the lower half-plane \(\fQ^a|_{\es}=\bP^a\)  and, after a short computation, that \(\fQ^{\es}|_i=\bQ_i\). Hence we see that exactly the same main objects, \(\bP\) and \(\bQ\), are used to construct both \(\fQ_{A|J}\) and \(\fQ_{A}|^{J}\). However, they are combined differently, both analytically (in the upper half-plane versus the lower half-plane) and algebraically (generated from \(\bP_a\) with \(\bQ_i\) versus \(\bP_a\) with \(\bQ^i\)).

\begin{figure}
  \centering
 \ifpicture{
  \includegraphics{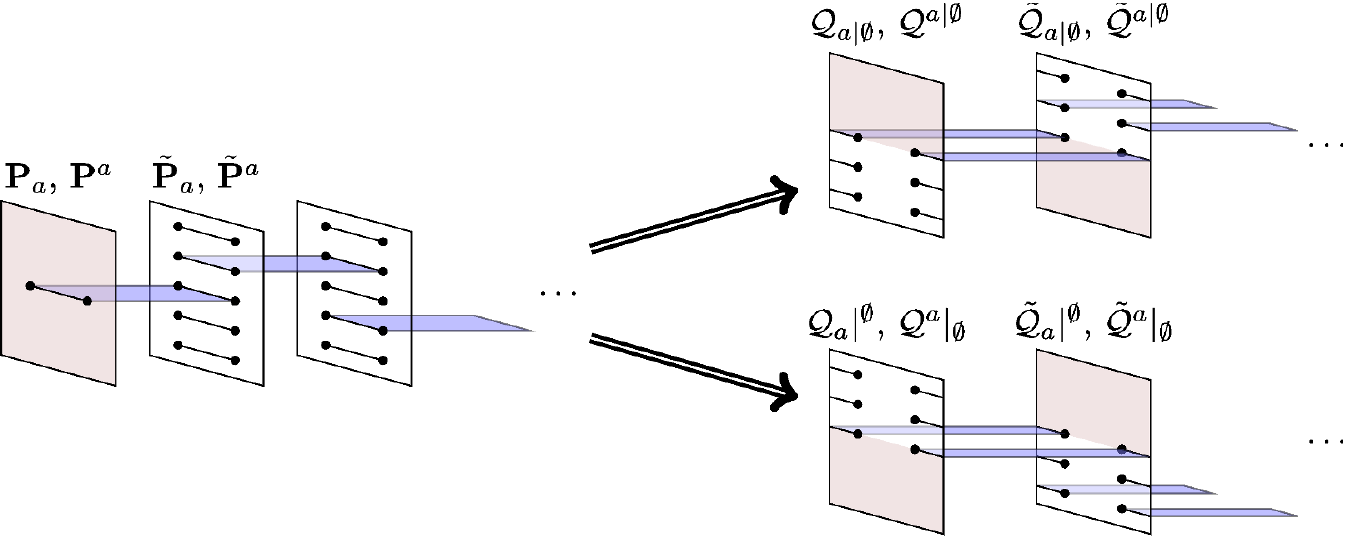}
}  \caption{UHPA and LHPA \(\fQ\)-systems: in their main Riemann sheet
  in the physical kinematics, the functions $\bP_a$ are analytic
  except on a short Zhukovsky cut on the real axis. There are two ways
  to define a mirror-Q-system from these functions: one option
  is to identify them with Q-functions on the upper half plane (for
  instance $\fQ_{a|\emptyset}=\bP_a$ on the upper half plane, hence
  $\fQ_{a|\emptyset}=\tilde \bP_a$ on the lower half plane). The other
  option is to identify them on the lower half plane (for instance
  $\fQ_{a}|^{\emptyset}=\bP_a$ on the lower half plane). The first
  option defines the UHPA \(\fQ\)-system, while the second one
  defines the LHPA \(\fQ\)-system.}
  \label{fig:Sheets}
\end{figure}

One may easily guess now that the \(\mu\)-function defines the H-rotation relating two systems throughout the mirror (long cuts) kinematics:
\be\label{muULA}
\check\fQ^{A}|_{I}=\left(\mu^{AB}\right)^{[\,|A|+|I|-1\,]}\check\fQ_{B|I}\,,
\ee
where \(\mu^{AB}=\mu^{a_1b_1}\ldots\mu^{a_kb_k}\), \(k=|A|=|B|\).

Hence we arrived at a new interpretation of \(\mu\) and \(\omega\) as H-transformations from one Q-system to another.

\subsection{A different point of view: from analytic Q-system to QSC }
\label{sec:difpoint}
The derivation of QSC presented in this paper has an advantage of being based on a relatively solid and well tested ground of TBA, or equivalently, of analytic Y- and T-systems. But a nice form of the resulting \(\bP\mu\) or \(\bQ\o\) equations defining the QSC and a relatively simple analytic structure of the functions entering there hint on the existence of  a  more general structure which might eventually lead to a simpler derivation based on a few physically transparent assumptions ({\it a-la} Zamolodchikovs S-matrix bootstrap) and certain symmetries, like those described earlier in this section.
Such a point of view on the QSC should be based on a good understanding of analyticity and symmetry properties of the underlying Q-system.

Here we first summarize the main properties of QSC
and then speculate  how they might be viewed as a beginning of such simplified derivation.

The  fundamental Q-system  can be recast into four different Q-bases which we denote by $\fQ_{A|I}$, $\fQ^{A|I}$, $\fQ^{A}|_{I}$, $\fQ_{A}|^{I}$. Although all of them describe the same QSC    we will colloquially  name them as 4 different Q-systems. Each of them has the same algebraic structure, i.e. the same QQ relations are satisfied, but the Q-functions with similar sets of indices are differently labeled and  they have different analyticity properties (different positions of branch-points and different large \(u\) asymptotics).   The first two Q-systems are UHPA and the last two are LHPA.  All of them are pairwise related to each other:
\begin{itemize}
\item $\fQ_{A|I}$ is a Hodge dual of $\fQ^{A|I}$ \eqref{Hodgedef}, and $\fQ^{A}|_{I}$ is a Hodge dual of $\fQ_{A}|^{I}$ \eqref{HodgedefLHP}.
\item In the mirror kinematics: $\fQ^{A}|_{I}$ is the H-rotation of $\fQ_{A|I}$, which explicitly means the following: consider $\fQ_{A|I}$ as functions with long cuts, rotate them using \(\mu^{-1}\), as in  \eqref{muULA}, and analytically continue to the lower half-plane. The result will be $\fQ^{A}|_{I}$.

The H-rotation in the mirror kinematics also relates  $\fQ_{A}|^{I}$ and $\fQ^{A|I}$. But, in accordance with the position of upper and lower indices, this rotation is done with $\mu$ instead of $\mu^{-1}$.

As a shorthand notation we write these relations  as
\begin{equation}\label{PmuGENERAL}
\fQ^{A}|_{I}=\CA(\mu^{-1})\cdot \fQ_{A|I}\,,\qquad \fQ_{A}|^{I}=\CA(\mu)\cdot \fQ^{A|I}\,,
\end{equation}
where the symbol $\CA$ reminds us that  these equalities include the analytic continuation between upper and lower half-planes.

\item In the physical kinematics: $\fQ_{A}|^{I}$ is the H-rotation of $\fQ_{A|I}$ \eqref{omULA}, with \(\o^{-1}\); and $\fQ^{A}|_{I}$ is the H-rotation of  $\fQ^{A|I}$, with \(\o\). The short-hand notation is
\begin{equation}\label{QomegaGENERAL}
\fQ_{A}|^{I}=\CA(\omega^{-1})\cdot \fQ_{A|I}\,,\quad \fQ^{A}|_{I}=\CA(\omega)\cdot \fQ^{A|I}\,.
\end{equation}

\end{itemize}
\begin{figure}
  \begin{center}
  \ifpicture{
    \includegraphics{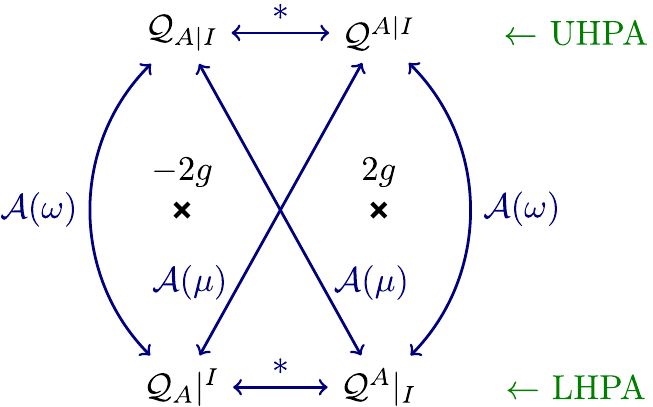}
}  \caption{\label{fig:ULHPA} Summary of transformations between the UHPA and LHPA
    \(\fQ\)-systems. The position of blue arrows with respect to the
    branch points $\pm2g$ indicates whether the analytic continuation
    is performed in the mirror or the physical kinematics. The
    commutativity of this diagram is the key \equref{Hd}.}
   \end{center}
\end{figure}
These relations can be nicely summarized as in \figref{fig:ULHPA}. In particular, they imply the pairwise relations between the single-indexed functions. Depending on the choice of the pair and of the kinematics, the relation can be a direct identification via analytic continuation,
\begin{align}\label{identification}
\hspace{3em}&& \bP_a &\equiv \hat\fQ_{a|\es}=\hat\fQ_{a}|^{\es}\,,
&
\bP^a &\equiv \hat\fQ^{a|\es}=\hat\fQ^{a}|_{\es}\,,
&
a &= 1,2,3,4\,,
&&\hspace{3em}
\no\\
\hspace{3em}
&& \bQ_i &\equiv \check\fQ_{\es|i}=\check\fQ^{\es}|_{i}\,,
&
\bQ^i &\equiv \check\fQ^{\es|i}=\check\fQ_{\es}|^{i}\,,
&
i &=1,2,3,4\,,
&&\hspace{3em}
\end{align}
or a simple relation involving $\mu$ and $\omega$:
\begin{align}\label{identification2}
\hspace{3em}&&&& \check\fQ_{a|\es}&=\mu_{ab}\,\check\fQ^{b}|_{\es}\,,
&
\check\fQ_{a}|^{\es}&=\mu_{ab}\,\check\fQ^{a|\es}\,,
&
a &= 1,2,3,4\,,
&&\hspace{3em}
\no\\
\hspace{3em}
&&&& \hat\fQ_{\es|i}&=\o_{ij}\,\hat\fQ_{\es}|_{j}\,,
&
\hat\fQ^{\es}|_{i}&=\o_{ij}\,\hat\fQ^{\es|j}\,,
&
i &=1,2,3,4\,.
&&\hspace{3em}
\end{align}

The single-indexed functions play a special role in this construction, we have even assigned a special notation for them, $\bP$ and $\bQ$. For one thing, they have the simplest analytic properties among all Q-functions:
\begin{itemize}
\item
On the Riemann sheets, where the identifications \eqref{identification} are  made, $\bP$'s have only one short cut and $\bQ$'s have only one long cut.
\end{itemize}
From this property it is clear that equations \eqref{identification2} are of course the same  as the basic equations of \(\bP\mu\) and \(\bQ\o\) systems of section~\ref{sec:QSCY}, namely  $\tilde\bP_a=\mu_{ab}\bP^b$ and $\tilde\bQ_i=\omega_{ij}\bQ^j$.

Note that on all other sheets $\bP$'s and $\bQ$'s have an infinite ladder of Zhukovsky cuts along the imaginary axis spaced by \(i\), as any generic function of spectral parameter in the AdS/CFT spectral problem.

Furthermore, the single-indexed functions define $\mu$ and $\omega$ through their discontinuities:
\begin{itemize}
\item $\tilde\mu_{ab}-\mu_{ab}=\bP_a\tilde\bP_b-\bP_b\tilde\bP_a\,,\ \ \ \tilde\o_{ij}-\o_{ij}=\bQ_i\tilde\bQ_j-\bQ_j\tilde\bQ_i\,.$

The antisymmetry of $\disc\mu_{ab}$ and $\disc\omega_{ij}$ strongly suggests that $\mu_{ab}$ and $\omega_{ij}$ are antisymmetric themselves and this is indeed the case.
  \end{itemize}
Other important and natural properties of the Q-system are:
\begin{itemize}
\item \(\fQ_{\bar\es|\bar\es}=1\), which is a quantum version of unimodularity of the superconformal group. We chose to normalize $\mu$ and $\omega$ to $\Pf(\omega)=\Pf(\mu)=1$ and, as a consequence, we also  get $\fQ_{\bar\es}|^{\bar\es}=\fQ^{\bar\es}|_{\bar\es}=\fQ^{\bar\es|\bar\es}=1$ as well.
\item Absence of poles for all Q-functions, anywhere on their Riemann surfaces except possibly at \(u=\infty\);
\item The asymptotic behavior of Q-functions at \(u\to\infty\) should be power-like.  The powers are fixed by 6 Cartan charges of the superconformal symmetry, as is given by \eqref{largeu2}. Each   solution for QSC obtained on the basis of this data  should correspond to a physical observable. This construction should describe all single trace local
operators.\end{itemize}
The listed properties fully determine the QSC. And in fact, this list is slightly over-complete. For example, we can check that the relations for discontinuities of \(\mu\) and \(\o\) written above follow from the other outlined properties, etc.  Below we propose a possible point of view on the construction with less amount of assumptions and we will try to give first hints how it can emerge from a certain bootstrap strategy.

Let us start  a derivation of QSC by assuming that there are certain "first principles" which imply the existence of a \(\gl(4|4)\) Q-system. Such existence should be a reflection of the (widely believed) quantum integrability
and the $\psu(2,2|4)$ symmetry, although  the physical origins of Q-system in integrable sigma models are still unclear and the "first principles" are yet to be identified.

We will need to know only a limited set of statements about this Q-system to restore it in full.  Assume that the basic Q-functions with one index have simple analytic properties:
\(\fQ_{a|\es}\) have only one short cut, and \(\fQ_{\es|i}\) have only one long cut on its defining Riemann sheet.  We can use these
$8$ one-indexed Q-functions to generate all the Q-functions, by solving the QQ-relations, and we can choose to solve the QQ-relations so as to make all Q-functions analytic in a half-plane. There is a simple test to uniquely decide whether to generate the UHPA or the LHPA Q-system from the one-indexed Q's: we demand that $\fQ_{\bar\es|\bar\es}=1$ (the property of quantum unimodularity, as we mentioned already). Consider the large-$u$ behaviour of Q-functions and check whether it is compatible with $\fQ_{\bar\es|\bar\es}=1$  property.  Since $\fQ_{\es|i}$ are the functions with long cuts and they do not have the same large-$u$ behaviour in different half-planes, it is unlikely that compatibility with $\fQ_{\bar\es|\bar\es}=1$ will be realized simultaneously in both UHPA and LHPA cases. For definiteness, let us assume that we  construct the UHPA Q-system with \(\fQ_{\bar\es|\bar\es}=1\).

On the other hand,  such UHPA Q-system constructed from the "first principles",  is unlikely to have any fundamental  reason to be preferred to the similar  LHPA Q-system which can be constructed from similar principles. It is expected that a conjugated construction exists which results in the LHPA and a certain symmetry should connect the two cases. But the only symmetries of the Q-system, apart from the  gauge adjustment, are H-rotations and Hodge duality. Hence we will rely on them. Since Hodge operation only relabels the objects, we can always choose a notation for a LHPA system in which it is related to the UHPA system only by the H-rotation, i.e. by \(\CA(\omega)\) or \(\CA(\mu)\), where $\omega$ and $\mu$ are so far arbitrary $4\times 4$ matrices which are periodic, respectively, in physical and mirror kinematics. We choose  to denote UHPA Q-functions by $\fQ_{A|I}$, the LHPA functions obtained by \(\CA(\omega)\) as \(\fQ_{A}|^{I}\) and the LHPA functions obtained by \(\CA(\mu)\) as \(\fQ^{A}|_{I}\), so that the H-transformations are explicitly realized as in \eqref{omULA} and \eqref{muULA}. Note that \(\fQ_{\bar\es}|^{\bar\es}=\det\omega^{-1} \fQ_{\bar\es|\bar\es}=\det\omega^{-1}\). Since \(\fQ_{\bar\es}|^{\bar\es}\) is analytic in the lower half-plane and $\det\omega^{-1}$ is periodic, the equality between them is only possible if both functions are analytic everywhere. Therefore, we can always normalize  \(\fQ_{\bar\es}|^{\bar\es}=\det\omega=1\) without altering the cut structure of the LHPA Q-system. By the same argument, we normalize \(\fQ^{\bar\es}|_{\bar\es}= \det\mu=1\).

At this stage there is no established algebraic relation  between \(\fQ^{A}|_{I}\) and \(\fQ_{A}|^{I}\).

Consider the following composite operation:  \(\CA^{-1}(\mu)\CA(\omega)\). By the symmetry argument, it should produce an UHPA Q-system, not necessarily identical to the original one but with the basic Q-functions having only one cut. Now we demand that this new UHPA Q-system and the original one are also related by the symmetry. Since  both systems are analytic in the same half-plane, the H-matrices relating them can be only constants, hence they can be absorbed in the redefinition of $\mu$ and $\omega$. Therefore we have only two options: either these two systems are equal or Hodge-dual to one another.

The situation which realizes in our physical system is of course the Hodge-duality, so explicitly we have:
\be\label{Hd}
\CA^{-1}(\mu)\CA(\omega)\cdot \fQ=*\fQ\,,
\ee
where \(*\) denotes taking the Hodge dual. Note that this automatically implies that $\fQ^{a|\es}$ and $\fQ^{\es|i}$ have only one cut, which is not necessary to assume originally as an axiom. We also self-consistently get that \(\fQ^{\es|\es}=1\), as follows on the one hand from the definition of $\CA^{-1}(\mu)\CA(\omega)$ and on the other hand as the Hodge dual of \(\fQ_{\bar\es|\bar\es}=1\) which is our original assumption. One can also deduce that $\fQ^{A}|_{I}$ and $\fQ_{A}|^{I}$ are Hodge-dual to one another and this explains our choice of notations. It is now easy to restore all the arrows in \figref{fig:ULHPA}.

The relation \eqref{Hd} is remarkable in many ways. First, loosely speaking,
it tells us that, up to  H-rotations, the Hodge duality transformation is a monodromy around the branch point: \(\tilde Q=*Q\). Then, we can derive \((\CA^{-1}(\mu)\cdot\CA(\omega))^2=1\) reflecting the square root nature of the branch point. The \equref{Hd} is reminiscent to the crossing symmetry and we can show that it is nothing but a summary of \(\mathbb{Z}_4\) symmetry properties from \fip. We believe it originates from a symmetry under the group of outer automorphisms of \(\psu(2,2|4)\) which is \(\mathbb{Z}_2\simeq \mathbb{Z}_4/\mathbb{Z}_2\). The relation to outer automorphisms is clear at strong coupling, see \secref{sec:QCA} and appendix A of \fip, while the finite coupling derivation still awaits to be done.

At this stage, we got a closed system of symmetry transformations between Q-systems related by analytic continuation. And, to our satisfaction, these symmetries encode all the monodromy data of the QSC! Indeed, let us show how the \(\bQ\omega\)-system follows from  this relation between UHPA and LHPA Q-systems. \Equref{Hd}  explicitly gives for \(\bQ_i\) the monodromy \(\tilde\bQ^i=\omega^{ij}\bQ_j\). The equation for the discontinuity of $\omega$ comes from the request that in \eqref{defahi}, $\fQ_a|^i$ is LHPA and and $\fQ_{a|j}$ is UHPA. Indeed, $(\fQ_a|^i)^-$ is analytic on the real axis only if \eqref{discomegadef} is satisfied, and the latter is satisfied only if
$\omega^{ij}-\tilde\omega^{ij}=\bQ^i\tilde\bQ^{j}-\bQ^{j}\tilde\bQ^i$, so that this last relation of the \(\bQ\omega\)-system is also a consequence of  \eqref{Hd}.     The equivalent arguments are applicable for the $\bP\mu$-system, with just replacing the long cuts by short ones.

From the discontinuity relation, we constrain \(\mu\) and \(\omega\) to be antisymmetric matrices\footnote{Symmetric parts of \(\mu\) and \(\omega\) have no cuts. A little more input is needed to show that they are actually zero, for instance if we know that these periodic functions should have no poles and decrease at infinity.} and as we learned before we can always normalize them to \(\Pf(\mu)=\Pf(\omega)=1\). The  inverse of any \(4\times 4\) antisymmetric matrix with the unite Pfaffian is  (minus) the  Hodge dual of the original matrix, e.g. for \(\mu\):
\begin{equation}
\label{muinv}
\mu^{ab}=(\mu^{-1})^{ab}=-\frac 12\e^{abcd}\mu_{cd}\,,
\end{equation}
and similarly for \(\o\). The role of \(n=m=4\) in \(\gl(n|m)\) is therefore very important, allowing us to keep the QSC equations linear in the components of \(\mu\) and \(\omega\). It allows one to pass from the formulae like \(\tilde \bP_a=\mu_{ab} \bP^b\) to the ones like \(\mu^{ab} \tilde\bP_b=\bP^a\) by a simple linear transformation which is rising and lowering the indices, in perfect harmony with the Hodge duality transformation as well. This hints on the exceptional role of the superconformal \(\psu(2,2|4) \) symmetry in the entire construction.

Finally, if we add a typical for integrability request of absence of poles and the power-like behaviour  at infinity, we will focus on the physical solutions for the quantum spectral curve. As is pointed out in \appref{app:unitarityfromanalyticity}, the analytic properties impose  constraints on the large $u$ behaviour of the Q-functions which can be identified with the unitarity constraints on the weights of typical representations of the \(\psu(2,2|4) \) algebra, suggesting to identify    the powers in the large \(u\) behavior of Q's  with particular linear combinations of global charges. The energy of a state can thus be read off from the large \(u\) asymptotics and the formula for this energy  thus follows from the analyticity of the Q-system.

\subsection{Conventions about the choice of the basis and asymptotics of \texorpdfstring{$\omega_{ij}$}{omega\_{ij}}}

The algebraic and analytic structure of the fundamental Q-system is invariant under any H-rotations with constant H-matrices from
 ${\rm GL}(4)\times {\rm GL}(4)$, assuming that $\mu$ and $\omega$ are also transformed in the covariant way. One can partially use this symmetry to introduce several convenient constraints on the QSC functions as we discuss here.
\subsubsection{Orderings conventions and asymptotics of \texorpdfstring{$\omega$}{omega}}
\label{omega-ass}

We partially use the freedom of H-rotations to insure that all $\bP$'s (and $\bQ$'s)
have different asymptotics at $u\to\infty$. Moreover we choose to arrange the order of magnitudes for these functions according to the rule
\begin{eqnarray}\label{magnitudeorderingD}
|\bP_1|<|\bP_2|<|\bP_3|<|\bP_4|\ {\rm and}\ |\bQ_{2}|>|\bQ_{1}|>|\bQ_{4}|>|\bQ_{3}|\ {\rm for}\ \Re(u)\gg1\ {\rm and}\ \Im(u)>0\,,
\no\\
\end{eqnarray}
which is the most convenient for comparison with representation theory in \appref{app:unitarityfromanalyticity}. This is the same ordering as in \eqref{largeu2}.

After the magnitude ordering was fixed, we still have a freedom in adding smaller {\Qfcts} to larger ones, by means of H-rotations. For \(\bQ\)'s, we will use it to constrain the behaviour of $\omega$ at infinity. First we note that $\omega$ should approach a constant matrix at $u\to\pm\infty$ because it is periodic and cannot increase exponentially fast, as this will lead to non-power like asymptotics of various Q-functions.
Next, once the choice \eq{magnitudeorderingD} is done we use the residual H-transformation to bring the asymptotic of $\omega$ to our favorite form.
For that we need some minor analytic input   to fix the physical solutions of QSC. For example, we can easily deduce from TBA (see \appref{subsec:Delta}) that asymptotically $\mu_{12}\sim u^{\Delta-J_1}$. This behaviour is however only possible if $\omega^{12}=-\omega_{34}$ is nonzero asymptotically.
Assuming that $\omega_{34}\neq 0$ asymptotically one can prove that the residual H-transformations allows us to set to
zero $\omega_{13},\;\omega_{14},\;\omega_{23}$ and $\omega_{24}$ and also set $\omega_{34}$ to our favorite non-zero value at infinity.
The remaining components are fixed uniquely by antisymmetry of $\omega$ and by $\Pf(\omega)=1$.
A possible choice is $\omega_{34}=1$ and hence $\omega_{12}=1$ which leads to \eq{largeuomega}.

Curiously, once the asymptotics of $\omega_{ab}$ is fixed at $u\to+\infty$
it gets also fixed at $-\infty$, but to the values differ by a phase factor.
To see this we write (for short cuts!):
\begin{equation}\label{QomegaQup}
\tilde{\hat\bQ}^{i}=\omega^{ij}\hat \bQ_j
\end{equation}
and compare its asymptotics at  \(u\to+\infty\) and \(u\to-\infty\). At \(u\to+\infty\), we see from \eqref{largeu2}
that \(\hat\bQ_{i}\sim \,u^{\hat M_i-1}\)  which implies, due to \eq{QomegaQup} and \eq{largeuomega},
that $\tilde{\hat\bQ}^{i}\sim \eta_{ij}\,u^{\hat M_j-1}$.
Next, we use  the fact that  $\hat \bQ_{j}$ is analytic in the upper half-plane so that its \(u\to-\infty\) asymptotics is determined by the analytic continuation from positive
 \(u\)  along a big upper semi-circle, and since $\tilde{\hat\bQ}^{i}$ is analytic in the lower half-plane, its  \(u\to-\infty\) behavior is determined by the
  analytic continuation along the lower semi-circle. As the result, at minus infinity we get an extra phases in \eq{QomegaQup}
which must be compensated by extra phase in $\omega$. More precisely at $u\to-\infty$
\begin{eqnarray}\label{large-u}
\hat\bQ_{i}\sim\,|u|^{\hat M_i-1}e^{i\pi(\hat M_i-1)}\quad \text{and}\qquad
\tilde{\hat\bQ}^i\sim\eta_{ij}\,|u|^{\hat M_j-1}\,e^{-i\pi(\hat M_j-1)}\;.
\end{eqnarray}
Plugging these asymptotics into the eq.\eqref{QomegaQup} we obtain:

\begin{eqnarray}\label{omega-infty}
&&\omega^{12}(-\infty)=e^{-2\pi i\hat M_1}\omega^{12}(\infty)\;\;,\;\;
 \omega^{34}(-\infty)=e^{-2\pi i\hat M_4}\omega^{34}(\infty)\,.
\end{eqnarray}
Since $\hat M_1=\frac{1}{2}\left(\Delta-S_1-S_2+2\right)$ and $\hat M_4=\frac{1}{2}\left(-\Delta+S_1-S_2\right)$
and $S_1$ and $S_2$ are integer charges we indeed reproduce  \eqref{largeuomegab}.

We also note that for non-even integer charges there could be an additional sign ambiguity due to
a branch cut at infinity (see \cite{Gromov:2014bva} for some examples).

\subsubsection{Complex conjugation and reality}\label{app:reality-1}

It is expected that the energy value for a generic state in AdS/CFT is real.
Even more, at the level of TBA equations not only the energy is real but also the Y-functions.
Hence the complex conjugation at the level of Q-system should not change the physical content of a particular QSC solution
and thus it is expected to be equivalent to a symmetry transformation of \secref{subsec:symmetry}.
Note that the complex conjugation affects the signs in QQ relations \eqref{QQbb}-\eqref{QQbf}, so to make it a true symmetry of the Q-system
one should supplement  the complex conjugation by the following sign change
\beq\la{singc}
{\cal Q}_{a_1\dots a_n|i_1\dots i_m}\;\;\to\;\;(-1)^{
\frac{(m+n)(m+n-1)}2}
\bar{\cal Q}_{a_1\dots a_n|i_1\dots i_m}\;.
\eeq
When discussing the complex conjugation for the functions with possible cuts on the real axis one should consistently fix the conventions for the cuts. For definiteness we consider the choice of physical kinematics when all Zhukovsky cuts are short. Then, we use that the T-functions are real and conclude from \eqref{Tp1p2} and \eqref{eq:15} that the complex conjugation does not raise the indices of Q-functions
(i.e. does not exchange the wings of the T-hook). Hence, Hodge transformation should be excluded from our consideration and we are left only with H-transformations to describe the conjugation properties of the Q-system.

\paragraph{Conjugation and reality of $\bP$ and $\mu$.} As we agreed to work in the physical kinematics, we have one short cut after conjugation of $\bP_a$ and thus the H-transformation could  only be given by a constant matrix, to preserve the power-like
asymptotics of $\bP$'s.
Furthermore, due to the ordering \eq{magnitudeorderingD} it should be a triangular matrix. In particular,
$\bP_1$ after conjugation, as the smallest of \(\bP_a\)'s, could only mix with itself, etc. This gives:
\beq\label{PbHP}
\bar \bP_a={(H_b)}_{a}{}^b\bP_b={\left(
\bea{cccc}
e^{i\phi_1}&0&0&0\\
t_{21}&e^{i\phi_2}&0&0\\
t_{31}&t_{32}&e^{i\phi_3}&0\\
t_{41}&t_{42}&t_{43}&e^{i\phi_3}
\eea\right)_{ab}}\;\bP_b\;.
\eeq
We note that the constant matrix $H_b$ should have
the property $H_b\bar H_b=1$, which is derived by applying complex conjugation once more to \eqref{PbHP}. This immediately implies that the diagonal elements of $H_b$
must be pure phases.

Let us show now that, as a consequence of existence of \({(H_b)_a}^b\),
there exists another H-transformation which makes all $\bP_a$ real.
We can easily see that $H_b^{1/2}$ will work as required, indeed
\beq
\bP\to H_b^{1/2}\bP\;\;\Rightarrow\;\;\bar\bP\to \bar H_b^{1/2}\bar\bP=\bar H_b^{1/2}H_b\bP=(\bar H_b H_b)^{1/2}H_b^{1/2} \bP=H_b^{1/2} \bP\;.
\eeq
As $H_b$ is a triangular matrix $H_b^{1/2}$ is well defined. We thus have shown  that without the loss of generality we can assume all $\bP$
to be real. It is however more common to take
\begin{eqnarray}\label{real1}
 {\bar{\bP}}_{a}=(-1)^{a}\bP_{a}\,,
\end{eqnarray}
which can be easily achieved  by the transformation $\bP_1\to i\bP_1,\;\bP_3\to -i \bP_3$.

One can deduce the conjugation property of $\bP^a$. Under H-rotations, this object should transform contravariantly to $\bP_a$. Also,  since $\bP^a$ is a  Q-function with 3 bosonic and 4 fermionic covariant (lower) indices, one acquires an extra sign factor, according to \eq{singc}. Therefore the complex conjugation should read $\bar\bP^a=-\bP^b\,(H_b^{-1})_b{}^a$, which reduces to
\be\label{real2}
\overline{\bP}^a=-(-1)^{a}{\bP}^a\,
\ee
for the convention \eq{real1}.

To justify the choice of \eq{real1} we note that it leads to real ${\wT}_{a,s}$ as one can easily see from \eqref{Tp1p2} and \eqref{eq:15}.

Let us now   find the conjugation property of $\mu_{ab}$. For that we first observe that the complex conjugation and tilde commute with one another: $\overline{\tilde \bP}=\tilde{\overline \bP}$. A simple way to see this feature is to think about $\bP$ as a function of Zhukovsky variable $\hat x(u)$ defined with short cuts. Since $\hat x(u)$ is a real function of $u$ and $\tilde{\hat x}=\frac 1{\hat x}$, one has  $\tilde{\overline \bP}(u)=\tilde{\overline \bP}[\hat x(u)]={\overline \bP}[1/\hat x(u)]=\overline{\tilde \bP}(u)$, which proves the suggested commutativity. Therefore, $\tilde\bP_a$ and $\tilde\bP^a$ obey the same conjugation rule as in \eqref{real1} and \eqref{real2},  correspondingly.
Then, from $\tilde\mu_{ab}-\mu_{ab}=\bP_a\,\tilde\bP_b-\bP_b\,\tilde\bP_a$ and $\tilde\bP_a=\hat\mu_{ab}\bP^b$ we see that
\be\label{realmu}
\overline{\check\mu}_{ab}=-(-1)^{a+b}\check\mu_{ab}\;.
\ee
Note that we derived the conjugation property of $\mu$ as a function with long cuts. We can consider \eqref{realmu} slightly below real axis and rewrite it for $\mu$  as a function with short cuts: $\overline{\hat\mu}_{ab}=-(-1)^{a+b}\tilde{\hat\mu}_{ab}$. But we also know that $\tilde{\hat\mu}=\hat\mu^{[2]}$, therefore one has $\overline{\hat\mu}_{ab}=-(-1)^{a+b}\hat\mu_{ab}^{[2]}$ or
\be\label{realmuph0}
\overline{\hat\mu^+}_{ab}=-(-1)^{a+b}\hat\mu_{ab}^+\,.
\ee

\paragraph{Conjugation and reality of $\bQ$ and $\omega$.}
In a similar way, we can deal with the ``fermionic" counterpart of the QSC, i.e. with $\bQ_i$ and $\bQ^i$.
Again, the complex conjugation must be equivalent to an H-transformation with a sign adjustment \eqref{singc}, but we have to use
its fermionic counterpart governed by a periodic matrix $(\hat H_f)_i{}^j$, such that $\det H_f=1$, with short cuts (as we decided to consider short cuts for Q-functions in this section):
\beq
\overline{\hat\bQ}_i=(\hat H_f)_i{}^j\hat\bQ_j\;\;,\;\;
\overline{\hat\bQ}{}^i=-\hat\bQ^j\,(\hat H_f^{-1})_j{}^i\;.
\eeq
Note that as a function with short cuts $\bQ_i$ has an infinite ladder of cuts in the lower half-plane.
As a result its complex conjugate is analytic below the real axis, but has a ladder of cuts above it.
Consequently, $H_f$ cannot be simply a constant matrix as it was the case of $\bP$'s.
A little trick is in order here: we define a new periodic matrix $h$ such that $H_f=h\,\omega^{-1}$. Then we get
\beq\la{qcon}
\overline{\hat\bQ}_i=\hat h_{ik}\,\omega^{kj}\hat\bQ_j=\hat h_{ik}\tilde{\hat\bQ}^k\quad{\rm and}\quad
\overline{\hat\bQ}{}^i=\tilde{\hat\bQ}_k\,(\hat h^{-1})^{ki}\;.
\eeq
Now we see that the matrix $h$ entangles
$\overline{\hat\bQ}_i$ and $\tilde{\hat\bQ}^k$
which are both analytic in the lower half-plane, and thus, like for the case with $\bP$'s, $h$ is simply a constant matrix.

From \(\IM u<0\) we can pass to the long-cut version of \eq{qcon}, which is more natural for fermionic Q-functions
\beq\label{eq:uniQ}
\overline{\check\bQ}_i=h_{ik}\,{\check\bQ}^k\quad{\rm and}\quad
\overline{\check\bQ}{}^i={\check\bQ}_k\,( h^{-1})^{ki}\;.
\eeq
We should have $\bar h= h^T$ for self-consistency, hence $h$ is a Hermitian matrix.

The explicit form of $h$ can be partially fixed from the large-$u$ asymptotic of $\bQ_i$
and $\omega_{ij}$. For that we use the first equality in \eq{qcon}
for large $u$:
\beq\la{largeu}
\overline{\hat\bQ}_i\simeq-h_{ik}\eta^{kj}\hat\bQ_j\;
\eeq   where we have used the large \(u\) asymptotics   \eqref{largeuomega} of \(\o\).
The situation here is very similar to the one with $\bP$: taking into account the ordering \eq{magnitudeorderingD} we see that $\bar{\hat\bQ}_2$
is the smallest one and thus the terms with other $\bQ$'s in the r.h.s. could not appear.
This type of reasoning constrains $h$ to the following
\beq
-h\,\eta=\left(
\bea{cccc}
+e^{i\psi_1}& *& 0 & 0\\
0& -e^{i\psi_2}& 0 & 0\\
*& *& +e^{i\psi_3} & *\\
*& *& 0 & -e^{i\psi_4}
\eea
\right)
\quad \Rightarrow\quad
h=\left(
\bea{cccc}
*& e^{i\psi_1}& 0 & 0\\
e^{i\psi_2}& 0& 0 & 0\\
*& *& * & e^{i\psi_3}\\
*& *& e^{i\psi_4} & 0
\eea
\right)\,,
\eeq
where $*$ represent some arbitrary coefficients and $\psi_i$ should be real for the same reason  as for the reality of $\phi_i$  in \eqref{PbHP}.
Since $h$ must be a Hermitian matrix we should have $\psi_2=-\psi_1$ and $\psi_4=-\psi_3$
and \(h_{31}=h_{32}=h_{41}=h_{42}=0\), i.e.\beq
h=\left(
\bea{cccc}
r_1& e^{+i\psi_1}& 0 & 0\\
e^{-i\psi_1}& 0& 0 & 0\\
0& 0& r_2 & e^{+i\psi_3}\\
0& 0& e^{-i\psi_3} & 0
\eea
\right)\;,
\eeq
where $r_1$ and $r_2$ are real.
We argue now that by doing a suitable H-transformation with constant $H$ we can bring $h$ to some standard form. We have the following transformation rules
\beq
Q_i\to H_i{}^jQ_j\;\;\Rightarrow\;\; h\to \bar H\, h\, H^T\;\;,\;\;\omega\to H\,\omega\,H^{T}
\eeq
and we would like to keep asymptotics of $\omega\sim \eta$, which constrains $H$ to obey $\eta=H\,\eta\,H^T$. It is easy to see that the following $H$ sets $r_1=r_2=0$
and $\psi_1=\psi_3=0$:
\beq
H=\left(
\begin{array}{cccc}
 0 & -\frac{e^{-i  {\psi_1}}+1}{2}  & 0 & 0 \\
 \frac{2}{e^{-i  {\psi_1}}+1} & -\frac{2 e^{i  {\psi_1}}
    {r_1}}{e^{2 i  {\psi_1}}+1} & 0 & 0 \\
 0 & 0 & 0 & -\frac{e^{-i  {\psi_3}}+1}{2}  \\
 0 & 0 & \frac{2}{e^{-i  {\psi_3}}+1} &
 -\frac{2 e^{i  {\psi_3}}{r_2}}{e^{2 i  {\psi_3}}+1} \\
\end{array}
\right)\quad {\rm gives}\quad h\to
\left(
\bea{cccc}
0&-1&0&0\\
-1&0&0&0\\
0&0&0&-1\\
0&0&-1&0
\eea
\right)\equiv -|\eta|\,.
\eeq
Therefore, without loss of generality we can always choose the following conjugation property
\beq\label{realfermions}
\overline{\bQ}_1=-{\bQ}^2\;\;,\;\;
\overline{\bQ}_2=-{\bQ}^1\;\;,\;\;
\overline{\bQ}_3=-{\bQ}^4\;\;,\;\;
\overline{\bQ}_4=-{\bQ}^3\;
\eeq
which is valid in the mirror kinematics.

This conjugation property also implies that $\bar {\hat\omega}=|\eta|\hat\omega^{-1}|\eta|$ i.e.
\beq
\bar\omega_{12}=\omega_{34}\;\;,\;\;\bar\omega_{13}=\omega_{13}\;\;,\;\;\bar\omega_{24}=\bar\omega_{24}\;\;,\;\;\bar \omega_{14}=-\omega_{14}
\;\;,\;\;\bar \omega_{23}=-\omega_{23}\;,
\eeq
which is valid in the physical kinematics.

It is instructive to discuss the difference between \eqref{real1},\eqref{real2} and \eqref{realfermions}. Complex conjugation maps the upper half-plane to the lower half-plane, and it also preserves the magnitude of functions. As bosons are naturally analytic in the physical kinematics, their analytic continuation from the upper to the lower half-plane does not change their magnitude at large $u$, hence we can choose $\bP$'s to be real (up to a constant phase) functions. The fermions, on the contrary, are naturally analytic in the mirror kinematics. Their analytic continuation from the upper to the lower half-plane, avoiding long cuts, changes their magnitude. Moreover, with our prescription of $\omega{(\pm\infty)}$, the functions $\bQ_3,\bQ_4$ which are small in the upper half-plane become large in the lower half-plane, and the opposite happens with $\bQ_1,\bQ_2$. Hence fermions should come in the conjugated pairs, and this happens indeed.

\paragraph{Complex conjugation of arbitrary Q-functions.} Complex conjugation of any Q-function which is analytic in the upper half-plane produces a function which is analytic in the lower half-plane. Hence, it is clear that it should relate UHPA and LHPA Q-systems. From the explicit conjugation rules for $\bP$'s and $\bQ$'s and recalling that complex conjugation results in an H-rotation plus a sign adjustment \eq{singc}, we easily restore the general conjugation rule
\be
\overline\fQ_{a_1\ldots a_m|i_1\ldots i_n}=(-1)^{\frac{(m+n)(m+n-1)}2}(-1)^{\sum a_i}(-1)^n|\eta_{i_1j_1}\ldots \eta_{i_nj_n}|\,\ \fQ_{a_1\ldots a_m}|^{j_1\ldots j_n}\,,
\no\\
\overline\fQ^{a_1\ldots a_m|i_1\ldots i_n}=(-1)^{\frac{(m+n)(m+n+1)}2}(-1)^{\sum a_i}(+1)^n|\eta^{i_1j_1}\ldots \eta^{i_nj_n}|\,\ \fQ^{a_1\ldots a_m}|_{j_1\ldots j_n}\,,
\ee
which is valid as written in the lower half-plane.

\subsubsection{Particular case of the Left-Right symmetric states}
\label{sec:particular-case-left}
 {
The left-right (LR) symmetry transformation  corresponds to the
exchange of the left and right \(\su(2|2)\) subalgebras of the full
superconformal symmetry \(\psu(2,2|4)\).   The LR-symmetric operators/states are characterized on the level of mirror Y-system  by the condition
\(Y_{a,-s}=Y_{a,s}\), and they include such important examples as  twist-\(L\) operators of the type
\(\tr (Z^{L-1}\nabla_+^S Z)+\dots\) from the \(\frak{sl}(2)\) sector of the theory. For T-functions the condition of LR symmetry depends on the gauge,
but for the distinguished \(\bT\)-gauge it is simply
\(\bT_{a,-s}=\bT_{a,s}\). We also have \(\wT_{1,s}=\wT_{1,-s}\), hence
one can always put
\(\bP^4=\bP_1\) and \(\bP^3=-\bP_2\).
 Then the following relations hold:

\begin{equation}
\bP_{a}={\chi_{ab}\bP^{b}}\,,\qquad  \bQ_{i}={\chi_{ij}\bQ^{j}}\,,\qquad  \chi_{ij}=\left(\begin{array}{cccc}
0 & 0 & 0 & 1 \\
0 & 0 & -1 & 0 \\
0 & 1 & 0 &0 \\
-1 & 0 & 0 & 0 \\
\end{array}
\right)\,.
\end{equation}

The anti-symmetric matrix \(\chi\) can be viewed as a symplectic metric on \(\mathbb{C}^4\)  rising or lowering both bosonic and fermionic indices of Q-functions in the case of a LR-symmetric state. So for a general LR-symmetric \(\fQ\)-function we  have  \begin{equation}
\fQ_{a_1\cdots a_k|i_1\dots i_m}=\chi_{a_1b_1}\dots \chi_{a_m b_m}\,\,\chi_{i_1j_1}\cdots \chi_{i_m j_m}\fQ^{b_1\cdots b_k|j_1\cdots j_m}\,.
\end{equation}}
Let us note that even in the case when LR-symmetry is absent, the transformation
\be
{\rm LR}: \fQ_{A|I}\mapsto \chi_{AB}\,\chi_{IJ}\,\fQ^{B|J}\,,
\ee
 being a combination of Hodge-duality and the special H-rotation, still preserves all algebraic and analytic relations of the fundamental Q-system. It is therefore always a symmetry of the equations (not the functions though), and this is the precise meaning of how the Hodge-transformation is interpreted as the LR-transformation. Obviously, on the level of T-functions, this transformation acts as ${\rm LR}: \bT_{a,s}\mapsto \bT_{a,-s}$.

\subsection{Exact Bethe equations}
\label{subsec:exactBethe}
The Asymptotic Bethe Ansatz (ABA)  of Beisert-Eden-Staudacher \cite{Beisert:2005fw,Beisert:2006ez} is a well-known and efficient approach for computing the spectrum at weak coupling and also for asymptotically large \(J_1\equiv L\). The ABA is a set of algebraic equations on a set of complex numbers, or Bethe roots, which
then give directly the anomalous dimensions in this approximation.
In this section we go beyond this asymptotic regime and discuss the role of  Bethe roots and Bethe equations in our new formalism, at finite coupling and $J_1$.
We will see that there is a natural definition of the exact Bethe roots which however do not play
the same crucial role as in the asymptotic case.

The situation we observe is in a strict analogy
with the Baxter equation formulation of the XXX Heisenberg spin chain spectral problem,
where the Bethe ansatz equation is replaced by an analyticity requirement.
The Baxter equation for the $\frak{sl}(2)$ spin chain of the length $L$ reads
\begin{equation}
\label{eqsl2}
T(u)Q(u)=(u+i/2)^L Q(u+i)+(u-i/2)^L Q(u-i)\,,
\end{equation}
where the analyticity requirement states that $T$ and $Q$ are polynomials.
The degree of  the polynomial $Q$ is the spin of the state.
For a given spin, this analyticity requirement has a finite number of solutions
corresponding to the highest weight states of the
spin chain. When a solution of \eqref{eqsl2} is found, the Bethe roots can be defined as the roots of $Q$,
which can also be shown to satisfy the Bethe equations as a consequence of \eq{eqsl2}:
\be
\left(\frac{u+i/2}{u-i/2}\right)^L\frac{Q^{[+2]}}{Q^{[-2]}}=-1\,,\ \ \text{at zeros of}\ \ Q\,.
\ee
Curiously, \eq{eqsl2} is precisely what the QSC reduces
to at weak coupling in the $\frak{sl}(2)$ sector where the role of $Q(u)$ is played by $\hat\mu_{12}(u+i/2)$,
which also becomes a polynomial at weak coupling \cite{Gromov:2013pga}.

Notably,
 the condition $\hat\mu_{12}(u_{4,j}+i/2)=0$ also gives exact Bethe roots
 defined in the TBA formalism as we are going to show now.
 In the case of TBA equations one defines the exact Bethe roots
 as $Y_{1,0}^*(u_{4,j})+1=0$, where the star denotes the physical
 branch of the Y-function defined as an analytic continuation through the first branch cut at $-i/2$, see e.g. \cite{Gromov:2009zb}\footnote{Note that not all the roots
 of $Y_{1,0}^*(u)+1$ give the Bethe roots. To identify correctly the right zeros one should follow them starting from the weak coupling limit,
 controlled by the ABA.}.
 To show that the zeros of $\mu_{12}^+$ produce the zeros of $Y_{1,0}^*+1$ we write, for short cuts,
\beq
1+\frac{1}{Y_{1,0}}=
\frac{\bT_{1,0}^+\bT_{1,0}^-}{\bT_{1,1}\bT_{1,-1}}
=
-\frac{\tilde{\hat\mu}^+_{12}\,\hat\mu^-_{12}}{\left(\bP^{3-} \bP^{4+}-\bP^{4-}
   \bP^{3+}\right) \left(\bP_2^{-} \bP_1^{+}-\bP_1^{-}
   \bP_2^{+}\right)}\la{Y10a}
\eeq
where we use  (\ref{eq:95}-\ref{eq:15}) and \eq{T10} to express $\bT_{a,s}$ explicitly in terms of $\bP$ and $\mu$.
We have to analytically continue $Y_{1,0}$ under the cut at $-i/2$. Using the explicit expression
\eq{Y10a} it is trivial to perform such continuation to get
\beq
1+\frac{1}{Y^*_{1,0}}
=
-\frac{\mu^+_{12}\,(\widetilde{\hat\mu^{--}_{12}})^{+}}{
\left(\bP^{3-} \tilde\bP^{4+}-\bP^{4-}
   \tilde\bP^{3+}\right) \left(\bP_2^{-} \tilde\bP_1^{+}-\bP_1^{-}
   \tilde\bP_2^{+}\right)}\la{Y10sim}\;.
\eeq
Note that
 \((\widetilde{\hat\mu^{--}_{12}})^{+}\) is not singular as  a consequence of our assumption of QSC regularity:
 all Q-functions and also $\mu_{ab}$ have no poles on any sheet of their Riemann surface.
Assuming also that the denominator does not go to zero, which can be verified at weak coupling,
we see that the condition $\mu_{12}(u_{4,j}+i/2)=0$ indeed implies $Y^*_{1,0}(u_{4,j})=-1$ exactly at any coupling.

\paragraph{Auxiliary Bethe roots.}
There exists no good definition within the TBA formalism for the auxiliary Bethe roots (i.e. those which are not momentum carrying).
But in the formalism of QSC we can develop a further analogy with the spin chains.
For generalized $\frak{su}(n)$  Heisenberg spin chains one defines $2^n$ different Q-polynomials.
Their polynomiality and the QQ-relations imposed on them generalize
the Baxter equation \eq{eqsl2} and also give a discrete set of solutions,
corresponding to the states of the spin chain. The Bethe roots are again zeros of these Q-functions,
which can be shown to satisfy some Bethe ansatz equations, as a consequence of the polynomiality and the QQ-relations.
Acting in the same way for the QSC at any coupling, we define a set of exact Bethe equations which, as we will see in \secref{sec:ABA}, coincide in the asymptotic limit  with the
familiar BES Bethe roots\footnote{See \cite{Gromov:2010kf} for the derivation of BES equations from the Y-system and from the generating functional, under natural analyticity and symmetry conditions.}.
For instance, let us first take the fermionic QQ-relation \eqref{QQbf}.
 Considering it at zeros of $\fQ_{Aa|I}$ and requiring the absence of poles in $Q_{A|Ii}$ one gets
\be\label{fermBetheE}
\frac{\fQ_{Aa|Ii}^+\fQ_{A|I}^-}{\fQ_{Aa|Ii}^-\fQ_{A|I}^+}=-1\,,\ \  \text{at zeros of}\  \ \ \fQ_{Aa|I}\,.
\ee
Applying the same logic as around equations \eqref{tmp1} and \eqref{shiftedtmp}, but now for arbitrary $A,I,i,j$ of \eqref{QQff}, one gets
\be\label{bosBetheE}
\frac{\fQ_{A|Ii}^{[+2]}\fQ_{A|I}^{-}\fQ_{A|Iij}^{-}}{\fQ_{A|Ii}^{[-2]}\fQ_{A|I}^{+}\fQ_{A|Iij}^{+}}=-1\,,\ \  \text{at zeros of}\  \ \ \fQ_{A|Ii}\,.
\ee
For the success of this procedure, it is important that $\fQ_{A|Ij}^{[\pm 2]}$ is regular at zeros of $\fQ_{A|Ii}$.
This equation and the regularity property are trivial consequences of our present formalism,
however, in our previous FiNLIE
description \cite{Gromov:2011cx} of the spectrum it was far from being obvious and for
a particular case $\fQ_{A|Ii}=\fQ_{12|12}$
it was observed in \cite{Leurent:2013mr} for an explicit perturbative solution.

For the spin chain case one can determine Q-functions simply by knowing their zeros. For instance, for the compact rational case, the Q-functions are simply polynomials. Hence solving the Bethe equations for the zeros of Q-functions is equivalent to finding the Q-system. One typically chooses a nesting path connecting \(Q_\es\) and \(Q_{\bar\es}\) on the Hasse diagram (which also defines a Dynkin diagram) and consider only the Bethe equations that involve the Q-functions on this path. This allows one to fix enough Q-functions to restore other ones by QQ-relations.

In the present case the sole knowledge of the roots of Q-functions does not allow
for a complete description as our Q-functions also have a complicated cut structure
and thus the exact Bethe roots defined here are in general much less restrictive  for
the exact solutions. Nevertheless, they could be important to identify particular solutions of QSC, for the numerical calculations
and their comparison with the ABA formalism at weak coupling. They could also help
with the classification of all physical solutions of QSC.

It is quite interesting that, at least in principle,
we can choose virtually any paths on the Hasse diagram,
not only  those  giving a ``Zhukovsky-polynomial'' solution of Beisert-Staudacher in the large volume limit.
It could happen that these Q-functions  become polynomial in some other physically
interesting regimes.

Let us emphasize that the definition of  exact Bethe roots as zeros of Q-functions
is not universal. For example, the exact momentum carrying Bethe roots can be alternatively defined as zeros of ${\cal Q}_{12|12}$, as suggests the logic of writing Bethe equations along a Hasse diagram.
This definition coincides with the TBA motivated definition through zeros of $\mu_{12}^+$ only asymptotically, but differ from the latter when the finite size effects are included, which was confirmed by the explicit  computation in \cite{Leurent:2013mr}.
The possibility for alternative and inequivalent definitions demonstrates the fact  that at the finite coupling the relevance of the Bethe roots is diminishing.

\section{Large Volume Limit}
\label{sec:ABA}

In this section we will establish the large length/charge Asymptotic Bethe Ansatz (ABA) limit of the quantum spectral curve.
We will  establish the quantities entering the QSC equations in this limit and restore the Beisert-Staudacher ABA equations.

\subsection{Conventions}\label{sec:ABAconventions} We should expect significant simplifications in the asymptotic limit of operators with large charges.  It is known that in this, particularly important, limit the spectral problem is governed by a system of algebraic   Beisert-Staudacher equations (ABA) \cite{Beisert:2005fw}
which describe the spectrum  with exponential precision in \(J_1\sim\Delta\).

To understand the correct scaling of various QSC quantities in the asymptotic limit, we can look at their large \(u\) asymptotics. We see for example that \(\bP_1\) scales as \(u^{-J_1/2}\) which becomes exponentially
small in \(J_1\) for large enough $u$ (or \(x(u)\)). At the same time \(\bP_4\) scales as \(u^{J_1/2}\) and thus becomes exponentially large.
To keep track of this scaling we introduce a formal expansion parameter \(\epsilon\sim u^{-J_1/2}\sim u^{-\Delta/2}\) and we assign a particular scaling guided by the large \(u\) asymptotics, as follows:
\begin{equation}\label{eq:scaQP}
\bQ^{\alpha}\sim\bQ_{\dot \alpha}\sim \bP_{\alpha}\sim\bP^{\dot \alpha}\sim\epsilon\;\;,\;\;
\bQ_{\alpha}\sim\bQ^{\dot \alpha}\sim\bP^{\alpha}\sim\bP_{\dot \alpha}\sim 1/\epsilon\,,
\end{equation}
where \(\alpha=1,2\) and \(\dot \alpha=3,4\).
Similarly
\begin{equation}\label{eq:scamu}
\mu_{\alpha\beta}\sim 1\;\;,\;\;\mu_{\alpha\dot \beta}\sim \epsilon^{-2}\;\;,\;\;\mu_{\dot \alpha\dot \beta}\sim \epsilon^{-4}
\end{equation}
for \(\beta=1,2\) and \(\dot \beta=3,4\). And following the same principle we assume
\(\omega_{ij}\sim \omega^{ij}\sim 1\).

To see from where the main simplification comes   we also have to deduce the scaling of \({\cal Q}_{a|i}\)
from its asymptotics, or simply from \eq{QPQ}:
\begin{equation}
{\cal Q}_{\alpha|\beta}\sim {\cal Q}_{\dot \alpha|\dot \beta} \sim 1\;\;,\;\;
{\cal Q}_{\alpha|\dot \beta}\sim \epsilon^{+2}\;\;,\;\;
{\cal Q}_{\dot \alpha|\beta}\sim \epsilon^{-2}\;.
\end{equation}
Then from \eq{muomega} we simply have
\begin{eqnarray}\label{muomega3}
\mu_{12}\simeq\omega^{12}\left({\cal Q}^-_{1|1}{\cal Q}^-_{2|2}
-{\cal Q}^-_{2|1}{\cal Q}^-_{1|2}\right)=\omega^{12}\cQ_{12|12}^-\,,
\end{eqnarray}
i.e. only the term with \(\omega^{12}\) in the r.h.s. survives in this
limit. For  $\tilde\bP_a$ we have in this asymptotic limit  a
similar simplification:
\begin{equation}\label{eqPta0}
\tilde \bP_\alpha=\mu_{\a b}\bP^b={\cal Q}_{\a |i}^-{\cal Q}_{b|j}^-\omega^{ij}\bP^b=
-{\cal Q}_{\a |i}^-\bQ_j\omega^{ij}\simeq
\left(-{\cal Q}_{\a |1}^-\bQ_2+{\cal Q}_{\a |2}^-\bQ_1\right)\omega^{12}=\omega^{12}\,\fQ_{\alpha|12}\,.
\end{equation}
Until the end of this section, we will consider all the cuts being short, if the otherwise is not specified. This means in particular that the periodicity conditions for \(\omega\) and \(\mu\) are
\begin{equation}\label{per}
\omega^{12[+2]}=\omega^{12}\;\;,\;\;\mu_{12}^{[+2]}=\tilde \mu_{12}\;.
\end{equation}
Also, we should consider the reality of $\mu$ in the physical kinematics \eq{realmuph0}:
\begin{equation}\label{realmuph}
\overline{\mu_{12}^+}=\mu_{12}^+\;.
\end{equation}
The conjugation condition for fermionic Q's should be also considered with short cuts, i.e. we  consider \eqref{qcon} which simplifies in the asymptotic limit to
\begin{align}\label{Qbarom}
&&&&\overline{\bQ_1}=&\omega^{12}\bQ_1\,,&\overline{\bQ_2}=-\omega^{12}\bQ_2\;,&&&&
\end{align}
prescription.
This relation implies
\begin{equation}\label{com}
\overline{\omega^{12}}=\frac{1}{\omega^{12}}=\omega_{34}\,.
\end{equation}

Finally, it will be convenient to introduce a notation ``\(\propto\)'',
where \(f_1\propto f_2\) means that \(f_1/f_2\) is an irrelevant constant
 multiplier.

 Let us point out an important subtlety.
 In general the analytic continuation and the expansion in $\epsilon$ are not expected to commute.
A simple example of the phenomenon is the function $1+1/x^L$ which with our precision is simply $1$. However, its analytic
continuation to the second sheet is $1+x^L$ which is $x^L$ with exponential precision.
Nevertheless, we see that such non-commutativity implies a drastic modification of
analytic properties of the function on another sheet -- namely we should hide extra
$L$ poles on the second sheet
(including the poles are  at infinity of the second sheet).
If the singularity structures of the expanded continuation and the continuation of  expansion coincide
there should not be any problem with  non-commutativity. In such a case it should be safe to do such analytic continuation
provided the analytic properties are well controlled. We will see some explicit examples below.

\subsection{Asymptotic solution of the QSC equations}
\subsubsection{Finding \texorpdfstring{$\mu_{12}$}{mu\_12} and \texorpdfstring{$\omega^{12}$}{omega\^{}12}}\label{sec:find-texorpdfstr-tex}

Using the above simplified relations, we now find the explicit form of \(\mu_{12}\) and  \(\omega^{12}\).
For that we introduce more of new notations. First, we label the zeros of \(\mu_{12} ^+\) on the physical sheet as $u_j$:
\begin{equation}
\mu_{12}(u_j+i/2)=0\;\;,\;\;j=1,\dots,N\,.
\end{equation}
We will see that \(\mu_{12}\) and \(\omega^{12}\) can be expressed in terms of $u_j$.
Furthermore, these zeros are the exact Bethe roots for the central node of TBA (\(Y_{10}^{\rm phys}(u_j+i/2)=0)\), as is discussed in \secref{subsec:exactBethe}. However,  we do not yet have to assign any physical meaning to $u_j$.

Since \(\mu_{12}^+\) is a real function, its zeros must be either real or
come in  complex conjugated pairs, which then implies that the Baxter
polynomial
\begin{equation} \label{eq:21}
\mathbb{ Q}\equiv \prod_{j=1}^N(u-u_j)
\end{equation}
is also real.
At the same time we see from \eq{com} that \(\omega^{12}\) cannot have zeros as they would inevitably generate poles in the complex conjugate points,
which contradicts the regularity of \(\omega^{12}\).

Next, we define a function $F$ by
\beq
F^2=\frac{\mu_{12}}{\mu_{12}^{++}}\frac{{\wQ}^+}{{\wQ}^-}=\frac{\mu_{12}}{\tilde\mu_{12}}\frac{{\wQ}^+}{{\wQ}^-}\,,
\eeq
with the choice of sign $F(+\infty)=+1$.
As we see it is defined in such a way that there are no poles or zeros on its defining sheet.
We note that even without the large volume approximation we have directly from the definition of \(F\)
\beq\la{FtF}
F\tilde F =\frac{{\wQ}^+}{{\wQ}^-}\;.
\eeq
Furthermore, in the large volume limit we can show that the discontinuities on all cuts are exponentially small,
except for the cut on the real axis. Indeed, using \eq{muomega3} we rewrite $F^2$ in terms of $\omega^{12}$,
which cancels due to periodicity
\beq\la{eqQ1212}
F^2=\frac{{\cal Q}_{12|12}^-}{{\cal Q}_{12|12}^+}\frac{{\wQ}^+}{{\wQ}^-}\,.
\eeq
As ${\cal Q}_{12|12}^-$ is analytic everywhere in the upper half-plane we know that $F^2$ is also analytic there.
Moreover, due to the reality of $\wQ$ and $\mu^+$ we see that $\bar F=1/F$ which implies the absence of cuts below the
real axis as well. We thus see that $F$ is a double valued function, which tends to $1$ at infinities
and satisfies \eq{FtF} with no poles or zeros on the main sheet.
Thus we can take log of \eq{FtF} and find $F$ by Hilbert transformation\footnote{Strictly speaking, the r.h.s. is equal to $\log\left(e^{\frac{i}{2}\mathfrak{p}}\frac{ B_{(+)}}{ B_{(-)}}\right)$, where $e^{\frac{i}{2}\mathfrak{p}}=\prod_{k=1}^{N}\frac{\sqrt{x_k^+}}{\sqrt{x_k^-}}$. This extra phase insures $F(+\infty)=1$. As eventually we will identify $u_k$ with asymptotic Bethe roots, and derive the cyclicity condition \eqref{cyclicitycond}, we know that $e^{i\mathfrak{p}}=1$, hence $e^{\frac{i}{2}\mathfrak{p}}=\pm 1$. To simplify the rest of the derivation, we write the equations only for the $+1$ case. Restoring the extra sign, if needed, is an easy task and in most of cases it is immediately evident.}
\beq
\log F=H\cdot\log\frac{\wQ^+}{\wQ^-}=\log\left(\pm \frac{ B_{(+)}}{ B_{(-)}}\right)\,,
\eeq
where the argument of the log in the r.h.s. is a rational function of the Zhukovsky variable $x$
\begin{equation}
B_{(\pm)}\equiv \prod\limits_{k=1}^{N}\sqrt{\frac{g}{x_k^\mp}}\left(\frac
1{x}-{
x_k^{\mp}}\right)
\;\;,\;\;
x_k^{\mp}=x(u_k\pm i/2)\,.
\end{equation}
The normalization factor $\sqrt{\frac{g}{x_k^\mp}}$ is chosen so as to have
\(
R_{(\pm)}B_{(\pm)}=(-1)^N{\wQ}^\pm\)\,,
where
\beq\la{RBdef}
R_{(\pm)}\equiv \tilde B_{(\pm)}= \prod\limits_{k=1}^{N}\sqrt{\frac{g}{x_k^\mp}}\left(x-{
x_k^{\mp}}\right)\,.
\eeq
Knowing $F$ we can reconstruct $\mu_{12}$. We should also remember that $\mu_{12}^+$ is real and it asymptotics is polynomial which
uniquely fixes the solution of the first order finite difference equation on $\mu_{12}$
\beq
\frac{\mu_{12}}{\mu_{12}^{++}}=\left(\frac{B_{(+)}}{B_{(-)}}\right)^2\frac{\wQ^-}{\wQ^+}\;\;\Rightarrow\;\;
\mu_{12}\propto
\wQ^-\prod_{n=0}^\infty
\frac{ B^{[+2n]}_{(+)}}{ B^{[+2n]}_{(-)}}
\prod_{n=1}^\infty
\frac{ B^{[-2n]}_{(-)}}{ B^{[-2n]}_{(+)}}\;.
\eeq
It is convenient to introduce a notation:
\begin{equation}\label{f-explicit}
\boxed{f\propto \prod_{n=0}^\infty\frac{ B^{[2n]}_{(+)}}{ B^{[2n]}_{(-)}}\,.}
\end{equation}
We assume that the infinite product is regularized in some way (for example by imposing $f(0)=1$). Alternatively, one can think of $f$ as a solution of
\be
\frac{f}{f^{++}}=\frac{B_{(+)}}{B_{(-)}}
\ee
which is analytic in the upper half-plane and which is normalised, for example, by $f(0)=1$.

Next, we use \eq{eqQ1212} to find ${\cal Q}_{12|12}$ by finding a solution, analytic
in the upper half-plane, which is uniquely given by
\beq
{\cal Q}_{12|12}\propto\wQ\, (f^+)^2\;.
\eeq
This allows us to find $\omega^{12}$ from \eq{muomega3}.
We summarize the main asymptotic relations from this section as follows
\begin{equation}\label{muomegaexplicitly}
\boxed{\mu_{12}\propto -\frac{ B_{(+)}}{ B_{(-)}}\bar f^{[-2]}f^{[+2]}{\mathbb Q}^-\;\;,\;\;
{\cal Q}_{12|12}\propto \wQ (f^+)^2\;\;,\;\;
\omega^{12}\propto \frac{ B_{(-)}}{ B_{(+)}}\frac{\bar f^{[-2]}}{f^{[+2]}}\;.}
\end{equation}
In conclusion we note again
that zeros of ${\cal Q}_{12|12}$ and that of $\mu_{12}$ coincide
in the large volume limit as was already mentioned in the \secref{subsec:exactBethe}.

\subsubsection{Finding \texorpdfstring{$\bP_{\alpha}$}{P\_alpha}}
The strategy here is essentially the same as before:
We build a function which becomes double-valued in our approximation and then fix it by zeros and poles.
Before that let us define a real function $\sigma$ with only one short cut on the defining sheet and with no poles or zeros there,  such that
\beq\la{sigmadef}
\sigma\tilde\sigma\propto \bar f^{[-2]}f^{[+2]}\;\;,\;\;\sigma(+\infty)= 1\;.
\eeq
Then, we introduce\beq
g_\alpha\equiv \bP_\alpha/\sigma\;.
\eeq
We check now that this $g_\alpha$ is a double valued function, i.e it has a double-sheeted Riemann surface, with two sheets connected through a single cut.
First, it is obvious that on the upper, defining sheet  there are no other cuts except the one on the real axis.
To get to the next sheet we use \eq{eqPta0}
\beq
\tilde g_\alpha\equiv
\tilde \bP_\alpha/\tilde\sigma
=\fQ_{\alpha|12}\omega^{12}\frac{\sigma}{f^{[+2]}\bar f^{[-2]}}
=\fQ_{\alpha|12}\frac{B_{(-)}\bar f^{[-2]}}{B_{(+)} f^{[+2]}}\frac{\sigma}{f^{[+2]}\bar f^{[-2]}}
\;
\eeq
from where we see that $\bar f^{[-2]}$ cancels and all other factors in the r.h.s. are
regular in the upper half-plane. As also from \eq{real2} $\bar g_\alpha=\pm g_\alpha$,  we see that there could not be any cuts or poles
in the lower half-plane either. Thus we conclude that $g_\alpha$ is simply a regular function of $x(u)$ for all finite $u$. It still can have pole at $x=0$ which corresponds to $u\to\infty$ on the second sheet. Given the power-like behaviour of $\bP_\alpha(u)$ at infinity, we see that the most general expression for $g_{\alpha}$ is
\beq\label{eqga}
g_\alpha \propto
R_{\alpha|\emptyset}B_{\alpha|12}\frac{1}{x^{L/2}}\,,
\eeq
where \(R^{}_{1|\emptyset},R^{}_{2|\emptyset}\) are some real polynomials in \(x\) containing zeros only outside the unit circle, and
\(B^{}_{1|12},B^{}_{2|12}\) are real polynomials containing zeros only inside the unit circle, similar to those defined in  \eqref{RBdef}. Their indices label the nodes on Hasse diagram  to which correspond the auxiliary Bethe roots defined by their zeros. $L$ can be though of as an arbitrary number so far, but it will be eventually  linked to the value of charges and number of Bethe roots. We will also show that $L$ coincides with the length of a spin chain that emerges in the weak coupling limit.

For $\bP_\alpha$ and $\fQ_{\alpha|12}$, equation \eqref{eqga} gives
\beq
\bP_\alpha \propto
R_{\alpha|\emptyset}B_{\alpha|12}\frac{\sigma}{x^{L/2}}\;\;,\;\;
\fQ_{\alpha|12}\propto B_{\alpha|\emptyset}\, R_{\alpha|12}ff^{[+2]}
\frac{x^{L/2}}{\sigma}\,.
\eeq

Finally, let us give an explicit form of the function
$\sigma$ by relating it to the dressing phase of
Beisert-Eden-Staudacher \cite{Beisert:2006ib,Beisert:2006ez}. We can recognise in \eq{sigmadef} an analytic continuation \cite{Volin:2009uv} of the crossing equation \cite{Janik:2006dc}, hence we can immediately write the solution:
\be\la{reltobes}
\frac{\sigma^+}{\sigma^-}=\prod_{k=1}^{N}\sigma_{\rm BES}(u,u_k)\,,
\ee
where sign conventions for $\sigma_{\rm BES}$ are the same as in \cite{Vieira:2010kb}.

\subsubsection{Finding \texorpdfstring{${\cal Q}_{\alpha|\beta}$}{Q\_alpha|beta}}
Recall that \({\cal Q}_{a|j}^+\omega^{jk}=-(\fQ_{a}|^k)^+\) belongs to the
  Q-system analytic in the lower half-plane, as is discussed in \secref{sec:discussionaxioms}. In our scaling
  only one term with $\omega^{12}$ survives and thus the following functions
should be analytic in the lower half-plane
\begin{equation}
{\cal Q}_{a|1}^+\omega^{12}\;\;,\;\;{\cal Q}_{a|2}^+\omega^{21}\;\;,\;\;a=1,\dots,4,
\end{equation}
As ${\cal Q}_{a|\alpha}^-$ themselves are analytic in the upper half-plane
we conclude that any ratio \({\cal Q}_{a|\alpha}/{\cal Q}_{b|\beta}\) for arbitrary \(a,b=1,\dots,4\) and \(\alpha,\beta=1,2\)
has no cuts on the whole complex plane and thus
is simply a rational function of $u$, which allows for the following parameterization
\begin{equation}\label{Qij}
{\cal Q}_{a|\beta}={\mathbb Q}_{a|\beta} \,\,q^+\,,
\end{equation}
where \({\mathbb Q}_{a|\beta}\) are polynomials of \(u\) and \(q\) is a so far  unknown function
analytic above the real axis. We find \(q\) by comparison to
\eq{muomega3}
 \begin{eqnarray}\label{muomega4}
\frac{\mu_{12}}{\omega^{12}}={\mathbb Q}^- f^2=q^2\left({\mathbb Q}^-_{1|1}{\mathbb Q}^-_{2|2}
-{\mathbb Q}^-_{2|1}{\mathbb Q}^-_{1|2}\right)\,.
\end{eqnarray}
This  tells us that, in a suitable normalization of the polynomials,  \(q\) defined in \eq{Qij}
coincides with  \(f\) defined by  \eqref{f-explicit} since by definition neither \(f\) nor \(q \) have zeros or poles. Thus we get
\begin{equation}\label{Qabf}
\boxed{{\cal Q}_{a|\beta}={\mathbb Q}_{a|\beta} f^+}
\end{equation}
and
\begin{equation}
\boxed{{\mathbb Q}_{1|1}{\mathbb Q}_{2|2}
-{\mathbb Q}_{2|1}{\mathbb Q}_{1|2}={\mathbb Q}\;.}
\end{equation}

\subsubsection{Finding \texorpdfstring{$\bQ_{\alpha}$}{Q\_alpha}}
We introduce \(r\equiv \frac{\bQ_1}{\bQ_2}\) which by the UHPA construction
has no short cuts above the real axis, and furthermore it is purely imaginary due to \eq{Qbarom}:
\begin{equation}
\bar r=\frac{\bar\bQ_1}{\bar\bQ_2}=\frac{\omega^{12}\bQ_1}{-\omega^{12} \bQ_2}=-\frac{\bQ_1}{\bQ_2}=-r\,,
\end{equation}
which implies that it has only one short cut on the defining sheet.
The reality property automatically propagates to the next Riemann
sheet. But  this ratio has no cuts there in the lower half-plane, because $\bQ_\alpha$ has the only cut in the mirror kinematics. Hence, because of these analyticity and reality properties, $\tilde r$ has no cuts in the upper half-plane as well. So $r$ is a rational function of \(x\), similarly to \(F\). We split its zeros and poles into those inside and outside of the unit circle and denote
\begin{equation}
\frac{\bQ_1}{\bQ_2}=r\propto\,\frac{R^{}_{\emptyset|1}B^{}_{12|1}}{R^{}_{\emptyset|2}B^{}_{12|2}}\;,
\label{Q1Q2ratio}\end{equation}
where \(R^{}_{\emptyset|1},R^{}_{\emptyset|2}\) are real polynomials in \(x\) containing zeros outside the unit circle, and
\(B^{}_{12|1},B^{}_{12|2}\) are real polynomials containing zeros inside the unit circle, similar to those defined in  \eqref{RBdef}.
 In order to have these polynomial well defined from their ratio,  we demand that \(R^{}_{\emptyset|1}\) and \(B^{}_{12|1}\) have the same zeros as $\bQ_1$ on the corresponding sheets, and that correspondingly  \(R^{}_{\emptyset|2}\) and \(B^{}_{12|2}\) have the same zeros as $\bQ_2$. The notations of indices suggest the place of these functions, and their zeros - auxiliary Bethe roots, in the asymptotic Q-system  (i.e. in Hasse diagram).

\Equref{Q1Q2ratio} suggests the following parameterization:
\begin{equation}\la{QAlower}
\bQ_\alpha=d_\alpha
R^{}_{\emptyset|\alpha}B^{}_{12|\alpha}\frac{S f^{[+2]}}{B_{(-)}}\;\;,\;\;\alpha=1,2\,,
\end{equation}
where $d_\alpha$ is a numerical constant. The factor \(f^{[+2]}/B_{(-)}\) can be absorbed into redefinition of \(S\) but it is convenient to keep it.
Indeed, due to this factor $\bar S=S$ as one can see from \eq{Qbarom}. Thus again $S$ has only one cut.

\paragraph{Fixing $S$} We will use QQ-relations to deduce $S$.
Starting from \eq{QPQ} we get
in the scaling limit \begin{equation}
f^{[+2]}{\mathbb Q}_{\alpha|\beta}^+-f\,{\mathbb Q}_{\alpha|\beta}^-\propto
R_{\alpha|\emptyset}\,B_{\alpha|12}
\,R^{}_{\emptyset|\beta}\,B^{}_{12|\beta}\frac{f^{[+2]}}{B_{(-)}}\,.
\end{equation}
Dividing by \(f^{[+2]}/B_{(-)}\) we get
\begin{equation}\label{fermionic0}
{{\mathbb Q}_{\alpha|\beta}^+ B_{(-)}-{\mathbb Q}_{\alpha|\beta}^- B_{(+)}\propto
R_{\alpha|\emptyset}B_{\alpha|12}
R^{}_{\emptyset|\beta}B^{}_{12|\beta}}S\sigma x^{-L/2}\;.
\end{equation}
We see from the last identity that the combination
$S\sigma$
must be a rational function of $x$.
The potential poles of this function could occur only
due to the zeros of $B_{\alpha|12}$ and $B^{}_{12|\beta}$,
since on the main sheet there are  for sure no poles, because of \eqref{QAlower} and the definition of $\sigma$. We will show now that the poles are also impossible at zeros of $B_{\alpha|12}$ and $B^{}_{12|\beta}$.

First, suppose there is a pole in $S$, on the second sheet, from a zero of $B^{}_{12|\beta}$. But than this pole and zero will simply cancel each other in \eqref{QAlower}. Hence $\bQ_{\beta}$ has no zero at the root of $B^{}_{12|\beta}$, which contradicts the very definition of $B^{}_{12|\beta}$.

Second, suppose there is a pole from a zero of $B_{\alpha|12}$. In subsection \eqref{subsec:QQdualities} we will construct $\fQ_{12|\alpha}$. From the way it is constructed it is clear that it is proportional to $R_{12|\alpha}\tilde S$, and these two terms are the only ones that can lead to zeros and poles of $\fQ_{12|\alpha}$. Hence such a pole in $S$ becomes a pole of a Q-function on the main sheet, which is impossible by our regularity assumption.

Thus $S\sigma$ could only have singularities at infinity
which we can fix  from the asymptotics to be
\beq
S=\frac{x^{L/2}}{\sigma}\;.
\eeq

\subsection{Exploring the results}
In the previous section we have found various Q-functions with one common property -- they are all of the form $\fQ_{A|I}$ where $A,I$ are multi-indices from $\{1,2\}$, hence these Q-functions form an $\su(2|2)$ Q-system. The distinguished role of the $\su(2|2)$ sub-algebra in the asymptotic limit of the AdS/CFT spectrum is well-known; for instance, it was extensively used in the approach of factorized scattering and nested Bethe ansatz \cite{Staudacher:2004tk,Beisert:2005tm}.

In this subsection, we complete the derivation of all still missing
asymptotic $\su(2|2)$ Q-functions, in addition to those derived above, and
demonstrate the dualities between them based on QQ-relations. Then we restore the whole set of BES ABA equations together with the expression for the energy of a state and the cyclicity condition.
\subsubsection{\texorpdfstring{$\su(2|2)$}{su(2|2)} Q-functions explicitly}
\label{sec:qfunctionsexplicitly}
We summarize the derived explicit QSC solution and also add the missing $\su(2|2)$ Q-functions $\fQ_{12|\alpha}$:
\begin{align}\label{su22Qsummary}
\bP_{\alpha}\equiv \fQ_{\alpha|\emptyset}&\propto x^{-\frac{L}2}\, R_{\alpha|\emptyset}\, B_{\alpha|12}\,{\sigma^{+1}}\,,\ \
&
\fQ_{\alpha|12}&\propto x^{+\frac L2}\, B_{\alpha|\emptyset}\, R_{\alpha|12}\,{\sigma^{-1}}\,f\,f^{[+2]}\,,
\no\\
\bQ_{\alpha}\equiv \fQ_{\emptyset|\alpha}&\propto x^{+\frac L2}\, R_{\emptyset|a}{}\, B_{12|a}\,{\sigma^{-1}}\,\frac{f^{[+2]}}{{}
B_{(-)}}\,, \
&
\fQ_{12|\alpha}&\propto{} x^{-\frac L2}\, B_{\emptyset|a}{}\, R_{12|\alpha}\,{\sigma^{+1}}\,f^{[+2]}{}\,B_{(+)}\,,
\no\\
\fQ_{\alpha|\beta}&\propto{\mathbb Q}_{\alpha|\beta}\,f^+\,,\ \
&
\fQ_{12|12}&\propto\mathbb Q\,(f^+)^2\,,
\end{align}
the derivation of $\fQ_{12|\alpha}$ will be done in the next subsection.

In these expressions, $R$'s and $B$'s are the above-defined polynomials in Zhukovsky variable $x$, and $\wQ$'s are the polynomials in the spectral parameter $u$.
The only functions which contain ladders of Zhukovsky cuts are $f$ and $\sigma$. They are defined
in \eq{f-explicit}, \eq{sigmadef} and \eq{reltobes}.
The relevant functions $\omega$ and $\mu$ are found in \eqref{muomegaexplicitly} and can be rewritten simply as:
\begin{equation}
\omega^{12}=\frac{\bar f^{[-2]}}f\,,\ \ \  \mu_{12}=f{\bar
f^{[-2]}}\,{\mathbb Q}^-\,.
\end{equation}

There is also the second (left) \(\su(2|2)\) Q-system which is treated absolutely in the same way and can be written in the full analogy in the Hodge-dual notations. For $\dot\alpha,\dot\beta\in\{3,4\}$: one has
\begin{align}
\bP^{\dot\alpha}\equiv\fQ^{\dot\alpha|\emptyset}&\propto x^{-\frac L2}{} R^{\dot\alpha|\emptyset}{} B^{\dot\alpha|34}\,{\sigma^{+1}}\,,
&
\fQ^{\dot\alpha|34}&\propto x^{+\frac L2}{} B^{\dot\alpha|\emptyset}{} R^{\dot\alpha|34}\,{\sigma^{-1}}f\,f^{[2]}\,,
\no\\
\bQ^{\do\alpha}\propto\fQ^{\emptyset|\dot\alpha}&\propto x^{+\frac L2}{} R^{\emptyset|\dot\alpha}{} B^{34|\dot\alpha}\,{\sigma^{-1}}\frac{f^{[+2]}}{{}
B_{(-)}}\,,
&
\fQ^{34|\dot\a}&\propto x^{-\frac L2}{} B^{\emptyset|\dot\alpha}{} R^{34|\dot\alpha}\,{\sigma^{+1}}f^{[+2]}{}
B_{(+)}\,,
\no\\
\fQ^{\dot\alpha|\dot\beta}&\propto{\mathbb Q}^{\dot\alpha|\dot\beta}f^+\,,
&
\fQ^{34|34}&\propto\mathbb
Q\,(f^+)^2\,.
\end{align}
The two $\su(2|2)$ Q-systems are actually interrelated through the central Q-function:
\be
\fQ_{12|12}=\fQ^{34|34}\,.
\ee

\subsubsection{QQ-relations and dualities}\label{subsec:QQdualities}
The Q-functions should satisfy various QQ relations which impose constraints on  possible zeros of Zhukovsky-Baxter polynomials $R$, $B$, and $Q$. In the context of ABA, these constraints are also known as dualities among the Bethe roots.

Let us demonstrate how to use these dualities to fix both the structure of $\fQ_{12|\alpha}$ and the position of its zeros. Consider the defining fermionic QQ-relation  \eq{QPQ}  written for two particular instances:
\be
&&\fQ_{\alpha|\beta}^+-\fQ_{\alpha|\beta}^-=\fQ_{\alpha|\es}\fQ_{\es|\beta}\,,
\no\\
&&\fQ_{\alpha|\beta}^+\fQ_{12|12}^--\fQ_{\alpha|\beta}^-\fQ_{12|12}^+=\fQ_{\alpha|12}\fQ_{12|\beta}\,.
\ee
The first of these relations leads to \eq{fermionic0} which can be written as
\begin{equation}\label{fermionic}
{{\mathbb Q}_{\alpha|\beta}^+ B_{(-)}-{\mathbb Q}_{\alpha|\beta}^- B_{(+)}\propto
R_{\alpha|\emptyset}B_{\alpha|12}
R_{\emptyset|\beta}B_{12|\beta}}\;,
\end{equation}
The second one leads to
\be\label{fermionic22}
{\mathbb Q}_{\alpha|\beta}^+ R_{(-)}-{\mathbb Q}_{\alpha|\beta}^- R_{(+)}\propto
B_{\alpha|\emptyset}R_{\alpha|12}\,\left(\frac{x^{L/2}\fQ_{12|\alpha}}{\sigma\,f^{[+2]}B_{(+)}}\right)\,,
\ee
where we used explicit expressions from \eqref{su22Qsummary},  $\frac f{f^{[+2]}}=\frac{B_{(+)}}{B_{(-)}}$, and $\wQ^\pm\propto B_{(\pm)}R_{(\pm)}$.

Now we note that the tilde applied to the l.h.s. of \eqref{fermionic} produces the l.h.s. of \eqref{fermionic22}. Hence the same should be true for the r.h.s. of these equations, which defines for us $\fQ_{12|\alpha}$ precisely as in  \eqref{su22Qsummary}.

\begin{figure}
  \centering
\includegraphics[scale=0.25]{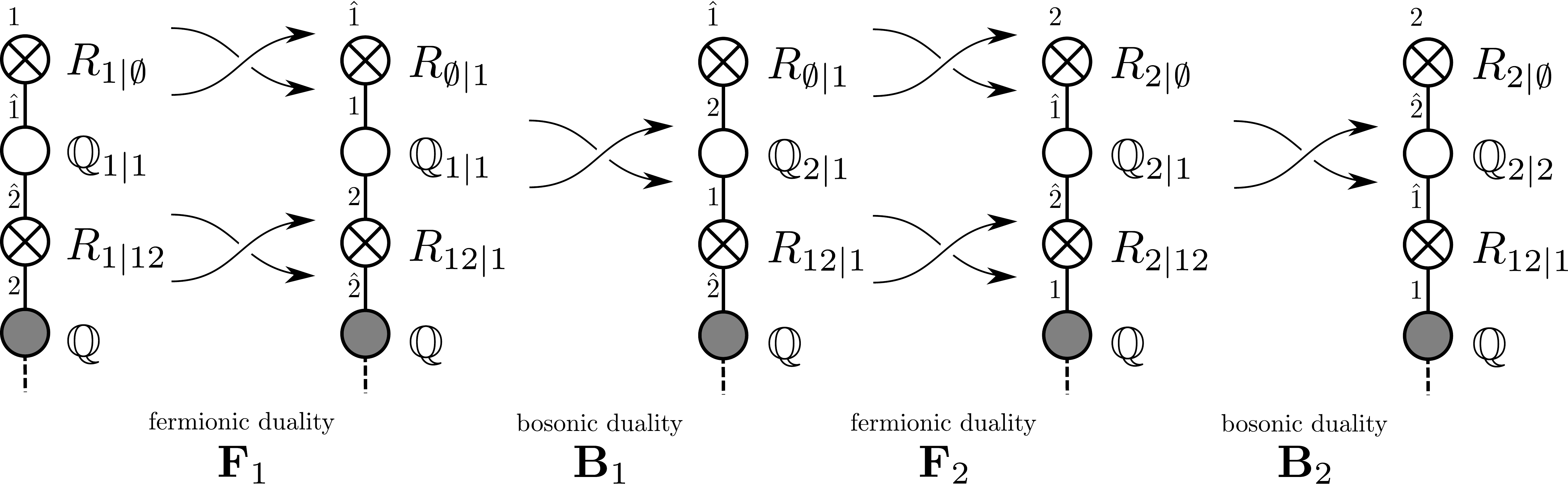}
  \caption{\textbf{Chain of dualities}:
  ABA for one of the wings exists in \(8\) different equivalent variants related by a chain of fermionic and bosonic dualities \cite{Beisert:2006ez,Pronko:1998xa,Gromov:2007ky}
  some of them are depicted on the picture. In the classical limit each of these possibilities corresponds to a particular ordering of the sheets
  denoted by \(1,2\) for the sheets in \(S^5\) and \(\hat 1,\hat 2\) for the sheets in \(AdS_5\).}
  \label{fig:mdual}
\end{figure}

The \equref{fermionic}, together with its tilde,  \eqref{fermionic22}, is nothing but the fermionic duality relation \cite{Beisert:2005fw}.
The dualities \({\bf F}_1\) and \({\bf F}_2\) in  \figref{fig:mdual} are two examples of it.

An example of the bosonic duality\cite{Pronko:1998xa}, like \({\bf B}_1\), follows from the QQ-relation $\fQ_{1|\alpha}^+
\fQ_{2|\alpha}^-
-
\fQ_{1|\alpha}^-
\fQ_{2|\alpha}^+
= \fQ_{\es|\alpha}\fQ_{12|\alpha}$
which leads in the large volume limit to
\begin{equation}\label{bosonic2}
{\mathbb Q}_{1|\alpha}^+
{\mathbb Q}_{2|\alpha}^-
-
{\mathbb Q}_{1|\alpha}^-
{\mathbb Q}_{2|\alpha}^+
\propto {\mathbb Q}_{\es|\alpha}{\mathbb Q}_{12|\alpha}\;.
\end{equation}
\subsubsection{Asymptotic Bethe ansatz equations}
Asymptotic Bethe Ansatz equations are  algebraic equations on zeros of Baxter polynomials.
An important feature of  ABA equations is that they contain explicitly only those Baxter polynomials
which belong to a given path on the Hasse diagram.  Each path is in one-to-one correspondence with the ordering of  sheets of the classical algebraic curve.
There are \(4\) sheets in AdS\(_5\) which we denote by numbers \(\hat 1,\hat 2,\hat 3,\hat 4\) and another \(4\) sheets in S\(^5\)
denoted as \(1,2,3,4\). From the point of view  of representation theory, it  is possible to choose different Borel subalgebras to describe the same highest weight representation.

The orderings of sheets which do not spoil the polynomial  (apart from the dressing factor) nature of Beisert-Staudacher equations,
are those where \(1,2,\hat 1,\hat 2\) are ordered before \(3,4,\hat 3,\hat 4\);
moreover, only the following two patterns are allowed \(*\hat *\hat * *\) or
\(\hat *\!*\!* \hat *\) for the first \(4\) and for the last \(4\) sheets. In total there are \(64\) possibilities
and, correspondingly, there are \(64\) versions of ABA equations related to each other via duality transformations.
 We discussed these duality transformations in the previous section, and here we pick one canonical choice
 \(1\hat1  \hat2 23 \hat3 \hat 4 4\). It corresponds to the ABA-type Kac-Dynkin-Vogan diagram in \figref{fig:abadiagram} of \appref{app:unitary} and to the standard, with respect to this diagram, choice of the Borel subalgebra.

We will obtain the set of nested Bethe ansatz equations from the QQ-relations, in a way similar to the case of Heisenberg super-spin chains \cite{Kazakov:2007fy}, using certain analytic properties of the involved Q-functions: in this case it will be their polynomiality (or sometimes "Zhukovsky polynomiality" in the current model, apart from $\sigma$ and $f$). The  choice
\(1\hat1  \hat2 23 \hat3 \hat 4 4\) means that we have to relate to each other only the Q-functions corresponding to this specific path on the Hasse diagram.
These relevant Q-functions are
\begin{equation}\label{relevQ}
{\cal Q}_{\emptyset|1},\;
{\cal Q}_{1|1},\;
{\cal Q}_{12|1},\;
{\cal Q}_{12|12}={\cal Q}^{34|34},\;
{\cal Q}^{34|4},\;
{\cal Q}^{4|4},\;
{\cal Q}^{4|\emptyset}\;.
\end{equation}
We see from here that we have to relate by Bethe equations the following Baxter-Zhukovsky polynomials
\begin{equation}
R_{\emptyset|1},\;
{\mathbb Q}_{1|1},\;
R_{12|1},\;
{\mathbb Q},\;R^{34|4},\;{\mathbb Q}^{4|4},\;R^{\emptyset|4}\;.
\end{equation}

Let us  denote the corresponding roots as \(u_{1,k},\;u_{2,k},\dots,\;u_{7,k}\). Note that $u_{4,k}= u_k$,  this is however true only in the asymptotic limit.
We start by taking \eq{fermionic} for \(\alpha=1,\beta=1\), and setting \(u=u_k^{\emptyset|1}\) which will set to zero the r.h.s.,
and we get
\begin{subequations}
\label{eq:BSa}
\begin{equation}
\boxed{\frac{{\mathbb Q}_{1|1}\left(u_{1,k}+\tfrac{i}{2}\right)B_{(-)}\left(u_{1,k}\right)}
{
{\mathbb Q}_{1|1}\left(u_{1,k}-\tfrac{i}{2}\right)B_{(+)}\left(u_{1,k}\right)
}=1\;\;,\;\;\forall k\;.}
\end{equation}
Similarly, we analytically continue \eq{fermionic} for \(A=1,B=1\) to another sheet, by replacing \(R\leftrightarrow B\)
and evaluate it at \(u=u_k^{12|1}\) so that again the r.h.s. vanishes and we get
\begin{equation}
\boxed{\frac{{\mathbb Q}_{1|1}\left(u_{3,k}+\tfrac{i}{2}\right)R_{(-)}\left(u_{3,k}\right)}
{
{\mathbb Q}_{1|1}\left(u_{3,k}-\tfrac{i}{2}\right)R_{(+)}\left(u_{3,k}\right)
}=1\;\;,\;\;\forall k\;.}
\end{equation}

To get an equation for \(u_{2,k}\) we use \eq{bosonic2} with $\alpha=1$. Evaluating first at \(u=u_{2,k}-\frac{i}{2}\)
we get rid of the first term as \({\mathbb Q}_{1|1}(u_{2,k})\equiv 0\).
Dividing the result by the same equation evaluated at \(u=u_{2,k}+\frac{i}{2}\) we get the following Bethe equation
\begin{equation}\label{Betheu2}
\boxed{
-\frac{
{\mathbb Q}_{1|1}(u_{2,k}-i)
}{
{\mathbb Q}_{1|1}(u_{2,k}+i)
}
=
\frac{{\mathbb Q}_{1|\emptyset}(u_{2,k}-\tfrac{i}{2}){\mathbb Q}_{12|1}(u_{2,k}-\tfrac{i}{2})}
{
{\mathbb Q}_{1|\emptyset}(u_{2,k}+\tfrac{i}{2}){\mathbb Q}_{12|1}(u_{2,k}+\tfrac{i}{2})
}\;\;,\;\;\forall k\;.}
\end{equation}
\end{subequations}

\paragraph{Middle node equation.}
This equation explicitly contains \(x^L\)
and thus its derivation might require the knowledge of the next order in \(\epsilon\) expansion. Fortunately, we can avoid this difficulty. Consider a special instance of the exact QQ relation \eq{QQff}:
\begin{equation}\label{tmp1}
\cQ_{12|1}\cQ_{12|123}=\cQ_{12|12}^+\cQ_{12|13}^--\cQ_{12|12}^-\cQ_{12|13}^+\;.
\end{equation}
Note now that $\cQ_{12|123}=\cQ^{34|4}$, so the only function that we do not know in the large volume approximation is $\fQ_{12|13}$. We, however, do not need to know it explicitly. Similar to derivation of \eqref{Betheu2}, we rewrite \eqref{tmp1} twice, shifted by $+i/2$ and $-i/2$,
\begin{eqnarray}\label{shiftedtmp}
\cQ_{12|1}^+(\cQ^{34|4})^+&=\cQ_{12|12}^{[+2]}\,\cQ_{12|13}-\cQ_{12|12}\,\cQ_{12|13}^{[+2]}\;,
\no\\
\cQ_{12|1}^-(\cQ^{34|4})^-&=\cQ_{12|12}\,\cQ_{12|13}^{[-2]}-\cQ_{12|12}^{[-2]}\,\cQ_{12|13}\;,
\end{eqnarray} and evaluate each of the equations at $u=u_j$, so that the terms proportional to $\fQ_{12|12}$ cancel out. Then we divide the obtained results to get
\begin{equation}\label{almostexactBethe}
-1=\frac{\cQ_{12|12}^{[+2]}\cQ_{12|1}^-(\cQ^{34|4})^-}{\cQ_{12|12}^{[-2]}\cQ_{12|1}^+(\cQ^{34|4})^+}
\end{equation}
for $u=u_j$. It remains to substitute the explicit expression for Q-functions in the asymptotic limit and to use the properties of $\sigma$ and $f$ to get
\begin{equation}\label{eq:BSm}
\frac{\wQ^{[+2]}}
{\wQ^{[-2]}}\left(\frac{B_{(-)}^-}{B_{(+)}^+}\right)^2\prod_{k=1}^N\sigma_{\rm BES}^{-2}(u,u_k)\times\frac{ {B}^-_{\emptyset|1}{R}^-_{12|1} {B}^{\emptyset|4-}{R}^{34|4-}}{ {B}^+_{\emptyset|1}{R}^+_{12|1} {B}^{\emptyset|4+}{R}^{34|4+}}=-\left(\frac{x^-}{x^+}\right)^{L}\,\ \ {\rm for}\ \ \ u=u_j\,,\quad j=1,\dots,N\,,
\end{equation}
which is the middle-node Beisert-Staudacher equation with the BES dressing phase.

In the weak coupling limit the Bethe equations simplify to Bethe equations for a $\psu(2,2|4)$ rational spin chain in which the parameter $L$  plays the role of the length of the chain. Moreover, we can deduce from the below-derived relations \eqref{powersinHasse}  that  $L$ is constrained to the range 
\be\label{ineqnew}
2J_1-\Delta+S_1+|S_2|\leq L\leq \Delta-S_1-|S_2|\,,
\ee
 hence the limit of large charges also typically means large length.
\subsubsection{Equation for the energy and cyclicity condition}
The simplest way to deduce the energy is from the asymptotics of \(\mu_{12}\sim u^{\Delta-J_1}\), or even better, using
\begin{equation}
\log\frac{\mu_{12}^{[+2]}}{\mu_{12}}=
\log\frac{\tilde\mu_{12}}{\mu_{12}}=
\log\frac{R_{(+)}B_{(-)}}{B_{(+)}R_{(-)}}\sim \frac{i(\Delta-J_1)}{u}\;\;,\;\;u\to\infty\,.
\end{equation}
At the same time
\begin{equation}
\log\frac{R_{(+)}B_{(-)}}{B_{(+)}R_{(-)}}\simeq \sum_{k=1}^N\log\frac{x^+_k}{x^-_k}
+\frac{i}{u}\sum_{k=1}^N\left(1+\frac{2gi}{x_k^+}-\frac{2gi}{x_k^-}\right)\,.
\end{equation}
The first term must be zero, which gives the trace cyclicity (zero total momentum) condition \begin{eqnarray}\label{cyclicitycond}
\sum_{k=1}^N\log\frac{x^+_k}{x^-_k}=0
\end{eqnarray}  and the second term gives  the familiar expression for the energy \cite{Santambrogio:2002sb}
\begin{eqnarray}
\;\;\Delta=J_1+\label{eq:enABA}
\sum_{k=1}^N\left(1+\frac{2gi}{x_k^+}-\frac{2gi}{x_k^-}\right)\;.
\end{eqnarray}
We hence accomplished derivation of the Beisert-Staudacher equation -- the auxiliary equations  \eqref{eq:BSa} and the middle node equation \eqref{eq:BSm}) -- as well as of the dispersion relation \eqref{eq:enABA} from the quantum spectral curve in the asymptotic limit. It still remains to prove that our main assumption about scalings \eqref{eq:scaQP} and \eqref{eq:scamu} is self-consistent, i.e. that the corrections to the obtained expressions in this section are indeed \(\epsilon\)-small. We postpone this question for further works.

\subsection{Numbers of roots and the conserved global charges}
\label{sec:NRCGC}
It is explicitly known how the number of Bethe roots is related to the global charges \cite{Beisert:2005fw}. Hence, comparison with the Beisert-Staudacher equations is the best way to determine the large $u$ behaviour of Q's in terms of global charges and to confirm our conjecture formulated in \secref{sec:asymptotics-at-large}.  We want to confirm this conjecture {\it at any value} of the charges, not only in the asymptotic large $\Delta\sim J_1$ regime discussed so far in this section. The basic argument is the physical well-known assumption that at sufficiently small coupling the asymptotic approximation is valid even at finite value of charges\footnote{This should be confirmed, in principle, directly from mathematical properties of QSC. However, this involves analysing \(\epsilon\)-corrections which is not performed in this paper.}. Hence our  strategy is to confirm the relation between global charges and the large $u$ behaviour of Q's at sufficiently small coupling when the asymptotic Bethe ansatz approximation is valid and then provide the arguments why this relation should hold even at finite coupling.

The necessary background about the representation theory and explanations about notations we use is given in \appref{app:unitary}.

Denote by $K_\alpha$ the number of Bethe roots $u_{\alpha,k}$ for $\alpha=1,2,\ldots, 7$. From \cite{Beisert:2005fw}, formula (5.3) there, the explicit relation to the $\su(2,2)$ and $\su(4)$ Dynkin labels is known. Respectively:
\begin{subequations}\label{dyn1}
\be
q_1&=&-K_1-K_3+K_4\,,
\\
q_2+\gamma&=&-L+K_3-2\,K_4+K_5\,,
\\
q_3&=&-K_7-K_5+K_4\,,
\ee
\end{subequations}
where $\gamma=\sum_{k=1}^N\left(\frac{2gi}{x_k^+}-\frac{2gi}{x_k^-}\right)$, and
\begin{subequations}\label{dyn2}
\be
r_1&=&K_1-2K_2+K_3\,,
\\
r_2&=&L+K_2-K_3-K_5+K_6\,,
\\
r_3&=&K_7-K_6+K_5\,.
\ee
\end{subequations}
We used that our choice of Kac-Dynkin-Vogan diagram corresponds to $\eta_1=\eta_2=-1$ in \cite{Beisert:2005fw}.

Now, on the one hand, the large-$u$ behaviour of Q-functions can be  written in terms of $K_i$ and $\gamma$. As a preparatory work, one should find the large $u$ expansion of $f$. Given that $\frac{f}{f^{[2]}}=\frac{B_{(+)}}{B_{(-)}}$, one gets
\be
\log\frac{f}{f^{[2]}}\simeq\frac 1x\sum_{k=1}^N\left(\frac{1}{x_k^+}-\frac{1}{x_k^-}\right)\simeq \frac{\gamma}{2iu}\,,
\ee
from where one finds $f\simeq u^{\gamma/2}$.

On the other hand, we can recall that, in notations \eqref{tranformulae},    $\bP_a\sim u^{-\hat\lambda_a}$, $\bQ_i\sim u^{-\hat\nu_i-1}$, $\bP^a\sim u^{\hat\lambda_a-1}$, $\bQ^{i}\sim u^{\hat\nu_i}$, and asymptotics for all other Q-functions follow from the QQ relations. Hence we can write the large-$u$ behaviour of Q-functions in terms of $\hat\lambda$'s and $\hat\nu$'s.

Therefore we can compare the two ways of computing the large-$u$ behaviour and fix in this way, using also \eqref{dyn1} and \eqref{dyn2},  the values of $\hat\lambda$'s and $\hat\nu$'s in terms of Dynkin labels. To that end, we compute the large-$u$ behaviour of Q-functions along the Hasse diagram, i.e. those listed in \eqref{relevQ}:
\begin{align}\label{powersinHasse}
-1-\hat\nu_1&=K_1+\frac \gamma 2+\frac L2\,,
\no\\
-\hat\lambda_1-\hat\nu_1&=K_2+\frac \gamma 2\,,
\no\\
-\hat\lambda_1-\hat\lambda_2-\hat\nu_1&=K_3+\frac \gamma 2-\frac L2\,,
\no\\
-\hat\lambda_1-\hat\lambda_2-\hat\nu_1-\hat\nu_2&=K_4+\gamma\,,
\no\\
-1+\hat\lambda_3+\hat\lambda_4+\hat\nu_4&=K_5+\frac \gamma 2-\frac L2\,,
\no\\
+\hat\lambda_4+\hat\nu_4&=K_6+\frac \gamma 2\,,
\no\\
+\hat\nu_4&=K_7+\frac \gamma 2+\frac L2\,.
\end{align}
On the l.h.s. we wrote the powers of large-$u$ asymptotics of Q-functions as it follows from \eqref{largeu2},\eqref{M-ass}. On the r.h.s. we wrote the same powers as it follows from the explicit solution \eqref{su22Qsummary}.

Expressing $\hat\lambda$ and $\hat\nu$ (shifted weights) in terms of $\lambda$ and $\nu$ (ordinary weights) according to \eqref{muhll}, and using \eqref{powersinHasse} to determine $K_i$, we arrive at conclusion that $r_i=\lambda_i-\lambda_{i+1}$ and $q_i=\nu_i-\nu_{i+1}$, as expected. To get this conclusion, we also used the zero charge condition $\sum\hat\lambda_i+\hat\nu_i=0$.
All the weights can be expressed through 6 Cartan charges \((J_1,J_2,J_3|S_1,S_2,\Delta)\) of \(\psu(2,2|4)\) by the  formulae   \eqref{tranformulae}.

Hence we confirmed that the large-$u$ asymptotic of Q-functions is defined by the global charges, based on the comparison with ABA. This comparison is valid only at sufficiently small coupling constant. For the finite coupling case, our arguments are as follows. In the representation theory, the charges $J_1,J_2,J_3,S_1,S_2$ are quantized and hence do not depend on the coupling constant. In \appref{app:unitarityfromanalyticity} we prove, using only analytic properties of QSC, that the corresponding powers in the large-$u$ asymptotics of Q-functions are also quantized and hence also do not depend on the coupling constant. Therefore, the established equivalence for these 5 charges is exact. For the remaining charge, the conformal dimension $\Delta$, we are able to show that this quantity, when defined from \(\mu_{12}\sim u^{\Delta-J_1}\), is the same as the  one defined in TBA. This is done in \appref{subsec:Delta}, which accomplishes our proof for identification of the global charges and large-$u$ asymptotics of Q-functions at finite coupling.

Also, let us note that global charges should satisfy the unitarity constraints. In \appref{app:unitary} we show that these constraints follow mostly from the analytic structure of QSC, which is another solid support for the proposed link between the asymptotics of Q-functions and the group-theoretical data.

\section{Quasi-classical approximation}
\label{sec:QCA}
In this section we discuss the classical limit of our construction. We compute various quantities in this limit and speculate about their physical meaning. We also establish various links to some results known in the literature.

The quasi-classical approximation in  Quantum Mechanics applies
in the limit \(\hbar\to 0\) when the quantum numbers of the state are large and scale as \(1/\hbar\).
In this case one can approximate the wave function \(\psi(x)\) by
\begin{equation}
\psi(x)\sim \exp\left({-\frac i\hbar\int^x p^{\rm cl}(z) dz}\right)\,,
\end{equation}
where \(p^{\rm cl}(z)\) is the classical momentum of the particle as a function of its coordinate.

Very similarly, in the Metsaev-Tseytlin sigma model the role of  \(\hbar\) is played by \(1/g\)
and, as we argue in this section, the role of the wave function in the above example is played by \(\bP_a\) and \(\bQ_i\).
More precisely, we will see that
\begin{eqnarray}\la{qclp}
{\bf P}_a= {\cal P}_{a|\emptyset}(u) \exp\left(-g\int_{0+i0}^{u/g} p_{\tilde a}(z)dz\right)\;\;,\;\;
{\bf Q}_i= {\cal P}_{\emptyset|i}(u) \exp\left(+g\int_{0+i0}^{u/g} p_{\hat i}(z)dz\right)\;.
\end{eqnarray}
where $p_{\tilde a}$ and $p_{\hat i}$
are the classical quasimomenta which are defined so that
$(e^{ip_{\tilde 1}},\dots,e^{ip_{\tilde 4}}|e^{ip_{\hat 1}},\dots,e^{ip_{\hat 4}})$ are eigenvalues of the classical $(4+4)\times(4+4)$
monodromy super-matrix. We also defined finite pre-exponents \({\cal P}_{a|\emptyset}(u)\) and \({\cal P}_{\emptyset|i}(u)\) as we are going to reconstruct
some partial information about them soon.

To see where \eq{qclp} comes from one can simply use our results in the large $L$ asymptotic (ABA) limit from \secref{sec:ABA}.
It is known that the ABA correctly reproduces the classical limit, as the wrapping corrections, not captured by ABA,
become relevant only at one loop.
Thus simply taking \eq{QAlower} we find
\begin{align}\la{qoq}
  \frac{{\bf Q}^+_1}{{\bf
      Q}^-_1}=&  \left(\frac{\hat x^+}{\hat x^-}\right)^{L/2}\!\!\!\!\frac{1}{\sigma_{\rm
      BES}} \frac{R_{\emptyset|1}^+}{R_{\emptyset|1}^-}
  \frac{B_{(-)}^-}{B_{(+)}^+}\frac{B_{12|1}^+}{B_{12|1}^-}
  \simeq e^{ i \left(\frac{4\pi {\cal J} x}{x^2-1}+\bar H_{3}+H_1-\bar H_4+\frac{{\cal Q}_2 x}{x^2-1}\right)}
  = e^{{ i p_{\hat 1}}}\,,
\end{align}
where $H_a$ and $\bar H_a$ are resolvents and ${\cal Q}_2$\footnote{not to be confused with Q-functions}
is a local conserved charge (energy) defined in terms of the Bethe roots by
\beqa
H_a(x)=\sum_j\frac{4\pi}{\sqrt\lambda}\frac{x^2}{x^2-1}\frac{1}{x-x(u_{a,j})}\;\;,\;\;\bar H_a(x)\equiv H_a(1/x)\;\;,\;\;
{\cal Q}_n=\sum_j\frac{4\pi}{\sqrt\lambda}\frac{x^{2-n}(u_{4,j})}{x^2(u_{4,j})-1}\;.
\eeqa
The last equality in \eq{qoq} is obtained by comparing with the expressions for the classical quasimomenta
in terms of the Bethe roots known from \cite{Beisert:2005fw}.

The natural variable for the classical limit is
$z=u/g$. Taking log of $\eq{qoq}$ and expanding the l.h.s. in $g\to\infty$ limit (with $z$ fixed) we get
\beq
\frac{i}{g}\partial_z\log \bQ_1= i p_{\hat 1}
\eeq
which leads to one of the relations \eq{qclp}.
 The rest of them can be obtained similarly, thus confirming the quasi-classical approximation \eq{qclp}.

In the following, we will discuss several properties and extend our intuition about QSC to the quasi-classical limit. We will show how the monodromy data of the QSC naturally fits the analytic continuation of the quasi-momenta, and hence  the $\mathbb Z_4$ symmetry of the coset sigma-model. We will also derive the expressions for T-functions in the character limit which are already known in the literature  \cite{Gromov:2009tq}. For that, in particular, we will get further information about the pre-exponents \({\cal P}\) and hence gain a certain bit of knowledge about QSC at one loop.

\subsection{Constraining pre-exponents}
To constrain the pre-exponents we use like before  the fact that \(\bP_a\) and \(\bQ_j\) are not completely independent, but rather related to each other
by \({\cal Q}_{a|j}\) defined by \eq{QPQ}.
From that equation we see that \({\cal Q}_{a|j}\) in our limit should scale as \(\sim \bP_a\bQ_j\) so we define
\begin{equation}
{\cal Q}_{a|j}={\cal P}_{a|j}(u)\exp\left(-g\int_{+0}^{u/g} \left[p_{\tilde a}(z)-p_{\hat j}(z)\right]dz\right)\;.
\end{equation}
Next, we relate ${\cal P}_{a|j}(u)$ to ${\cal P}_{\emptyset|j}(u)$ and  ${\cal P}_{a|\emptyset}(u)$ using \eq{QPQ}. For that we have to find
the difference ${\cal Q}_{a|j}^+-{\cal Q}_{a|j}^-$ which can be easily evaluated for large $g$ to be
\begin{equation}
{\cal Q}_{a|j}^+-{\cal Q}_{a|j}^-\simeq
{\cal P}_{a|j}
\exp\left(-g\int_{+0}^{u/g} \left[p_{\tilde a}(z)-p_{\hat j}(z)\right]dz\right)
\left(
e^{-\frac{i}{2} p_{\tilde a}+\frac{i}{2}p_{\hat j}}
-
e^{+\frac{i}{2} p_{\tilde a}-\frac{i}{2}p_{\hat j}}
\right)\,.
\end{equation}
It is convenient to introduce notation $x_a\equiv e^{-ip_{\tilde a}}\;,\;
y_i\equiv e^{-ip_{\hat i}}$. Next, comparing with the r.h.s. of \eq{QPQ} we see that
\begin{equation}\la{Pab}
{\cal P}_{a|i}=\frac{{\cal P}_{a|\emptyset}{\cal P}_{\emptyset|i}}{\frac{\sqrt{x_a}}{\sqrt{y_i}}-\frac{\sqrt{y_i}}{\sqrt{x_a}}}\;.
\end{equation}
At the same time, from \eq{QP} and \eq{QP2}:
\begin{equation}\la{QmPQ}
\bQ_i=-\bP^a{\cal Q}_{a|i}^+\;\;,\qquad\;\;\bP_a=-\bQ^i{\cal Q}_{a|i}^+\;,
\end{equation}
from where we see that in analogy with \eq{qclp} we  should have for the upper indices
\begin{eqnarray}\la{qclp2}
{\bf P}^a= {\cal P}^{a|\emptyset}(u) \exp\left(+g\int_{+0}^{u/g} p_{\tilde a}(z)dz\right)\;\;,\qquad\;\;{\bf Q}^i= {\cal P}^{\emptyset|i}(u) \exp\left(-g\int_{+0}^{u/g} p_{\hat i}(z)dz\right)\;
\end{eqnarray}
In addition, due to \eq{QmPQ} the pre-exponents are constrained by
\begin{equation}
{\cal P}_{\emptyset|i}=-{\cal P}^{a|\emptyset}{\cal P}_{a|i}\frac{\sqrt{x_a}}{\sqrt{y_i}}\;\;,\;\;
{\cal P}_{a|\emptyset}=-{\cal P}^{\emptyset|i}{\cal P}_{a|i}\frac{\sqrt{x_a}}{\sqrt{y_i}}.
\end{equation}
In combination with \eq{Pab} this gives
\begin{subequations}
  \begin{align}
    \label{eq:23}
1=&-\sum_a\frac{{\cal P}^{a|\emptyset}{\cal P}_{a|\emptyset}}
{1-y_i/x_a}\,,&i=&1,\dots,4\,,\\
1=&-\sum_a\frac{{\cal P}^{\emptyset|i}{\cal P}_{\emptyset|i}}
{1-y_i/x_a}\,,&a=&1,\dots,4\,.
  \end{align}
\end{subequations}
These equations, considered as a set of linear equations on \({\cal P}^{a_0|\emptyset}{\cal P}_{a_0|\emptyset}\) and
 \({\cal P}^{\emptyset|i_0}{\cal P}_{\emptyset|i_0}\) give
 \begin{subequations}\la{PPrel}
   \begin{align}
     \label{eq:24}
{\cal P}^{a_0|\emptyset}{\cal P}_{a_0|\emptyset}=&-\frac{\prod_{i}(x_{a_0}-y_i)}{x_{a_0}\prod_{b\neq a_0}(x_{a_0}-x_b)}\,\,,&a_0=&1,\dots,4\,,\\
{\cal P}^{\emptyset|i_0}{\cal P}_{\emptyset|i_0}=&+\frac{\prod_{a}(y_{i_0}-x_a)}{y_{i_0}\prod_{j\neq i_0}(y_{i_0}-y_j)}\,\,,&i_0=&1,\dots,4\,,
   \end{align}
 \end{subequations}
where we use the unimodularity condition of the classical monodromy matrix  which in our notations reads as
$\la{unicl}
\prod_{a=1}^4 \frac{x_a}{y_a}=1
$.
We can also check that due to \eq{PPrel} one has \(\sum_a {\cal P}^{a|\emptyset}{\cal P}_{a|\emptyset}=\sum_i {\cal P}^{\emptyset|i}{\cal P}_{\emptyset|i}=0\)
as it should be from \(\bP_a\bP^a=\bQ_i\bQ^i=0\) (see \eq{Qomegafinal},\eq{Pmufinal}).

\subsection{Quasi-classical limit of the discontinuity relations}
\begin{figure}
  \centering
{\includegraphics{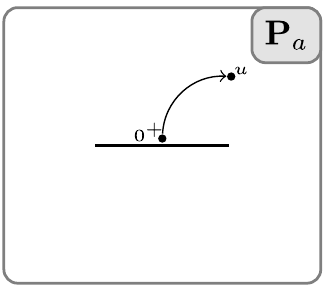}~\includegraphics{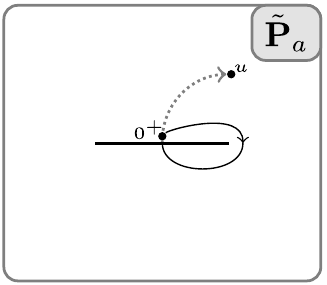}}
 \caption{Integration contour to define $\bP_a$ and $\tilde\bP_a$ from \eqref{qclp}.}
 \label{fig:Pcc}
\end{figure}

Here we briefly discuss the classical limit of the discontinuity relations
\eq{Pmufinal}  and their relation to the ${\mathbb Z}_4$ symmetry of the coset model.
To that end, consider \(\mu_{ab}\). In terms of \(z\) the branch points are fixed at \(\pm 2+n/g\) and, quasi-classically, all of them are squeezed to the points \(\pm 2\).
At the same time, \(\mu_{ab}(z)\) is a periodic function with \(i/g\) period which goes to zero. Assuming \(\mu_{ab}\) has a sensible quasi-classical limit
we see that in the domain \(-2g<\Re u<2g\) \(\mu_{ab}\) must be simply a constant.
This simple observation leads to the essential simplification of the discontinuity relation of $\bP_a$ \eq{Pmufinal}, which we repeat here for convenience:
\begin{eqnarray}
&&\bPt_1=\mu_{12}\bP^2+\mu_{13}\bP^3+\mu_{14}\bP^4\,,\\
&&\bPt_2=\mu_{21}\bP^1+\mu_{23}\bP^3+\mu_{24}\bP^4\;.
\end{eqnarray}
We note that \(\bP^3\) and \(\bP^4\)
are exponentially small compared to \(\bP^2\), so, assuming \(\mu\)'s are all of the same order we get simply
\(
\bPt_1=+\mu_{12}\bP^2,\;\bPt_2=-\mu_{12}\bP^1
\)
where \(\mu_{12}\) is in this limit is just a constant which we set to \(-1\) by a suitable rescaling of \(\bP\)'s.
Thus we simply should have
\begin{equation}\la{discqq}
\bPt_1=-\bP^2\;\;,\;\;\bPt_2=\bP^1\;.
\end{equation}
We can see that this equation is perfectly consistent with \eq{qclp} and \eq{qclp2}. Indeed, to analytically continue
\eq{qclp} under the cut we will have to integrate \(p_{\tilde a}(u)\)
around the cut and then under the cut, as illustrated in \figref{fig:Pcc}. An important property of the classical curve, which is related to \({\mathbb Z}_4\)
automorphism of \(\psu(2,2|4)\) algebra is that the analytic continuation of the quasi-momenta \(p_1(u)\) (or \(p_2(u)\))
is \(-p_2(u)\) (or \(-p_1(u)\)). In particular
\begin{equation}
\tilde{\bf P}_1= \tilde{\cal P}_1(u) \exp\left(-\int_{0^+}^{0^-} p_{\tilde 1}(u)du\right)
 \exp\left(+\int_{0^+}^{u} p_{\tilde 2}(u)du\right)\sim \bP^2\;.
\end{equation}
We see that, as a consequence of the ${\mathbb Z}_4$ symmetry of the classical theory reflected in this specific property of the quasi-momenta,
we indeed reproduce correctly the nontrivial leading exponential factor in the discontinuity relation coming from the QSC.

\subsection{Quasi-classical limit of T-functions and characters of monodromy matrix}
In this section we establish a link between the quasi-classical limit of $\bP$ functions, found in this section,
and the quasi-classical limit of T-functions used earlier for one-loop test of TBA/Y-system equations in \cite{Gromov:2009tq,Gromov:2010vb}.
More precisely, we show that the quasi-classical limit of T-functions defined in \eq{Tp1p2},\eq{eq:12} as \begin{eqnarray}\la{rea}
{\mathbb T}_{1,s}(u)&=&
\bP_1(u+\tfrac{is}{2})
\bP_2(u-\tfrac{is}{2})
-
\bP_2(u+\tfrac{is}{2})
\bP_1(u-\tfrac{is}{2})\,,\\
{\mathbb T}_{2,+s}(u)&=&{\mathbb T}^{[+s]}_{1,1}(u){\mathbb T}^{[-s]}_{1,1}(u)
\end{eqnarray}
coincides  with the \(\frak{psu}(2,2|4)\) characters of the classical monodromy Matrix
in ``rectangular'' unitary representations, characterized by two labels $(a,s)$ (see \cite{Gromov:2010vb}).
The characters are explicit functions of eigenvalues of the monodromy Matrix {\ie} \(x_a\) and \(y_a\)
 \cite{Gromov:2010vb}:
\begin{align}\nn
\chi_{1,s}=&
\frac{x_1^{s-1} \left(x_1-y_1\right) \left(x_1-y_2\right) \left(x_1-y_3\right)
   \left(x_1-y_4\right)}{\left(x_1-x_2\right) \left(x_1-x_3\right)
   \left(x_1-x_4\right)}+\frac{x_2^{s-1} \left(x_2-y_1\right) \left(x_2-y_2\right)
   \left(x_2-y_3\right) \left(x_2-y_4\right)}{\left(x_2-x_1\right)
   \left(x_2-x_3\right) \left(x_2-x_4\right)}\,,\\
\chi_{2,s}=&\la{char}
\frac{x_1^{s-2} x_2^{s-2} \left(x_1-y_1\right) \left(x_2-y_1\right)
   \left(x_1-y_2\right) \left(x_2-y_2\right) \left(x_1-y_3\right)
   \left(x_2-y_3\right) \left(x_1-y_4\right)
   \left(x_2-y_4\right)}{\left(x_1-x_3\right) \left(x_2-x_3\right)
   \left(x_1-x_4\right) \left(x_2-x_4\right)}\;.
\end{align}
According to \cite{Gromov:2010vb} the characters in \eq{char} should match  the corresponding T-functions
in the domain of the spectral parameter \(-2<{\rm Re~ } z<+2\) because the initial T-system is formulated in the mirror kinematics. Hence all cuts
should be taken to be long. In the classical limit the cuts merge together (as they are separated by $\sim i/g$) and
make the limit much more complicated outside the region \(-2<{\rm Re~ } z<+2\), where we have to use different analytic expressions.
In addition for definiteness we take $z$ slightly above the real axis or \({\rm Im~ u}=A\)
such that \(g\gg A\gg 1\sim s\). This means that $\bP_2(u-is/2)$ gets analytically continued under its single cut
\begin{equation}
\check{\mathbb T}_{1,+s}(u)=
\bP_1(u+\tfrac{is}{2})
\tilde\bP_2(u-\tfrac{is}{2})
-
\bP_2(u+\tfrac{is}{2})
\tilde\bP_1(u-\tfrac{is}{2})
\end{equation}
which due to \eq{discqq} becomes
\begin{equation}
\check{\mathbb T}_{1,+s}=
\bP_1(u+\tfrac{is}{2})
\bP^1(u-\tfrac{is}{2})
+
\bP_2(u+\tfrac{is}{2})
\bP^2(u-\tfrac{is}{2})\,.
\end{equation}
Using \eq{qclp} and \eq{qclp2} we have \(\bP_a(u+is/2)= \bP_a(u+i0) x_a^{s/2}\)
and \(\bP^a(u+is/2)= \bP^a(u+i0) x_a^{-s/2}\) so that quasi-classically
\begin{equation}
\check{\mathbb T}_{1,+s}(u)=x_1^s{\cal P}_1{\cal P}^1+x_2^s{\cal P}_2{\cal P}^2\;.
\end{equation}
We know explicitly the combinations ${\cal P}_1{\cal P}^1$  in terms of $x_a,\;y_i$ from \eq{PPrel}. Plugging these expressions into the last  equation we
 indeed reproduce  precisely  the character \(-\chi_{1,s}\)\footnote{The minus sign could is due to our gauge choice.}. Similarly, we treat \({\mathbb T}_{2,+s}\) as
\begin{eqnarray}\nn
\check{\mathbb T}_{2,+s}
&=&
\left(\bP_1^{[+s+1]}\bP_2^{[+s-1]}-\bP_2^{[+s+1]}\bP_1^{[+s-1]}\right)
\left({\tilde\bP_1^{[-s+1]}\tilde\bP_2^{[-s-1]}}-\tilde\bP_2^{[-s+1]}\tilde\bP_1^{[-s-1]}\right)\\
&=&-{\cal P}_{1|\emptyset}{\cal P}^{1|\emptyset}
{\cal P}_{2|\emptyset}{\cal P}^{2|\emptyset}
\left(x_1-x_2\right)^2 x_1^{s-1}x_2^{s-1}
\end{eqnarray}
which together with \eq{PPrel} gives precisely the character \(\chi_{2,s}\).

Finally, we recall that it is enough to know $\wT_{1,\pm s}$, $\wT_{2,\pm s}$ and $\mu_{12}$ to restore the whole T-hook, as was discussed in \secref{sec:emergence}. One can find $\mu_{12}$ from the fact that  $\bT_{0,s}=(\mu_{12}^{[s+1]})^2\simeq 1$  in the character limit since $\bT_{0,s}$ should be the character of the trivial representation. Hence, $\mu_{12}=\pm 1$, which coincides with the normalisation introduced previously in this section. Therefore, by verifying  the expressions for $\wT_{1,\pm s}$, $\wT_{2,\pm s}$, and $\bT_{0,s}=1$, which we did, we confirm that the whole T-hook is reproduced from the QSC in the quasi-classical approximation.

\section{Conclusions and discussion}

\label{sec:conclusions}
In this paper we achieved the long-standing goal  of  formulation of a concise and mathematically transparent system of
Riemann-Hilbert equations for anomalous dimensions of an arbitrary local operator in planar \neqfour{} SYM theory or,
alternatively, for the energy of the dual closed superstring state. The equations
generalize the classical spectral curve represented by a specific Riemann surface~\cite{Beisert:2005bm},
to the full quantum case, describing the spectrum for arbitrary value of the `t Hooft coupling \(\lambda=16\pi^2 g^2\).

A particularly important set of functions in this construction are
4+4 functions of a natural spectral parameter \(u\),
having only a single Zhukovsky cut with the branch points at \(\pm2g\)
on the real axis of their defining sheet:
\(\bP_a(u),\,\,a=1,2,3,4\) having a short cut
and \(\bQ_{j}(u),\,\,{j}=1,2,3,4\) having a long cut.
These functions with  simple and neat analytic properties happened to be
 deeply hidden inside the AdS/CFT TBA/Y-system. Remarkably,  not only they lead
to much simpler than TBA and  analytically transparent Riemann-Hilbert-type equations, but they also
allow to describe the planar spectrum in
complete generality, overcoming a well known problem of TBA/Y-system formulation,
limited mainly to the simplest states obeying certain additional symmetries (such
as Konishi-like states).
On the other hand, these fundamental functions also have a very intuitive
physical meaning. For example,
in the semi-classical limit of the underlying superstring sigma model~\cite{Beisert:2005bm}
we find
\(\bP_a\sim \exp\left({-g\int^{u/g} dz\,\, \tilde p_a(z)}\right),
\quad \bQ_{j}\sim
\exp\left({+g\int^{u/g} dz\,\, \hat p_j(z)}\right) \) related to the  quasi-momenta \((\tilde p_1,\tilde p_2,\tilde p_3,\tilde p_4|\hat p_1,\hat p_2,\hat p_3,\hat p_4)  \) of the classical spectral curve, also having only one pair of fixed branch points on their defining sheets.
This formula is of course familiar as a quasi-classical
limit of a wave function in a one dimensional quantum mechanics! This striking analogy
suggests that these objects should be also building blocks
for an exact wave function in separated Sklyanin variables.
We also have identified the weak coupling limit of these functions where they happen to be
simply related to Baxter Q-functions of the $\psu(2,2|4)$ XXX Heisenberg spin chain.
This identification allows for a concrete classification of
physical solutions of the QSC
based on the analytic continuation in $g$ from weak coupling
where the Riemann-Hilbert-type equations reduce to a system of algebraic equations.

At the same time, \(\bQ_i,\bP_a\)
give rise to a powerful integrability
setting -- the Q-system of 256 various Q-functions
representing  Grassmannians' coordinates with an additional self-consistent and intriguing
analyticity structure described in \secref{sec:YTQ}. The  analytic continuation
of various Q-functions under their Zhukovsky-type cuts can be described in terms of a certain isomorphism of the Q-system relating the upper-half-plane and lower-half-plane analytic Q-functions.
One can even inverse the logic here and derive the QSC from the
requirement of existence of such an isomorphism. This observation reveals a remarkable mathematical beauty of the spectral problem, not immediately seen from its former formulations.
This  combination of the algebraic, grassmanian construction of the
Q-system with the underlying analytic  structure w.r.t. spectral
parameter can be called the {\it analytic Q-system}.

\
The actual method, first announced by the authors  in \cite{Gromov:2013pga}  and  presented in detail in the current paper,   has already found a few powerful applications.
It was used to get the most accurate perturbative calculation at weak \cite{Dima-talk,Dima-toappear}
and at strong coupling in \cite{Gromov:2014bva,Gromov:2014eha}, where also a pomeron intercept was found up to the 6-th order.
The method was proven to be very powerful for exact analytic
nonperturbative calculations for the generalized cusp anomalous
dimension \cite{Gromov:2012eu,Gromov:2013qga,Gromov:2013pga}, for slope and curvature function \cite{Gromov:2014eha,Gromov:2014bva}.
Recently, following our methods, the QSC was also build in the ABJM theory
\cite{Bombardelli:2009xz,Gromov:2009at,Cavaglia':2013hva,Cavaglia:2014exa}. Exact slope function
in ABJM theory computed using the QSC methods has lead to a well justified conjecture for the interpolation
function $h(\lambda)$, entering into all integrability based calculations in this theory \cite{Gromov:2014eha}.
Further development of our methods should also allow for a comprehensive
description of the open strings spectrum with various
types of integrable
boundary conditions, which cover such physically relevant cases
as quark--anti-quark potential \cite{Correa:2012hh,Drukker:2012de}, $D\bar D$-systems \cite{Bajnok:2013wsa}.
Also the QSC approach may help to scan for all possible integrable boundary conditions
and may even lead to their complete classification.
Even more straightforward generalizations are
$\beta,\gamma,\eta-$deformed theories \cite{Gromov:2010dy,Arutynov:2014ota}.
Finally, we hope that some dualities of ${\rm AdS}_3/{\rm CFT}_2$ type \cite{Babichenko:2009dk,Borsato:2014exa} could be
studied by our exact methods.

The QSC equations presented here seem to be an ideal approach for attacking a set of
longstanding complicated problems.
In particular the
BFKL pomeron spectrum seems already to be within the reach of our methods.

More speculative possible domains of application of QSC
are the calculations of form-factors and correlation functions.
Indeed, as we discussed it is very appealing to associate $\bQ's$ and $\bP's$
with the wave functions which thus could be very useful building blocks
for the exact formulae for these more general observables in the theory \cite{Negro:2013wga}.

It would be interesting to construct a similar analytic Q-system in other integrable quantum sigma models, such as the principal chiral field or the Gross-Neveu model. Apart from a deeper understanding of related mathematical structures  of  integrable  finite volume 2d QFT's, it could help to clarify various unsolved problems, such as the  description and classification of various excited states in these models.\footnote{Such a general classification is achieved for the \(SU(N)\times SU(N)\) principal chiral field at \(N=2\) \cite{Gromov:2008gj} but little is known beyond the vacuum and the mass gap states for \(N>2\) \cite{Kazakov:2010kf}.}   It would be good to work out  a physical bootstrap procedure, generalizing the Zamolodchikovs' S-matrix bootstrap to the  finite volume case, which leads directly to the corresponding Y-system, or  Q-system, omitting the TBA procedure.
Indeed, as it is done in this paper -- through the TBA procedure for the mirror theory, starting from the S-matrix bootstrap -- seems to be way too complicated regarding the simplicity and naturalness of the final equations based on Y-system or Q-system.

Another important open problem which is left to solve is the
operatorial formulation of the Q-system.
At weak coupling a locality makes the problem well posed and
solvable. It reduces to the one for a non-compact supersymmetric
Heisenberg spin chain. Q-operators are well understood and explained
for non-compact
\cite{Derkachov:1999pz,Derkachov:2002tf,Belitsky:2006cp,Frassek:2011aa}
and supersymmetric \cite{Belitsky:2006cp,Frassek:2010ga}.
Even though it is clear that at a finite coupling this problem literally is not very well posed  due to a scheme dependence,
 in a relevant formulation, it still should contain a certain rational bit of information.
A possible way to it might be similar to the one applied  for the twisted spin chains in \cite{Kazakov:2007na,Kazakov:2010iu} where the operatorial form of Q-functions was constructed
by application  to characters of co-derivatives w.r.t. twists~\cite{Kazakov:2007na,Kazakov:2010iu}.
This approach may also shed some light on the meaning of Q-functions
from the gauge theory point of view.

\section*{Acknowledgements}   We thank M.Alfimov,  J.Balog, A.Cavagli\`a, D.Fioravanti, I. Kostov, F.Levkovich-Maslyuk, V.Pestun, A.Sever,  G.Sizov, R.Tateo and Z.Tsuboi for discussions.
The work of V.K. is supported by the ANR
grant StrongInt (BLANC- SIMI- 4-2011)  and by the ESF grant HOLOGRAV-09-
RNP- 092.   The research
of N.G., V.K., and D.V. leading to these results has received
funding from the People Programme (Marie Curie Actions)
of the European Union's Seventh Framework Programme
FP7/2007-2013/ under REA Grant Agreement
No 317089. The research of V.K.  leading to these results has received
funding from the European Research Council under the European
Community's Seventh Framework Programme (FP7/2007-2013 Grant Agreement
no.320769).  V.K. is also very grateful to Yukawa Institute for Theoretical Physics (Kyoto) for hospitality, where a part of this work was done. V.K. also thanks the Ambrose Monell Foundation, for the generous support.  A substantial part of the research of
S.L. leading to these results was performed in Imperial College
London, with support from the ERC Advanced grant No.290456. S.L. also thanks Nordita for hospitality, where a part of the work was done.

\newpage
\appendix

\section{{\it Mathematica} code to check derivations}
\subsection{
  Details of derivation for section 2}\la{apA}
All the steps in section 2 can be done by hands, but to avoid needless waste of time we suggest to follow
elementary steps using \(\it Mathematica\).

First we enter the definition of the \(\wT\) -functions (denoted by \verb"t") in terms of \(\bP_a\)'s (denoted as \verb"P[a]")
and \(\bP^a\)'s (denoted as \verb"P[-a]"). $\mu_{1,2}$ is denoted as
$\mu$\verb+[1,2]+, but for the simplicity of notation, when the notation
$\mu$\verb+[1,2]+ is used in an expression involving {\Pfcts} with odd
shift, $\mu$\verb+[1,2]+ denotes $\mu_{1,2}^+$; by contrast, when it
appears in an expression involving {\Pfcts} with even shifts, it
denotes $\mu_{1,2}$. We also define \(\bT\) and
denote it by \verb+T+.
\begin{verbatim}
  (*Right strip:*)
t[1,s_][u_] := P[1][u+I s/2] P[2][u-I s/2] - P[2][u+I s/2] P[1][u-I s/2]/;s>0;
t[2, s_][u_] := t[1,1][u+I s/2] t[1,1][u-I s/2] /; s > 1
t[0, s_][u_] = 1;
  (*Left strip:*)
t[1, s_][u_] := P[-4][u+I s/2] P[-3][u-I s/2] -  P[-3][u+I s/2] P[-4][u-I s/2]/;s<0;
t[2, s_][u_] := t[1,-1][u+I s/2] t[1,-1][u-I s/2] /; s < -1
  (*black gauge - more convenient for upper strip*)
T[a_, s_][u_] := (-1)^(a s) t[a, s][u]/\[Mu][1, 2]^(a - 2) /; Abs[s] >= a;
T[a_, +2][u_] := T[+2, a][u];
T[a_, -2][u_] := T[+2, -a][u];
\end{verbatim}
Next, we will have to be able to find others \(\bT\) using Hirota identity.
For that we should remember that the shift operator in the Hirota identity is defined
with long cuts, whereas  \(\bP\) are the functions with short cuts.
To correctly deal with this situation we introduce \verb+up+ and \verb+dn+
operations which are compositions of the shift and analytic continuation \verb+toPtilde+.
We also define a function \verb+Disc+ which computes discontinuity
\begin{verbatim}
  (*Define monodropy of P's*)
toPtilde = {P[a_][u] :> Pt[a][u], Pt[a_][u] :> P[a][u]};
  (*Computes discontinuity*)
Disc = #-(#/.{\[Mu][1,2]->\[Mu]t[1,2],Pt[a_][u+b_]:>Ptt[a][2b/I][u+b]}/.toPtilde)&;
  (*Shift operators, which take into account multivaluedness of P*)
up = (# /. u -> u + I/2 /. toPtilde) &;
dn = (# /. toPtilde /. u -> u - I/2) &;
\end{verbatim}
After that we are ready to use Hirota to derive one by one all
necessary \(\bT\) functions:
\begin{verbatim}
Hir[T_][a_,s_]:=(up[T[a,s][u]]dn[T[a,s][u]]-T[a+1,s][u]T[a-1,s][u]
                                           -T[a,s+1][u]T[a,s-1][u]);
   (*Finds T from Hirota*)
FindT[{a1_, s1_}, {a2_, s2_}] := Block[{ee}, Quiet[T[a1, s1][u_] =.];
ee=T[a1, s1][u]/.Solve[Hir[T][a2, s2]==0,T[a1, s1][u]][[1]];T[a1,s1][u_]=ee;]
\end{verbatim}
This short code provides us with all necessary tools we need.
For example by running
\begin{verbatim}
FindT[{2, 1}, {2, 2}]; Disc[T[2, +1][u]] // Factor
\end{verbatim}
we find \(\bT_{2,1}\), compute its discontinuity reproducing \eq{formath1}!
It is not much harder now to get \eq{mastereq} for \(n=1\). We just find one by one \(\bT_{2,-1},\bT_{1,0},\bT_{2,0}\) and \(\bT_{3,1}\)
and impose absence of discontinuity on \(\bT_{2,0}^+\) and \(\bT_{3,1}^+\):
\begin{verbatim}
FindT[{2,-1},{2,-2}];FindT[{1,0},{1,1}];FindT[{2,0},{1,0}];FindT[{3,1},{2,1}];
sltp=Solve[0 == Disc[T[2, +1][u]] // Simplify, \[Mu]t[1,2]][[1]];
sl12=Solve[{Disc[up[T[2, 0][u]]] == 0,
            Disc[up[T[3, 1][u]]] == 0},
{Ptt[1][2][u+I],Ptt[2][2][u+I]}][[1]] /. sltp // Simplify
\end{verbatim}
Similar equation for the left wing is obtained using
\begin{verbatim}
FindT[{1, 0}, {1, -1}]; FindT[{2, 0}, {1, 0}]; FindT[{3, -1}, {2, -1}];
sltm=Solve[0 == Disc[T[2, -1][u]] // Simplify, \[Mu]t[1,2]][[1]];
sl34 = Solve[{0 == Disc[up[T[2, 0][u]]],
              0 == Disc[up[T[3, -1][u]]]},
{Ptt[-3][2][u+I],Ptt[-4][2][u+I]}][[1]] /. sltm // Simplify
\end{verbatim}
Testing \eq{mastereq} for \(n=2\) does not take more then \(6\) extra lines of code.
For that we assume \eq{mastereq} to hold for  this test of analyticity for \(\bT_{3,0}^{[+2]},
\bT_{4,1}^{[+2]},\bT_{4,-1}^{[+2]}\). Indeed we proof then  \eq{mastereq} for \(n=2\):
\begin{verbatim}
eq[n_]=Flatten[{sl12, sl34}/.Ptt[a_][2][u + I]:>Ptt[a][2 n][u+I]/.I->I n];
FindT[{3, 0}, {2, 0}];FindT[{4, 1}, {3, 1}];FindT[{4, -1}, {3, -1}];
  (*exclude Pt[-4][u] using constraint*)
noP4 = Table[Solve[0==Disc[T[2,-1][u]]/.sltp//Simplify,
       Pt[-4][u]][[1]]/. u->u-I n/2,{n, -10, 10}] // Flatten;
Disc[up[up[T[3, 0][u]]]]/.sl12/.sl34/.sltp/.eq[2]/.noP4//Factor
Disc[up[up[T[4,+1][u]]]]/.sl12/.sl34/.sltp/.eq[2]/.noP4//Factor
Disc[up[up[T[4,-1][u]]]]/.sl12/.sl34/.sltp/.eq[2]/.noP4//Factor
\end{verbatim}

\subsection{QQ-relations}

\paragraph{Solution to the QQ-relations}

The following lines of code define the QQ-relations
(\ref{definingQQ}):
\begin{verbatim}
QQ[A_List,a_,b_,J_List]:=Q[A|J,u]Q[Join[A,{a,b}]|J,u]==Q[Join[A,{a}]|J,u+I/2]*
     Q[Join[A,{b}]|J,u-I/2]-Q[Join[A,{a}]|J,u-I/2]Q[Join[A,{b}]|J,u+I/2]
QQ[A_List,J_List,i_,j_]:=Q[A|J,u]Q[A|Join[J,{i,j}],u]==Q[A|Join[J,{i}],u+I/2]*
     Q[A|Join[J,{j}],u-I/2]-Q[A|Join[J,{i}],u-I/2]Q[A|Join[J,{j}],u+I/2]
QQ[A_List,a_,J_List,j_]:=Q[Join[A,{a}]|J,u]Q[A|Join[J,{j}],u]==Q[A|J,u-I/2]*
     Q[Join[A,{a}]|Join[J,{j}],u+I/2]-Q[Join[A,{a}]|Join[J,{j}],u-I/2]Q[A|J,u+I/2]
\end{verbatim}
After running the above code, 
\verb+QQ[+$A$\verb+,+$a$\verb+,+$b$\verb+,+$I$\verb+]+ returns the
QQ-relation (\ref{QQbb}), whereas
\verb+QQ[+$A$\verb+,+$I$\verb+,+$i$\verb+,+$j$\verb+]+ (resp
\verb+QQ[+$A$\verb+,+$a$\verb+,+$I$\verb+,+$i$\verb+]+) returns the
relation (\ref{QQff}) (resp (\ref{QQbf})) -- provided $A$ and $I$ are
entered as lists. For instance ~
\verb+QQ[{},1,2,{1}]+  returns
\textsf{Q[\{\}$|$\{1\},u]Q[\{1,2\}$|$\{1\},u]==Q[\{1\}$|$\{1\},$\mathbbm i$/2+u]Q[\{2\}$|$\{1\},-$\mathbbm i$/2+u]\\-Q[\{1\}$|$\{1\},-$\mathbbm i$/2+u]Q[\{2\}$|$\{1\},$\mathbbm i$/2+u]}.

Let us now define a function \verb+Develop+ which substitutes
Q-functions according to (\ref{Qes}-\ref{symdet},\ref{sport}):
\begin{Verbatim}[commandchars=\\\#\&]
Develop[expr_]:=expr//.Q[_[A_,J_],u_]:>Block[{n},Which[A==J=={},1,\hspace#\stretch#1&&(*use \eqref#Qes&*)
    (n=Length@A-Length@J)==0,Det[Table[Q[{a}|{i},u],{a,A},{i,J}]],\hspace#\stretch#1&&(*use \eqref#symdet0&*)
    J=={},Det[Table[Q[{a}|{},u+I k/2],{a,A},{k,n-1,1-n,-2}]],\hspace#\stretch#1&&(*use \eqref#eq:3&*)
    A=={},Det[Table[Q[{}|{i},u+I k/2],{i,J},{k,-n-1,1+n,-2}]],\hspace#\stretch#1&&(*use \eqref#eq:5&*)
    n>0,Sum[Product[Signature@s,{s,{A,A1~Join~A2}}]*Q[A1|{},u]*Q[A2|J,u+I n/2]
            ,{A1,Subsets[A,{n}]},{A2,{Complement[A,A1]}}],\hspace#\stretch#1&&(*use \eqref#eq:125&*)
    n<0,(-1)^(n Length@A) Sum[Product[Signature@s,{s,{J,J1~Join~J2}}]*Q[A|J1,u+I n/2]*
           Q[{}|J2,u],{J2,Subsets[J,{-n}]},{J1,{Complement[J,J2]}}]\hspace#\stretch#1&&(*use \eqref#eq:126&*)
        ]]//.{Q[_[{},{}],_]->1,\hspace#\stretch#1&&(*uses \eqref#Qes&*)
            q:Q[_[{a_},{i_}],v:u+b_]:>Which[b/I>1/2,
                       Q[{a}|{i},v-I]+Q[{a}|{},v-I/2]Q[{}|{i},v-I/2],\hspace#\stretch#1&&(*uses \eqref#sport&*)
            b/I<=-1/2,Q[{a}|{i},v+I]-Q[{a}|{},v+I/2]Q[{}|{i},v+I/2],\hspace#\stretch#1&&(*uses \eqref#sport&*)
            True,q]}
\end{Verbatim}

One can check that it solves the QQ-equations
(\ref{definingQQ}): for instance \verb+Develop[QQ[{},1,2,{}]]+
returns
\textsf{True} whereas \verb+Develop[QQ[{},1,2,{1,3}]]+ returns a
large expression\footnote{Depending on the speed of your computer,
  \texttt{Develop[QQ[\{2,3\},1,\{1,2,3\},4]]} may return \textsf{True} at
  once without having to ask for a \texttt{FullSimplify}.}; this expression simplifies to \textsf{True} as
one can check by running
\verb+FullSimplify[Develop[QQ[{},1,2,{1,3}]]]+.

Using \verb+FullSimplify+ to simplify long expressions can sometimes be
very slow (and it sometimes even doesn't succeed), hence one can as
well substitute random values to each expression of the form
\verb+Q[{+$a$\verb@}|{},u+@$p$\verb+I/2]+,
\verb+Q[{}|{+$i$\verb@},u+@$p$\verb+I/2]+,
\verb+Q[{+$a$\verb@}|{@$i$\verb@},u]@ or
\verb+Q[{+$a$\verb@}|{@$i$\verb@},u+I/2]@. The outcome will be
\textsf{True} if and (almost)only if the expression simplifies to
\textsf{True}. This is what the function \verb+CheckEq+ (defined
below) does:
\begin{verbatim}
CheckEq[e_]:=Block[{ff},ff[n_,p_,q_]:=ff[n,p,q]=Rationalize[Random[],0];
         Develop[e]/.Q[_[a_,i_],b_]:>ff[a,i,FullSimplify[(b-u)/(I/2)]]]
\end{verbatim}
With this definition, \verb+CheckEq[Develop[QQ[{2,3,4},1,{},1]]]+ will
quite quickly return \textsf{True} on a small modern computer, whereas
\verb+FullSimplify[Develop[QQ[{2,3,4},1,{},1]]]+ does not succeed to
return \textsf{True} in a reasonable time.

\paragraph{Hodge transformation}

The \equref{Hodgedef} is implemented as follows:
\begin{verbatim}
Qup[_[A_,J_],u_]:=Block[{Ap=Complement[Range@4,A],Jp=Complement[Range@4,J]},
    (-1)^(Length@Ap*Length@J)*Signature[Join[Ap,A]]*Signature[Join[Jp,J]]Q[Ap|Jp,u]]
\end{verbatim}
It is easy to check that the upper-indexed Q-functions obey the same
QQ-relations as the original one, by evaluating for instance
\verb+CheckEq[Develop[QQ[{2},1,3,{4,3}]/.Q->Qup]]+ (which returns \textsf{True}).

We are now almost ready to check equations such as (\ref{QQorto1}):
the only missing thing is that the equations (\ref{QQorto1}) assume
that $\fQ_{\emptyset|\emptyset}=\fQ^{\emptyset|\emptyset}=1$. While
the above definitions rely on (\ref{Qes}-%
\ref{symdet},\ref{sport}), which ensures that
$\fQ_{\emptyset|\emptyset}=1$, the condition
$\fQ^{\emptyset|\emptyset}=1$ has to be specifically enforced by thefunction \verb+Qforced+ (below) which uses these
relations to express $\fQ_{4|4}$ and $\fQ_{\emptyset|4}$:
\begin{verbatim}
Qforced[_[{},{4}],v_]:=Qforced[_[{},{4}],v]=Solve[Develop[Q[{1,2,3,4}|{1,2,3,4},v+I/2]
             ==Q[{1,2,3,4}|{1,2,3,4},v-I/2]],Q[{}|{4},v]][[1,1,2]]
Qforced[_[{4},{4}],uu__]=Solve[Develop[Q[{1,2,3,4}|{1,2,3,4},uu]==1],
             Q[{4}|{4},uu]][[1,1,2]];
Qforced[A___]=Q[A];FSimp[expr_]:=(Develop[expr]//.Q->Qforced)
\end{verbatim}

With the above code, it is now easy to check all relations derived
from the QQ-relations. For instance,
\verb+CheckEq[FSimp[Q[{}|{2},u]==-Sum[Qup[{a}|{},u]Q[{a}|{2},u-I/2],{a,1,4}]]]+
evaluates to \textsf{True}, as expected from (\ref{QQrel1}).

\section{Details of the relations between the Q-, T- and {\Ysys}s}
\label{app:relation-to-TBA}
Historically, development of the finite size AdS/CFT spectrum solution roughly followed the left-to-right direction in the following diagram.
\begin{equation}\label{YTQ}
 \raisebox{-1em}{\includegraphics[width=0.4\textwidth]{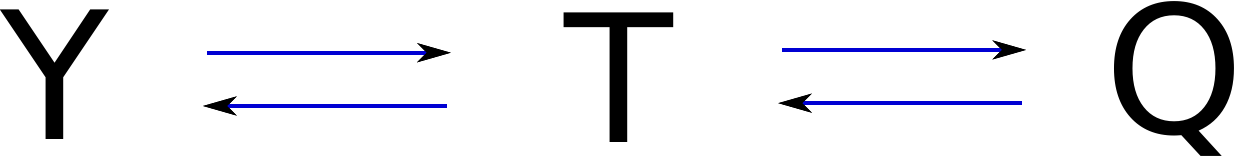}}
\end{equation}
The algebraic Y-T-Q equivalence was realized quite a while ago: both Y- and T-systems were written down already in the work \cite{Gromov:2009tv}, and soon after that the Wronskian solution for T's in terms of Q's was proposed   \cite{Gromov:2010km}. This algebraic representation took a good deal of inspiration from the similar constructions in integrable quantum spin chains with \(\su(n)\) symmetry \cite{Krichever:1996qd}  (see also \cite{Bazhanov:1996dr} for a similar, even operatorial construction for an \(\su(n)\) CFT) and turned out to be a natural generalization to the non-compact and super-symmetric case.

The analytic structure, at the contrary, turned out to be very specific to this particular integrable model. Although the basic features of this structure  were already sketched out in  \cite{Gromov:2009tv} it took a  long effort to properly understand it. Originally it was encoded through the TBA equations \cite{Bombardelli:2009ns,Gromov:2009bc,Arutyunov:2009ur} and then, step by step it reduced to the analytic Y-system \cite{Cavaglia:2010nm,Balog:2011nm}, a distinguished \(\bT\)-gauge \cite{Gromov:2011cx}, and to the mixture of T- and Q- functions in the FiNLIE  formulation  \cite{Gromov:2011cx}. While moving from left to right on the diagram \eq{YTQ}, the analytic structure was becoming more and more transparent.    The AdS/CFT quantum spectral curve, which can be also called  the analytic Q-system, proposed in the current paper (see also the already published short version \cite{Gromov:2013pga}) gives  the clearest, and in many respects the ultimate insight into the analytic structure of the underlying Q-functions.

Of course, it is important to demonstrate explicitly that this QSC -- the analytic Q-system -- is equivalent to the original TBA equations, at least for certain well-understood cases. \Secref{sec:QSCTBA} already contains the basic ideas of this equivalence. But to keep things short there, we omitted some important steps of the proof which might make this construction looking mysterious. The goal of this appendix is to systematically review the Y-T-Q equivalence which greatly relies on the classical integrability of the Hirota dynamics and hence on the machinery of the {\Qsys}.

The due remark is that we cannot fully rigorously demonstrate  the
equivalence with TBA for {\it arbitrary} state of the AdS/CFT
spectrum. The back-up of our analysis is in our explicit studies of
the \(\sl(2)\) sector that were thoroughly done up to, and including the
double wrapping orders at weak coupling
\cite{Leurent:2012ab,Leurent:2013mr} and non-perturbatively, in the
small \(S\) regime \cite{Gromov:2012eu,Gromov:2013qga,Gromov:2014bva} as
well as in a few examples of numerical solution of TBA
equations. Though the derivations below are written as if for generic
state,  they might have unaccounted subtleties in each particular case
due to state-dependent singularities, typically in the {\Yfcts}. We have enough evidence, however, to believe that the main discontinuity properties of the Y-, T-, Q- functions are  not sensible to these potential subtleties.

Also, the reader should understand that a state-by-state comparison with TBA is impossible and in principle impractical  because the TBA equations are  not even  written explicitly for an arbitrary state. Even for the known cases, these equations are only conjectured and  based on the contour deformation trick \cite{Dorey:1996re}, still unproven even for simpler sigma-models. The QSC has its own universal requirement  addressing the arbitrary state --  the condition of absence of poles  which, for instance, can be immediately recast into the exact Bethe equations along Hasse diagram as discussed in \secref{subsec:exactBethe}. Hence, as concerns arbitrary state, we believe that QSC should be thought of as a framework for multiple approaches to the AdS/CFT spectral problem. We demonstrated its viability by deriving the asymptotic Bethe equations in the large volume approximation, see \secref{sec:ABA}.

The \secref{app:fundamentalT} discusses the relation $Y\rightleftarrows T$ in \eqref{YTQ}.
It discusses the properties of the fundamental {\Tsys}, which
is equivalent to the TBA and the discontinuity relations of
\cite{Cavaglia:2010nm,Balog:2011nm}. This equivalence was demonstrated
in \cite{Gromov:2011cx} for LR-symmetric case and  we generalize it
for arbitrary state.

The next step is to relate
analytic properties of T- and {\Qfcts} in appropriate gauges. First, \secref{sec:solution-t-system} describes the direction $\fQ\to{\bT}$, and
shows how to construct the {\Tsys} from the fundamental Q-system. For the reader familiar with Wronskian ansatz, we emphasize right away that a bold application of this ansatz to the fundamental Q-system {\it will not} immediately reproduce the mirror T- and Y-systems. The proper course of action also requires  understanding of the available symmetries and proper usage of them.

Finally, the opposite direction $\bT\to\fQ$ is discussed in \secref{sec:quant-sprect-curve}, where we show how to derive the full QSC
from the analytic properties of the {\bTfcts}.

\subsection{\texorpdfstring{Y$\leftrightarrow \bT$}{Y<->T}}
\label{app:fundamentalT}
\begin{figure}
\begin{center}
\includegraphics{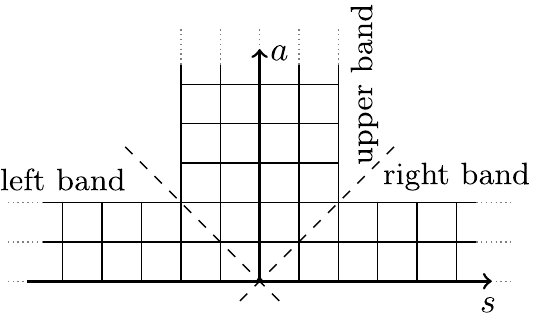}%
\caption{\label{fig:Thook1}
  T-hook: in the ``right'' and ``left'' band, the \fcts{\(\wT\)} are
  analytic whereas the \fcts{\(\bT\)} are
  analytic in the upper band.
}
\end{center}
\end{figure}The equivalence between TBA equations and the Y-system with extra
analyticity constraints (analytic Y-system) was successfully
demonstrated in \cite{Cavaglia:2010nm,Balog:2011nm}, and we will
assume it as the established one. Our goal would be to translate the
analyticity constraints on Y's  to the language of T-functions. This
exercise is  however not immediate because T's are not uniquely defined
objects. Indeed, the relation \(Y_{a,s}=\frac{T_{a,s+1}T_{a,s-1}}{T_{a+1,s}T_{a-1,s}}\) defines T's only up to the gauge transformation
\be\label{gaugeapp}
T_{a,s}\to g_1^{[a+s]}\,g_2^{[a-s]}\,g_3^{[-a+s]}\,g_4^{[a+s]}\,T_{a,s}
\ee
which is also a symmetry of the Hirota bilinear identities \eqref{Hir} constraining T's.

An achievement of  \cite{Gromov:2011cx} was to demonstrate the existence,
for the case of left-right (LR) symmetric solutions, of a special
gauge \(\bT\) in which the analyticity of T-functions can be clearly
formulated  and has a  natural physical
interpretation. Here we will proceed in a similar way  but the \(\bT\)-gauge will be this time extended
 to the case of non-LR-symmetric states, i.e we will no longer use the
 condition \(Y_{a,-s}=Y_{a,s}\).  {\it A priori}, one could expect two different
 "distinguished" gauges, left and right. This is
 actually not the case and the right and left part of the T-hook have the same special gauge, due to the property
\begin{eqnarray}
\label{eq:9}Y_{1,+1}Y_{2,+2}=Y_{1,-1}Y_{2,-2}\,,
\end{eqnarray}
as will be shown below. The property \eqref{eq:9}  is true even for the case without LR-symmetry and can be immediately deduced from the TBA equations (e.g. in \cite{GromovKKV}). In later sections, \eqref{eq:9} will propagate to the ``quantum unimodularity'' of the QSC.

 Our derivation extensively uses the appendix C of \cite{Gromov:2011cx} and in the cases when the proof goes without modification we simply refer to the corresponding place in \cite{Gromov:2011cx}. Finally let us note that the  derivation of the \(\bT\)-gauge can be done directly from TBA using the ``telescoping'' procedure \cite{Gromov:2011cx}   and the backwards compatibility from the \(\bT\)-gauge analyticity to the analytic Y-system and TBA is straightforward, as explained in \cite{Gromov:2011cx}. This will prove the equivalences of  \({\rm TBA}\leftrightarrow {\rm Y}\leftrightarrow \bT\).

\subsubsection{Statements}

\begin{statement}[existence]\label{prop:existenceT} There exists a  \(\bT\)-gauge with the following properties:\footnote{See comments on the notations in the rest of this subsection.}
\begin{subequations}\label{eq:2}
\begin{equation}
  \label{eq:1} \renewcommand{\arraycolsep}{0cm}
  \begin{array}{c@{~}|@{~}r@{\:}l@{~~}r@{\:}l}
\textrm{reality}&\bT_{a,s}=&\overline {\bT}_{a,s}\\\hline
\multirow{3}{*}{\textrm{analyticity}}&\bT_{a,0}\in& {\cal A}_{a+1}\\
&\bT_{a,\pm 1}\in& {\cal A}_{a}\\
&\bT_{a,\pm 2}\in& {\cal A}_{a-1}\\\hline
\multirow{4}{*}{\textrm{group theoretical}}
&\bT_{n,2}=&
\bT_{2,n}\,,&n\geq& 2\\
&\bT_{n,-2}=&
\bT_{2,-n}\,,&n\geq& 2
\\
&\bT_{0,0}^+=&\bT_{0,0}^-\\
&\bT_{0,s}=&\bT_{0,0}^{[-s]}\\\hline
\textrm{\({\mathbb Z}_4\) symmetry}&\hat\bT_{a,s}^c=&(-1)^s \hat\bT_{-a,s}^c
  \end{array}
\end{equation}
In addition, the gauge \(\wT\) defined by\footnote{The sign in
  \equref{eq:Fnice} differs from the sign in
  \cite{Gromov:2011cx}, but this corresponds to the irrelevant
  ambiguity in the choice of a sign in the definition
  $\CF\equiv\sqrt{\bT_{0,0}}$.}
\begin{align}
\label{eq:Fnice}
 \wT_{a,s}=&(-1)^{a\,s}
\bT_{a,s}(\CF^{[a+s]})^{{a-2}},& \CF\equiv& \sqrt{\bT_{0,0}}\,,
\end{align}
obeys the following properties
\begin{equation}
  \label{eq:111} \renewcommand{\arraycolsep}{0cm}
  \begin{array}{c@{~}|@{~}r@{\:}l@{~~}r@{\:}l}
    \textrm{reality}&\wT_{a,s}&=\overline {\wT}_{a,s}\\\hline%
\multirow{3}{*}{\textrm{analyticity}}
&\wT_{0,\pm s}&=1\\
&\wT_{1,\pm s}&\in {\cal A}_{s},\ \ s\geq 1\\
&\wT_{2,\pm s}&\in {\cal A}_{s-1},\ \ s\geq 2\\\hline%
\multirow{2}{*}{\textrm{\({\mathbb Z}_4\) symmetry}}
&\hat\wT_{a,-s}^{c_r}&=(-1)^a \hat\wT_{a,s}^{c_r}\,,\\
&\ \hat\wT_{a,-s}^{c_l}&=(-1)^a \hat\wT_{a,s}^{c_l}\,\\
&\hat\wT_{1,\pm s}&, s\geq 1,\ \text{is analytic for}&u\in&\mathbb{C}\setminus \hbZ_s \setminus \hbZ_{-s}
\end{array}
\end{equation}
\end{subequations}
\end{statement}
The \(\bT\)-gauge can be rightfully called ``physical'' as it was done in \cite{Gromov:2011cx}, however to avoid confusion with the physical kinematics we will refer to it as to the ``distinguished'' one\footnote{We did not have enough of evidence in \cite{Gromov:2011cx} to call this kinematics as physical,
instead it was called  ``magic''. In this work we got a convincing demonstration of it by deriving the asymptotic Bethe ansatz.}. In addition, we conjecture the following property:
\begin{property}[regularity]
  \(\bT_{0,0}\) has no poles and it has a power-like behaviour at infinity.
\end{property}

Although we gathered a significant evidence for the regularity of \(\bT_{0,0}\), we could not  complete a proof of  the absence of poles for a generic state.  It is quite clear that the  regularity is directly linked to the exact Bethe equations, see \secref{subsec:exactBethe}.  In this appendix, we will only use the regularity to prove the following
\begin{statement}[uniqueness] If \(\bT_{0,0}\) satisfies the
  regularity conjecture, the distinguished gauge is unique up to an
  overall normalization\footnote{We call ``overall normalization'' a
  redefinition of  normalization of the T-functions which
  leaves the Hirota equation trivially invariant. An example is the
  transformation \(T_{a,s}\leadsto {\rm const}\times (-1)^{a+s} T_{a,s}\).
} and up to  an overall functional rescaling
\begin{align}\label{xrescaling}
\bT_{a,s}\to&
\left(\prod_{k=-(|s|-1)/2}^{(|s|-1)/2}\frac{x^{[+a+2k]}}{x^{[-a+2k]}}\right)^{\Lambda\,
\mathrm{sgn}(s)}\
  \bT_{a,s}\,,
  \end{align}
  where $\Lambda$ is some constant.
\end{statement}

\paragraph{Notations:}
In this section, all functions without explicit hat are defined in the mirror kinematics (long cuts) because  the original Y-system equations are valid only there. In particular, the reality properties are defined in the mirror kinematics.
Only the functions with explicit hat, like \(\hat\bT\) and \(\hat\wT\), are
defined in the physical kinematics (short cuts).
In agreement with \secref{sec:spectr-param-riem}, the analytic
continuation between mirror and physical is done slightly above the
real axis.

The notation \(f\in\CA_n\) means
that the function \(f(u)\) has no cuts for \(-n/2<\Im(u)<n/2\). One says, with some abuse of terminology,
that \(f\) is analytic in this domain, although it might have poles
there. If we want to emphasize in addition that \(f\) has no poles, we
say that \(f\) is regular. The notation \(\hat Z_{s}\) denotes the
support of a short Zhukovsky cut at \([-2g+i s/2,+2g+is/2]\),  and the notation \(\check Z_{s}\)
denotes the support of long Zhukovsky cut at \([-\infty+i s/2,-2g+is/2]\cup[2g+is/2,+\infty+is/2]\).

In (\ref{eq:1}), the relation \(\bT_{0,0}^+=\bT_{0,0}^-\) means that
\(\bT_{0,0}\) is  \(i\)-periodic in the mirror
kinematics. Furthermore, the conditions~\eqref{eq:2} impose that
\(\cF=\sqrt{\bT_{0,0}}\) only has branch points at positions $\pm 2
g+i\mathbf Z$, hence \(\bT_{0,0}\) has only double zeroes, to avoid
extra branch points in \(\cF\). Moreover, \(\cF\) is also real. By analyzing the {\Tsys} at weak coupling one concludes that
\begin{equation}
\CF^+=\CF^-\,,
\end{equation}
but not \(\CF^+=-\CF^-\) as it could have happened in principle  when taking square root.

Throughout this appendix, we will refer to the domain $a\ge|s|$ as to
the ``upper band'' (see \figref{fig:Thook1}), and similarly the domain
$s\ge a$ (resp $s\le-a$) will be called the right band (resp the left
band) of the T-hook.

In (\ref{eq:1}), we use an additional object denoted as \(\bT_{a,s}^c\): we define \(\bT_{a,s}^c\equiv\bT_{a,s}\) if \(a\geq|s|\), whereas elsewhere \(\bT_{a,s}^c\) is defined as the analytic continuation of
\(\bT_{a,s}\) in \(a\) from \(a\geq |s|\) to negative \(a\) at fixed \(u\)\footnote{
One way to define an analytic continuation from integer values of the parameter \(a\) is through the Wronskian solution, {\ie} using the second line of
\eqref{TfromQ} disregarding \(a\geq |s|\) constraint. The Wronskian solution operates with such objects as \(Q^{[a]}=Q(u+\frac{i\,a}2)\), so it explicitly  depends on \(a\) in a continuous way.

It is also possible to define this  analytic  continuation
  without reference to {\Qfcts}: as a solution of the Hirota equation
  in the \(\su(4)\) band \(\{(a,s)|a\in\mathbb
  Z,\,s\in[-4,4]\cup\mathbb Z\}\), see Fig.~7 in
  \cite{Gromov:2011cx}. To be fully accurate, Hirota equation is not sufficient to
uniquely define \(\hat\bT^c_{-2,\pm1}\) because
\(\hat\bT^c_{0,\pm1}=0\). So we should restrict the \(\mathbb{Z}_4\) symmetry relation \(\hat\bT_{a,s}^c=(-1)^s \hat\bT_{-a,s}^c\) to the cases of \(\bT_{a,\pm 2}\), \(\bT_{0,s}\), \(\bT_{\pm 1,s}\), \(\bT_{\pm 2,0}\). This restriction, however, does not weaken our constraints, they still fully determine the {\bT}-gauge \cite{Gromov:2011cx}.}. Note that \(\bT_{a,s}^c\) does not
coincide with  the actual {\Tfct}  \(\bT_{a,s}\)  of the
\(\wT\)-hook for \(a<|s|\).
One should also clearly distinguish two different cases: notation \({\bf T}^c\) means that \(u\) is fixed with \(|\Re(u)|<2g\) when analytically continuing in \(a\), while notation \(\hat\bT^c\) means that \(u\) is fixed with \(|\Re(u)|>2g\). Generically, \({\bf T}^c\) and \(\hat{\bf T}^c\) are not related by any analytic continuation in \(u\). For instance, \(\hat\bT_{0,\pm 1}^c=0\)
according to \(\mathbb{Z}_4\) symmetry, but \(\bT_{0,\pm 1}^c\neq0\).

In the same way, one defines  \(\wT_{a,s}^{c_r}\) as the analytic
continuation in \(s\) from the right band of the \(\wT\)-hook, so that \(\wT_{a,s}^{c_r}=\wT_{a,s}\)
for \(s\geq a\). Similarly,  \(\wT_{a,s}^{c_l}\) is the analytic
continuation in \(s\) from the left band, so that \(\wT_{a,s}^{c_r}=\wT_{a,s}\)
for \(s\leq -a\).  For instance, generically all three functions \(\wT_{2,0},\wT_{2,0}^{c_r},\wT_{2,0}^{c_l}\) are different. And again, one should distinguish the cases
\(\wT^c\) and \(\hat\wT^c\). The \(\mathbb{Z}_4\) symmetry in
(\ref{eq:111}) is realized for \(\hat\wT^{c_r}\) and \(\hat\wT^{c_l}\).

\subsubsection{Proof of uniqueness of the distinguished gauge}
\label{sec:bTproof}
The proof is
   given for the sake of completeness. It does not contain insights needed to understand the fundamental {\Qsys}. The reader may skip
   it if he or she understands the statements in (\ref{eq:2}) and (\ref{xrescaling}).

\paragraph{Group-theoretical constraints plus analyticity.}The departing point is to ensure that there exists a  gauge \(\sT\) with real T's   having a proper analyticity
strips  in the upper band.  The most straightforward way to prove its
existence is to use a gauge  freedom and fix \(\sT_{a,\pm 2}=1\) for
\(a\geq 2\) and then to reverse relations \(1+Y_{a,s}^{-1}=\frac
{\sT_{a,s}^+\sT_{a,s}^-}{\sT_{a,s+1}\sT_{a,s-1}}\) separately for each
given value of \(a\) using analyticity and reality of {\Yfcts}, as was
done in \cite{Balog:2011nm}. The
analyticity strips and reality of {\Yfcts} are obvious from the TBA equations.

For any analytic gauge \(\sT\), the discontinuity relation (1.7) of \cite{Cavaglia:2010nm}  is reduced to the statement that
\begin{eqnarray}\label{defB}
{\bf B}=\frac{1}{Y_{1,{\color{mblue}\pm1}}Y_{2,{\color{mblue}\pm 2}}}\frac{\sT_{1,0}}{\sT_{0,0}^-}
\end{eqnarray}
is analytic in the upper half-plane, see appendix C.2 of \cite{Gromov:2011cx}. Already at this stage it is crucial to use \eqref{eq:9}, so that \(\bB\) is the same for the right and the left part of the T-hook.

Consider now an arbitrary gauge transformation which preserves reality:
\begin{eqnarray}\label{bTsT}
\bT_{a,s}=f_1^{[a+s]}f_2^{[a-s]}\bar f_1^{[-a-s]}\bar f_2^{[-a+s]}\sT_{a,s}\,.
\end{eqnarray}
We would  also like to preserve analyticity, so we restrict ourselves to transformations such that \(f_{1}^-\) and \(f_2^-\) are analytic in the upper half-plane.

Firstly, we constrain the product \(f_1\) and \(f_2\) by the requirement:
\begin{eqnarray}\label{prodf1f2}
{\bf B}=\frac{(f_1\,f_2)^-}{(f_1\,f_2)^+}\,.
\end{eqnarray}
This requirement implies that, from \eqref{defB}, \eqref{bTsT} and \(Y_{a,s}=\frac{T_{a,s+1}T_{a,s-1}}{T_{a+1,s}T_{a-1,s}}\),
 we have
 \begin{subequations}
 \label{Y11Y22cons}
 \begin{eqnarray}
  \label{Y11Y22consa}
\frac{1}{Y_{1,{\color{mblue}\pm1}}Y_{2,{\color{mblue}\pm 2}}}\times\frac{\bT_{1,0}}{\bT_{0,0}^{\color{mred}-\color{black}}}
=\frac{\bT_{3,{\color{mblue}\pm2}}{\bT_{0,{\color{mblue}\pm1}}}}{\bT_{2,{\color{mblue}\pm3}}}\times{\frac
  1 {\bT_{0,0}^{\color{mred}-\color{black}}}}=1\,,
\end{eqnarray}
and, by taking complex conjugation and using reality of T's and Y's,
\begin{eqnarray}
\frac{1}{Y_{1,{\color{mblue}\pm1}}Y_{2,{\color{mblue}\pm 2}}}\times\frac{\bT_{1,0}}{\bT_{0,0}^{\color{mred}-\color{black}}}
=\frac{\bT_{3,{\color{mblue}\pm2}}{\bT_{0,{\color{mblue}\pm1}}}}{\bT_{2,{\color{mblue}\pm3}}}\times{\frac
  1 {\bT_{0,0}^{\color{mred}+\color{black}}}}=1\,,
\end{eqnarray}
\end{subequations}
One immediate consequence is
\begin{eqnarray}\label{per1}
\bT_{0,0}^+=\bT_{0,0}^-\,.
\end{eqnarray} Secondly, let us constrain the ratio of \(f_1\) and \(f_2\) by imposing
\begin{eqnarray}\label{T01T01}
\bT_{0,1}=\bT_{0,-1}\,.
\end{eqnarray}
Such constraint is possible to satisfy. Indeed, one has to solve
\begin{eqnarray}\label{tofixf}
1=\frac{\bT_{0,+1}}{\bT_{0,-1}}=\left(\frac{(f_1/f_2)^+}{(f_1/f_2)^-}\times
{\rm c.c.}\right)\frac{\sT_{0,+1}}{\sT_{0,-1}}\,.
\end{eqnarray} \(\frac{(f_1/f_2)^+}{(f_1/f_2)^-}\) is analytic in the upper half-plane and its complex conjugate is analytic in the lower half-plane. Any real function  can be decomposed as a product of two complex-conjugated functions analytic in the upper and lower half-planes, respectively, so this equation  always has a solution.

We therefore can always find \(f_1\) and \(f_2\) such that \(\bT\)-gauge is real, having proper analyticity strips in the upper band (\(a\geq |s|\)) and satisfying  \(\bT_{0,1}=\bT_{0,-1}\), \(\bT_{0,0}^+=\bT_{0,0}^-\). Using the  Hirota equation at the boundary \(\bT_{0,s}^+\bT_{0,s}^-=\bT_{0,s+1}\bT_{0,s-1}\) and \eqref{per1}, \eqref{T01T01}, one concludes then that
\begin{eqnarray}\label{bottomline}
{\bT_{0,s}}=\left(\cF^{[+s]}\right)^2,\ \ \CF^{+}=\CF^{-}\,.
\end{eqnarray}

From (\ref{Y11Y22cons}) we can now  conclude
\begin{eqnarray}
        \bT_{2,3}=\bT_{3,2}\,,\ \ \ \bT_{2,-3}=\bT_{3,-2}\,.
\end{eqnarray}
This derivation almost coincides with the construction of a gauge
solvable by Wronskian ansatz \cite{Gromov:2010km,Tsuboi:2011iz}.
Two things are added: first, an extra property \(Y_{1,1}Y_{2,2}=Y_{1,-1}Y_{2,-2}\)
was needed to derive \(\CF^{+}=\CF^{-}\) of (\ref{bottomline}), which is not
necessarily true for an arbitrary {\Tsys}. Second, we managed to preserve the analyticity
strips and the reality of {\Tfcts} while constructing this gauge.

\paragraph{\(\mathbb{Z}_4\) symmetry of the right/left bands.} Consider  only the
right band (\(s\geq a\)) for a moment. In the same way as above, one
can show the existence of a gauge \(\cT\) where the {\Tfcts} are
real and have proper analyticity strips in the right band (\(s\geq a\)). In
any such gauge the condition \(\hat\cT^{c_r}_{1,0}=0\) is satisfied and its derivation from TBA is explained in appendix C.1 of  \cite{Gromov:2011cx}.

To demonstrate the full \(\mathbb{Z}_4\) symmetry of \eqref{eq:111}, one should show that
it is possible to perform a gauge transformation from  the gauge \(\cT\) to a new gauge \(\wTr\)%
\footnote{We will prove later that it coincides with the gauge
denoted as \(\wT\) in \eqref{eq:Fnice}, but at the moment we should not
assume that they are the same gauge.
} that preserves reality and analyticity, but also ensures \(\wTr_{0,s}=1\) and \(\hat\wTr_{1,-1}^{c_r}=-\hat\wTr_{1,1}\). Then we can use
section 4.2 of  \cite{Gromov:2011cx} to prove the complete \(\mathbb{Z}_4\) symmetry
\(\hat\wTr_{a,-s}^{c_r}=(-1)^a\hat\wTr_{a,s}^{c_r}\) %
, and the finiteness
of the number of cuts for T- and {\Qfcts} in the physical kinematics.

The gauge transformation which we use to define the gauge \(\wTr\) has to
preserve the reality, hence it has to be of the form
\begin{eqnarray}\label{arbgaugeg}
\wTr_{a,s}=g_1^{[a-1+s]}g_2^{[-a+1+s]}\bar g_1^{[-a+1-s]}\bar g_2^{[a-1-s]}\cT_{a,s}\,,
\end{eqnarray}
with \(g_1\) and \(g_2\) analytic above real axis. On the lower boundary, one has  \(\cT_{0,s}=1\). Hence we impose \(g_1\,g_2^{++}=1\) to also have \(\wTr_{0,s}=1\)\,. To proceed further, we write the
\(\mathbb{Z}_4\) condition which we wish to achieve:
\begin{eqnarray}\label{eqong1g2}
1=-\frac{\hat \wTr_{1,-1}^{c_r}}{\hat\wTr_{1,+1}}=-\left(\frac{\hat g_1^-\hat g_2^-}{\hat
g_1^+\hat g_2^+}\times c.c\right)\frac{\hat{\cT}^{c_r}_{1,-1}}{\hat{\cT}_{1,+1}}\,.
\end{eqnarray}

Similarly to \eqref{tofixf}, (\ref{eqong1g2})
has a solution giving  us the product  \(g_1g_2\).
By knowing \(g_1g_2\) and \(g_1g_2^{++}\) we restore \(g_1\) and \(g_2\). Hence we
can always  find a  \(\wTr\)-gauge which has all the \(\mathbb{Z}_4\) properties and the proper analyticity in the right band.

We do the same procedure in the left band (\(s\leq -a\)) and construct
a \(\wTl\)-gauge there. A priory, the gauges \(\wTr\) and \(\wTl\)
constructed for
the right and the left bands  are
 different.
 But below we will show that one can choose the unique
 \(\wT\)-gauge for both  bands.
  This will be possible due to a freedom remaining in the
choice of \(g\)'s.
\paragraph{Relation between \(\bT\)- and \(\wT\)-gauges}
First, we will show that the gauge \(\wT_{a,s}=\frac{(-1)^{a\,s}\bT_{a,s}}{(\CF^{[a+s]})^{{2-a}}}\)
has proper analyticity strips in both right and left bands. For this we use the following analyticity
condition coming from the discontinuity condition (F.5) of \cite{Cavaglia:2010nm} (see appendix C.2 of \cite{Gromov:2011cx}):
\begin{align}
\left(\frac{\bT_{2,1}}{\wTr_{1,2}}\frac{\wTr_{1,1}^-}{\bT_{1,1}^-}\right)^2\frac{\bT_{0,0}^-}{\bT_{1,0}}\frac{Y_{1,1}}{Y_{2,2}}&&\textrm{is analytic in the upper half-plane.}
\end{align}
 By using the first equality in (\ref{Y11Y22consa})
and \(Y_{2,2}=\frac{\bT_{2,1}}{\bT_{1,2}}\) we conclude that also
\begin{eqnarray}
{\bf D}=\frac{\bT^{}_{1,2}}{\wTr^{}_{1,2}}\frac{\wTr_{1,1}^-}{\bT_{1,1}^-}=\sm\frac{
\wT^{}_{1,2}}{\wTr^{}_{1,2}}\frac{\wTr_{1,1}^-}{
\wT_{1,1}^-}
\end{eqnarray}
is analytic in the upper half-plane.

The gauge transformation between  \(\wT\)-s and \(\wTr\)-s
is of the form (\ref{arbgaugeg}) because we know that
both gauges are
real, but \(g_1\) and \(g_2\) might not have
nice analytic properties. Now
we put restrictions on \(g\)-s. First, since \(\wT_{0,s}=\wTr_{0,s}=1\), one gets \(g_1g_2^{++}=1\)\,.
Then, since both \(\wT_{2,s}=\bT_{2,s}=\bT_{s,2}\) and \(\wTr_{2,s}\)
belong to the class \(\CA_{s-1}\), the ratio
\(g_1^{[+2]}/g_1^{[-2]}\)  is analytic in the upper half-plane. Finally, from analyticity of \(\bf D\) in the upper half-plane one gets analyticity
of \(\frac{g_1^{[+2]}g_1^{[-2]}}{g_1^2}\) in the upper half-plane. Hence we
conclude that \(g_1/{g_1^{[-2]}}\) is analytic in the upper half-plane. Therefore, since
\begin{eqnarray}\label{arbgaugeg2}
\wTr_{1,s}=\left(\left(\frac{g_1}{g_1^{[-2]}}\right)^{[s]}\times c.c\right)
\wT_{1,s}\,,
\end{eqnarray}
we conclude that \(\wT_{1,s}\) has proper analyticity strips in the right band. By
repeating the same argument with the gauge \(\wTl\), we conclude that  \(\wT_{1,s}\) has proper analyticity strips in the left
band as well.

Now we have to show that  \(\hat{
  \wT}_{1,-1}^{c_r}=-\hat{
  \wT}_{1,+1}\) (resp  \(\hat{
  \wT}_{1,-1}=-\hat{
  \wT}_{1,+1}^{c_l}\)),
which will finally mean that \(\wT\) is indeed a \(\mathbb{Z}_4\)-symmetric
gauge for the
right (resp left) band%
. For this it is enough to show that \(\hat{%
  \wT}_{1,1}\) and
\(\hat{%
  \wT}_{1,-1}\) have only two cuts, \(\hat Z_{\pm 1}\), because then
we can use the logic of appendix C.4
in \cite{Gromov:2011cx} which  proves both the
\(\mathbb{Z}_4\)-symmetry of \(\bT\) in the upper band and the \(\mathbb{Z}_4\) symmetry of \(
\wT\)-s  in the right and left bands.
To this end, we use the fact  that the discontinuity
condition (1.6) of \cite{Cavaglia:2010nm} is equivalent  to the statement that
the product\begin{eqnarray}\label{middlenode}
\hat{%
  \wT}_{1,1}\hat{%
  \wT}_{1,-1}
\end{eqnarray}
has only two  cuts \(\hat Z_{\pm 1}\)  (this is obtained by straightforward generalization
of the logic in C.3 of \cite{Gromov:2011cx} for non LR-symmetric
cases).

We can force  \(\hat{%
  \wT}_{1,1}\) and
\(\hat{%
  \wT}_{1,-1}\) to separately have only two cuts. For this we notice that the  so far derived group-theoretical constraints on  the  \(\bT\)-gauge \eqref{eq:1}  do not not constrain fully the gauge freedom but the following gauge transformation is still possible
\begin{eqnarray}\label{gaugephi}
\bT_{a,s}\to \left(\prod_{k=-(|s|-1)/2}^{(|s|-1)/2}\frac{e^{i\,\phi^{[+a+2k]}}}{e^{i\,\phi^{[-a+2k]}}}\right)^{\mathrm{sgn}(s)}\bT_{a,s}\,.
\end{eqnarray}
To also
preserve reality and analyticity,  \(\phi\) should be a mirror-real function with only one long cut \(\check Z_0\). We
can use this function to change \(%
\wT_{1,1}\):
\begin{eqnarray}
\hat{%
  \wT}_{1,1}\to e^{{\rm i}(\hat\phi^+-\bar{\hat\phi}^-)}\hat{%
  \wT}_{1,1}\,.
\end{eqnarray}
Hence, to enforce $\hat\wT_{1,1}$ to have only two short cuts, we require that
\begin{eqnarray}
&&{\rm i}\ {\rm disc\ }\hat\phi^{[-2n]}={\rm -disc\ }\log \hat{%
  \wT}_{1,1}^{[-1-2n]}\,,\
\ n=1,2,\ldots\,,\no\\
&&{\rm i}\ {\rm disc\ }\bar{\hat\phi}^{[+2n]}={\rm +disc\ }\log \hat{%
  \wT}_{1,1}^{[+1+2n]}\,,\
\ n=1,2,\ldots\,.\label{eq:10}
\end{eqnarray}
Let us show how to construct such a function: first construct a function \(\hat
e\) with short cuts which is analytic on the real axis and has specific discontinuities across the other cuts \(\hat Z_n\):
\begin{eqnarray}
&&{\rm i}\ {\rm disc\ }\hat e^{[-2n]}={\rm -disc\ }\log \hat{%
  \wT}_{1,1}^{[-1-2n]}\,,\
\ n=1,2,\ldots\,,\no\\
&&{\rm i}\ {\rm disc\ }{\hat e}^{[+2n]}={\rm +disc\ }\log \hat{%
  \wT}_{1,1}^{[+1+2n]}\,,\
\ n=1,2,\ldots\,.
\end{eqnarray}
Then \(\phi\) is a function with one mirror cut which is a solution of the
Riemann-Hilbert problem
\begin{eqnarray}\label{eqonphi}
\phi^{[+0]}+\phi^{[-0]}=\hat e,\ \ u\in\check Z_0\,.
\end{eqnarray}
Note that \(\hat e\) is magic-real by construction, hence \(\phi\) will be  mirror-real,
as it should. Finally note that the solution of \eqref{eqonphi} is fixed up to a term \(\delta\phi=\Lambda\log x\) which is responsible for the remaining functional rescaling symmetry \eqref{xrescaling}.

Hence, by adjusting \(\phi\), we can force \(\hat{%
  \wT}_{1,1}\) to have
only two cuts, and then, due to (\ref{middlenode}),  \(\hat{%
  \wT}_{1,-1}\)
has only two cuts. Now by repeating the logic of C.4 in \cite{Gromov:2011cx}
we conclude that  all the  properties of the \(\bT\)-gauge listed in the statement are satisfied even
if there is no LR symmetry.
\paragraph*{Uniqueness.}
The proof of uniqueness is already contained in the derivation above, we need only to summarize it. By demanding the group-theoretical constraints we fixed all but one gauge freedom. The remaining one \eqref{gaugephi} is determined by the function \(\phi\) which should preserve analyticity in the mirror kinematics and also analyticity in the physical kinematics (which exists in the right and left bands due to \(\mathbb{Z}_4\) symmetry). The only transformation preserving these analyticities is \eqref{xrescaling}. Note that in the LR-symmetric case we demand \(\bT_{a,s}=\bT_{a,-s}\) and even the transformation \eqref{xrescaling} is forbidden.

Finally, there is always a gauge freedom in multiplication by an \(i\)-periodic function. But such a function should not have branch points to preserve analyticity and, due to conjecture of regularity, it should also preserve the  condition of  absence of poles and of a polynomial behaviour at infinity. Hence it should be a constant. \qed
\ \\

The presented  proof is a constructive one. It shows us how to build  the \(\bT\)-gauge starting from any analytic gauges \(\sT\) and \(\cT_r\), \(\cT_l\)  in the corresponding bands. This construction was practically used in the derivation of the FiNLIE  \cite{Gromov:2011cx} for LR-symmetric states.  The generalization presented here allows, in principle,   to repeat the logic of \cite{Gromov:2011cx} and to derive the FiNLIE for arbitrary state, departing from the proper \(\sT\)- and \(\cT\)-gauges that are suitably parameterized in terms of resolvents.

\subsection{\texorpdfstring{$\fQ\to{\mQ}\to{\bT}$}{Q->T}}
\label{sec:solution-t-system}
\subsubsection{Wronskian parameterization}\label{sec:wronparam} The Hirota bilinear relations \eqref{Hir} are an infinite system of equations for an infinite set of functions $T_{a,s}$. This system appears to be integrable and the net result of this integrability is that T-functions can be expressed  from determinants of a finite set of {\Qfcts}. This is the so-called Wronskian parameterization. For the AdS/CFT case, it was written in \cite{Gromov:2010km} and is constructed as follows: consider $\gl(4|4)$ {\Qfcts} satisfying algebraic relations from \secref{sec:qq-relations}. We demand $Q_{\es|\es}=1$ but we impose no restrictions on $Q_{\bar\es}$ for the moment.
Let us split bosonic indices \(\{1,2,3,4\}\) into two sets: the ``right'' set \(\{1,2\}\), its elements to be labeled by \(\alpha\), and
the ``left'' set \(\{3,4\} \), its elements to be labeled by \(\dot\alpha\).
{\Tfcts} that are expressed as
 \begin{eqnarray}\label{TfromQ}
  T_{a,s}=
  \begin{cases}  \frac{(-1)^{as+1}}
{a!(2-a)!} \eta^{\alpha_1\alpha_2} Q_{\alpha_1\ldots \alpha_a|\es}^{[+s]}Q_{\alpha_{a+1}\ldots
\alpha_234|1234}^{[-s]}\,, \ \ \ \ \ \ \ &s\geq a\,,\\
\frac{%
  1%
}%
{(2-s)!(2+s)!}\e^{i_1\ldots i_4} {Q_{12|i_1\ldots
    i_{2-s}}^{[+a]}Q_{34|i_{3-s}\ldots
    i_{4}}^{[-a]}}
  ,&a\geq|s|\,,\\
\frac{(-1)^{as+a+1}}%
{a!(2-a)!}\eta^{\dot\alpha_1\dot\alpha_2}Q_{\dot\alpha_1\ldots \dot\alpha_a|\es}^{[+s]}Q_{12\dot\alpha_{a+1}\ldots \dot\alpha_2|1234}^{[-s]}\,,&-s\geq a\,,
\end{cases}
 \end{eqnarray}
automatically satisfy  Hirota equations
\cite{Gromov:2010km,Tsuboi:2011iz,KLV-Qsystem}.

The Wronskain parameterization constructs T-functions in a gauge with \(T_{n,\pm 2}=T_{2,\pm n}\), \(n\geq 2\) and \(T_{0,s}=f^{[-s]}\). Any solution of Hirota equation can be brought to a gauge of this type, construction of the \(\bT\)-gauge is a good example. Moreover, it is possible to show that any ``smooth enough'' solution in this gauge class has a Wronskian parameterization.

The goal of this section is to depart from the fundamental Q-system and to construct the Q-functions that reproduce, through the Wronskian parameterization, all the above-described analytic properties of the \(\bT\)-gauge  and hence the \(\bT\)-gauge itself, due to its uniqueness property. We stress that the fundamental Q-system may not be boldly substituted into \eqref{TfromQ} for achieving this goal, but it will become clear very soon what is the correct procedure.

\subsubsection{Mirror Q-system}
\label{sec:mirrorQtoTBA}
Splitting of bosonic indices between two sets (right and left) in \eqref{TfromQ}
brakes the \(\Gl(4)\) symmetry (\ref{Htransform}) of bosonic H-rotations
to \(\Gl(2)\otimes\Gl(2)\). It is certainly important which bosonic Q's
will be called \(Q_{1|\es}\) and \(Q_{2|\es}\) and which will be called
\(Q_{3|\es}\), \(Q_{4|\es}\): for instance the analyticity strips of the
T-functions will follow from the analyticity of \(Q_{1|\es}\) and
\(Q_{2|\es}\) in the upper half-plane (because these Q-functions appear
with a positive shift in \eqref{TfromQ}) and from the analyticity of \(Q_{3|\es}\) and
\(Q_{4|\es}\) in the lower half-plane.
 Hence, we have to choose a proper basis in the Q-system (the Q-basis) prior to constructing the Wronskian solution.

The AdS/CFT {\Ysys} is defined in the mirror kinematics (long cuts), hence the
{\Qfcts} that solve it should obey QQ-relations in the mirror
kinematics. On the other hand, we have the fundamental {\Qsys} from \secref{sec:analyticQ} at our
disposal, with QQ-relations defined in the upper half-plane. Hence we
continue these QQ-relations to the mirror kinematics from the upper
half-plane and restrict ourselves to this kinematics. Now we can do
H-rotations from the fundamental {\Qsys},
keeping in mind that the rotations should be \(i\)-periodic in the
mirror kinematics and due to the  possible cuts  they might be not
periodic in the physical kinematics. We pose the following question:
How to lift the ambiguity in choosing of the Q-basis and fix the right one, by the appropriate mirror H-rotations, so as to reproduce the \(\bT\)-gauge?

This correct Q-basis will be called {\it mirror} and  denoted by the bold
font \(\mQ\), to reflect that it should reproduce the mirror  {\Tsys} in
the distinguished gauge \(\bT\).
For the fermionic Q's, there is no subtlety in choosing the mirror
Q-basis because the whole construction is invariant under H-rotations of
fermions. It is particularly nice to keep the fundamental \(\fQ_{\es|i}\)  as the basis functions since they already have simple analyticity in the mirror kinematics. Hence we just choose \(\mQ_{\es|i}=\fQ_{\es|i}\)  and also use simplified notations \(\mQ_{i}\equiv\mQ_{\es|i}\) which explains our notation choice for \(\bQ\)'s in the {\Qosys}.

For bosonic \(Q\)'s, we will use the notations \(\bosQ{a}\equiv \mQ_{a|\es}\). We recall that the mirror {\Tfcts} have
certain analyticity strips according to \eq{eq:1}, and hence, for the right band, \(\bosQ{1},\bosQ{2}\) are expected to be analytic in the upper half-plane. Clearly, they should be identified with \(\bP\)'s. But how to choose among possible \(\bP\)'s? There is one distinguished choice: \(\bP_1\) and \(\bP_2\) are smaller in magnitude than \(\bP_3\), \(\bP_4\) at large \(u\), see \secref{sec:asymptotics-at-large}, and in this sense they are defined unambiguously. So we will identify
\begin{equation}
\label{eqQ1}
\bosQ{1}=\bP_1\,,\ \ \bosQ{2}=\bP_2\,, \ \  \Im(u)>0\,,
\end{equation}
Similarly, for the left band, \(\bosQ{3},\bosQ{4}\) are expected to be analytic in the lower half-plane. On the other hand, they should be a certain mirror H-rotation of $\bP_a$'s. This hints us to use $\bP^a$'s to define \(\bosQ{3},\bosQ{4}\) because they can be considered as LHPA objects obtained from UHPA $\bP_a$'s through a rotation with $H=\mu$, see \secref{sec:discussionaxioms}. \(\bP^3,\bP^4\) are the two smallest functions among \(\bP\)'s with upper indices and in this sense they are also unique. Hence we will identify:
\begin{equation}
\label{eq3443}
\bosQ{3}=\hat\bP^4\,,\ \ \bosQ{4}=-\hat\bP^3\,, \ \  \Im(u)<0\,.
\end{equation}
the sign choice is for further convenience.

\Equref{eq3443} is written as follows for \(\Im(u)>0\):   \(\bosQ{3}=\tilde\bP^4=\mu^{4a}\bP_a\), \(\bosQ{4}=-\tilde\bP^3=-\mu^{3a}\bP_a\)\,, so, in summary, the desired mirror H-transformation from the fundamental to the mirror {\Qsys} is
\begin{align}\label{eq:funtomir1}
&&&&&&
\begin{pmatrix}
\bosQ 1\\ \bosQ 2\\ \bosQ 3\\ \bosQ 4
\end{pmatrix}
&=
\begin{pmatrix}
1&0&0&0\\
0&1&0&0\\
+\mu_{23}&-\mu_{13}&\mu_{12}&0\\
+\mu_{24}&-\mu_{14}&0&\mu_{12}
\end{pmatrix}
\begin{pmatrix}
\check\bP_1\\\check\bP_2\\\check\bP_3\\\check\bP_4
\end{pmatrix}\,,
&&&&&&
\end{align}
where the check on top of \(\bP\)'s reminds us that in this particular formula \(\bP\)'s are considered in the mirror kinematics and are defined by analytic continuation from the upper half-plane where they are free from cuts.

More generally, the functions \(\mQ_{A|I}\) and \({\cal Q}_{A|I}\) are related by the
transformation (\ref{Htransform}) where \(H_f=1\) and where \(H_b\) is the
matrix of equation (\ref{eq:funtomir1}). This matrix has the
determinant \(\mu_{12}^2\), hence
\begin{equation}
\mQ_{\bar\es}=(\mu_{12}^+)^2%
\,.
\end{equation}
Since \(\bT_{0,s}=\mQ_{\bar\es}^{[-s]}\), we recover, by comparing with \eqref{bottomline}, \(\CF=\mu_{12}^+\).

Since \(\mQ_{\bar\es}\neq1\), one cannot treat Hodge-dual Q-functions defined by \eqref{Hodgedef} on the equal footing with the original functions. However, we can improve situation by adjusting the definition of Hodge-dual to the case \(\mQ_{\bar\es}\neq1\):
\begin{equation}
\begin{aligned}\label{HodgedefM}
 \mQ^{A|I}\equiv&
(\CF^{|A|-2})^{[\,|A|-|I|\,]}
 {(-1)^{|A'||I|}\e^{AA'}\e^{I'I}\mQ_{A'|I'}}{}\,,&\textrm{where
    }\{A'\}=&\{1,2,3,4\}\setminus \{A\},\\&& \{I'\}=&\{1,2,3,4\}\setminus
    \{I\}\,.
  \end{aligned}
\end{equation}
One should update correspondingly the normalization in \eqref{QQrel1}
\begin{eqnarray}\label{QQrel12}
&&\mQ_{\emptyset|i}=
-\frac{1}{\CF^+}
\, \mQ^{a|\emptyset}\mQ_{a|i}^\pm\,,\qquad \mQ^{\emptyset|i}=
\frac 1{\CF^+}
\mQ_{a|\emptyset}(\mQ^{a|i})^\pm\,,
\end{eqnarray}
and \eqref{QQorto2}
\begin{eqnarray}\label{ortoQij2}
\mQ^{a|i}\mQ_{a|j}=-\delta^{i}{}_j\,\CF\;\;,\;\;\mQ^{a|i}\,\mQ_{b|i}=-\delta^a{}_b\,\CF\,,
\end{eqnarray}
but the  other properties of the Q-system remain the same.

In this way one has again \(\mQ^{\es|\es}=1\) and Hodge-duality remains indeed the symmetry of the system. In particular, one can deduce that, in full analogy with the lower-index formulae, \(\mQ^{i}=\fQ^{\es|i}\) and
\begin{align}\label{eq:funtomir2}
&&&&&&
\begin{pmatrix}
\bosQup 1\\ \bosQup 2\\ \bosQup 3\\ \bosQup 4
\end{pmatrix}
&=
\begin{pmatrix}
\mu_{12}&0&-\mu_{23}&-\mu_{24}\\
0&\mu_{12}&+\mu_{13}&+\mu_{14}\\
0&0&1&0\\
0&0&0&1
\end{pmatrix}
\begin{pmatrix}
\check\bP^1\\\check\bP^2\\\check\bP^3\\\check\bP^4
\end{pmatrix}\,,
&&&&&&
\end{align}
which can be rewritten also as
\be
&&\bosQup 3=\bP^3\,,\ \ \bosQup 4=\bP^4\,, \ \ \Im(u)>0\,,
\\
&&\bosQup 1=-\hat\bP_2\,,\ \ \bosQup 2=\hat\bP_1\,, \ \ \Im(u)<0\,.
\ee
We can now write down T-functions more explicitly, with the help of Hodge-dual \(\bQ\)'s:
\begin{subequations}
\label{wronskian}
\begin{align}
  \label{eq:6}
        \bT_{0,s}&=\bQ_{\bar\es}^{[-s]},\\
\label{eq:6b1}       \bT_{1,s}&=-(-1)^s\CF^{[s+1]}\left(\bQ_{1|\emptyset}^{[+s]}(\bQ^{1|\es})^{[-s]}+\bQ_{2|\emptyset}^{[+s]}(\bQ^{2|\es})^{[-s]}\right)\,,
s\geq 1\,,\\
\label{eq:6b2}        \bT_{1,s}&=+(-1)^s\CF^{[s+1]}\left(\bQ_{3|\emptyset}^{[+s]}(\bQ^{3|\es})^{[-s]}+\bQ_{4|\emptyset}^{[+s]}(\bQ^{4|\es})^{[-s]}\right)\,,s\leq
-1\,,\\
        \bT_{2,s}&=+\bQ_{12|\emptyset}^{[+s]}(\bQ^{12|\es})^{[-s]}\,,\ s\geq
2\,,\ \ \bT_{2,s}=+\bQ_{34|\emptyset}^{[+s]}(\bQ^{34|\es})^{[-s]}\,, s\leq-2\,,
\\  %
\label{bTa2}
        \bT_{a,2}&=+\bQ_{12|\emptyset}^{[+a]}(\bQ^{12|\es})^{[-a]}\,,\ \
\bT_{a,-2}=+(\bQ^{34|\es})^{[+a]}\bQ_{34|\emptyset}^{[-a]}\,,\ \ a\geq 2
\\  %
\label{bTa1}
        \bT_{a,1}&=-\bQ_{12|i}^{[+a]}(\bQ^{12|i})^{[-a]}\,,\
\bT_{a,-1}= (\bQ^{34|i})^{[+a]}\bQ_{34|i}^{[-a]}\,,\ \ a\geq 1
\\  %
 \bT_{a,0}&=+\frac 12 \bQ_{12|ij}^{[+a]}(\bQ^{12|ij})^{[-a]}\,,\ \  a\geq 0\,.\label{eq:7}
\end{align}
\end{subequations}

It is straightforward to check that  T-functions constructed from the
mirror Q-system satisfy all the properties of the distinguished
\(\bT\)-gauge. Indeed, analyticity is immediate from the half-plane
analyticities of the Q-functions. Reality comes from conjugation
properties of Q's given in \secref{app:reality-1}. Absence of poles and power-like asymptotics is a property of fundamental $\fQ$'s, hence of mirror $\mQ$'s, and hence of $\bT$'s.
Finally, for \(\mathbb Z_4\) symmetry it is enough to demonstrate it for \(\wT\)-functions, as the \(\mathbb Z_4\) symmetry for the \(\bT\)-functions follows from it \cite{Gromov:2011cx}. Since \(\CF=\mu_{12}^+\), we can write down
\begin{equation}
\wT_{1,s}=\frac{(-1)^{s}}{{\CF}^{[s+1]}}\bT_{1,s}=%
-%
\left(\bosQs{1}{[+s]}(\bosQup{1})^{[-s]}+\bosQs 2 {[+s]}(\bosQup {2})^{[-s]}\right)=
\left(\hat\bP_{1}^{[+s]}\hat\bP_{2}^{[-s]}-\hat\bP_{2}^{[+s]}\hat\bP_{1}^{[-s]}\right)\,,
\end{equation}
which is nothing but (\ref{Tp1p2})  and which is explicitly \(\mathbb Z_4\)-symmetric.

To conclude, we demonstrated how to reconstruct the T-functions in the \(\bT\)-gauge from the fundamental Q-system. A non-trivial step was to first perform the H-rotation in the mirror kinematics so as to map fundamental Q's to the mirror Q's. It is crucial to apply the Wronskian ansatz to the latter ones but not to the fundamental Q's. Hence the procedure can be summarized as the \(\fQ\to{\bQ}\to{\bT}\) mappings.
\subsection{\texorpdfstring{$\bT\to\bQ\to\fQ$}{T->Q}}
\label{sec:quant-sprect-curve}
In this section we derive the fundamental Q-system from the distinguished T-gauge. The overall logic is the opposite to the previous section, {\ie} we will first derive the properties of the mirror Q-basis and then explain how one could ``guess'' the existence of the fundamental basis. In this section, the properties of  neither mirror nor fundamental Q-system are assumed to be known.  Note that decoding analyticity of Q-functions from T's is a daunting task as opposed to the \(Q\to T\) direction. It will take us quite an effort to perform it.

First we note that since \(\bT_{0,s}=\mQ_{\bar\es}^{[-s]}\) and, on the other hand, one has \eqref{bottomline}, we should identify \(\bQ_{\bar\es}=\CF^2\). Therefore, it is advantageous to normalize Hodge-dual functions as in \eqref{HodgedefM}. Then the Wronskian parameterization \eqref{wronskian} only flips left and right wings of the T-hook when exchanging Q-functions with their Hodge duals. Hence, Hodge-dual objects should enter in a symmetric way in the whole construction.

Let us remind that the \(\bT\)-gauge is fully constrained, up to a minor \eqref{xrescaling}. However, Q-functions of the Wronskian parameterization are not firmly fixed by the choice of the gauge. One can still perform H-rotations. Our first step is to find, by performing H-rotations, a Q-basis with as simple as possible analytic properties of \(\bQ\)'s solving the Hirota equations.  In performing this step, we will heavily use the results of \cite{Gromov:2011cx}, in particular of appendix D there, and generalize it to the non LR symmetric case. However, some conceptually new findings will be presented as well.
\subsubsection{Analyticity of \fcts{\texorpdfstring{$\mQ$}{Q}} in the mirror basis}\label{sec:TBA2Q}
Most of the basic analytic structure can be read off from
\fip. Indeed, the Wronskian solution was already studied in detail in
that paper, but separately for the left/right and the upper bands of
the T-hook solvable by \(\gl(2)\)- and \(\gl(4)\)-symmetric Wronskian
ans\"atze, correspondingly. In the distinguished gauge these solutions
are allowed to be glued together into \eqref{wronskian} since both
conditions \(\bT_{2,\pm n}=\bT_{n,\pm 2}\) and \(\bT_{0,s}=\bT_{0,0}^{[-s]}\)
are satisfied. So at this point we just have to compare the notations
of {\fip} and of the current paper.

\paragraph{One cut for bosonic Q's.}

It was proved in {\fip} that one can always find a Q-basis, by means
of H-rotation, which solves the right and left bands in the following
way\footnote{We adjusted the notation for the non-LR-symmetric case by
  introducing \(\wQ_3\) and \(\wQ_4\). In {\fip},
  \(\bT_{1,s}=\bT_{1,-s}\) was satisfied, so \(\wQ_1\) and \(\wQ_2\)
  were enough. Apart for this remark, LR symmetry is not used in the
  derivation of \eqref{rightleft}, hence we rely on
  {\fip}.}$^,$\footnote{The reader should be warned that in this
  section, the symbol $\wQ$ denotes Q-functions in the
  $\wT$-gauge. This should not be confused with the polynomials
  denoted by the same symbol in \secref{sec:ABA}.}.%
\begin{eqnarray}\label{rightleft}
 \bT_{1,s}
&=&(-1)^{s}\CF^{[1+s]}(\hat\wQ_{1}^{[+s]}\hat \wQ_{2}^{[-s]}-\hat\wQ_{2}^{[+s]}\hat \wQ_{1}^{[-s]})\,,\ \ s\geq +1,\no\\
\bT_{1,s}
&=&(-1)^{s}\CF^{[1+s]}(\hat\wQ_{3}^{[+s]}\hat \wQ_{4}^{[-s]}-\hat\wQ_{4}^{[+s]}\hat
\wQ_{3}^{[-s]})\,,\ \ s\leq -1\,,
 \
\end{eqnarray}
where \(\hat\wQ_i\) have only one short cut in the physical kinematics. \(\wT\)-gauge is explicitly \(\mathbb Z_4\)-symmetric in this parameterization.

\begin{figure}
  \begin{center}
    \includegraphics[width=0.2\textwidth]{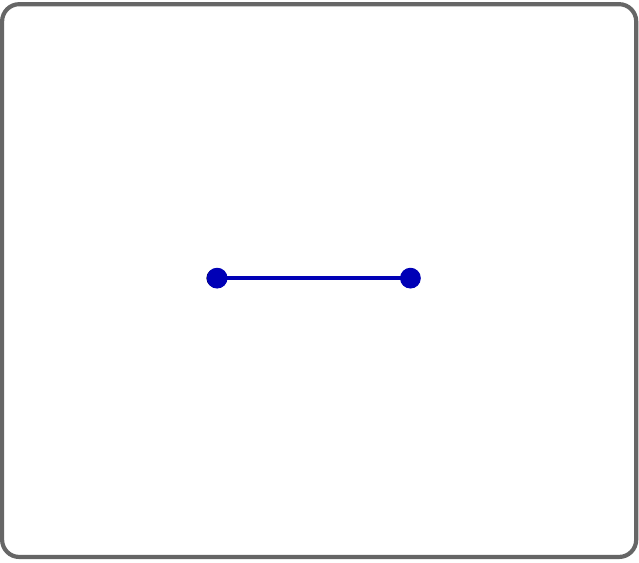}
  \caption{\label{fig:one cut} \(\hbosQ{i}\) and \(\hbosQup{i}\) have only one short cut.}
   \end{center}
\end{figure}

By direct comparison with \eqref{eq:6b1} and \eqref{eq:6b2} one gets
\begin{align}\label{bQwQ}
\bosQ{1}&=\hat\wQ_{1}\,, &  \bosQ{2}&=\hat\wQ_{2}\,, & \bosQup{3}&=-\hat\wQ_3\,, & \bosQup{4}&=\hat\wQ_4\,,  & \Im(u)&>0\,,
\no\\
\bosQup{1}&=-\hat\wQ_{2}\,, &  \bosQup{2}&=\hat\wQ_{1}\,, & \bosQ{3}&=\hat\wQ_4\,, & \bosQ{4}&=\hat\wQ_3\,,  & \Im(u)&<0\,;
\end{align}
And we arrive to the following essential conclusions. Firstly, \(\hbosQ{i}\) and \(\hbosQup{i}\) have only one short cut, see \figref{fig:one cut}.
We should not forget that \(\bQ\)'s without hat are functions in the mirror kinematics. They are analytic only in a half-plane but connected to the physical kinematics with one-cut structure by the continuation from a natural domain of analyticity. From \eqref{bQwQ} it is clear that this natural domain is the upper half-plane for \(\bosQ{1},\bosQ{2},\bosQup{3},\bosQup{4}\) and the lower half-plane for \(\bosQup{1},\bosQup{2},\bosQ{3},\bosQ{4}\).

Secondly, we see that
\begin{gather}\label{Qbos}
        \ensuremath{\hbosQ{a}={\eta_{ab}\hbosQup{b}}}
\,.\end{gather}
Both properties do not follow solely from the Wronskian solution but essentially rely on \(\mathbb Z_4\) symmetry.

\paragraph{One cut for fermonic Q's.}
In {\fip}, the upper band was solved with the help of Q-functions \(\qs\) and \(\ps\):
\begin{align}\label{upper}
&&\bT_{a,s}&=\frac{\epsilon^{i_1,i_2,i_3,i_4}}{(2-s)!(2+s)!}\qs_{i_1,\dots,i_{2-s}}^{[+a]}\ps_{i_{2-s+1},\dots,i_4}^{[-a]}\,,&a\geq |s|\,.
\end{align}
The multi-indexed \(\qs\)'s and \(\ps\)'s are expressed as combination of \(\qs_i,\ps_i\) and \(\qs_\es,\ps_\es\) as follows
\be\label{detexplicit}
\qs_{ij}=\frac 1{\qs_{\es}}\left|
  \begin{matrix}
    \qs_{i}^{+}&\qs_{i}^{-}\\
    \qs_{j}^{+}&\qs_{j}^{-}
  \end{matrix}\right|\,,\ \ \ \ \qs_{ijk}=%
\frac 1{\qs_{\es}^+\qs_{\es}^-}
\left|
  \begin{matrix}
    \qs_{i}^{[2]}&\qs_{i}&\qs_{i}^{[-2]}\\
    \qs_{j}^{[2]}&\qs_{j}&\qs_{j}^{[-2]}\\
    \qs_{k}^{[2]}&\qs_{k}&\qs_{k}^{[-2]}
  \end{matrix}\right|%
  \,,\ \ \
  \qs_{1234}=\frac{\det\limits_{1\leq i,j\leq 4}\qs_{i}^{[5-2j]}}{\qs_{\es}^{[2]}\qs_{\es}\qs_{\es}^{[-2]}}
  \,,
\ee
and the same for \(\ps\). The relations \eqref{detexplicit} are nothing but the determinant QQ-relation \eqref{eq:3} adjusted to the case \(Q_{\es}\neq 1\).

\begin{figure}
 \begin{center}
 \begin{tabular}{cccccc}
 \(\qs_{\es},\ \qs_{1234}\)&\(\qs_i,\ \qs_{ijk}\)&\(\qs_{ij}\)\hspace{1.8em}
 &
 \hspace{1em}\(\ps_{ij}\)&\hspace{0.5em}\(\ps_i,\ \ps_{ijk}\)&\hspace{0.9em}\(\ps_{\es},\ \ps_{1234}\)
 \\
 \\
 \includegraphics[height=.15\textwidth]{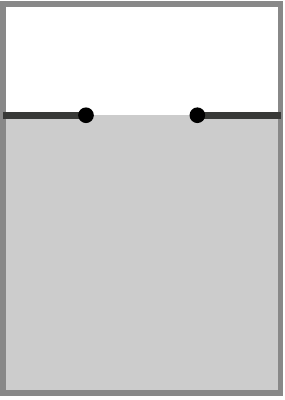}&
 \hspace{.8em}\includegraphics[height=.15\textwidth]{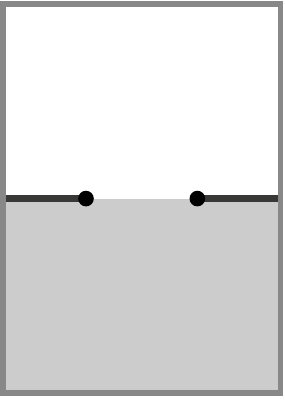}&
 \includegraphics[height=.15\textwidth]{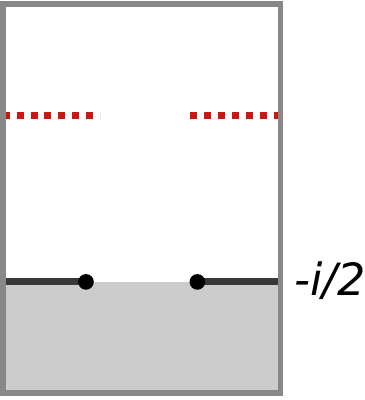}
 \hspace{-0.6em}
 &
\hspace{-0.6em} \includegraphics[height=.15\textwidth]{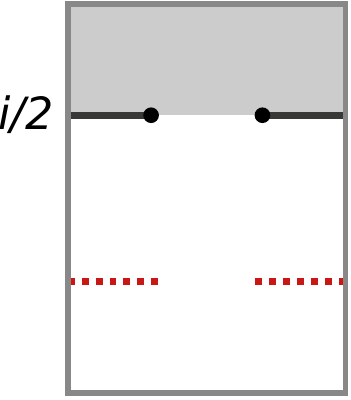}&
 \includegraphics[height=.15\textwidth]{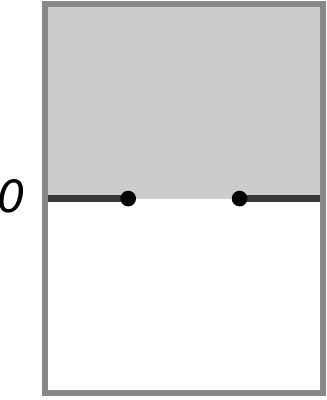}&
 \hspace{.7em}
 \includegraphics[height=.15\textwidth]{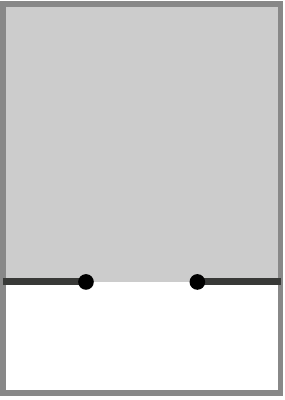}
 \\
 \\
 &\(\mQ_i\)&\(\mQ_{1|i},\ \mQ_{2|i}\)\hspace{1.8em}
 &
 \hspace{1em}\(\mQ_{3|i},\ \mQ_{4|i}\)&\hspace{0.5em}\(\mQ_i\)&
 \\
 \vspace{-0.5em}
 \\
  &\(\mQ^i\)&\(\mQ^{3|i},\ \mQ^{4|i}\)\hspace{1.8em}
 &
 \hspace{1em}\(\mQ^{1|i},\ \mQ^{2|i}\)&\hspace{0.5em}\(\mQ^i\)&
 \end{tabular}
 \caption{\label{fig:analyticity}Domains of analyticity of various Q-functions. White region is the place where a function is analytic. Above each figure we denoted  corresponding \(\ps\)'s and \(\qs\)'s in the notations of \cite{Gromov:2011cx}. Below each figure we denoted some of \(\mQ\)'s in the mirror basis with the same analyticity. Dotted line addresses \(\mQ\)'s only. It denotes the mirror cut which is possible in principle if only algebraic QQ-relations are used. However, with the help of \(\mathbb Z_4\) symmetry we show that this cut is actually absent.}
 \end{center}
 \end{figure}

In contradistinction to \(\hat\wQ\)-functions having one cut only,  \(\qs_i\) and \(\ps_i\) are only analytic in a half-plane, both in the physical and mirror kinematics. One can even show that a better than half-plane analyticity would render Y-system trivial  {\fip}. The precise statement is: we can only achieve the following domains of analyticity by choosing a proper basis via certain H-rotation:
\begin{itemize}
  \item \(\qs_{\es}\) and \(\qs_{1234}\) are analytic (have
no cuts) for \(\Im(u)>1/2\),
 \(\qs_{i}\) and \(\qs_{ijk}\) are analytic for \(\Im(u)>0\),
 \(\qs_{ij}\) are analytic for \(\Im(u)>-1/2\),

  \item \(\ps_{\es}\) and \(\ps_{1234}\) are analytic for \(\Im(u)<-1/2\),
 \(\ps_{i}\) and \(\ps_{ijk}\) are analytic for \(\Im(u)<0\),
\(\ps_{ij}\) are analytic for \(\Im(u)<1/2\),
\end{itemize}
these properties are depicted in \figref{fig:analyticity}.

The proof is based on the TQ-relations
and it is given in appendix D.6 of \cite{Gromov:2011cx}. For non
LR-symmetric case the proof requires slight generalization which is
however obvious and we leave it as an exercise for a curious
reader.

Note the following nontrivial property: from the determinant
expression \eqref{detexplicit} for \(\qs_{ijk}\), one would naively expect that \(\qs_{ijk}\) has ``less analyticity'' than \(\qs_{i}\), because \(\qs_{i}^{[-2]}\) is analytic only for \(\Im(u)>1\) whereas \(\qs_{i}\) is analytic for  \(\Im(u)>0\). However, a certain cancellation of cuts happens and eventually \(\qs_{ijk}\) and \(\qs_i\) have the same analyticity half-plane: \(\Im u>0\). This a direct consequence of the \(\bT\)-gauge being the same for the left and right parts of the T-hook. Here we see that the most analytic Q-basis is also universal, there is no separately "left" and "right" most analytic bases. It will be eventually crucial in the derivation of the quantum spectral curve.

\begin{wrapfigure}{r}{.5\textwidth}
\begin{center}
\includegraphics{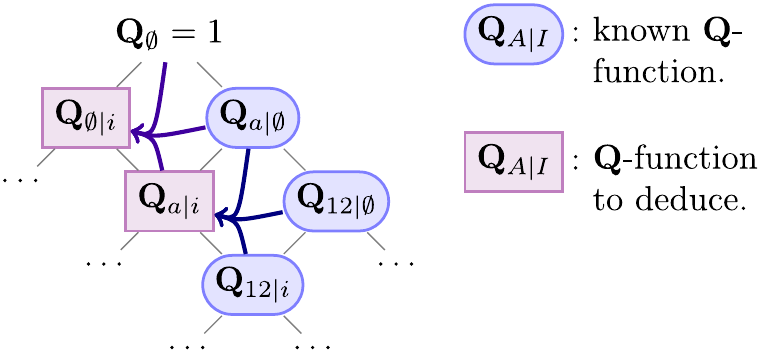}
\caption{\label{fig:strategy}Use defining QQ-relations \eqref{QQbf}, first to determine \(\mQ_{a|i}\) and then \(\mQ_{\es|i}\).}
\end{center}
\vspace{-6em}
\end{wrapfigure}

At this point we should ask: why the analyticity of the upper band of
the T-hook is ``worse'' than the one from the right/left bands? The
answer is impressively simple but at the same time it was not
understood before: because \(\qs\)- and \(\ps\)-functions are
composite objects! We have to just compare \eqref{upper} and
(\ref{TfromQ}) (or equivalently \eqref{bTa2}-\eqref{eq:7}) to conclude%
\begin{align}
\label{upperQq} &%
  \mQ_{12|I}\equiv\e_{II'}\mQ^{34|I'}=\mathbf{q}_I\,,%
&%
   \mQ_{34|I}\equiv\e_{II'}\mQ^{12|I'}=\mathbf{p}_I\,.%
\end{align}
We know that for instance \(\mQ_{12|I}\) is derivable from the basic one-indexed Q-functions \(\mQ_{i|\es}\) and \(\mQ_{\es|i}\). The properties of the bosonic \(\mQ_{i|\es}\) were identified above, and they are simple. So our next goal is to dig up the properties of \(\mQ_{\es|i}\). Will they be simple as well? To approach the answer, we will use QQ-relations \eqref{definingQQ} to systematically reduce the number of indices, as outlined in \figref{fig:strategy}.
Our first step is to compute the Q-functions with two indices,
\begin{align}\label{QQforonecut}
\mQ_{a|i}=&\frac{\mQ_{ab|i}^+\mQ_{a{|\emptyset}}^--\mQ_{ab|i}^-\mQ_{a{|\emptyset}}^+}{\mQ_{ab{|\emptyset}}}\nonumber\\=&\frac{\mQ_{12|i}^+\mQ_{a{|\emptyset}}^--\mQ_{12|i}^-\mQ_{a{|\emptyset}}^+}{\mQ_{12{|\emptyset}}}\,,
\end{align}
for \(\{a,b\}=\{1,2\}\). Analytic properties of Q-functions in the r.h.s. of  \eqref{QQforonecut} were already identified. Using them, we deduce that \(\mQ_{a|i}\) is analytic in the upper half-plane for \(\Im u>1/2\). But analyticity seems to be even better than that! A determinant of \(\bQ_{a|i}\),
\be
\qs_{ij}=\mQ_{12|ij}=
\left|
\begin{matrix}
\mQ_{1|i} & \mQ_{1|j} \\
\mQ_{2|i} & \mQ_{2|j}
\end{matrix}
\right|\,,
\label{qsmQ}
\ee
has a bigger analyticity domain, \(\Im u>-1/2\). Can it be that the
upper cut of \(\mQ_{a|i}\) is absent, {\ie} that \(\mQ_{a|i}\) is
analytic for \(\Im u>-1/2\), exactly like the combination
$\mQ_{12|ij}$? The answer is positive, however the reason is more
sophisticated than an extra analyticity of \eqref{qsmQ}.

\begin{statement}
\(\disc\mQ_{a|i}^+=0\) for any \(a=1,2\) and \(i=1,2,3,4\).
\end{statement}
The proof is given below. It is technical and can be harmlessly skipped by the reader.
\begin{proof} Let us start by showing that all the functions
\begin{eqnarray}
\label{eq:16}f_{a,i}\equiv\frac{\mQ_{a|i}^+}{\bosQs{a}{[2]}\mQ_{12|i}^{[2]}}=\frac 1{\mQ_{12}^+}\left(\frac{\bosQ{a}}{\bosQs{a}{[2]}}-\frac{\mQ_{12|i}}{\mQ_{12|i}^{[2]}}\right)
\end{eqnarray}
have the same discontinuity on the real axis for any \(a,b\in\{1,2\}\) and
\(i,j\in\{1,2,3,4\}\).
It follows from noting that the last equality in (\ref{eq:16}) is nothing but a
rewriting of the first equation in  \eqref{QQforonecut}, hence one can get
\begin{eqnarray}
&&\disc\left(f_{ai}-f_{bi}\right)=\disc\left(\frac
  1{\mQ_{12}^+}\left(\frac{\bosQ{a}}{\bosQs{a}{[2]}}-\frac{\bosQ{b}}{\bosQs{b}{[2]}}\right)\right)=\disc\left(\frac{-\e_{ab}\mQ_{12}^+}{\mQ_{12}^+\bosQs{a}{[2]}\bosQs{b}{[2]}}\right)=0\,,\ \ \forall\, a,b\,,
\end{eqnarray}
and similarly we compute
\begin{eqnarray}
&&\disc\left(f_{ai}-f_{aj}\right)=\disc\left(\frac {-1}{\mQ_{12}^+}\left(\frac{\mQ_{12|i}}{\mQ_{12|i}^{[2]}}-\frac{\mQ_{12|j}}{\mQ_{12|j}^{[2]}}\right)\right)=\disc\left(\frac{-\mQ_{12}^+\mQ_{12|ij}^+}{\mQ_{12}^+\mQ_{12|i}^{[2]}\mQ_{12|j}^{[2]}}\right)=0\,,\ \ \forall\, i,j\,,
\no\\
\label{eq:B52}
\end{eqnarray}
as requested. Since the denominator in \(f_{ai}\) is
analytic on the real axis, we have actually shown that
\begin{align}
\label{eq:18}
  \disc\left(\mQ_{a|i}^+\right)=&\frac{\bosQs{a}{[2]}\mQ_{12|i}^{[2]}}{\bosQs{b}{[2]}\mQ_{12|j}^{[2]}}\disc\left(\mQ_{b|j}^+\right)\,.
  \end{align}
So, if the discontinuity of any non-trivial linear combination of
\(\mQ_{a|i}^+\) is 0, so will be the discontinuity of all \(\bQ_{a|i}^+\). Let us consider the following combination%
\footnote{
The idea of constructing \(\mathcal{C}\) is to supplement \(\mQ_{a|i}\) with additional Q-functions having enough indices to contract them with Levi-Civita symbols. In this way we ensure that \(\mathcal{C}\) is easily related to T-functions for which we know the analytic properties in detail. The supplementary (shifted) Q-functions should be analytic on the real axis since we want to prove analyticity of \(\mQ_{a|i}^+\) there. Then we tinker with several variations of this construction until we find \(\mathcal{C}\) that  produces the desired proof.
}
\begin{align}
\mathcal{C}=\frac 16\e^{ab}\e^{ijkl}\mQ_{a|i}^+\hbosQs{b}{[-2]}\mQ_{34|jkl}^{[-2]}\,.
\end{align}

The fact that \(\mathcal{C}\) is non-trivial for the purpose of computing discontinuity is confirmed by the property
\(\disc(\mathcal{C})=\wT_{1,2}\bT_{2,1}\frac{\disc\left(\mQ_{1|1}^+\right)}{\bosQs{1}{[2]}\mQ_{12|1}^{[2]}}\),
so that  \(\disc(\mathcal{C})=0\) is only possible if
\(\disc%
  (\mQ_{1|1}^+%
)=0\).

Rewrite now \(\mathcal{C}\) in terms of T- and {\Yfcts} and
show that it has no cut on the real axis.
Slightly above real axis, we get
\begin{eqnarray}
\mathcal{C}&=&\frac
16\e^{ab}\e^{ijkl}\frac{1}{\mQ_{12}^+}\left(\mQ_{12|i}^{[2]}\mQ_{34|jkl}^{[-2]}\times\bosQ{a}
\hbosQs b{[-2]}-\mQ_{12|i}^{}\mQ_{34|jkl}^{[-2]}
\times\bosQs{a}{[2]}\hbosQs b{[-2]}\right)\\
&=&\frac 1{\hat\wT_{1,1}^{[1+0]}}\left(\bT_{2,1}\wT_{1,1}^--\bT_{1,1}^-\wT_{1,2}\right)=\frac{\bT_{2,1}}{\hat\wT_{1,1}^{[1+0]}}\left(\wT_{1,1}^-+\frac{\bT_{1,1}^-}{\frac{\bT_{2,1}}{\bT_{1,2}}\cF^{+}}\right)=\bT_{2,1}\frac{\wT_{1,1}^-}{\hat\wT_{1,1}^{[1+0]}}\left(1+\frac{1}{Y_{2,2}}\right)\,.
\no\end{eqnarray}
Here we used the explicit Wronskian parameterization
\eqref{wronskian}, the relation \(Y_{2,2}=\frac{\bT_{2,1}}{\bT_{1,2}}\)\,, and relation \eqref{eq:Fnice} between \(\bT\)-s and \(\wT\)-s.

Now we compute
\begin{eqnarray}
\frac{\mathcal{C}}{\tilde{\mathcal{C}}}=\frac{\wT_{1,1}^-\wT_{1,1}^+}{\hat\wT_{1,1}^{[1+0]}\hat\wT_{1,1}^{[-1-0]}}\times\frac{1+\frac 1{Y_{2,2}}}{1+Y_{1,1}}\,,
\label{magrat}
\end{eqnarray}
where we used that \(\bT_{2,1}\) is analytic on the real axis and that \(\tilde Y_{2,2}=Y_{1,1}^{-1}\).

The combination on the r.h.s. is equal to 1, because the second
term, known as the magic ratio, is the inverse of the first term
\cite{Gromov:2011cx}. Hence \(\disc\,{\mathcal{C}}=0\) which proves
that \(\disc\mQ_{a|i}^+=0\).
\end{proof}
In the proof above we did use the extra analyticity of the combination \eqref{qsmQ} but also we heavily used \(\mathbb Z_4\) symmetry.

Our output is that \(\mQ_{1|i}\) and \(\mQ_{2|i}\) are analytic for \(\Im(u)>-1/2\) and similarly one can show that \(\mQ_{3|i}\) and \(\mQ_{4|i}\) are analytic for \(\Im(u)<1/2\), see \figref{fig:analyticity}.

\vspace{1em}

We can continue with further reduction of the number of indices. The next and the final step is to write down
\begin{align}
\mQ_{i}=&\frac{\mQ_{a|i}^+-\mQ_{a|i}^-}{\mQ_{a{|\emptyset}}}\,.
\end{align}
If \(a=1\) or \(2\), we conclude that \(\mQ_i\) is analytic for \(\Im(u)>0\). But taking \(a=3\) and \(4\) implies that \(\mQ_i\) is analytic \(\Im(u)<0\). So \(\mQ_i\) is analytic everywhere except on the real axis where, being a function in the mirror kinematics, it can have only the long mirror cut!

Exactly the same chain of arguments applies for \(\mQ^i\). Indeed, we can solve the T-hook by Hodge-dual objects exactly in the same manner as we did with the lower-index ones. Both solutions are algebraically equivalent. What is the most important, we do not need to change the Q-basis to engineer Hodge-dual Q-functions with the maximal possible analyticity. The mirror basis is universal! For the left
and right bands it is obvious, see \eqref{bQwQ}. For the upper band the observation follows from simultaneous analyticity of \(\qs_{ijk}\) and \(\qs_i\) for \(\Im(u)>0\) (and \(\ps_{ijk}\) and \(\ps_i\) for \(\Im(u)<0\)), the non-triviality of this property was discussed shortly after \eqref{detexplicit}.

\begin{wrapfigure}{r}{0.3\textwidth}
  \begin{center}
    \includegraphics[width=0.2\textwidth]{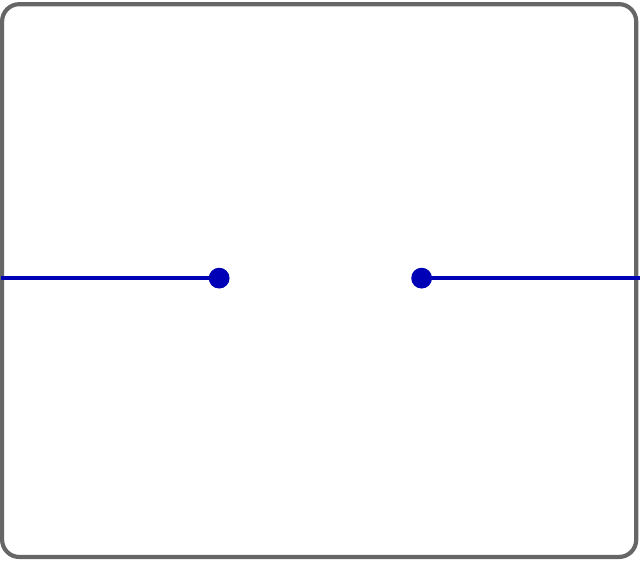}
  \caption{\label{fig:long cut} \(\mQ_{i}\) and \(\mQ^{i}\) have only one long cut.}
   \end{center}
\end{wrapfigure}

Now we can formulate the main conclusion of this subsection: \(\mQ_{\es|i}\) and \(\mQ^{\es|i}\) have only one long cut, see \figref{fig:long cut}, so the mystery of ``less analyticity'' of the upper band is resolved. We see  a very nice interplay in comparison with bosonic \(\hat\mQ\)'s which have only one short cut. An interesting question is why one has short cuts in the one case and long cuts in the other. Apparently, this cut structure is intimately connected to the representation theory as is discussed in \appref{sec:Quantization of charges}.

\subsubsection{\texorpdfstring{$\mQ\omega$-system}{Q-omega system}}\label{Z symmetry}
A sharp reader might have noticed that the analogy between bosons and
fermions is not complete. Apart from having a simple cut structure, we know from \eqref{Qbos} that bosonic \(\bQ\)'s are related to one another through the physical kinematics. We can also rephrase   \eqref{Qbos} for the functions in the mirror kinematics:
\begin{eqnarray}\label{tildebos}
 \tbosQ{a}=\eta_{ab}\bosQup{b}\,.
\end{eqnarray}
Is there an analog of \eqref{tildebos} for fermions? The answer was almost given in {\fip} as we explain below, however it should be decoded for the case of one-indexed Q's. As we now understand, this decoding is likely to simplify the formulae.

From the \(\mathbb Z_4\) property \(\hat\bT_{0,1}^c=0\) and \eqref{upper} it follows \(\hat\qs_i=\alpha\,\hat\ps_i^{[2]}+\beta\,\hat\ps_i+\gamma\,\hat\ps_i^{[-2]}\), where the coefficients \(\alpha,\beta,\gamma\) are some functions of the spectral parameter which are the same for all \(i\in\{1,2,3,4\}\). Equivalently, from \(\hat\bT_{0,-1}^c=0\), we see that \(\hat\ps_i=\alpha'\,\hat\qs_i^{[2]}+\beta'\,\hat\qs_i+\gamma'\,\hat\qs_i^{[-2]}\). By constructing certain determinants, similarly to how it was done in equations (D.2) of {\fip}, we can identify all the coefficients \(\alpha,\ldots,\gamma'\) with T-functions, so that the linear dependence between shifted \(\qs_i\) and \(\ps_i\) reads explicitly in a Baxter-like form:
\begin{subequations}
\label{halfbaxter}
\be
\label{halfbaxtera}
&&\hat\qs_i\, \hat\bT_{1,-2}^c+\hat\ps_i\,\hat\bT_{2,-1}=\hat\ps^{[2]}_i\hat\bT_{1,-1}^-+\hat\ps^{[-2]}_i\hat\bT_{1,-1}^+\,,\\
\label{halfbaxterb}
&&\hat\ps_i\, \hat\bT_{1,+2}^c+\hat\qs_i\,\hat\bT_{2,+1}=\hat\qs^{[2]}_i\hat\bT_{1,+1}^-+\hat\qs^{[-2]}_i\hat\bT_{1,+1}^+\,.
\ee
\end{subequations}
These relations are written in the physical kinematics.%

We should recall the following useful relation before proceeding\footnote{It is derived as follows. Since \(\bT_{a,2}=\bT_{2,a}\) for \(a\geq 2\),  \(\bT_{2,s}=\wT_{2,s}=\hat\wT_{1,1}^{[+s]}\hat\wT_{1,1}^{[-s]}\), and \((\hat\bT_{a,2}^c)^+(\hat\bT_{a,2}^c)^-=\hat\bT_{a+1,2}^c\hat\bT_{a-1,2}^c\) for \(a\in\mathbb{Z}\), we can conclude that \(\hat\bT_{a,2}^c=\hat\wT_{1,1}^{[+a]}\hat\wT_{1,1}^{[-a]}\) for \(a\in\mathbb Z\). Similarly \(\hat\bT_{a,-2}^c=\hat\wT_{1,-1}^{[+a]}\hat\wT_{1,-1}^{[-a]}\).}
\be\label{relationT}
\hat\bT_{a,2}^c=\hat\wT_{1,1}^{[+a]}\hat\wT_{1,1}^{[-a]}\,,\ \ {\rm in\ particular}\ \ \ \hat\bT_{1,\pm 2}^c=\hat\wT_{1,\pm 1}^{+}\hat\wT_{1,\pm 1}^{-}.
\ee
Also, since \(\hat\bT_{a,2}^c=\hat\qs_{\es}^{[+a]}\hat\ps_{1234}^{[-a]}\), we identify \(\hat\qs_{\es}=\hat\ps_{1234}=\hat\wT_{1,1}\) and similarly \(\hat\ps_{\es}=\hat\qs_{1234}=\hat\wT_{1,-1}\). Finally we remind that \(\bT_{1,\pm 1}=\CF\,\wT_{1,\pm 1}\).

\vspace{1em}
Should the \(\ps\)-term be absent in \eqref{halfbaxterb}, it would be the standard Baxter equation, a functional equation of the second order. For such equations we know that the following Wronskian combination of any two solutions \(\qs_i\) and \(\qs_j\) is \(i\)-periodic:
\(
\frac 1{\hat\bT_{1,1}}\,
\left|
\begin{matrix}
\qs_i^+ & \qs_j^+
\\
\qs_i^- & \qs_j^-
\end{matrix}
\right|
=\frac {\hat\qs_{\es}}{\hat\bT_{1,1}}\hat \qs_{ij}=\frac 1{\hat\CF}\hat \qs_{ij}\,,
\)
where we used relations summarized after \eqref{relationT}.

Should the \(\qs\)-term be absent in \eqref{halfbaxtera}, it would be another standard Baxter equation, with the Wronskian \(i\)-periodic combination \(
\frac 1{\hat\bT_{1,-1}}\,
\left|
\begin{matrix}
\ps_i^+ & \ps_j^+
\\
\ps_i^- & \ps_j^-
\end{matrix}
\right|
=\frac {\hat\ps_{\es}}{\hat\bT_{1,-1}}\hat \ps_{ij}=\frac 1{\hat\CF}\hat \ps_{ij}\).

When both equations \eqref{halfbaxter} are considered in the full generality, the following Wronskian-type combination
\footnote{We shifted the definition of \(\omega\) by \(i/2\) and
  flipped its sign in comparison with {\fip}.}
\footnote{In the case of LR symmetry another $i$-periodic combination \(\chi\) is possible, see \secref{sec:particular-case-left}. }
\be
\omega_{ij}^+=-\frac {1}{\hat\CF}\left(\hat\qs_{ij}+\hat\ps_{ij}\right)
\ee
turns out to be periodic in the physical kinematics. By default, \(\omega\) is considered as a function in the physical kinematics, hence we do not write an explicit hat on top of it. There are 4 linearly independent solutions to \eqref{halfbaxter}, hence \(\omega\) is a \(4\times 4\) antisymmetric matrix.

Existence of periodic \(\omega\) is a result of \fip. We have only generalized it to the non LR symmetric case. The proof of periodicity is rather standard. In the expression below we substitute \(\qs^{[2]}\) and \(\ps^{[2]}\) with the linear combinations dictated by \eqref{halfbaxter}, and several cancellations occur. We explicitly show only the cancellation coming from \(\hat\bT_{1,\pm 2}^c\) terms:
\be
\omega_{ij}-\omega_{ij}^{[2]}&=&\frac 1{\hat\bT_{1,1}^+}\,
\left|
\begin{matrix}
\qs_i^{[2]} & \qs_j^{[2]}
\\
\qs_i & \qs_j
\end{matrix}
\right|-\frac 1{\hat\bT_{1,1}^-}\,
\left|
\begin{matrix}
\qs_i & \qs_j
\\
\qs_i^{[-2]} & \qs_j^{[-2]}
\end{matrix}
\right|
+
\frac 1{\hat\bT_{1,-1}^+}\,
\left|
\begin{matrix}
\ps_i^{[2]} & \ps_j^{[2]}
\\
\ps_i & \ps_j
\end{matrix}
\right|-\frac 1{\hat\bT_{1,-1}^-}\,
\left|
\begin{matrix}
\ps_i & \ps_j
\\
\ps_i^{[-2]} & \ps_j^{[-2]}
\end{matrix}
\right|
\no\\
&=&\frac{\hat\bT_{1,+2}^c}{\hat\bT_{1,+1}^+\hat\bT_{1,+1}^-}
\left|
\begin{matrix}
\hat\ps_i & \hat\ps_j
\\
\hat\qs_i & \hat\qs_j
\end{matrix}
\right|
+\frac{\hat\bT_{1,-2}^c}{\hat\bT_{1,-1}^+\hat\bT_{1,-1}^-}
\left|
\begin{matrix}
\hat\qs_i & \hat\qs_j
\\
\hat\ps_i & \hat\ps_j
\end{matrix}
\right|
=\left(\frac{\hat\wT_{1,+1}^+\hat\wT_{1,+1}^-}{\hat\bT_{1,+1}^+\hat\bT_{1,+1}^-}-\frac{\hat\wT_{1,-1}^+\hat\wT_{1,-1}^-}{\hat\bT_{1,-1}^+\hat\bT_{1,-1}^-}\right)
\left|
\begin{matrix}
\hat\ps_i & \hat\ps_j
\\
\hat\qs_i & \hat\qs_j
\end{matrix}
\right|
\no\\
&=&0,
\ee
where on the last two steps we used \eqref{relationT} and \(\bT_{1,\pm 1}=\CF\,\wT_{1,\pm 1}\).

Now we rewrite \(\omega\) in terms of \(\mQ\)'s:
\be
\omega_{ij}^+=-\frac{\hat\mQ_{12|ij}+\hat\mQ_{34|ij}}{\hat\CF}=\frac {1}{\hat\CF}\hat\mQ_{a|i}\,\hat\mQ_{b|j}\,\eta^{ab}\,.
\ee
In this form it is especially simple to demonstrate that \(\Pf(\omega)=1\):
\be
\Pf(\omega^+)&=&\frac 18\e^{ijkl}\,\omega_{ij}\,\omega_{kl}=\frac 1{8\,\hat\CF^2}\e^{ijkl}\hat\mQ_{a|i}\,\hat\mQ_{b|j}\,\hat\mQ_{c|k}\,\hat\mQ_{d|l}\,\eta^{ab}\,\eta^{cd}
\no\\
&=&\frac 1{\hat\CF^2}\left(\det\limits_{1\leq a,i\leq 4}\hat\mQ_{a|i}\right)\Pf(\eta^{-1})=\frac{\mQ_{\bar\es}}{\CF^2}=1\,.
\ee
The last equality implies the following one \(\omega^{ij}\equiv(\omega^{-1})^{ij}=-\frac 12\e^{ijkl}\o_{kl}\).

\(\omega\) is known to relate \(\hat\qs\)'s and \(\hat\ps\)'s in the physical kinematics {\fip}, which in our new interpretation means for certain functions that it relates \(\tilde\mQ\) and Hodge-dual \(\mQ\) in the mirror, precisely in the spirit of \eqref{tildebos}!  It is the simplest to demonstrate the statement for \(\mQ_{a|j}\). Consider the following relations
\be\label{eqd:1}
\omega_{ij}^+\hat\mQ^{a|j}=\frac {1}{\hat\CF}\eta^{bc}\hat\mQ_{b|i}\hat\mQ_{c|j}\hat\mQ^{a|j}=\eta^{ab}\hat\mQ_{a|i},
\ee
where we used \eqref{ortoQij2}. It is tempting to transfer \eqref{eqd:1} to the mirror kinematics. The answer depends for which value of the spectral parameter we will write it. Consider \(\alpha\in\{1,2\}\), then slightly {\bf below} real axis one has
\be\label{eqd:2}
\bQ_{\alpha|i}^+=\eta_{\alpha b}\,\omega_{ij}(\bQ^{b|j})^+\,,\ \ \
{\rm and}\ \ \ \tilde\bQ_{\alpha|i}^-=\eta_{\alpha b}\,\omega_{ij}(\bQ^{b|j})^-\,,
\ee
where we used the analyticity structure of \(\bQ_{\alpha|j}\), see
\figref{fig:analyticity}. To proceed further, one can write the
difference of the two equations in \eqref{eqd:2} and use the defining
QQ relation \(\bQ_{\alpha|i}^+-\bQ_{\alpha|i}^-=\bosQ{\alpha}\bQ_{i}\): for the
current value of \(\alpha\) and the spectral parameter it reads \(\bQ_{\alpha|i}^+-\tilde\bQ_{\alpha|i}^-=\tbosQ{\alpha}\tilde\bQ_{i}\). The bosonic \(\bosQ{\alpha}\) will conveniently cancel out due to \eqref{tildebos} and we get
\begin{subequations}
\label{QtoQapp}
\be\label{eqd:3}
\tilde\bQ_{i}=\omega_{ij}\bQ^{j}\,,\ \ \  u\in{\mathbb{R}-i\,0}\,.
\ee
This is precisely like \eqref{tildebos}, except for the restriction \(u\in\mathbb{R}-i\,0\). But, in fact, this restriction is not needed, and the analogy is complete. Indeed, one should consider \({\dot \alpha}\in\{3,4\}\)  and perform a similar analysis to get
\be\label{eqd:4}
\tilde\bQ_{i}=\omega_{ij}\bQ^{j}\,,\ \ \  u\in{\mathbb{R}+i\,0}\,,
\ee
\end{subequations}
as desired.

At this point, the reader may wonder how %
\eqref{eqd:3} and \eqref{eqd:4} can %
simultaneously hold. \eqref{eqd:3} tells us that \(\tilde\bQ_{i}=\tilde\omega_{ij}\bQ^{j}\) for \(u\in{\mathbb{R}+i\,0}\), because \(\omega\) is the function in the physical kinematics as opposed to \(\mQ\)'s. Hence the only possibility for \eqref{eqd:4} to hold is \((\omega_{ij}-\tilde\omega_{ij})\mQ^{j}=0\). Let us check whether this is indeed so by computing \(\disc\omega_{ij}\). To this end, introduce \(\check\omega\), the function in the mirror kinematics, and note that slightly above the real axis:
\be
\check\omega_{ij}=\omega_{ij}\ \ \  {\rm and}\ \ \ \check\omega_{ij}^{[2]}=\tilde\omega_{ij}\,,
\ee
the first one is just the definition of connection of mirror and physical kinematics, the second one should be clear from the diagram
\be
\begin{tabular}{ccc}
\(\check\omega\)&&\(\omega\)\\
\includegraphics[width=0.3\textwidth]{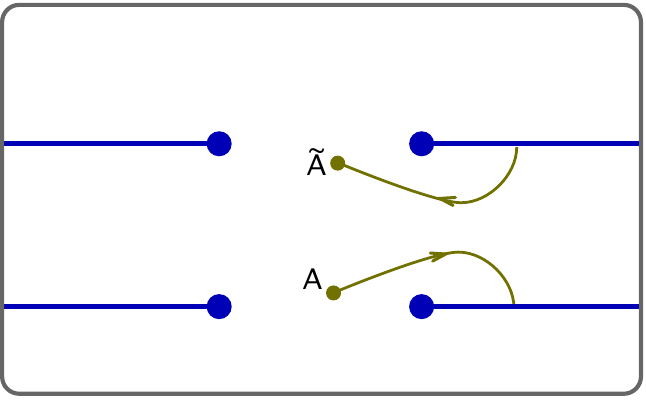}&\(\raisebox{3.5em}{\ensuremath{\leftrightarrow}}\)&\includegraphics[width=0.3\textwidth]{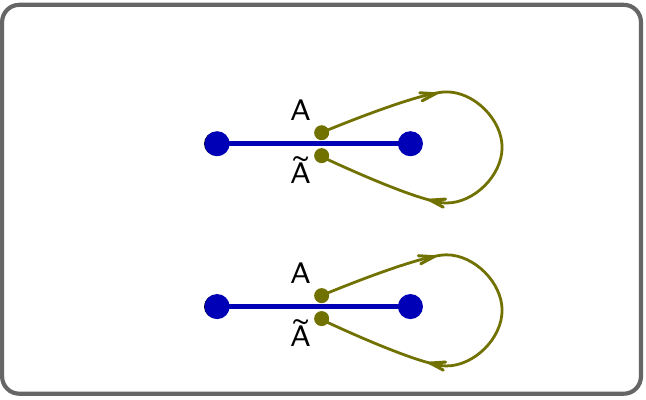}
\end{tabular}\,.
\ee
The jump of \(\omega\) across the cut is computed as follows
\begin{eqnarray*}
\disc\o_{ij}&\equiv&\o_{ij}-\tilde\o_{ij}=\check\o_{ij}-\check\o_{ij}^{[2]}=\frac{\eta^{ab}}{2\CF^-}\left(\bQ_{ab|ij}^--\bQ_{ab|ij}^+\right)=
\frac{\eta^{ab}}{\CF^-}\bosQ{a}
\left|
\begin{matrix}
\bQ_{b|i}^+ & \bQ_{i}
\\
\bQ_{b|j}^+ & \bQ_{j}
\end{matrix}
\right|
\\&=&-\frac{\eta^{ab}}{\CF^-}\bQ^{k}\bQ_{a|k}^-
\left|
\begin{matrix}
\bQ_{b|i}^- & \bQ_{i}
\\
\bQ_{b|j}^- & \bQ_{j}
\end{matrix}
\right|
=
\left|
\begin{matrix}
\omega_{ik} & \bQ_{i}
\\
\omega_{jk} & \bQ_{j}
\end{matrix}
\right|
\bQ^{k}
=
\left|
\begin{matrix}
\tilde\bQ_i & \bQ_{i}
\\
\tilde\bQ_j & \bQ_{j}
\end{matrix}
\right|.
\end{eqnarray*}
Hence we see that
\be
\omega_{ij}-\tilde\omega_{ij}=\bQ_j\tilde\bQ_i-\bQ_i\tilde\bQ_j\,,
\ee
and since \(\bQ_i\bQ^i=0\) (QQ relation \eqref{orto1}) and \(\tilde\bQ_i\bQ^i=0\) (follows from e.g. \eqref{eqd:3}), we conclude that indeed \((\omega_{ij}-\tilde\omega_{ij})\mQ^{j}=0\).

By now we actually derived all the properties of the \(\bQ\omega\) system \eqref{Qomegafinal}. \(\omega\) appears here with one more interesting interpretation -- as a Wronskian of certain Baxter-type equations, and this is how \(\omega\) was first discovered historically.

\subsubsection{Deeper look into analytic structure\label{sec:deeperlook}.}
We will now make the final step prior to building the fundamental {\Qsys}. In the previous sections we stressed several times that considering either \(\bQ\)'s or their Hodge-duals are two equivalent ways to solve the T-hook.  Even more than that, taking Hodge-dual does not require adjusting the mirror Q-basis by some H-rotations. All Q-functions have their best analyticity in the very same basis. It imposes further interesting constraints on the Q-functions beyond their natural domains of analyticity. The archetypal example is the relation
\begin{eqnarray}
 \bQ^{4|1234}=\frac 1{\cF^+}\bQ_{123|\emptyset}\,.
\end{eqnarray}
The l.h.s. is a determinant object from functions analytic in the
upper half-plane, hence it is also analytic there, and more precisely
for \(\Im(u)>1\). By contrast, the analyticity of the r.h.s. is far from being trivial. Indeed, explicitly
\begin{eqnarray}
  \frac 1{\CF^{+}}\bQ_{123|\emptyset}=\left|
\begin{array}{lll}
\bQ_{1|\es}^{[+2]} & \bQ_{2|\es}^{[+2]} & \frac 1{\CF^+}\bQ_{3|\es}^{[+2]} \\
\bQ_{1|\es}^{\hphantom{[+2]}} & \bQ_{2|\es}^{\hphantom{[+2]}} & \frac 1{\CF^+}\bQ_{3|\es}^{\hphantom{[+2]}}  \\
\bQ_{1|\es}^{[-2]} & \bQ_{2|\es}^{[-2]} & \frac 1{\CF^+}\bQ_{3|\es}^{[-2]} \\
\end{array}\right|\,.
\end{eqnarray}
The last column is not analytic in the upper half-plane as neither \(\bosQ{3}\) nor \(\cF\) are analytic there. But, since the full determinant is analytic, the discontinuity of the last column should be a linear combination of the first two columns:
\begin{eqnarray}\label{DeltaQoF}
\disc\left(\frac{\bQ_{3|\es}^{[2(n+k)]}}{\CF^+}\right)=\delta^{41}(n)\,\bQ_{1|\es}^{[2(n+k)]}+\delta^{42}(n)\bQ_{2|\es}^{[2(n+k)]}\,,\ \  k=-1,0,1\,,\ \ n=2,3,\ldots\,,
\end{eqnarray}
where \(\delta\)-s are some functions of the spectral parameter.
By induction, \(\delta^{ij}(n)\) do not depend on \(n\).

Another interesting example is the second orthogonality condition in \eqref{ortoQij2} which can be written as
\begin{equation}
\label{eq:25}\left({\CF^{-1}}\bQ_{a|i}\right)\bQ^{b|i}=-\delta_a{}^b\,.
\end{equation}
The r.h.s. is analytic everywhere which cannot be said about either of
the functions in the l.h.s. Specializing (\ref{eq:25}) to \(a=3\),
$b=3,4$ and taking its discontinuity, we derive (by comparing with the
$a=1,2$, $b=3,4$ case of (\ref{eq:25}))
\be\label{ddiscQij}
\disc\left(\frac{\bQ_{3|i}^{[2n-1]}}{\CF^+}\right)=\gamma^{41}(n)\bQ_{1|i}^{[2n-1]}+\gamma^{42}(n)\bQ_{2|i}^{[2n-1]}\,,\ \ i={1,2,3,4}\,;\ \ n=1,2,\ldots\,.
\ee
Let us show that \(\gamma\)'s also do not depend on \(n\), by taking the difference of \eqref{ddiscQij} at two adjacent values of \(n\):

\be
0&=&\disc\left(\frac{\bQ_{3|i}^{[2n+1]}-\bQ_{3|i}^{[2n-1]}}{\CF^+}\right)-\sum_{a=1}^2\left(\gamma^{4a}(n+1)\bQ_{a|i}^{[2n+1]}-\gamma^{4a}(n)\bQ_{a|i}^{[2n-1]}\right)
\no\\
&=&\bQ_{\es|i}^{[2n]}\left[\disc\left(\frac{\bQ_{3|\es}^{[2n]}}{\CF^+}\right)-\sum_{a=1}^2\gamma^{4a}(n+1)\bQ_{a|\es}^{[2n]}\right]-\sum_{a=1}^{2}\left[\gamma^{4a}(n+1)-\gamma^{4a}(n)\right]\bQ_{a|i}^{[2n-1]}.
\ee
Each of the square brackets should vanish separately because their multipliers depend on \(i\) which can be arbitrary. First, we conclude that \(\gamma\) is indeed independent of \(n\). And second, by recalling \eqref{DeltaQoF} we see that actually \(\gamma=\delta\).

To our surprise, we instantaneously get control on the all semi-infinite ladder of cuts of the Q-functions. Indeed, the discontinuity at the leading cut is determined by \eqref{tildebos}, and all the rest are known if \(\delta\) is known! It will be very interesting indeed to find explicitly \(\delta\)'s. To compute say \(\delta^{41}\), we %
multiply \eqref{ddiscQij} at \(n=1\) by \((\bQ_{2|j})^+\) (and
antisymmetrize over \(i\leftrightarrow j\)) and get
\be
\delta^{41}\,\bQ_{12|ij}^+=-\disc\left(\frac{\bQ_{23|ij}^+}{\CF^{+}}\right)=\frac {\bQ_{23|ij}^{+}\disc\CF^{+}-\CF^+\disc\bQ_{23|ij}^{+}}{\CF^+\widetilde{\CF^+}}\,.
\no\\
\ee
Since \(\bQ_{2|j}^+\) is analytic on the real axis, one has
\(\disc\bQ_{23|ij}^+=
\left|
\begin{smallmatrix}
\bQ_{2|i}^+&\disc\bQ_{3|i}^+
\\
\bQ_{2|j}^+&\disc\bQ_{3|j}^+
\end{smallmatrix}
\right|=
\left|
\begin{smallmatrix}
\bQ_{2|i}^+&(\bosQ{3}\bQ_i-\tbosQ{3}\tilde\bQ_i)
\\
\bQ_{2|j}^+&(\bosQ{3}\bQ_j-\tbosQ{3}\tilde\bQ_j)
\end{smallmatrix}
\right|
\,,
\) therefore \((\disc\bQ_{23|ij}^+)\bQ^{i}\tilde\bQ^j=0\), so we can perform  a projection:
\begin{equation}
\delta^{41}\,\bQ_{12|ij}^+\bQ^{i}\tilde\bQ^j=\frac {\bQ_{23|ij}^{+}\bQ^{i}\tilde\bQ^j}{\CF^+\widetilde{\CF^+}}\disc\CF^+\,.
\end{equation}

Departing from \(\bQ^{1|i}\bQ_{1|i}=-\CF^{[+2]}\), one can compute the
discontinuity\footnote{To get the second line of (\ref{eq:26}), we
  use the relation
$\bosQup {1}=
\bQ_{i}(\bQ^{1|i})^\pm$
obtained from
\eqref{QQrel2}.
This equation is not affected by the normalization \eqref{HodgedefM},
unlike \equref{QQrel1} which had to be modified to \eqref{QQrel12}.
} of \(\CF^+\):
\begin{eqnarray}\label{eq:26}
\disc\CF^+\equiv\CF^+-\widetilde{\CF^+}&=&(\bQ^{1|i})^-(\tilde\bQ_{1|i}^--\bQ_{1|i}^-)=(\bQ^{1|i})^-\bQ_{i}\bosQ{1}-(\bQ^{1|i})^-\tilde\bQ_{i}\tbosQ{1}=\no\\
&=&\bosQup{1}\bosQ{1}-\tbosQup{1}\tbosQ{1}=\disc \bosQ{1}\bosQup{1}\,.
\end{eqnarray}

Using \(\mathbb{Z}_4\) symmetry in the form \eqref{tildebos}, we get \(\disc(\bosQ{1}\bosQup{1})=-\disc(\bosQ{1}\tbosQ{2})\). For the future relations, it will be also useful to derive \(\disc\CF^+\) departing from  \(\bQ^{3|i}\bQ_{3|i}=-\CF\). In summary, we find
\begin{eqnarray}\label{jumpF}
\disc\CF^+=-(\bosQ{1}\tbosQ{2}-\bosQ{2}\tbosQ{1})=\bosQ{3}\tbosQ{4}-\bosQ{4}\tbosQ{3}\,.
\end{eqnarray}

On the other hand,
\(\bQ_{12|ij}^+\bQ^i\tilde\bQ^j=\disc(\bosQ{1}\tbosQ{2})=-\disc\CF^+\)
and
\(\bQ_{23|ij}^+\bQ^i\tilde\bQ^j=\disc(\bQ_2\tilde\bQ_3)=\disc(\bosQup{1}\tbosQup{4})\). Therefore
\be
\delta^{41}=\frac{\disc(\bosQup{4}\tbosQup{1})}{\CF^+\widetilde{\CF^+}}\,.
\ee

Independence of \(\delta\) of \(n\) allows us to restate \eqref{DeltaQoF} in a very appealing form
\begin{eqnarray}
\frac{\bosQ{3}}{\CF^+}=-\CA^{4}+\sum_{a=1}^2\Omega^{4a}\bosQ{a}\,,\ \  \Omega^+=\Omega^-\,, \disc{\Omega^{4a}}=\delta^{4a},
\end{eqnarray}
where  \(\Omega\) is a mirror \(i\)-periodic function defined by its discontinuity\footnote{The ambiguity of defining \(\Omega\) by its discontinuity amounts in adding \(\bosQ{1}\) and \(\bosQ{2}\) to \(\CA^4\) and does not spoil its analyticity.}, and where \(\CA^4\) is a function analytic in the upper half-plane for \(\Im(u)>0\).

Clearly, the same type of reasoning can be applied for all \(\bosQ{a}\) and \(\bosQup{a}\). The general statement is the following: given the mirror \(i\)-periodic functions \(\Omega_{ab}=-\Omega_{ba}\) defined by their discontinuities
\begin{eqnarray}\label{defOmega}
\disc\Omega^{ab}=\frac{\disc (\bosQup{a}\tbosQup{b})}{\CF^{+}\widetilde{\CF^{+}}}\,,
\end{eqnarray}
so that \(\Omega^{12}=-\Omega^{34}=1/\CF^+\), and
\(\Omega_{ab}\equiv\eta_{ac}\eta_{bd}\,\Omega^{cd}\) \footnote{Note
  that in this case, \(\Omega^{cd}\neq\left(\Omega^{-1}\right)_{cd}\),
  by contrast with the definition of upper indices for other
  matrices. However, the antisymmetry of the matrix $\Omega$ ensures
  $\Omega_{ab}\Omega^{bd}=\mathrm{Pf}(\Omega)\delta_a{}^d$.},
the following combinations
\begin{eqnarray}\label{OHtransform}
 \CA_a\equiv\Omega_{ab}\bosQup{b}\,,\ \  \CA^a\equiv\Omega^{ab}\bosQ{b}\,
\end{eqnarray}
are analytic in a half-plane. Explicitly,
\(\CA^1\), \(\CA^2\), \(\CA_{3}\) and \(\CA_{4}\) are analytic in the lower half-plane,
whereas \(\CA^{3}\), \(\CA^{4}\), \(\CA_{1}\) and \(\CA_{2}\) are analytic in the upper half-plane.

And correspondingly, by considering \eqref{ddiscQij} we see that
\be
\CA_a|{}^{i}\equiv\Omega_{ab}\bQ^{b|i}\,,\ \ \ \CA^{a}|_{i}\equiv\Omega^{ab}\bQ_{b|i}
\ee
are analytic in half-planes precisely where, correspondingly, \(\bQ_{a|i}\) and \(\bQ^{a|i}\) are analytic.

Finally, let us compute \(\tilde\CA\) then.  Because of \eqref{defOmega}, \(\tilde\Omega^{ab}\bosQ{b}=\Omega^{ab}\bosQ{b}\), therefore
\be
\tilde\CA^a=\Omega^{ab}\tbosQ{b}=\Omega^{ab}\eta_{bc}\bosQup{c}=-\eta^{ab}\Omega_{bc}\bosQup{c}=-\eta^{ab}\CA_{b}\,,
\end{eqnarray}
which is an absolute equivalent of \eqref{tildebos} showing that \(\hat\CA_a=-\eta_{ab}\hat\CA^b\), hence \(\CA\)-s have only one cut!

Hence \(\CA\) is the full analog of bosonic \(\bQ\). It is tempting to write down \(\CA_a=\pm\bosQ{a}\) but we checked on explicit example that this is not the case. Hence \(\CA\) presents an extra hidden structure of the Q-system.

\subsubsection{Fundamental {\Qsys}}
\label{sec:fqsys}
Fermionic one-indexed Q-functions have only one cut in the mirror
kinematics. If they are continued through the long cut, the result is
controlled by the physical $i$-periodic matrix \(\omega\):
\(\tilde\bQ_i=\omega_{ij}\bQ^j\). Bosonic one-indexed Q-functions have
only one cut in the physical kinematics. In this kinematics they obey the
relation \(\hbosQ{a}=\eta_{ab}\hbosQup{b}\) with explicitly known
constant matrix, so there are actually \(8/2=4\) bosonic {\mQfcts}, a
half of the number of fermionic ones. %
Among \(\hat\CA_a\) and \(\hat\CA^a\) there are also only 4 independent functions, so in fact an ensemble of \(\hat\CA_a\)'s and \(\hbosQ{a}\)'s is a proper counterpart of fermionic \(\bQ_{a}\).

The properties of fermions and bosons are not fully symmetric because the QQ-relations
are consistently defined only in the mirror kinematics, as dictated by the T-hook, thus treating bosons and fermions differently. To reach the full symmetry, it would be ideal to have a Q-system which is equally good in mirror and physical kinematics. And in fact this is possible. All our single-index functions are analytic in a half-plane, and we can define the QQ-relations in either upper or lower half-plane. To achieve this goal, we have to rotate the bosonic functions. For the choice of the upper half-plane, the rotated basis  \(\check\bP_a=H_a{}^b\bQ_b\) is explicitly given by
\begin{subequations}
\label{eq:11}
\begin{align}
&&&&&&
\begin{pmatrix}
\check\bP_1\\ \check\bP_2\\ \check\bP_3\\ \check\bP_4
\end{pmatrix}
&\equiv
\begin{pmatrix}
\bQ_1\\ \bQ_2\\\CA^4\\ -\CA^3
\end{pmatrix}
&=
\begin{pmatrix}
1&0&0&0\\
0&1&0&0\\
-\O^{1,4}&-\O^{2,4}&\O_{1,2}&0\\
\O^{1,3}&\O^{2,3}&0&\O_{1,2}
\end{pmatrix}
\begin{pmatrix}
\bQ_1\\\bQ_2\\\bQ_3\\\bQ_4
\end{pmatrix}\,,
&&&&&&
\label{eq:11a}
\end{align}
and for the Hodge-dual quantities one has
\begin{align}
\label{eq:11b}
&&&&&&
\begin{pmatrix}
\check\bP^1\\ \check\bP^2\\ \check\bP^3\\ \check\bP^4
\end{pmatrix}
&\equiv
\begin{pmatrix}
-\CA_2\\\CA_1\\ \bQ^3\\ \bQ^4
\end{pmatrix}
&=
\begin{pmatrix}
\O_{1,2}&0&-\O_{2,3}&-\O_{2,4}\\
0&\O_{1,2}&\O_{1,3}&\O_{1,4}\\
0&0&1&0\\
0&0&0&1
\end{pmatrix}
\begin{pmatrix}
\bQ^1\\\bQ^2\\\bQ^3\\\bQ^4
\end{pmatrix}\,.
&&&&&&
\end{align}
\end{subequations}
Check over \(\bP\)-s reminds us that equations are written in the mirror kinematics. All \(\bP\)-s are equal to such bosonic \(\bQ\)'s or \(\CA\)'s that they are  analytic in the upper half-plane, and from there they are continued to physical kinematics where they have only one cut. The convention is that \(\bP\)-s without check are defined in the physical kinematics: \(\bP\equiv\hat\bP\).

In general, the above-defined matrix \(H\) rotates the full mirror Q-basis to a new Q-basis which we call fundamental:
\be\label{deq:10}
\fQ_{A|I}=H_A{}^{B}\bQ_{B|I}\,,
\ee
in particular \(\bP_a=\fQ_{a|\es}\) and \(\bQ_{i}=\fQ_{\es|i}=\bQ_{\es|i}\).

We can check that definitions \eqref{deq:10} and \(\bP^a\equiv\fQ^{a|\es}\) produce the transform \eqref{eq:11b}, so \(H\) is indeed a consistently defined H-transformation. It remains to assure that all \(\fQ\)'s are analytic in the upper half-plane as intended. For single-indexed \(\fQ\)'s this is so by construction. We should check only \(\fQ_{a|i}\), the analyticity of other Q-functions follows automatically through determinant QQ-relations. For \(\alpha=\{1,2\}\) one has \(\fQ_{\alpha|i}=\bQ_{\alpha|i}\) which is known to be analytic. Then we use  \(\fQ_{3|i}=\CA^{4}|_{i}\) and \(\fQ_{4|i}=-\CA^{3}|_{i}\) and recall analyticity of the corresponding \(\CA\)'s discussed in the previous subsection. Therefore we conclude that all \(\fQ_{a|i}\) are analytic in the upper half-plane, for \(\Im(u)>-1/2\).

Quite remarkably, we achieve as well  \(\fQ_{\bar\es}=1\) in the fundamental basis, since \(\det H=\CF^{-2}\). Though it was implicit, the property  \(\Omega_{12}=-\Omega_{34}\) was used which follow from \eqref{jumpF} and originally from \eqref{eq:9}.

From monodromies of  \(\bQ\)-s and \(\CA\)-s, it is now easy to compute the monodromies of \(\bP\)-s:
\be\label{A74}
\tilde\bP^a=\mu^{ab}\bP_b\,, \ \ \tilde\bP_a=\mu_{ab}\bP^b\,,
\end{eqnarray}
where \(\mu_{ab}=-\mu_{ba}\), \(\mu_{12}=\CF^-\),
\(\mu_{34}=-\CF^-\Pf(\O)\), and \(\mu_{\alpha\dot \alpha}=-\CF^-\,\O_{\alpha\dot \alpha}\)
for \(\alpha\in\{1,2\}\) and \(\dot \alpha\in\{3,4\}\). From these definitions
of $\mu$ it follows that \(\Pf(\mu)=1\) and then \(\mu^{ab}=(\mu^{-1})^{ab}=-\frac 12\e^{abcd}\mu_{cd}\). Finally, we compute \(\disc\mu\). The easiest way to do it is to notice that from \eqref{A74} it follows that \(\disc(\mu)\propto\bP\wedge\tilde\bP\). The coefficient of proportionality is restored from \(\disc(\mu_{12})=\disc(\CF^-)=\disc(\tilde\bQ_1\bQ_2)=\disc(\tilde\bP_1\bP_2)\) and similarly \(\disc(\mu^{34})=-\disc(\tilde\bP^3\bP^4)\)\,. So finally
\be\label{A76}
\disc(\mu_{ij})=\disc(\tilde\bP_i\bP_j)\,\ \ \disc(\mu^{ij})=-\disc(\tilde\bP^i\bP^j)\,.
\end{eqnarray}
The properties \eqref{A74} and \eqref{A76}, together with periodicity  of \(\mu\) is precisely the \(\bP\mu\) system. The orthogonality \(\bP^i\bP_i=0\) is just a QQ-relation.

Therefore, we  accomplished derivation of the fundamental Q-system and its main subsystems: \(\bP\mu\) and \(\bQ\omega\) and hence gave a total account of the path \(\bT\to\bQ\to\fQ\).

\subsection{Conclusions}
In this appendix we demonstrated that analytic structures of the analytic Y-system and the quantum spectral curve follow from one another, and in this sense the two systems are equivalent.

The first milestone of establishing the equivalence was to realize that there exist the unique gauge \(\bT\) and  the unique mirror Q-basis, despite the absence of LR symmetry. We showed that one-indexed \(\hat\bQ_{i|\es}\), \(\hat\bQ^{i|\es}\) have only one short cut while one-indexed \(\bQ_{\es|i}\), \(\bQ^{\es|i}\) have only one long cut. The periodic matrix \(\omega\) naturally appeared as a Wronskian of certain Baxter-type relations.

The second milestone was to recognize that it is possible to rotate the mirror Q-basis so as to make all Q-functions analytic in the upper half-plane. The possibility of this rotation is due to the hidden analyticity which was decoded from the universality of the mirror Q-basis. At this point the reader can recognize an ad-hoc ansatz \eqref{P3P4} as the rotation \eqref{eq:11a}.

Our choice of the upper half-plane was random. It is of course possible to choose the lower half-plane and the corresponding H-rotation instead. Two choices are  related by a certain H-rotation, and it is simple to deduce from \secref{sec:fqsys} that this rotation is defined by \(\mu\)! Another way to relate upper and lower half-plane choices is to use \(\omega\) as a fermionic H-rotation through the physical kinematics. This is a subject of \secref{sec:discussionaxioms}.

Apart of analytic structure we should also relate reality, regularity, and asymptotics at infinity of two systems. The question of reality was clarified:  conjugation properties were traced for the relations \(Y\leftrightarrow {\bT}\) and \(\fQ\rightarrow{\bT} \) in this appendix, whereas the direction \(\bT\to\fQ\) is presented in \secref{app:reality-1}. As was mentioned, we cannot fully derive the regularity of the quantum spectral curve from TBA, nevertheless we assume it based on a solid evidence from a handful of explicit examples. We discuss a link of regularity to the exact Bethe equations in \secref{subsec:exactBethe}. Finally, asymptotics at infinity is tightly related to the global charges, hence this aspect of the TBA $\leftrightarrow$ QSC relation will be discussed in the next appendix devoted to the QSC from the point of view of the representation theory.

\section{Unitarity and global charges}
\label{app:unitary}
This appendix consists of three subsections. The sub\secref{app:repr-theory-psu224} summarizes the classification of unitary representations of \(\psu(2,2|4)\) algebra. This classification is explicitly relevant for QSC because the global charges of a given state define the large-$u$ asymptotic behaviour of Q-functions. The latter fact was demonstrated in the main text, \secref{sec:NRCGC}, for sufficiently small coupling constant when the large volume approximation was applicable. In the same \secref{sec:NRCGC} it was explained how to generalize the argument for arbitrary  value of the coupling: for 5 charges \(J_1,J_2,J_3,S_1,S_2\), it is enough to show that they are quantized. It is the goal of sub\secref{app:unitarityfromanalyticity} to derive this quantization directly from analytic properties of QSC. Moreover we also discuss appearance of the main unitarity constraint \eqref{uniprincipal} from QSC analyticity, without exploiting the representation theory. The remaining 6th charge \(\Delta\) is not quantized, and we derive the way it appears in QSC through comparison with TBA in sub\secref{subsec:Delta}.
\subsection{Representation theory for \texorpdfstring{$\psu(2,2|4)$}{psu(2|2)}}\label{app:repr-theory-psu224}
\subsubsection{Kac-Dynkin-Vogan diagrams}\label{sec:kac-dynkin-vogan}
\(\gl(4|4)\) algebra is defined by the super-commutation relations
\be
\left[E_{ij},E_{kl}\right]=\delta_{jk}E_{il}-(-1)^{(p_i+p_j)(p_k+p_l)}\delta_{li}E_{kj}\,,
\ee
where \(p\) is a parity grading function. \(p_i\) can be either \(0\) or \(1\), depending on the grading of the index \(i\).  \(E_{ij}\) is an odd generator if \(p_i\neq p_j\) and even otherwise.

To  describe a unitarity representation, one should  define a choice of the complex conjugation which is conveniently done by
\be\label{rulestar}
E_{ij}^*=(-1)^{c_i+c_j}E_{ji}\,,
\ee
where \(c\) is a new grading function which, similarly to \(p\), has values either \(0\) or \(1\).

\(\psu(2,2|4)\) is obtained from the corresponding real form of \(\gl(4|4)\) by considering only super-traceless combinations, e.g.
\be
h_i=E_{ii}-(-1)^{p_i+p_{i+1}}E_{i+1,i+1}\,,
\ee
and by imposing the zero-charge condition \(\sum_i E_{ii}=0\).

\begin{figure}[t]
\begin{center}
\parbox{0.49\textwidth}{
\includegraphics[width=0.4\textwidth]{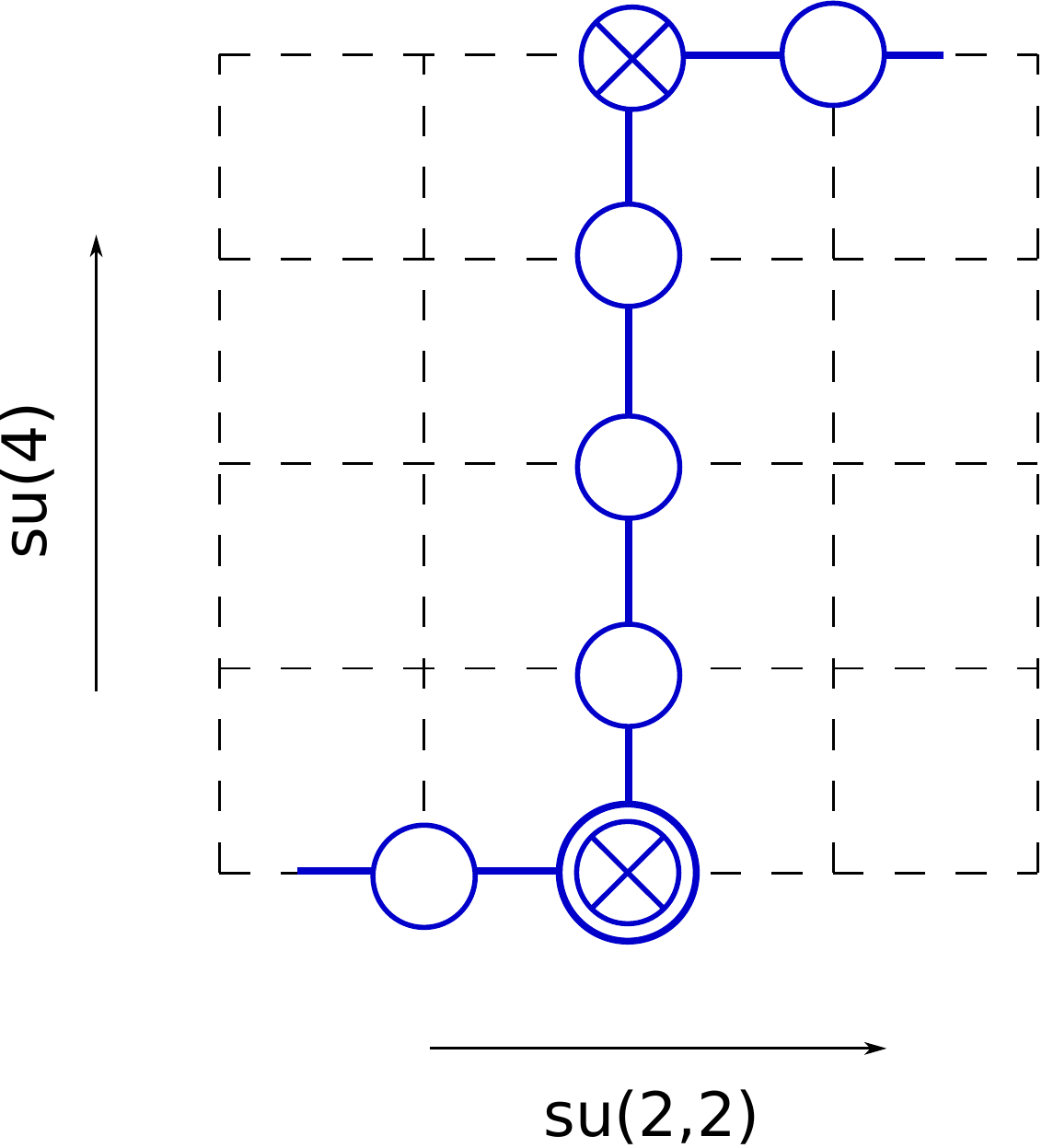}
}
\parbox{0.49\textwidth}{
\includegraphics[width=0.4\textwidth]{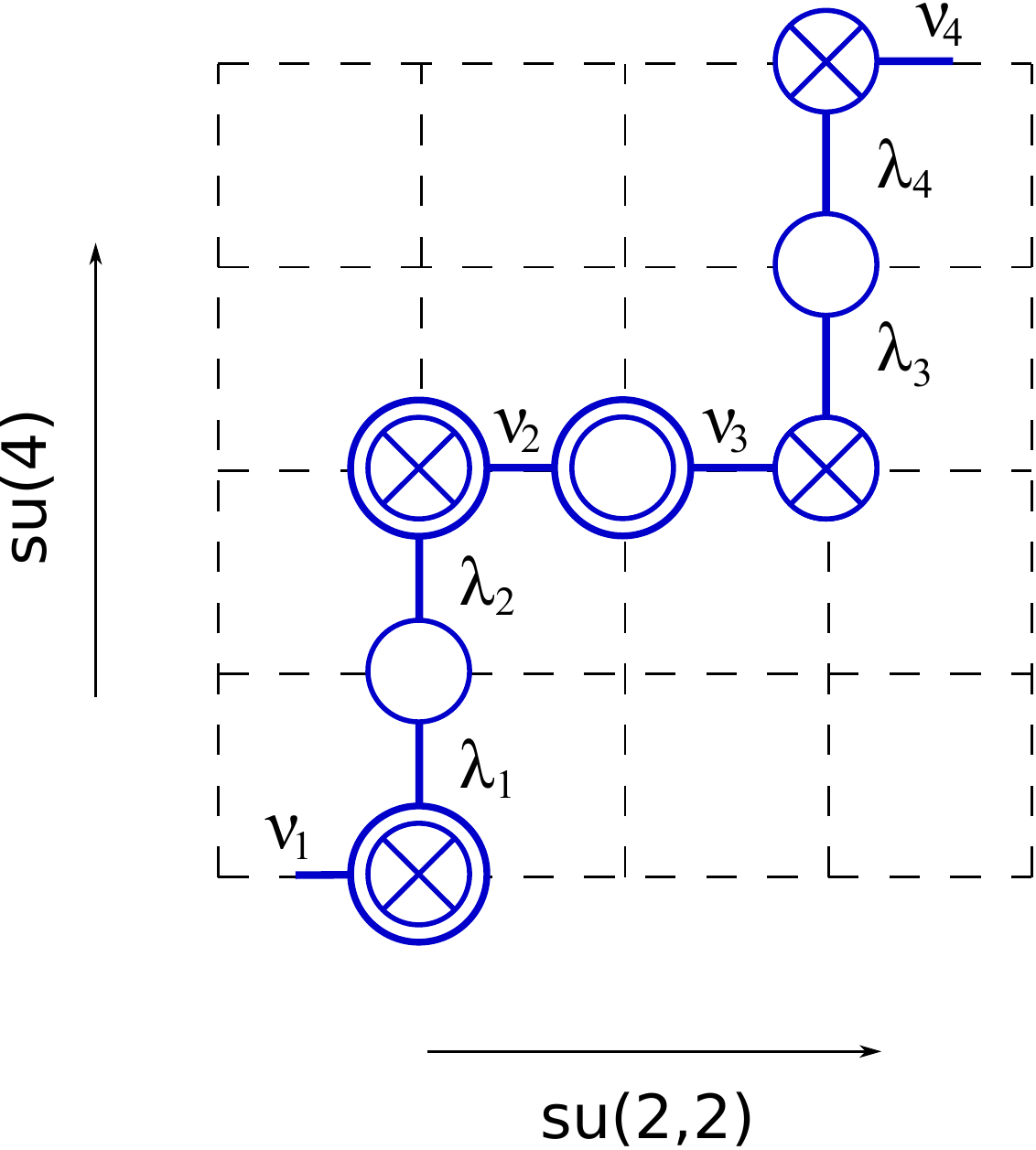}
}
\caption{
\label{fig:abadiagram}
Compact beauty (left) and non-compact \abadm{} (right) choice of \(p\)-gradings for \(\gl(4|4)\) algebra. The figure further precises the real form of the algebra by denoting \(c\)-odd nodes with an extra circle. For the \abadm{} diagram, the weights with respect to bosonic subalgebra are  shown: when acting on the highest weight, the generators of \(\uuu(4)\)'s Cartan subalgebra, \(E_{22},E_{33},E_{66},E_{77}\), have the eigenvalues respectively  \(\lambda_1,\lambda_2,\lambda_3,\lambda_4\), whereas the generators of \(\uuu(2,2)\)'s Cartan subalgebra, \(E_{11},E_{44},E_{55},E_{88}\), have  the eigenvalues respectively \(\nu_1,\nu_2,\nu_3,\nu_4\).
}
\end{center}
\end{figure}

It is convenient to depict both \(p\)- and \(c\)-gradings on the Kac-Dynkin-Vogan diagram. For our case, it consists of 7 nodes, each node represents the gradings of \(E_{i,i+1}\):
\be
\raisebox{20pt}
{
\begin{tabular}{c|c|c}
&\(p\)-even&\(p\)-odd\\
\hline
\(c\)-even &  \parbox[][][c]{24pt}{
                \begin{picture}(24,24)
                \put(12,12){\circle{15}}
                \end{picture}
                } & \parbox[][][c]{24pt}{
                \begin{picture}(24,24)
                \put(12,12){\circle{15}}
                \put(6.8,6.8){\line(1,1){10.2}}
                \put(6.8,17.2){\line(1,-1){10.2}}
                \end{picture}
                }  \\

\hline
\(c\)-odd  & \parbox[][][c]{24pt}{
                \begin{picture}(24,24)
                \put(12,12){\circle{15}}
                \put(12,12){\circle{19}}
                \end{picture}
                } & \parbox[][][c]{24pt}{
                \begin{picture}(24,24)
                \put(12,12){\circle{15}}
                \put(12,12){\circle{19}}
                \put(6.8,6.8){\line(1,1){10.2}}
                \put(6.8,17.2){\line(1,-1){10.2}}
                \end{picture}
                }\end{tabular}
}
                \,,
\ee
What is called the distinguished diagram is
{
\raisebox{-.25cm}
  {
\newcommand{\nee}[1]{\put(7.5,0){\circle{15}}\put(15,0){#1}}
\newcommand{\neo}[1]{\put(7.5,0){\circle{15}}\put(7.5,0){\put(5.3033,5.3033){\line(-1,-1){10.6066}}\put(-5.3033,5.3033){\line(1,-1){10.6066}}}\put(15,0){#1}}
\newcommand{\noe}[1]{\put(9.8,0){\circle{15}}\put(9.8,0){\circle{19}}\put(19.6,0){#1}}
\newcommand{\noo}[1]{\put(9.8,0){\circle{15}}\put(9.8,0){\circle{19}}\put(9.8,0){\put(5.3033,5.3033){\line(-1,-1){10.6066}}\put(-5.3033,5.3033){\line(1,-1){10.6066}}}\put(19.6,0){#1}}
\newcommand{\ml}[1]{\put(0,0){\line(1,0){12}}\put(12,0){#1}}
\newcommand{\lgl}[1]{\put(0,0){\line(1,0){14}}\put(14,0){#1}}
{\begin{picture}(165,15)
\put(-23,9.5){\nee{\ml{\noe{\ml{\nee{\lgl{\neo{\lgl{\nee{\lgl{\nee{\lgl{\nee{}}}}}}}}}}}}}}
\end{picture}}
}}.
However, it is long time known that this diagram, dubbed ``beast'' \cite{Beisert:2003yb}, is not convenient for describing  the unitary representations. Two convenient choices,  ``beauty'' and ``\abadm''\footnote{Actually, there are 4 different diagrams that are used in the ABA equations \cite{Beisert:2005fw}. We consider only the non-compact bosonic one for simplicity.}, are presented in \figref{fig:abadiagram} \footnote{These diagrams are introduced and explained for spin chains in \cite{Kazakov:2007fy}, but without specification of $c$-grading. The notation including $c$-grading is to be discussed in more detail in \cite{VM-Representations}, see also \cite{Volin:2010xz}.}. For the \abadm{} case, the asymptotic Bethe equations can be written in terms of rational (apart from  the dressing phase) functions of Zhukovsky variables  \cite{Beisert:2005fw}. With discovery of QSC, one can in principle write Bethe equations, exact and asymptotic, for other choices of \(p\)-gradings, see \secref{subsec:exactBethe}. However, these equations are not written in terms of rational functions and hence seem to be of less significance, at least for as what concerns the large volume regime.

Any unitary representations of any real form of \(\gl(n|m)\)-type superalgebras are of a highest-weight type\footnote{There is a simple explanation for this fact: let $f$ be a fermionic generator. Obviously, $f+f^*$ is real and hence $(f+f^*)^2$ is positive. But $(f+f^*)^2=\{f,f^*\}=b$, hence $b$ has a positive spectrum. We can choose $f$ in a way that $b$ is an element of the Cartan sub algebra. For instance, in the case of rule \eqref{rulestar} it is obvious. For a more formal proof see e.g. \cite{JakobUHW,Furutsu91}.}, and precisely these representations should be addressed by the quantum spectral curve. The highest weight \(|\Omega\rangle\) is defined by condition \(E_{ij}|\Omega\rangle=0\) for \(i<j\).

There are two common ways to parameterize a highest weight representation of \(\psu(2,2|4)\). The first one is by a set of 7 Dynkin labels which are the eigenvalues of \(h_i\)'s on the \(|\Omega\rangle\). For {\bf the beauty diagram}, they are denoted as \(\left[q_1,\omega_-,r_1,r_2,r_3,\omega_+,q_3\right]\) with \(\omega_{\pm}\) corresponding to \(p\)-odd roots. Because of the zero charge constraint, which reads
\be\label{zerocharge}
\frac 12(r_1-r_3)+\frac 32(q_1-q_3)=-\omega_{-}-\omega_{+}\,,
\ee
there are only 6 independent parameters.

The second way explicitly operates with \(6\) parameters:  3 Dynkin labels \([r_1,r_2,r_3]\) of the \(\su(4)\) subalgebra  together with 3 Dynkin labels \([q_1,q_2,q_3]\) of the \(\su(2,2)\), defined with respect to the highest weight of the supermultiplet. One has \(q_2=\omega_--\omega_+-r_1-r_2-r_3\). We also commonly use \(\lambda\)'s and \(\nu\)'s which are defined by
\be\label{deflanu}
\lambda_i-\lambda_{i+1}=r_i\,,\ \ \  \nu_i-\nu_{i+1}=q_i\,.
\ee
Due to the zero charge condition one should request
\be\label{zerocharge5}
\sum_{i=1}^4(\lambda_i+\nu_i)=0\,,
\ee
whereas the overall shift \(\lambda\to\lambda+\Lambda\), \(\nu\to\nu-\Lambda\) is physically inessential. We will see that it corresponds to a residual gauge symmetry of the quantum spectral curve \eqref{oneparameterxrescaling}. The Lorenz spins \(s_1,s_2\) of \(\su(2,2)\) are nothing but \(s_1=q_1/2\) and \(s_2=q_3/2\). The conformal dimension \(\Delta\)  and the charge \(J_1\), one of the angular momenta on \(S^5\), are given by
\begin{eqnarray}
\Delta=-\left(q_2+\frac 12\left(q_1+q_3\right)\right)\,,\ \ \  J_1=r_2+\frac 12(r_1+r_3)\,.
\end{eqnarray}
The other charges of \(\mathfrak{so}(2,4)\simeq\su(2,2)\) and \(\mathfrak{so}(6)\simeq\su(4)\) are given by
\begin{eqnarray}
r_1=J_2-J_3\,\ \ \  && q_{1}=S_1+S_2\,,\no\\
r_2=J_1-J_2\,\ \ \  && q_2=-\Delta-S_1\,,\no\\
r_3=J_2+J_3\,\ \ \  && q_3=S_1-S_2\,.
\end{eqnarray}
For convenience of the reader, we summarize various transition formulae
\begin{eqnarray}
\label{tranformulae}
\lambda_1=\frac {+J_1+J_2-J_3}2+\Lambda\,,\ \ \ \ \ \ \  &&\nu_1=\frac{-\Delta+ S_1+S_2}{2}-\Lambda\,,\no\\
\lambda_2=\frac {+J_1-J_2+J_3}2+\Lambda\,,\ \ \ \ \ \ \ &&\nu_2=\frac{-\Delta-S_1-S_2}{2}-\Lambda\,,\no\\
\lambda_3=\frac {-J_1+J_2+J_3}2+\Lambda\,,\ \ \ \ \ \ \ &&\nu_3=\frac{+\Delta+ S_1-S_2}{2}-\Lambda\,,\no\\
\lambda_4=\frac {-J_1-J_2-J_3}2+\Lambda\,,\ \ \ \ \ \ \  &&\nu_4=\frac{+\Delta-S_1+S_2}{2}-\Lambda\,,
\end{eqnarray}
and
\begin{eqnarray}
J_2=\frac 12(r_1+r_3)\,,\ \ \  && J_3=-\frac 12(r_1-r_3)\,,\no\\
S_1=\frac 12(q_1+q_3)\,,\ \ \  && S_2=\frac 12(q_1-q_3)\,.
\end{eqnarray}

\subsubsection{Long multiplets (typical representations)}
As demonstrated above, the same super-algebra admits different choices of a Kac-Dynkin-Vogan diagram. The weights of a representation {\it do depend} on this choice.  In particular, \(\su(4)\) and \(\su(2,2)\) charges {\it do depend} on this choice. Indeed, a supermultiplet is a collection of several bosonic multiplets. Depending on the choice of the Kac-Dynkin-Vogan diagram, the highest weight vector changes and it may become a part of a different bosonic multiplet, hence the weights \(\lambda\) and \(\nu\) will change by an integer amount.

The proper invariant objects are the so called shifted weights which we denote by \(\hat\lambda\) and \(\hat\nu\).  Already in the purely bosonic case, they play a significant role, in particular Casimirs of the representation are symmetric polynomials in them \cite{Okounkov96shiftedschur}. A similar statement holds also for the supersymmetric case, but only for the case of long multiplets \cite{MolevCapelli}: the Casimirs of a typical representation are supersymmetric polynomials in \(\hat\lambda\) and \(\hat\nu\).

Long multiplets are characterized by a property that for any vector \(|v\rangle\) of the representation module and for any \(p\)-odd generator \(E_{ij}\) either \(E_{ij}|v\rangle\) or \(E_{ij}^*|v\rangle\) is non-zero.
In this case, ordinary weights transform according to the rule
\be\label{eq:fermduality}
\parbox{0.4\textwidth}{\includegraphics[width=0.4\textwidth]{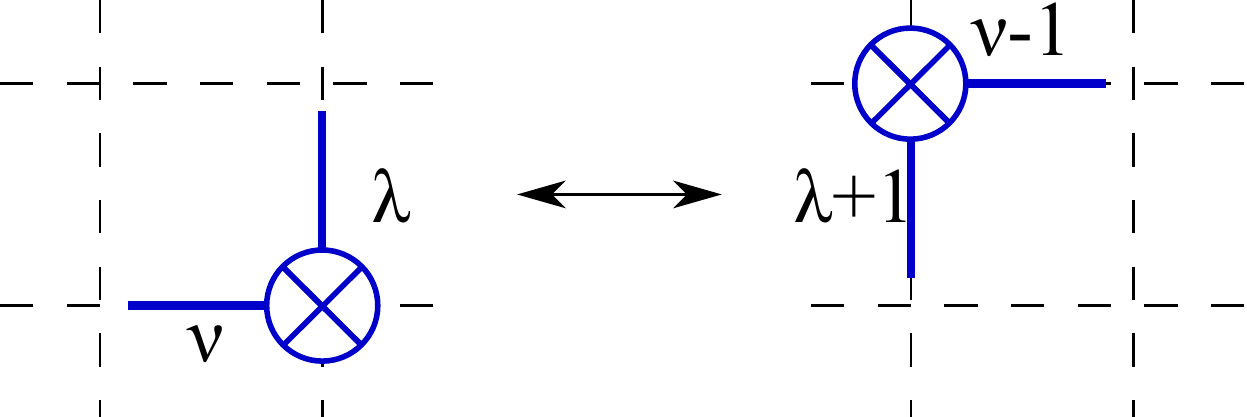}}
\ee
when changing from one Kac-Dynkin-Vogan diagram to another\footnote{All possible diagrams can be obtained by composition of this elementary move.}.

Relation \eqref{eq:fermduality} shows us how the shifted weight is defined. Roughly,  \(\hat\lambda=\lambda+\){horizontal displacement from the square's diagonal in \figref{fig:abadiagram}} and \(\hat\nu=\nu+\){vertical displacement from the square's diagonal in \figref{fig:abadiagram}}.

Precisely, for the case of beauty grading one has
\begin{equation}
\begin{tabular}{cccc}
\(\hat\lambda_1=\lambda_1+2\,,\) & \(\hat\lambda_2 =\lambda_2+1\,,\)&\(\hat\lambda_3=\lambda_3\,,\)&\(\hat\lambda_4=\lambda_4-1\,,\)
\\
\(\hat\nu_1=\nu_1-1\,,\)&\(\hat\nu_2=\nu_2-2 \,,\)&\(\hat\nu_3=\nu_3+1\,,\)&\(\hat\nu_4=\nu_4\,.\)
\end{tabular}
\Domain{\rm Beauty}
\end{equation}

For the case of \abadm{} grading one has
\begin{equation}
\label{muhll}
\begin{tabular}{cccc}
\(\hat\lambda_1=\lambda_1+1\,,\) & \(\hat\lambda_2 =\lambda_2\,,\)&\(\hat\lambda_3=\lambda_3+1\,,\)&\(\hat\lambda_4=\lambda_4\,,\)
\\
\(\hat\nu_1=\nu_1-1\,,\)&\(\hat\nu_2=\nu_2 \,,\)&\(\hat\nu_3=\nu_3-1\,,\)&\(\hat\nu_4=\nu_4\,.\)
\end{tabular}
\ \ \ \ \ \ \ \
\Domain{\rm \abadm}
\end{equation}
\vspace{.5em}
\begin{center}
{{\it In the main text we consider only the \abadm{} choice to make an easier comparison with well-established asymptotic Bethe equations.}}
\end{center}
\vspace{.5em}

The unitarity constraint for long multiplets is the easiest to formulate in terms of Dynkin labels for {\bf the beauty diagram}. It states that \(q_1,q_3,r_1,r_2,r_3\) are non-negative integers satisfying the inequalities
\be\label{unitconstraintomega}
\omega_{-}<-1-q_1\,,\ \ \ \omega_{+}>+1+q_3\,.\Domain{\rm Beauty}
\ee
When the equality is reached in \eqref{unitconstraintomega}, the long multiplet becomes reducible containing short multiplets as its irreducible part. This should happen only at zero 't Hooft coupling when all charges, including the eigenvalue \(\Delta\) of the dilatation operator  become integers.

We can rewrite these unitarity constraints as the constraints on bosonic Dynkin labels and with respect to ABA-diagram: all Dynkin labels except \(q_2\) are integers satisfying
\begin{equation}\label{constr5}
\ \ q_1\geq 2\,\ \ q_3\geq 2\,,\ \ r_1\geq 0,\ \ r_2\geq 2\,,\ \ r_3\geq 0\,,\Domain{\rm \abadm}
\end{equation}
while generically non-integer \(q_2\) satisfies
\begin{equation}\label{constru}
q_2\leq-(q_1+q_3+\sum_ir_i)-\frac
12|(r_1-r_3)+(q_1-q_3)|\,.
\ \ \ \ \
\Domain{\rm \abadm}
\end{equation}
The last one looks as follows in terms of \(\mathsf{so}(2,4)\) and \(\mathsf{so}(6)\) charges
\begin{equation}
\Delta\geq J_1+J_2+S_1+|J_3-S_2|\,.
\Domain{\rm \abadm}
\end{equation}
We provide the data for the important case of \(\sl(2)\) sector. In \abadm{} description it is characterized by \(S_2=J_2=J_3=0\) and \(S_1=S\) is then just the Lorentz spin. One immediately gets from \eqref{constr5} that \(J_1\geq 2\) and \(S\geq 2\). The case when the equalities are reached corresponds to  Konishi operator \(\Tr Z\,\D^2\,Z\). The most nontrivial unitarity constraint \eqref{constru} is written as \(\Delta\geq J_1+S\), where the equality is reached at zero coupling when \(\Delta\) becomes an integer engineering dimension $\Delta_0$. Combining the weak coupling equality $\Delta_0=J_1+S$ and constraint \eqref{ineqnew}, we can fix unambiguously the spin chain length in $\sl(2)$ sector  to be $L=J_1$.

It is  a matter of simple algebra to formulate the unitarity constraints in terms of invariant shifted weights:
\be\label{uniint}
\hat\lambda_i-\hat\lambda_{i+1}\in\mathbb{Z}\ \ \ \  &{\rm and}&\ \ \ \ \hat\lambda_i-\hat\lambda_{i+1}\geq 1\,;
\\
\hat\nu_1-\hat\nu_2\in\mathbb{Z}\ \ \ \ &{\rm and}&\ \ \ \ \hat\nu_1-\hat\nu_2\geq 1\,,
\label{uniintferm}\\
\hat\nu_3-\hat\nu_4\in\mathbb{Z}\ \ \ \ &{\rm and}&\ \ \ \ \hat\nu_3-\hat\nu_4\geq 1\,.
\no
\ee
and
\be\label{uniprincipal}
\hat\nu_1-\hat\nu_4+\hat\lambda_1-\hat\lambda_4+|\hat\lambda_1+\hat\lambda_4+\hat\nu_1+\hat\nu_4| \leq 0\,.\ \ \leftrightarrow\ \ \hat\nu_1+\hat\lambda_1\leq 0,\ \hat \nu_4+\hat\lambda_4\geq 0\,.
\ee
Recall also that the zero charge condition \eqref{zerocharge5} should be respected.

\subsubsection{Short multiplets (atypical representations)}
\label{sec:atypical}
Short multiplets are realized when one or more of the following conditions hold: \(\omega_-=-1-q_1\), \(\omega_-=q_1=0\), \(\omega_+=1+q_2\), \(\omega_+=q_3=0\). We remind that shifted weights are not invariant objects in this case since the transformation rule \eqref{eq:fermduality} generically does not hold.

The only atypical representation realized in planar \neqfour{} SYM's spin chains at finite coupling is the 1/2 BPS multiplet generated from the BMN vacuum \(\Tr\,Z^{J_1}\).  It can be considered as a state in the \(\sl(2)\) sector with \(S=0\) with protected conformal dimension \(\Delta=J_1\). The corresponding QSC solution seems to be a degenerate one. If one take the $S\to 0$ limit of the solution  in \cite{Gromov:2014bva}, which should correspond to the BMN vacuum, on finds \(\bP_i=\bP^i=0\). This limit also has other nonstandard features, in particular \(\mu_{24}\) has no longer a polynomial asymptotics. This is related to the need to analytically continue from unitarity region for \(S\geq 2\) to an isolated point \(S=0\) through non-integer values of \(S\).

All other short multiplets exist only at zero coupling. At finite coupling they combine into long multiplets and become non-protected ones. This phenomenon is captured by the Beisert-Staudacher equations \cite{Beisert:2005fw} and hence by the quantum spectral curve, since the former is derived from the latter in \secref{sec:ABA}.
\subsection{Unitarity constraints from analyticity of QSC}
\label{app:unitarityfromanalyticity}
The information about the  representation of a particular physical state is encoded in the large \(u\) asymptotic of the quantum spectral curve in the upper half-plane \eqref{largeu2}. In this appendix we use notations $\hat\lambda_a\equiv \tilde M_a$, and $\hat\nu_i\equiv -\hat M_i$, so that \eqref{largeu2} reads:
\begin{eqnarray}\label{largeu3}
\bP_a\simeq A_a\, u^{-\hat \lambda_a}\,,\ \ \bQ_{\hat i}\simeq B_i\,u^{-\hat\nu_i-1}\,,\ \  \bP^a\simeq A^a\, u^{\hat \lambda_a-1}\,,\ \ \bQ^{i}\simeq B^i\,u^{\hat\nu_i}\,,\ \ u\to\pm\infty+i\,0\,.
\end{eqnarray}
After the discussion in \secref{sec:atypical}, it is not surprising now that the asymptotics is given in terms of the invariant quantities -- shifted weights. Relation \eqref{largeu3} is derived in \secref{sec:NRCGC}. It is one of the main goals of this appendix to provide the necessary technical background for \secref{sec:NRCGC}. Relation \eqref{largeu3} is also in the full agreement with the quasi-classical approximation discussed in \secref{sec:QCA}.

We can use \eqref{largeu3} to impose unitarity restrictions \eqref{uniint},\eqref{uniintferm} and \eqref{uniprincipal} on the possible solutions of the QSC. However,  we can also consider a different approach. Suppose that we do not know about connection to representation theory but try to answer the question what generic restrictions can be imposed on the asymptotics of Q-functions solely from the analytic structure of QSC. We will answer this question below, derive in this way  \eqref{uniint},\eqref{uniintferm}, \eqref{uniprincipal} and hence demonstrate that analyticity of QSC naturally encodes the unitarity constraints. While the quantization conditions  \eqref{uniint},\eqref{uniintferm} will be demonstrated from scratch, for \eqref{uniprincipal} we will need certain bits of information from the large volume approximation and then use the continuity argument.

Therefore, our departing assumption is that we know all the properties of QSC but we do not know the physical interpretation of its large-\(u\) asymptotics. For this asymptotics, we will only assume that it is power-like. One additional assumption is that no \(\bP\)'s coincide and no \(\bQ\)'s coincide.

The first thing to do, is to use invariance of the fundamental {\Qsys} with respect to H-rotations to choose a convenient basis for \(\bP\)'s and \(\bQ\)'s . If
one wants to preserve analyticity in the upper half-plane, the
H-matrices can be only constants. By performing constant
H-transformations we can always achieve that no pair of
\(\bP\)-functions scale with the same power at infinity and the same for {\bQfcts}.  We parameterize the asymptotic behaviour in such
a basis by \eqref{largeu3}, where \(\hat\lambda\)'s and \(\hat\nu\)'s, following
the logic of this section, are just some numbers. It is also our free choice to prescribe the ordering of magnitudes of \(\hat\lambda\)'s and \(\hat\nu\)'s, and we make a choice as in \eqref{magnitudeorderingD} which we repeat here for clarity:
\begin{eqnarray}\label{magnitudeordering2}
\hat\lambda_1>\hat\lambda_2>\hat\lambda_3>\hat\lambda_4\ \ \ {\rm and}\ \ \ \hat\nu_3>\hat\nu_4>\hat\nu_1>\hat\nu_2\,.
\end{eqnarray}
In the following we will work only in such a basis.

\subsubsection{Unimodularity and projectivity} Unimodularity and projectivity are firmly encoded in the fundamental Q-system. The  Hodge-symmetry is possible in principle only when $\fQ_{\bar\es}$ is at least periodic, the property which has an interpretation of quantum unimodularity \fip. Since $\fQ_{\bar\es}$ should be free of cuts in the upper half-plane and of poles, it can be only a constant which we normalize to $\fQ_{\bar\es}=1$. The latter equality and $Q_{\bar\es}=\det\limits_{1\leq i,j\leq 4}Q_{i|j}$ force us to have $\sum_{n=1}^4(\hat\lambda_n+\hat\nu_n)=0$ which is the zero charge constraint, {\ie} the projectivity (${\mathfrak{p}}$ of $\psu(2,2|4)$) is respected. Projectivity of the representation manifests itself also in the following way. The gauge transformation \eqref{rescalingsym2} is a symmetry of the quantum spectral curve. If one wants to respect the analyticity of the latter, this symmetry is however  significantly constrained. It reduces to a one-parameter family of $x$-rescalings:
\begin{eqnarray}
\bP_a\to x^{+\Lambda}\,\bP_a\,,&&\ \ \  \bQ_i\to x^{-\Lambda}\,\bQ_i\,,\label{oneparameterxrescaling}
\no\\
\ \ \bP^a\to x^{-\Lambda}\,\bP^a\,,&&\ \ \  \bQ^i\to x^{+\Lambda}\,\bQ^i\,,
\no\\
\ \ \ \mu_{ab}\to\mu_{ab}\,,&&\ \ \ \omega_{ij}\to\omega_{ij}\,,
\end{eqnarray}
which translates into redefinitions \(\hat\lambda_a\to\hat\lambda_a+\Lambda\), \(\hat\nu_i\to\hat\nu_i-\Lambda\).  Since these shifts by $\Lambda$ originate from the symmetry of QSC, they  should not affect physical quantities, in complete agreement with \eqref{deflanu} and \eqref{zerocharge5}. Note that \eqref{oneparameterxrescaling} reads on the level of T-functions as \eqref{xrescaling}.

\subsubsection{Quantization of charges}
\label{sec:Quantization of charges}
\begin{figure}[t]
\begin{center}
\includegraphics[width=0.8\textwidth]{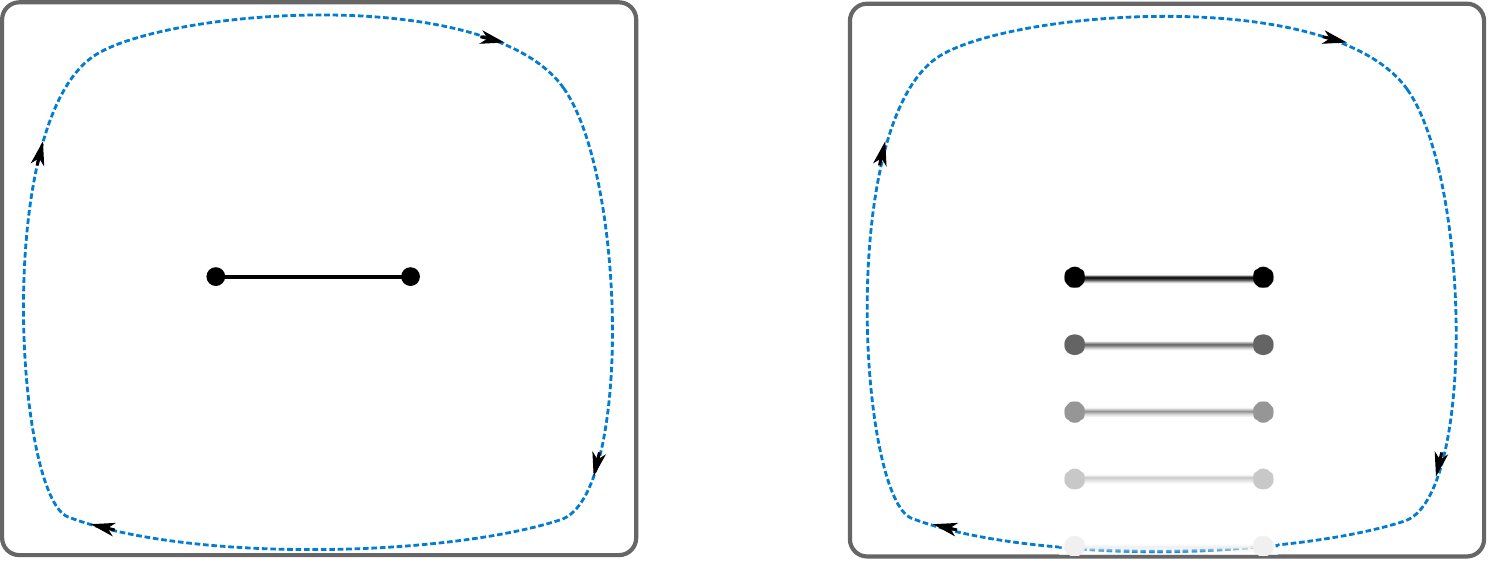}
\caption{\label{fig:monodromies}
Left: Monodromy of \(\bP\) around \(u=\infty\) should be trivial because \(\bP\) has only one cut. Right: Monodromy of \(\bQ\) around \(u=\infty\) is not trivial because in the set up when \(\bQ\)'s have short cuts, there are infinitely many of them. However, \(\bQ_1/\bQ_2\) and \(\bQ_3/\bQ_4\) are more regular at \(u\to-i\infty\) than one would naively expect, so eventually the mondromy of these combinations is also trivial.}
\end{center}
\end{figure}

The first, rather straightforward, observation is that \(\bP\)'s have only one cut, hence they should have trivial monodromy around infinity, and hence the \(\hat\lambda\)'s in \eqref{largeu3} should be integer \footnote{Sometimes it is convenient to use $x$-rescalings \eqref{oneparameterxrescaling} to impose $\sum\hat\lambda_a=\sum\hat\nu_i=0$, for instance this is the choice for the LR symmetric case, {\it cf.} \cite{Gromov:2013pga}. In this normalization $\lambda$'s can become fractional for certain states but their difference is still integer.}, see \figref{fig:monodromies}. Hence, in the ordering prescription \eqref{magnitudeordering2}, we immediately get the unitarity constraint \eqref{uniint}! Quantization of $\hat\lambda$'s reflects the compactness of $\su(4)$ algebra which is an R-symmetry of \neqfour{} SYM.

 Unlike this simple observation about quantization of asymptotics of  \(\bP\)'s, we should generically expect for  \(\bQ\)'s  that their asymptotics is not quantized. Indeed, they have only a long cut on the distinguished sheet, and the analytic continuation around infinity necessarily crosses it. As  the number of branch points is infinite on the other Riemann sheets, see \figref{fig:monodromies}, we should generically expect a non-trivial monodromy and hence the absence of charge quantization. On the one hand, it is fully acceptable, and even plausible, because the asymptotic of \(\bQ\)'s should reflect, after all, the representation of non-compact conformal symmetry algebra $\su(2,2)$, in particular it should contain a non-integer conformal dimension.  More than that: unlike the rational spin chains where Hamiltonian commutes with the symmetry algebra, the very fact that the AdS/CFT integrability includes energy as one of the symmetry charges imposes on us to consider Q-functions with infinitely many branch points.

On the other hand, the two other AdS charges, \(S_1\) and \(S_2\) are expected to be quantized. We will indeed confirm now this quantization  from analyticity of QSC.

Consider the quantity \(\bQ_{1}/\bQ_{2}\) and consider its analytic continuation  around a clockwise contour of very large radius, starting from \(\Im(u)>0\) domain, see \figref{fig:monodromies}. First, when we cross the real line from above for \(u>0\), we should apply analytic continuation formula: \(\tilde\bQ_{1}/\tilde\bQ_{2}=(\omega_{1j}\bQ^{j})/(\omega_{2j}\bQ^{j})\) and, since our contour is very far from the origin, we can use the magnitude ordering \eqref{magnitudeorderingD} and approximate this ratio by  \(-(\omega_{12}\bQ^{2})/(\omega_{12}\bQ^{1})\) or simply by \(-\bQ^{2}/\bQ^{1}\). We see that \(\omega_{ij}\) cancelled out from this ratio in the large-$u$ approximation which works in particular for \(u\to-i\infty\). Hence, while \(\tilde\bQ_i\)  has a semi-infinite ladder of cuts in the lower half-plane, these cuts are suppressed asymptotically  in the combination  \(\tilde\bQ_{1}/\tilde\bQ_{2}\). Hence, we can proceed with our analytic continuation through negative imaginary axis and then up to negative real axis. Finally, when crossing the negative real axis, we recover back \(\bQ_{1}/\bQ_{2}\) using the analytic continuation formula.

In conclusion, the leading term in large \(u\) expansion of \(\bQ_{1}/\bQ_{2}\) has a trivial monodromy, hence \(\hat\nu_1-\hat\nu_2\) should be integer. By applying the same argument to \(\bQ^{3}/\bQ^{4}\) we show that \(\hat\nu_3-\hat\nu_4\) is also integer. Then, using the ordering of magnitudes \eqref{magnitudeordering2}, we recover the expected quantization \eqref{uniintferm}.

The same logic cannot be applied to the ratio \(\bQ_2/\bQ_3\)%
. Indeed, \(\tilde\bQ_{2}/\tilde\bQ_{3}=(\omega_{2j}\bQ^{j})/(\omega_{3j}\bQ^{j})\simeq\omega_{12}/\omega_{13}\) for \(u\to-i\infty\), hence the infinite ladder of cuts is not suppressed and the monodromy is not expected to be trivial. As we discussed above, this is a healthy sign because \(\hat q_2=\hat\nu_2-\hat\nu_3\) is directly related to \(\Delta\) and should not be  integer.

\subsubsection{Main unitarity constraint}
The most nontrivial constraint is inequalities \eqref{uniprincipal}:  \(\hat\lambda_1+\hat\nu_1\leq 0\) and \(\hat\lambda_4+\hat\nu_4\geq 0\). The equality is realized only in the case when the long multiplet becomes reducible. This effect is reproduced by the quantum spectral curve: if either \(\hat\lambda_1+\hat\nu_1= 0\) or \(\hat\lambda_4+\hat\nu_4= 0\), the fundamental Q-system undergoes certain degeneration as one can see from \eqref{AABB0}. This simple test also shows us that the signs of \(\hat\lambda_1+\hat\nu_1\) and \(\hat\lambda_4+\hat\nu_4\) should remain the same at any value of the coupling constant, assuming the solution never degenerates. Hence we will discuss the signs of \(\hat\lambda_1+\hat\nu_1\) and \(\hat\lambda_4+\hat\nu_4\) at weak coupling.

At weak coupling, the discussion of \secref{sec:ABA} is expected to be applicable. One of the results of this discussion is \eqref{Qabf} which reads \(\fQ_{a|\alpha}\simeq \wQ_{a|\alpha}f^+\), where \(\wQ_{a|\alpha}\) is a polynomial of \(u\) and \(f\) is a function which has the  large-\(u\) asymptotics \(u^{\gamma/2}\) with \(\gamma=\sum_{k=1}^N\left(\frac{2gi}{x_k^+}-\frac{2gi}{x_k^-}\right)\simeq \sum_{k=1}^N\frac{2g^2}{u_k^2+1/4}\).

The asymptotics of $\fQ_{a|\alpha}$ reads $\fQ_{a|\alpha}\propto u^{-(\hat\lambda_a+\hat\nu_\alpha)}$, hence the smallest among \(\wQ_{a|\alpha}\) at large $u$ is \(\wQ_{1|1}\). If \(\wQ_{1|1}\) is a non-trivial polynomial (not constant) then one concludes that \(\hat\lambda_1+\hat\nu_1< 0\) at sufficiently small coupling. If \(\wQ_{1|1}\) is a constant then the sign of \(\hat\lambda_1+\hat\nu_1\) will be the opposite to the sign of $\gamma$. In this case, we should rely on an explicit Bethe ansatz solution to determine the sign of \(\gamma\) and we expect, though cannot prove in full generality, that \(\gamma>0\) and hence \(\hat\lambda_1+\hat\nu_1< 0\). At least this is so in the \(\sl(2)\) sector and other cases when all the Bethe roots are real. By the same reasoning one gets  \(\hat\lambda_4+\hat\nu_4> 0\).

Hence we reproduced the inequalities \eqref{uniprincipal}, at least for many interesting states described asymptotically by real Bethe roots and for any states for which \(\wQ_{1|1}\) and \(\wQ^{4|4}\) are non-trivial polynomials.
\ \\

Hence we showed that all unitarity constraints follow from analytic properties of the QSC, inequality \eqref{uniprincipal} is obtained however with certain assumptions. In the derivation, we did not used any reference to representation theory of \(\psu(2,2|4)\). We of course used that the Q-system is of \(\gl(4|4)\) type, but neither real form of \(\gl(4|4)\) was specified nor any Verma module was introduced. Besides using the parts of this result in deriving asymptotics of Q-functions, we can think about it as another strong justification for the overall viability of the QSC approach.

\subsection{Proof of exact \texorpdfstring{$\Delta$}{Delta}-dependence of asymptotics of Q-functions} \label{subsec:Delta}

 In TBA, the energy $E=\Delta-J_1$ is given by the expression
\be\label{ED}
E=\sum_{k=1}^N\hat\e_1(u_j)+\sum_{a=1}^\infty\int\frac{du}{2\pi i}\frac{\partial \check\e_a}{\partial u}\log(1+Y_{a,0})\,,
\ee
where $\e_a=a+\frac{2ig}{x^{[a]}}-\frac{2ig}{x^{[-a]}}$, see for instance \cite{Gromov:2009bc}. In \fip, section 3.7, it was shown that the energy defines the asymptotic behaviour of the $Y_{11}Y_{22}$. Indeed, the latter product satisfies the TBA equation
\be\label{Y11Y22D}
\log Y_{11}Y_{22}=\log\frac{R_{(+)}B_{(-)}}{R_{(-)}B_{(+)}}+\sum_{a=1}^\infty\int\frac{du}{2\pi i}\mathcal{Z}_a*\log(1+Y_{a,0})\,,
\ee
where the only thing we need to know about $\mathcal{Z}_a$ is that the large $u$ expansion of \eqref{Y11Y22D} results in $\log Y_{11}Y_{22}\sim \frac{i\,E}{u}$ with $E$ given precisely by \eqref{ED}.

It is a standard assumption that for considering various excited states one should perform the contour deformation trick, in particular the driving term in \eqref{ED} can be included into the integral if one deforms the contour of integration to surround points $u_j$ on a Riemann sheet where $Y_{a,0}(u_j)=-1$. It is also a common prescription to choose the same contour of integration both in \eqref{ED} and \eqref{Y11Y22D} and in general, in all TBA integrals involving $\log(1+Y_{a,0})$ as an integrand. Hence, conclusion about $\log Y_{11}Y_{22}\sim \frac{i\,E}{u}$ does not depend on the state we want to consider. On the other hand, we know from \cite{Gromov:2013pga} that $Y_{11}Y_{22}=\frac{\hat\mu_{12}^{[2]}}{\hat\mu_{12}}$, and therefore it is clear that $E$ is the same quantity both in TBA and QSC, independently of the physical state in question.

\bibliographystyle{hutphysc}
\bibliography{bibliography}

\end{document}